\documentclass[10pt,english,aps,rmp,twocolumn,amsmath,amssymb,floatfix,unsortedaddress,eqsecnum,longbibliography]{revtex4-2}

\usepackage[T1]{fontenc}
\usepackage{babel}
\usepackage{upgreek} 
\usepackage{amsmath}
\usepackage{amssymb}
\usepackage{graphicx}
\usepackage[unicode=true,pdfusetitle,
 bookmarks=true,bookmarksnumbered=false,bookmarksopen=false,
 breaklinks=false,pdfborder={0 0 1},backref=false,colorlinks=false]
 {hyperref}
\usepackage{xcolor}

\makeatletter
\usepackage{braket}
\DeclareMathOperator{\Tr}{Tr} 

\makeatother

\begin{document}
\title{Quantitative theory of magnetic interactions in solids}
\author{Attila Szilva}
\affiliation{Department of Physics and Astronomy, Division of Materials Theory, Uppsala University,
Box 516, SE-75120, Uppsala, Sweden}
\author{Yaroslav Kvashnin}
\affiliation{Department of Physics and Astronomy, Division of Materials Theory, Uppsala University,
Box 516, SE-75120, Uppsala, Sweden}
\author{Evgeny A. Stepanov}
\affiliation{CPHT, CNRS, {\'E}cole polytechnique, Institut Polytechnique de Paris, 91120 Palaiseau, France}
\author{Lars Nordstr{\"o}m}
\affiliation{Department of Physics and Astronomy, Division of Materials Theory, Uppsala University,
Box 516, SE-75120, Uppsala, Sweden}
\author{Olle Eriksson }
\affiliation{Department of Physics and Astronomy, Division of Materials Theory, Uppsala University, Box 516, SE-75120 Uppsala, Sweden and Wallenberg Initiative Materials Science for Sustainability, Uppsala University, 75121 Uppsala, Sweden}
\author{Alexander I. Lichtenstein}
\affiliation{Institut f{\"u}r Theoretische Physik, Universit{\"a}t Hamburg, Notkestra{\ss}e   9 , 22607 Hamburg, Germany}
\author{Mikhail I. Katsnelson}
\affiliation{Institute for Molecules and Materials, Radboud University, Heyendaalseweg 135, 6525 AJ, Nijmegen, The Netherlands}

\begin{abstract}
In this report we review the method of {\it explicit} calculations of interatomic exchange interactions of magnetic materials. This involves exchange mechanisms normally referred to as Heisenberg exchange, Dzyaloshinskii-Moriya interaction and anisotropic symmetric exchange. The connection between microscopic theories of the electronic structure, such as density functional theory or dynamical mean field theory, and interatomic exchange, is given in detail. The different aspects of extracting information for an effective spin Hamiltonian that involves thousands of atoms, from electronic structure calculations considering significantly fewer atoms (1-50) is highlighted. Examples of exchange interactions of a large group of materials is presented, which involves heavy elements of the 3d period, alloys between transition metals, Heusler compounds, multilayer systems as well as overlayers and adatoms on a substrate, transition metal oxides, 4f elements, magnetic materials in two dimensions and molecular magnets. Where possible, a comparison to experimental data is made, that naturally becomes focused on the magnon dispersion. The influence of relativity is reviewed for a few cases, as is the importance of dynamical correlations. Development to theories that handle out of equilibrium conditions is also described here. The review ends with a short description of extensions of the theories behind  {\it explicit} calculations of interatomic exchange, to non-magnetic situations, e.g. that describe chemical (charge) order and superconductivity.

\end{abstract}

\maketitle

\tableofcontents

\section{Introduction}
\label{intro}

\begin{figure}
\includegraphics[width=0.95\linewidth]{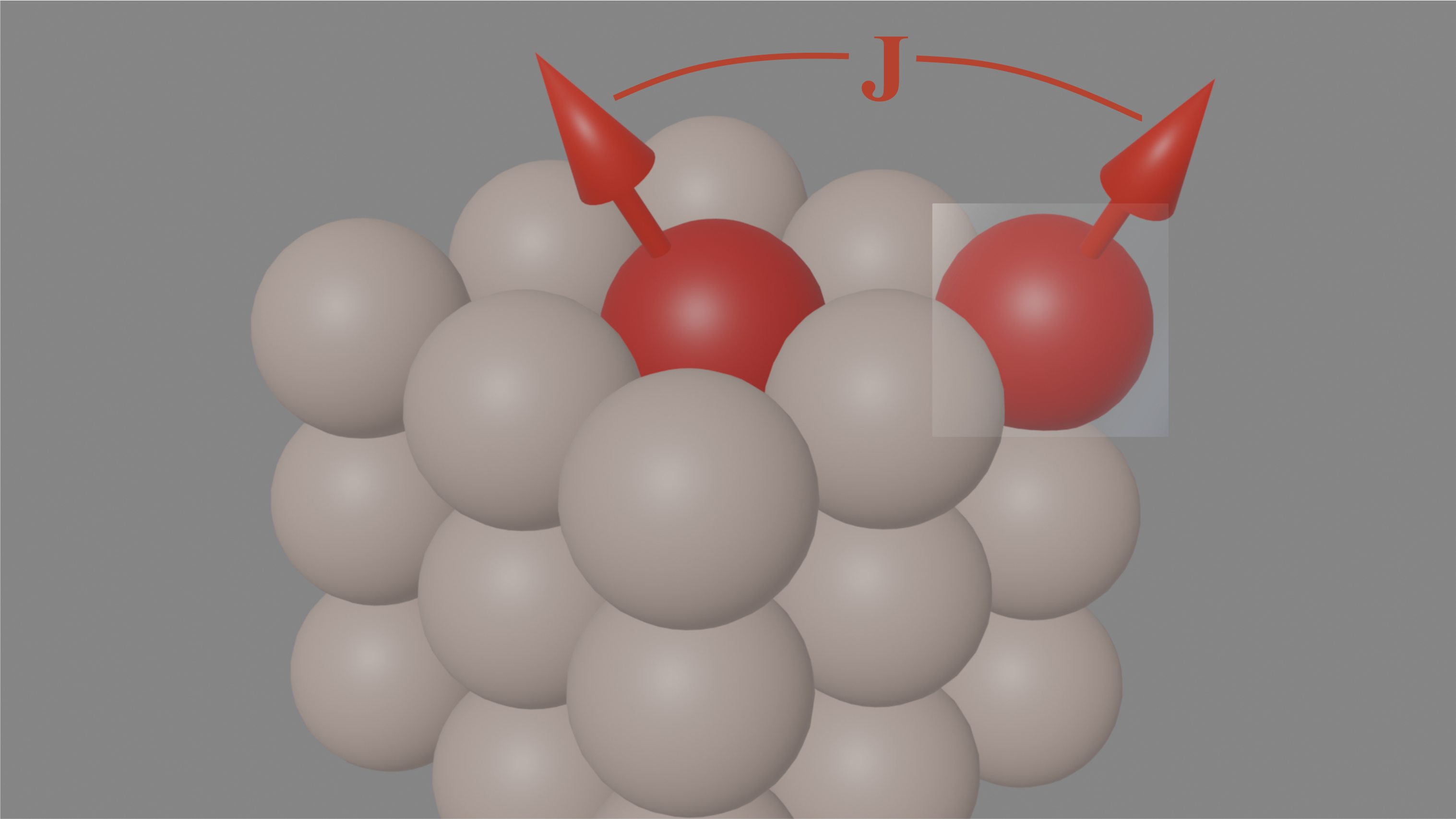}
\caption{(Color online) Schematic illustration of the multiscale step of using information from electronic structure calculations considering the primitive unit cell (bright square) to evaluate the exchange interaction, $J$, between two atomic spin-moments shown by red (dark grey) arrows. Note that atomic magnetic moments are only depicted for one pair of atoms, moments of other atoms are not shown.
\label{fig1}
}
\end{figure}

Magnetic phenomena are naturally of quantum nature. This follows from the success that quantum theory has had in describing magnetism, but it can also be ascribed to the discovery of a theorem of Bohr and van Leeuwen that demonstrates that a classical treatment fails in describing {\it any} magnetic properties at thermal equilibrium, with the magnetic susceptibility identically equal to zero~\cite{mohn2006magnetism}. Quantum mechanics has offered an excellent tool to analyze and interpret magnetic materials and since its birth, nearly one hundred year ago, the magnetism community has developed concepts as well as experimental and theoretical techniques to study magnetism. There are many textbooks covering the essentials of these techniques, as well as magnetic materials and magnetic phenomena~\cite{mohn2006magnetism, white1983quantum, kubler2017theory, stohr2006magnetism, buschow2003physics, getzlaff2008magnetism, jensen1991rare, coey2010magnetism, eriksson2017atomistic, vonsovsky1974magnetism, Yosida_Magnetism, goodenough1963magnetism, fazekas1999lecture,Skomski2021}. The purpose of this review article is by no means an attempt to cover what has already been described in detail in the references mentioned above. Instead, the main ambition of this work is to describe in detail how interatomic exchange interactions can be evaluated from {\it ab-initio} electronic structure theory, in a framework based on density functional theory (DFT)~\cite{hohenberg1964inhomogeneous, kohn1965self} and dynamical mean field theory (DMFT)~\cite{RevModPhys.68.13, PhysRevB.57.6884, kotliar-DMFT}. The pioneering work that this review focuses on, was published in 1984~\cite{liechtenstein1984exchange}, and since then many important contributions have been made to what is now a vibrant research field, that include both fundamental questions on the nature of the interatomic exchange interaction but also involve practical investigations in how to find functional materials with tailor-made properties. The latter studies involve green energy technologies,  e.g. the attempt to find permanent magnets that do not contain the costly and (from mining perspective) environmentally troublesome rare-earth metals, as well as to discover materials to be used in magneto caloric devices~\cite{gutfleisch2011magnetic, tegus2002transition}. As this review describes in detail, it is possible to evaluate the interatomic exchange interaction between any pair of magnetic atoms of a solid, from theoretical electronic structure calculations that considers atoms only within a primitive unit cell. This is illustrated schematically in Fig. \ref{fig1}, and the capability of extracting information from one scale (that of a conventional unit cell) to another scale (that involves thousands or even millions of atoms) is an important step in realizing approaches for an effective description of magnetism and magnetization dynamics.

This review hence describes how to calculate from electronic structure theory, the interaction term, $\mathcal{J}_{ij}$, of the celebrated Heisenberg Hamiltonian; 
\begin{equation}
\mathcal{H}_{H}= \sum_{<ij>} {\mathcal{J}}_{ij} {\vec{S}}_i \cdot {\vec{S}}_j,
\label{eqn1}
\end{equation}
where the summation is made over pairs of atomic spins, ${\vec{S}}_{i}$, and how its relativistic generalization~\cite{udvardi2003first} allows one to evaluate the Dzyaloshinskii-Moriya (DM) interaction (of vector form - ${\vec{\mathcal{D}}}_{ij}$)\footnote{See Eqs.~\eqref{tensor1} and~\eqref{tensor2} for a more precise generalization.} in; 
\begin{equation}
\mathcal{H}_{DM}= \sum_{<ij>} {\vec{\mathcal{D}}}_{ij} \cdot \left( {\vec{S}}_i \times {\vec{S}}_j \right).
\label{eqn2}
\end{equation}
Since this review is focused on methods to evaluate interatomic exchange interactions from electronic structure theory, a word on the nature of the electron states is relevant. In solids the electron states producing an atomic spin, that are mapped to describe low energy excitations by means of Eqns.\ref{eqn1} and \ref{eqn2}, are traditionally divided into localized electron states, where the electronic structure is described by atomic physics, or itinerant Bloch states. Traditionally the Heisenberg Hamiltonian was adopted primarily for the class of magnetic materials with localized electron states, but as this review outlines, many investigations have shown its success also for systems where the electron states are best described as Bloch states. The key aspect for this success is partially described in Section \ref{momentdescription}, that demonstrates that magnetism (and atomic spins) can be localized in space even though the electronic structure is completely itinerant. With modern developments in the theory of electronic structure, it is in fact quite possible to describe with equal accuracy the electronic structure of localized and itinerant electron systems, something we return to below in the review. The key question is actually not so much a question of localized versus itinerant electron states, but rather how configuration dependent the calculated parameters of Eqns.\ref{eqn1} and \ref{eqn2} are. This is discussed in Section \ref{detailsLKAG}.

The steps described in this review, that are used to derive an expression of interatomic exchange interactions, can be seen as the most robust argument (or derivation) for using the Heisenberg Hamiltonian (and its generalizations) to analyze magnetic phenomena, compared to the original argument of Heisenberg and Dirac (described in many textbooks, e.g. Refs.[1-3]), who considered a rather simple system, that of a two electron system and the energy difference between spin-singlet and spin-triplet states (a derivation which is covered in most textbooks in solid state physics). In fact, the connection between electronic structure information and interatomic exchange interactions, pioneered in Ref.~\onlinecite{liechtenstein1984exchange}, can be seen as the magnetic parallel to the quantum mechanical forces that are available from the Hellmann-Feynman theorem. The similarity extends also to their use; the interatomic exchange interactions can be used for torque minimization to find a ground state magnetic configuration, similar to the force minimization, to obtain the geometrical minimum of the nuclear position. Also, the use of a magnetic torque for studies of dynamics of magnets (in so-called spin dynamics simulations~\cite{antropov1995ab}) is completely analogous to the use of forces for molecular dynamics simulations. It should be noted that coupled spin-lattice dynamics simulations~\cite{antropov1995ab}, involving both interatomic forces and exchange, have also been described and used in practical simulations~\cite{PhysRevB.99.104302}. 

This review outlines {\it explicit} calculations of exchange parameters, where the term {\it explicit} implies that the parameters are obtained explicitly and directly once the solution to an electronic structure calculation is obtained~\cite{liechtenstein1984exchange, liechtenstein1985curie, liechtenstein1987local}. This can be compared to implicit approaches, where a Hamiltonian of the form used in Eqs.~\eqref{eqn1} and~\eqref{eqn2} is used to fit total energies obtained from electronic structure calculations, for a large number of magnetic configurations. A third method that is frequently employed, is to calculate in a DFT framework, the total energies  of spin-spiral configurations for several wavelengths of the spin-spiral. In this way one can obtain information of a reciprocal space representation of the exchange, and after a Fourier transform, the real space interatomic exchange paramaters are obtained~\cite{kubler1988density,Sandratskii_1991,Mryasov_1991,PhysRevB.58.293,PhysRevB.66.134435,doi:10.1080/000187398243573,jakobsson2015first}.

The implicit, cluster expansion approach, as well as the spin-spiral approach have been used with success, but they are outside of the scope of this review. We do however note a few key references that have outlined the cluster expansion approach~\cite{drautz2004spin,singer2011spin} and excellent treatises have covered the spin-spiral approach~\cite{kubler1988density, doi:10.1080/000187398243573, jakobsson2015first}. The focus of this review is, as mentioned, on the {\it explicit} method of extracting intraatomic exchange directly from a single electronic structure calculation, and details in how this is done both formally and practically are presented. We note that the method introduced in Ref.~\onlinecite{liechtenstein1984exchange} has its strengths in that it is a  universal way to calculate exchange parameters of Eqs.~\eqref{eqn1} and~\eqref{eqn2}, in the sense that systems with or without translation symmetry can be considered and that alloys and compounds can be treated on equal footing. It also offers an orbital decomposition of the interactions, that opens up for symmetry analysis from contributions between different irreducible representations of the electron states. It is also an excellent way to investigate general trends of the exchange interaction. 

It should be noted that similar approaches as the one reviewed here for the calculations of interatomic exchange, have been derived and used with success in other fields of solid state science. Examples involve for instance the chemical interaction between atomic species in alloys, in a method referred to as the ''Generalized perturbation theory'', first presented in Ref.~\onlinecite{ducastelle1976generalized} (for a review see Ref.~\onlinecite{PhysRevB.70.125115}), that is designed to calculate chemical interactions in alloys. We will touch briefly on this method in Section~\ref{Section10}.

\subsection{A short description of the early history of magnetism}
Before continuing these introductory remarks, we make here a short expose of the early historical discoveries of magnetic phenomena. In ancient times it was known that a type of stone which was found in northern Greece, close to a place called Magnesia, could attract iron. Thales of Milet (a Greek philosopher living in the 7'th century B.C.) is documented to be aware of the mysterious and invisible force these stones could have on iron~\cite{mohn2006magnetism,verschuur1996hidden}. Other philosophers of the past that were attracted to the mysterious properties of these magnetic minerals (later named Loadstone, where the magnetism stems from Fe$_3$O$_4$) involve Plinus the elder and Lucretius, both active in the first century A.D.~\cite{verschuur1996hidden}. The name of this first discovered magnetic mineral comes from the Lodestar (the pole star) which leads (or marks) the northern direction~\cite{verschuur1996hidden}. 
The first documented magnetic device used for establishing direction, the compass, is to be found in a Chinese manuscript dated to the 11$^{th}$ century~\cite{mohn2006magnetism}, and it is indeed curious that this technology is used widely even today, one thousand years later! Apart from its use in navigation, these first compasses were used for construction of buildings and their alignment, in the belief that they would be in harmony with the forces of nature~\cite{verschuur1996hidden}. 

Other historical breakthroughs in the science of magnetism and magnetic materials involve Peter Peregrinus (13$^{th}$ century) who undertook several experiments with Loadstone, and discovered that a magnet has poles. He in fact used the term ``polus'' to describe the north and south end of a magnet~\cite{verschuur1996hidden}. Curiously, he is known for a quote that ``experience rather than argument is the basis of certainty in science'', a principle most natural science lives by today, which he realized over half a millennium ago! Some three hundred years after the investigations of Peter Peregrinus, the first treatise of magnetism was published by William Gilbert~\cite{verschuur1996hidden}. In his book, with a title translated in English to ''On the Loadstone and Magnetic Bodies and on the Great Magnet the Earth; a New Physiology, Demonstrated by Many Arguments and Experiments'', he presented, among many things, his greatest realizations, that magnetism could be found to disappear when the material was heated and that the earth itself is magnetic~\cite{verschuur1996hidden}. 

The final major historical leap in the science of magnetism, before the development of quantum mechanics, is the discovery of electromagnetism, one of the greatest discoveries of the 19$^{th}$ century. This is something that is covered in almost all textbooks on physics, and is for this reason not discussed here further. We note however that from a practical point of view, several discoveries made in the 19$^{th}$ century, concerning magnetic materials, now form a firm basis for technologies used to propel our society. To mention concrete examples, Faradays induction law, which allows the conversion of mechanical energy to electricity, is used in all power plants. Also, the development of electrical motors, which becomes more and more a standard technology for motorized vehicles, has performance based on the magnetic field strength~\cite{gutfleisch2011magnetic,tegus2002transition}. A final example is that of magnetic refrigeration, and the principle of adiabatic demagnetisation, that is considerably less energy demanding compared to a compressor based technology of cooling~\cite{gutfleisch2011magnetic,tegus2002transition}.
Hence, many technologies that rely on magnetic materials are used in our society, being key components to the economy and to the well-being of household and private use. The functionality of these technologies is based on the performance of the magnetic materials they are constructed around. Hence, in a general aim of a more electrified society, and with the ambition to find greener technologies to generate electricity, e.g. in farms of wind power mills, the search of magnetic materials with tailored properties has become a very active field of science~\cite{gutfleisch2011magnetic,tegus2002transition}. 

We end this subsection with a comment on the coupling of magnetism and biology. The coupling of magnetism and living matter were discussed over the many centuries that magnetic phenomena were known. For instance, Bartholomew the Englishman (13$^{th}$ century) advocated its medicinal powers~\cite{verschuur1996hidden}, and the ideas of Franz Anton Mesmer (late 18$^{th}$ and early 19$^{th}$ century) around ``animal magnetism'' have escaped few. Although the ideas of Mesmer are now regarded as nonsense, the influence of magnetic fields on biological matter is well established, e.g. as demonstrated by levitating animals or fruit when subjected to strong magnetic fields\footnote{As can be seen here: https://www.youtube.com/watch?v=A1vyB-O5i6E.}. It is also established that birds use nano-particles of magnetite (Fe$_3$O$_4$) to navigate in the Earths magnetic field~\cite{wiltschko2006bird}. In addition, it is well known that magnetic materials, again in the form of magnetite, can be produced by bacteria, e.g. magnetotactic bacteria, and the bacterium GS-15 is known to produce magnetite~\cite{snowball2002bacterial}. It has even been speculated that this is one reason for large amounts of fine-grained magnetite found in ancient sediments. Single-domain magnetite produced in this way could then reveal the magnetic recording of the ancient geomagnetic field.

\subsection{On magnetic materials and magnetic phenomena}
\label{momentdescription}
Most elements have for the free atom a pairing of electron spins, due to intraatomic exchange interaction among the electrons. This leads to atomic moments, similar to what is illustrated schematically in Fig. \ref{fig1}, for almost all elements of the Periodic Table, provided they are isolated atoms. In the solid state things are more complicated since the spin-pairing energy competes with the kinetic energy, which is lowest for equal population of spin up and spin down electrons. Combined with the fact that band formation of electron states can make the kinetic energy rather significant, one ends up with a competition between two mechanisms, one favoring equal population of spin states, and one that favors spin pairing and local (or atom centered) moments. Stoner theory quantifies this competition and allows one to identify a simple rule for when magnetic order is to be expected (see e.g. Ref.~\onlinecite{mohn2006magnetism}). Among most elemental solids, it is in fact the kinetic energy and band formation that dominates, so that an equal amount of spin up and down electron states are populated, making these materials either Pauli paramagnetic or diamagnetic. Spontaneous magnetic order occurs only for a limited number of elements of the Periodic Table, and at (or just below) room temperature merely four are found to have spontaneous ferromagnetic (FM) order (bcc Fe, hcp Co, fcc Ni and hcp Gd). However, there are thousands of compounds and alloys that show significant magnetic moments, and there are plenty of materials to investigate with respect to the many interesting magnetic phenomena that have been reported. 

\begin{figure}
\includegraphics[width=\linewidth]{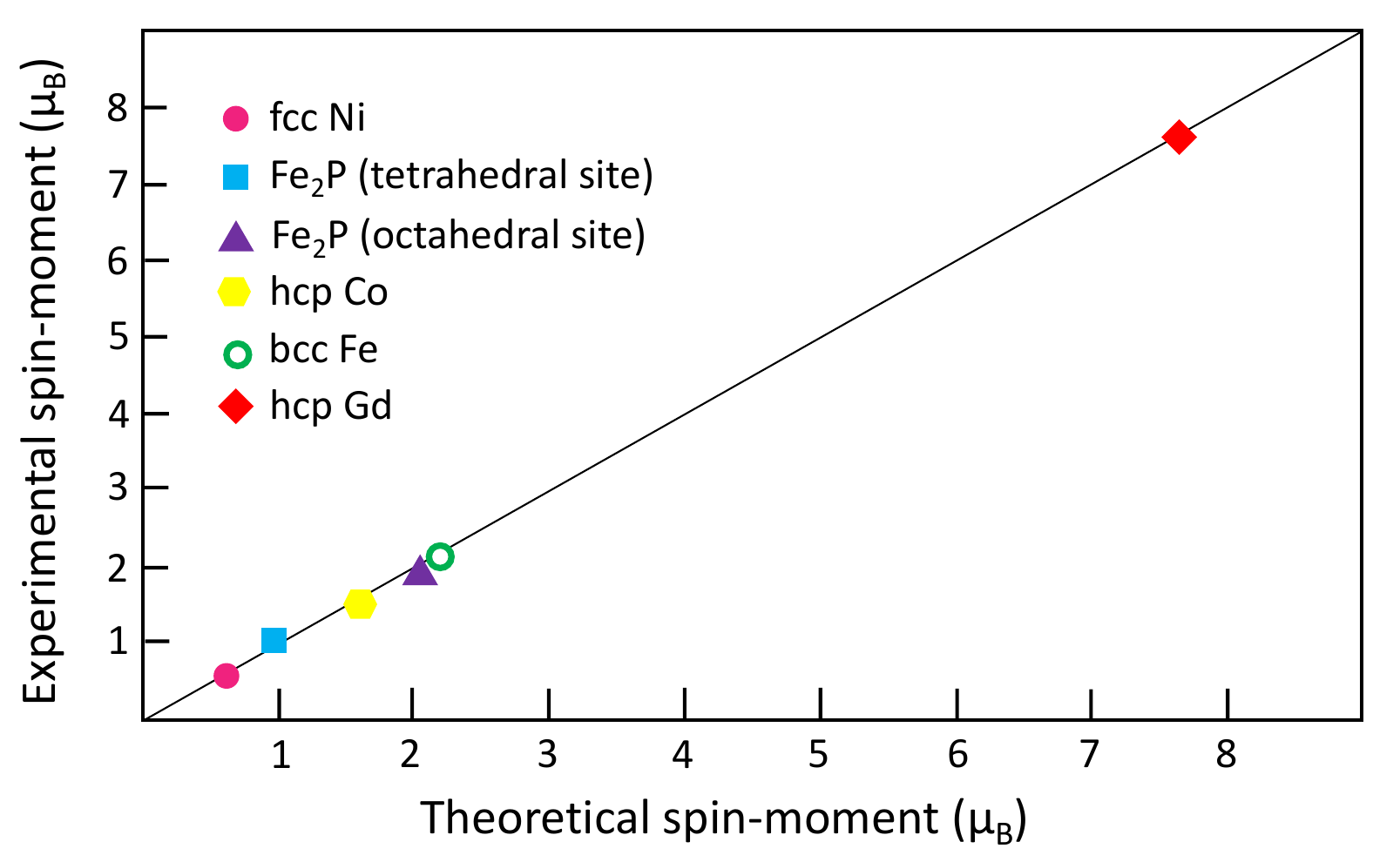}
\caption{(Color online) Comparison between measured magnetic spin moments and results obtained from theoretical calculations based on DFT (data taken from Ref.~\onlinecite{eriksson2017atomistic}). Data given in $\mu_B$ per atom.
\label{fig2}
}
\end{figure}

One of the more efficient ways to evaluate the delicate balance between band formation and spin-pairing, relies on DFT and the invention of efficient methods for solving the electronic structure of solids, so that measured magnetic moments can be reproduced with good accuracy. These calculations are often referred to as {\it ab-initio}, indicating that they are carried out without experimental input. Results from {\it ab-initio} theory are shown in Fig. \ref{fig2}, where a comparison is made to experimental results for the four ferromagnetic elements mentioned above, as well as for the ferromagnetic, hexagonal compound Fe$_2$P. This compound has Fe atoms occupying two distinct crystallographic sites, a tetrahedral and an octahedral site, and neutron scattering measurements have revealed that the magnetic moments of these sites are quite different. As Fig. \ref{fig2} shows, {\it ab-initio} theory reproduces the measured magnetic moments with good accuracy. The moments for Fe$_2$P are particularly interesting since they reveal a delicate balance between band formation and interatomic exchange energy, resulting in very different moments for Fe atoms situated on different crystallographic sites. The results shown in Fig. \ref{fig2} actually reveal a rather typical accuracy of theory based on DFT, at least when it comes to reproducing magnetic moments. Based on 2935 calculations, Ref. \cite{PhysRevX.11.011031} shows the predictive power of spin polarized density functional theory (combined with cluster-multipole expansion) by reproducing the experimental magnetic configurations with an accuracy of $\pm0.5\mu_{B}$. Notable difficulties of DFT based theory are however found for correlated electron systems, where multiconfiguration effects become important, something we will also discuss in this review. 

\begin{figure}
\includegraphics[width=0.83\linewidth]{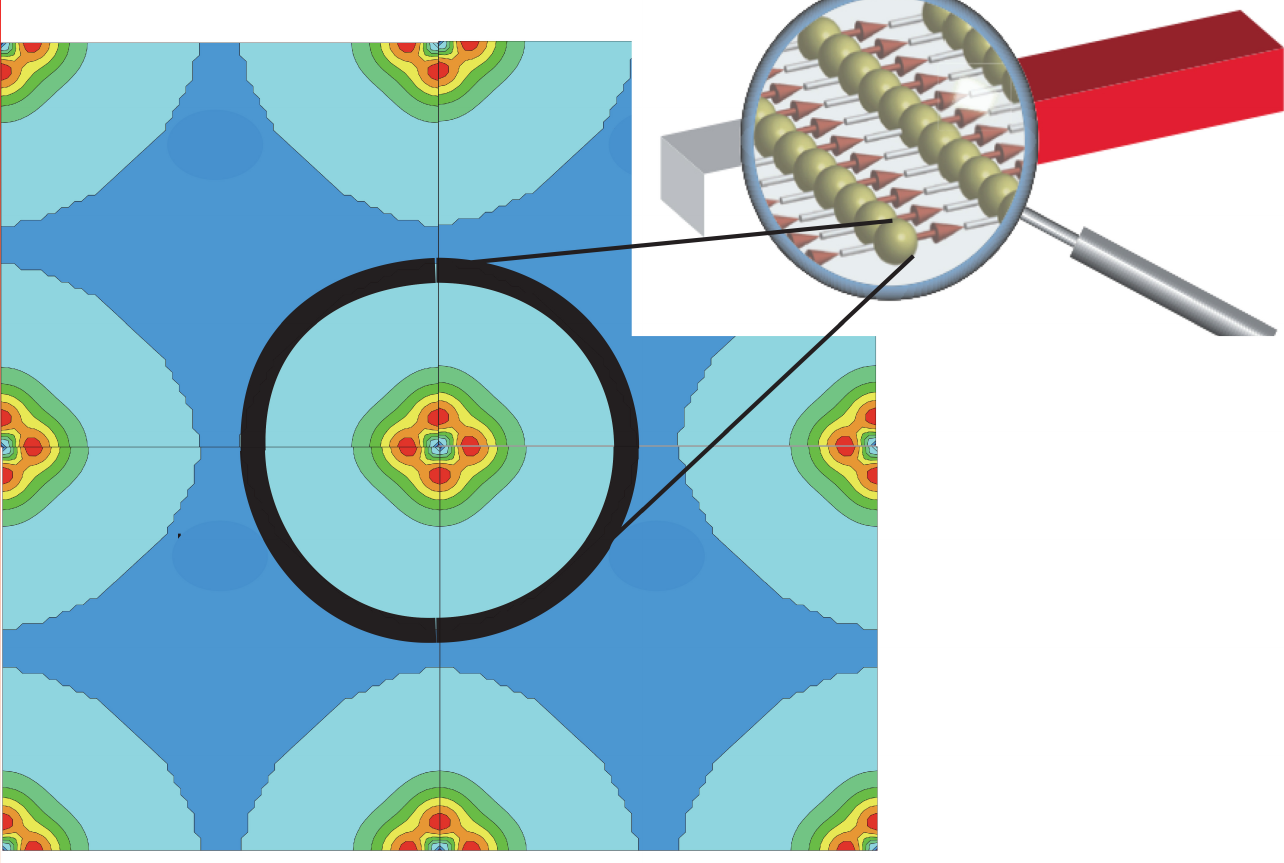}
\caption{(Color online) Magnetization density of bcc Fe, from theoretical calculations based on DFT (redrawn after Ref.~\onlinecite{eriksson2017atomistic}).
\label{fig3}
}
\end{figure}

{\it Ab-initio} theory provide another important piece of information; that provided by the magnetization density. This is illustrated in Fig. \ref{fig3} for bcc Fe, for a plane inside the crystal, spanned by vectors parallel to the axis of the conventional unit cell of the bcc structure. Note that red (dark grey) coloration indicate high magnetization density while blue (light grey) indicate low values of this density. As the figure shows, the magnetization density is high only in a small region that is located close to the atomic nuclei of the Fe atoms. This is a typical result and allows one, for almost all materials, to describe the magnetic state as being composed of atom centered (or atomic) moments, as illustrated in the upper right of the figure. The results shown in Fig. \ref{fig3} justifies a discussion based on atomic moments and the different types of phenomena such moments display.
These results also shine light on the dynamics of magnetism, and the distinction made between fast (electrons) and slow (site dependent magnetisation directions) variables, illustrating the concept of ''temporarily broken ergodicity'' as analysed in detail in Refs. \cite{Gyorffy_1985, Staunton_1985, STAUNTON198415}. This is the basic principle for performing atomistic spin-dynamics simulations, e.g as outlined in Ref.\cite{eriksson2017atomistic}, where the slow variables evolve under the influence of a local Weiss field.

The inset of Fig.~\ref{fig3} illustrates a specific arrangement of atomic moments, and as the figure shows, bcc Fe is a ferromagnet; all atomic moments point in the same direction. However, for other materials many different orderings of atomic moments have been reported, such as antiferromagnetism, where every other magnetic moment shown in the inset of Fig. \ref{fig3} would have its direction reversed. A majority of the materials that have finite atomic moment have either of these two types of collinear magnetic order. However, more complex magnetic orders exist in nature, where atomic moments form a noncollinear arrangement (see e.g. Ref.~\onlinecite{kubler2017theory}). Among the elements, such order is found predominantly among the lanthanides~\cite{jensen1991rare}. As highlighted with the Nobel Prize in Physics 2021, glass-like phenomena are also found for specific groups of magnetic materials, in which one singular magnetic ground state never is realized. Instead, the magnetism can be understood to reflect a multi-valley landscape, where very many different configurations of the atomic moments result in very similar energies~\cite{snowball2002bacterial}. Thermal fluctuations can make the system drift from one configuration to the next, and aging phenomena are a fingerprint of spin glasses. Dilute alloys (e.g. Mn impurities in a Cu matrix~\cite{cannella1972magnetic}) and more recently, elemental Nd~\cite{kamber2020self,benj2022verl}, are known spin glass systems. 

Each class of magnetic materials has its own characteristic in terms of ground state properties as well as fingerprints revealing its excited state. This involves quasiparticple properties of the collective excitations, referred to as magnons (see e.g. Ref.~\onlinecite{mohn2006magnetism}), the temperature dependence of the magnetic state, the value of the ordering temperature and the critical exponents used to characterize second order phase transitions, when the magnetic state vanishes with temperature. Many (if not all) of these phenomena are typically analyzed using Eqs. (1) (and (2)), and different forms of Heisenberg exchange interactions have been discussed to be responsible for the widespread list of magnetic properties found in nature. As reviewed in this communication, this involves direct exchange, super- and double exchange, RKKY interaction and interlayer exchange. In these investigations the dimensionality of the magnetic material is a natural component of the analysis, and the celebrated Mermin-Wagner theorem describes the connection between dimensionality and finite temperature effects of spontaneously broken symmetries of magnets (see e.g. Ref.~\onlinecite{PhysRevLett.17.1133, ruelle1999statistical, PhysRevB.39.2344, PhysRevB.60.1082}).

\subsection{Recent trends in magnetism}
Recent trends in magnetism often have focused on systems in the nano-scale. This relates for instance to magnetic multilayers and trilayers, where perhaps the most celebrated finding is the giant magneto resistance (GMR) effect and its applications for sensors~\cite{baibich1988giant, binasch1989enhanced}. Most applications in magnetic information storage currently rely on sensors based on the tunneling magneto resistance (TMR) effect~\cite{bowen2001large}, where interestingly the use of MgO as an optimal  tunneling layer was predicted~\cite{butler1985theory} by ab-initio theory before experimental verification. These investigations have focused on systems that are confined in one dimension, and thin film physics is reviewed in many textbooks on magnetism~\cite{kubler2017theory,stohr2006magnetism}. Such studies of quasi two-dimensional systems have now expanded to focus on toplogical magnetic states (e.g. skyrmions, merons and hopfions)~\cite{belavin1975metastable} as well as investigations of purely 2D materials. The latter class is particularly interesting, given the strong influence on geometrical dimensionality and magnetism, as stated by the Mermin-Wagner theorem. However, Cr trihalides have been synthesized and their magnetic properties are by now rather well known~\cite{Huang2017}. Other aspects of magnetism that currently are under investigation are coupled to questions on the ultra-fast dynamics, pioneered in Ref.~\onlinecite{beaurepaire1996ultrafast}, as well as magnonics~\cite{kruglyak2010magnonics} and spintronics~\cite{wolf2001spintronics}. As a final remark of this subsection we note the recent interest in spin-ice~\cite{bramwell2001spin} and spin-liquid states~\cite{norman2016colloquium}, as well as the so-called Kitaev systems~\cite{kitaev2006anyons}.

\subsection{Early theories of interatomic exchange}

This review has a starting point in the work of Ref.~\onlinecite{liechtenstein1984exchange}, but it is clear that the work published in Ref.~\onlinecite{liechtenstein1984exchange} has overlap with earlier works, that also attempted to find a formalism that allows one to extract interatomic exchange from information given by {\it ab-initio} electronic structure theory. We specifically mention the early works of Oguchi and coworkers~\cite{oguchi1983magnetism, PhysRevB.28.6443}, where a similar, but not identical method was presented. 
In this work an approach was used where the magnetic moments were rotated by 180 degrees in order to extract the exchange interaction strength, instead of the use of infinitesimally small rotations, which is the essence of the works in~\cite{liechtenstein1984exchange}.
The results of Refs.~\onlinecite{oguchi1983magnetism, PhysRevB.28.6443} were in fact similar to earlier works by Inoue and Moriya~\cite{inoue1967interaction} and by Lacour-Cyet and Cyrot~\cite{lacour1974magnetic}. We also mention here the early works of Gyorffy et al.~\cite{gyorffy1980momentum} and by Liu~\cite{liu1961exchange}, that inspired the works of~\cite{liechtenstein1984exchange}, and an early work from Wang et al.~\cite{wang1982magnetism} who studied fluctuating local band theory of itinerant electron ferromagnetism in nickel and iron. 

It should also be noted that exchange interactions in solids have bearing on many phenomena and theories of magnetism, that due to space limitations can not be covered in detail here. For instance, spin-fluctuation theories have been used with great success to analyse excited state properties of magnetic solids, including temperature dependence of magnetism, susceptibility and specific heat (see e.g. Ref.\cite{mohn2006magnetism}). These theories are typically connected to Landau or Ginzburg-Landau theories, which are not the topic of this review since they have been covered in many textbooks already \cite{mohn2006magnetism,white1983quantum,kubler2017theory,stohr2006magnetism}.

\subsection{A comment on nomenclature}
\label{nomenclature}

Before entering the main results of this review, we make a comment on the form of the Hamiltonian used in this text. In the derivations and the examples given below we will use the expressions,
\begin{equation}
\mathcal{H}_{H}= \sum_{<ij>} J_{ij} \vec{e}_i \cdot \vec{e}_j,
\label{neweqn1}
\end{equation}
and 
\begin{equation}
\mathcal{H}_{DM}= \sum_{<ij>} \vec{D}_{ij} \cdot \left( \vec{e}_i \times \vec{e}_j \right),
\label{neweqn2}
\end{equation}
where $\vec{e}_i$ is a unit vector describing the direction of the magnetic moment of the atom at site $i$.
In this review we will refer to the interaction parameters in Eq.~\eqref{neweqn1} either as interatomic exchange or as Heisenberg exchange ($J_{ij}$) and DM interaction ($\vec{D}_{ij}$), or simply as the $J_{ij}$'s or $\vec{D}_{ij}$'s. It should also be noted that the definition of the interatomic energy, used in this work, is with a plus sign in front of the summations in Eq.~\eqref{neweqn1}, where the summation is made over pairs of atoms $<ij>$. One sometimes uses a slightly different notation, where the summation is made such that $i \ne j$, but the indexes $i$ and $j$ run over all atoms considered in a calculation. In this case a factor $1/2$ appears in front of the summations in Eq.~\eqref{neweqn1}, to ensure that each pair interaction is calculated only once. Some authors choose to use a minus sign in front of Eq.~\eqref{neweqn1}. We also note that in the derivation of the interatomic exchange formulate a sum where the local interaction $i=j$ is also considered will be temporally needed as shown e.g. in Eq. (\ref{twoforallsites}. Importantly, in Section~\ref{examples}, where results of exchange parameters are given, the numerical values (in the main text and also in the figures) are consistent with the nomenclature given by Eq. (\ref{neweqn1}). 

A comparison between Eqs.~\eqref{eqn1} and~\eqref{neweqn1} gives that $J_{ij} = {\mathcal{J}_{ij}} S_i S_j$ where $S_i$ and $S_j$ stand for the lengths of the vectors $\vec{S}_i$ and $\vec{S}_j$. Similarly we obtain that ${\vec{D}_{ij}={\vec{\mathcal {D}}}_{ij}} S_i S_j$. This distinction is important when comparing interactions obtained from different theoretical methods and experiments. The different forms of Eqs.~\eqref{eqn1} and~\eqref{neweqn1} (and between Eqs.~\eqref{eqn2} and~\eqref{neweqn2}) also allows an important distinction between quantum- and classical spin Hamiltonians. We adopt here the nomenclature that Eqs.~\eqref{neweqn1} and~\eqref{neweqn2} allow for infinitesimal rotations of the direction of an atomic moment, and hence $\vec{e}_i$ can be treated as a classical vector. This is different from approaches when $\vec{S}_i$ (as in Eqs.~\eqref{eqn1} and~\eqref{eqn2}) is considered as a quantum mechanical operator. The latter is obviously preferable from a formal point of view, but it is in many cases impractical. In fact, all materials specific examples given in this review make use of Eqs.~\eqref{neweqn1} and~\eqref{neweqn2}. In the review, we will describe magnetic fields that are expressed in energy units. In other words, we consider a magnetic field as $\vec{B} = \frac{1}{2} g \mu_{B}  \vec{\tilde{B}} $ where $\tilde{B}$ is measured in Tesla, $\mu_{B}$ is the Bohr-magneton and $g$ approximately equals -2 for electrons. We also note that we will use bold symbols for vectors in real and reciprocal space while symbols with an arrow denotes vectors in spin space. Finally, we note that we use the dot ($\cdot$) symbol when components of a vector or a tensor are contracted (summed over), e.g., $\vec{A}\cdot \vec{B}=\sum_\mu A^\mu B^\mu$ or $\vec{D}\cdot\mathbb{C}\cdot\vec{E}=\sum_{\mu\nu} D^{\mu}{C}^{\mu\nu}E^{\nu}$, the cross ($\times$) symbol is used for cross product (or vector product) and the star ($\star$) symbol will be used when an equation continues on a new line.

\section{Linear response theory of the susceptibility}
\label{linearresp}

In this review we present a description of magnetic interactions of many-electron systems, via the separation of specific spin degrees of freedom (roughly, directions of localized magnetic moments) from a complete quantum description of all properties of the system starting from the Schr\"odinger equation. This cannot be done without approximations, due to a presence of strong interelectron interactions. Nevertheless, it makes sense to start with a formally rigorous scheme and then introduce these approximations step by step, something we do here.

For equilibrium properties, there are two main practical schemes: density functional theory based on the Hohenberg-Kohn theorem~\cite{hohenberg1964inhomogeneous} with the associated Kohn-Sham quasiparticles~\cite{kohn1965self} and Green function formalism based on Luttinger-Ward generating functional~\cite{luttinger1960ground,hedin1965new}. Spin dynamics deal with out-of-equilibrium properties, and, fortunately, both these main techniques can be generalized for this case. For the Green function functional, this is done in the most general form by Baym and Kadanoff~\cite{baym1961conservation} but in reality this method does not have any applications to the properties of real materials since it is computationally too demanding. Only for model systems is there a real progress~\cite{aoki2014nonequilibrium}. Since in this review we are focused on the applications to real materials, connecting calculated results to experimental observations, we will not consider the time-dependent Green function functionals here.

On the other hand, the time-dependent generalization of density functional theory has been realized. It is based on the Runge-Gross theorem~\cite{runge1984density} and its generalization to spin-polarized calculations~\cite{liu1989time}. There are numerous examples of the time-dependent density functional theory (TDDFT), applied to specific magnetic materials~\cite{cooke1985new,savrasov1998linear,buczek2011different,singh2019adiabatic,sharma2007first,gorni2018spin}. In principle, if one knows the exact time-dependent density functional and, in particular, the so-called exchange-correlation kernel~\cite{runge1984density}, one can calculate the dynamical magnetic susceptibility, and find the spin-wave spectrum as the poles of the dynamical susceptibility. A fitting of exchange parameters could even be done to the calculated spectrum. This method would be formally exact, but not very practical, at least at this stage, since the successes in building of reliable expressions for the exchange-correlation kernel are still very restrictive (note however the first attempts that have been made~\cite{castro2012controlling,thiele2008adiabatic}). In order to proceed with practical calculations, we introduce an approximation, that is, the so-called adiabatic approximation within TDDFT (ADA-TDDFT). According to this approximation, the exchange correlation kernel is equal to its equilibrium form. This is naturally a significant simplification. Indeed, whereas the full exchange correlation kernel depends on two times, in the adiabatic approximation it depends only on one time, via the time-dependence of the charge and spin densities only. After this approximation is made, one can proceed to the final expression for the exchange parameters~\cite{katsnelson2004magnetic}. We will follow here this derivation, which generalizes earlier theories~\cite{callaway1981magnetic}. 

We proceed with the master equation of density functional theory, the Kohn-Sham equation, that has the form of a single particle Schr\"{o}dinger equation. Within the self-consistent ADA-TDDFT approximation it has the form
\begin{eqnarray}
\label{schroed}
i\frac{\partial \psi }{\partial t} &=&H\psi  \nonumber \\
H &=&-\nabla ^{2}+V({\bf r})- \left(\vec{B}_{xc}({\bf r})+{\vec{B}}_{ext}( {\bf r}) \right) \cdot \vec{\sigma} \,
\label{firstKS}
\end{eqnarray}
where $V({\bf r})$ is the effective
potential, ${\vec{B}}_{ext}({\bf r})$ and ${\vec{B}}_{xc}({\bf r})$
are the external magnetic field and the
exchange-correlation field, respectively, that couple to the electrons spin, and $\vec{\sigma}$ stands for the Pauli spin matrices $\{\sigma_x,\sigma_y,\sigma_z\}$. Note that in this work we adopt the original formulation of density functional theory, that was formulated at $T=0$. The work by Mermin \cite{PhysRev.137.A1441}, and subsequent works \cite{PhysRevB.82.205120, PhysRevLett.107.163001}, showed that the power of density functional theory extends also to finite temperature. However, for the purposes of this review, it is sufficient to adopt the original formulation of density functional theory. Note also that Rydberg units are used here: $\hbar=2m=e^2/2=1$.

Next, we employ the adiabatic approximation, assuming that the functional dependencies of the exchange-correlation potential, and hence the field
of the charge and spin density, are the same as in the stationary case. In the local spin density approximation (LSDA) the effective potential depends on the values of charge and spin densities at the same spatial and temporal point only:
\begin{eqnarray}
\label{local}
V({\bf r}) &=&V_{ext}({\bf r})+\int d{\bf r}^{\prime }\frac{n({\bf r}%
^{\prime })}{\left| {\bf r-r}^{\prime }\right| }+\frac{\partial }{\partial n}%
[n\varepsilon _{xc}]  \nonumber \\
{\vec{B}}_{xc} ({\bf r}) &=&-\frac{{\vec{m}}}{m}\frac{\partial }{\partial m}[n\varepsilon
_{xc}],  
\end{eqnarray}
where $n({\bf r})$ and ${\vec{m} ({\bf r})}$ is the charge and spin density, ${m} ({\bf r})$ is the magnitude of ${\vec{m} ({\bf r})}$,
$\varepsilon _{xc}$ is the exchange-correlation energy density, and $V_{ext}({\bf r})$ is the external potential, that is, the electrostatic
potential of nuclei. Note that the spin-orbit interaction will be consider later in the review. We also note that in the expressions above, we have in some places omitted for simplicity the spatial argument ${\bf r}$, that enters all variables in Eq.~\eqref{local}. We will in some of the equations below also adopt this simplifying notation. 

The spin susceptibility which we are interested in is the linear-response function, therefore we 
consider the limit ${\vec{B}}_{ext}({\bf r})\rightarrow 0$. Then the
effective complete ``non-equilibrium'' field contains both an external field as well as an additional exchange correlation field, due to redistribution of the spin density, and the variation of this field can be expressed as:
\begin{equation}
\delta B_{tot}^{\alpha }=\delta B_{ext}^{\alpha }+\frac{\delta B_{xc}^{\alpha }}{%
\delta m^{\beta }}\delta m^{\beta },  \label{linear}
\end{equation}
where $\alpha \beta $ are Cartesian indices and a sum over
repeated indices is assumed.

The exact, non-local, frequency-dependent spin
susceptibility, $\widehat{\chi }^{\alpha \beta }$, is the kernel of the operator that connects the variation of
the spin density and the external magnetic field:
\begin{equation}
\delta m^{\alpha }=\widehat{\chi }^{\alpha \beta } \delta
B_{ext}^{\beta } \,. \label{linear1}
\end{equation}
We use here the standard definition of the operator product:
\begin{equation}
(\widehat{\chi }\varphi )({\bf r)=}\int d{\bf r}^{\prime }\chi ({\bf r,r}%
^{\prime })\varphi ({\bf r}^{\prime }) \,.  \label{def}
\end{equation}
A parallel consideration for the calculation of the spin-susceptibility follows from the Runge-Gross theorem~\cite{runge1984density} and its generalization to the spin-polarized case~\cite{liu1989time}, where in the
time-dependent density functional theory one has the exact relation
\begin{equation}
\delta m^{\alpha }=\widehat{\chi }_{0}^{\alpha \beta } \delta
B_{tot}^{\beta } \,, \label{linear2}
\end{equation}
where $\widehat{\chi }_{0}^{\alpha \beta }$ is the susceptibility
of an auxiliary system of one-electron, Kohn-Sham particles. Comparing the
equations~\eqref{linear}, \eqref{linear1}, and~\eqref{linear2}, we arrive at the result that
\begin{equation}
\widehat{\chi }^{\alpha \beta }=\widehat{\chi }_{0}^{\alpha \beta }+\widehat{%
\chi }_{0}^{\alpha \gamma }\frac{\delta B_{xc}^{\gamma }}{\delta m^{\delta }}%
\widehat{\chi }^{\delta \beta } \,,  \label{dyson}
\end{equation}
which is a particular case of the Bethe-Salpeter equation~\cite{salpeter1951relativistic}, with $\frac{\delta B_{xc}^{\gamma }}{\delta m^{\delta }}$ playing the role of the vertex, $\Gamma$. One may note that this equation turns out to be formally exact within ADA-TDDFT. Actually, even if one does not assume the local spin density approximation, equation~\eqref{dyson} is still exact, but the vertex, $\Gamma$, is then not local in spatial coordinates. The adiabatic approximation assumes however its locality in time.

The local spin density approximation~\eqref{local} leads to further simplifications. Indeed, one then obtains the expression
\begin{equation}
\frac{\delta B_{xc}^{\gamma }}{\delta m^{\delta }}=\frac{B_{xc}}{m}\left(
\delta _{\gamma \delta }-\frac{m^{\gamma }m^{\delta }}{m^{2}}\right) +\frac{%
m^{\gamma }m^{\delta }}{m^{2}}\frac{\partial B_{xc}}{\partial m} \,,
\label{local1}
\end{equation}
where the first term in Eq.~\eqref{local1} is purely transverse and the
second one is purely longitudinal with respect to the local
magnetization density (or the local magnetic moment) and $B_{xc}$ is the length of $\vec{B}_{xc}$.

As a next simplification, we restrict ourselves to the case of collinear magnetic ground states, with moments along the z-direction. Then, the coupling between the longitudinal and transverse components of the magnetic susceptibility vanishes. For the transverse spin susceptibility, which is commonly denoted by $\chi^{+-}$ and depends on the frequency $\omega$, we have an especially simple expression:
\begin{align}
&\chi ^{+-}({\bf r,r}^{\prime },\omega) = \chi _{0}^{+-}({\bf r,r}^{\prime
},\omega) \notag\\
&+\int d{\bf r}^{\prime \prime }\chi _{0}^{+-}({\bf r,r}^{\prime
\prime },\omega ) I_{xc}({\bf r}^{\prime \prime}) \chi ^{+-}({\bf r}^{\prime
\prime }{\bf ,r}^{\prime },\omega )  
\label{transverse}
\end{align}
where
\begin{equation}
I_{xc}=\frac{2B_{xc}}{m} \,,
\label{stoner}
\end{equation}
is an exchange-correlation, Stoner (or Hund) interaction. This is the standard RPA equation for the transverse susceptibility written for the spatially inhomogeneous case. As one may see, it follows directly from the adiabatic local spin-density approximations of TDDFT, without any further assumptions. The magnetic
and charge electron densities as well as bare magnetic susceptibility are related to the Kohn-Sham states in the usual way,
\begin{eqnarray}
\label{magn}
m=\sum_{\mu \sigma } \sigma f_{\mu \sigma }\mid \psi _{\mu \sigma }({\bf r)}
\mid ^{2} \,,
\end{eqnarray}
\begin{eqnarray}
\label{magnOE1}
n=\sum_{\mu \sigma } f_{\mu \sigma }\mid \psi _{\mu \sigma }({\bf r)} \mid ^{2} \,, 
\end{eqnarray}
and
\begin{align}
&\chi _{0}^{+-}({\bf r,r}^{\prime },\omega )= \notag \\
&\sum_{\mu \nu
}\frac{f_{\mu \uparrow }-f_{\nu \downarrow }}{\omega-\varepsilon
_{\mu \uparrow }+\varepsilon _{\nu \downarrow }}\psi _{\mu
\uparrow }^{\ast }({\bf r)}\psi _{\nu
\downarrow }({\bf r)}\psi _{\nu \downarrow }^{\ast }({\bf r}^{\prime }{\bf )}%
\psi _{\mu \uparrow }({\bf r}^{\prime }{\bf )} \,.
\label{empty}
\end{align}
In these expressions, $\psi _{\mu \sigma }$ and $\varepsilon _{\mu \sigma }$ are eigenstates
and eigenenergies for the time-independent Kohn-Sham equation,
\begin{align}
\left(H_{0} - \sigma B_{xc}\right) \psi _{\mu
\sigma} &= \varepsilon _{\mu \sigma }\psi _{\mu \sigma }  \nonumber \\
H_{0} &= -\nabla ^{2}+V({\bf r)} \,.  
\label{sham}
\end{align}
Here $\sigma$ (without a vector symbol) stands for the spin index $\pm1=\uparrow \downarrow$ and $f_{\mu \sigma }=f\left( \varepsilon _{\mu \sigma }\right)$ is the
Fermi distribution function and $\mu$ labels the Kohn-Sham states. 

The same approach leads to expressions for the longitudinal spin susceptibility, which turns out to be coupled to the charge density. Since these expressions are not necessary for the derivation of the values of exchange parameters we do not show them here, but refer to the work in Ref.~\onlinecite{katsnelson2004magnetic}.

Further transformations are needed to make the expressions for the spin wave spectrum more explicit.
First, when substituting Eq.~\eqref{stoner} into Eq.~\eqref{transverse} we have the product of exchange-correlation field and wave functions. According to Eq.~\eqref{sham} this can be transformed as
\begin{align}
2B_{xc}\psi _{\mu \uparrow}\psi _{\nu \downarrow }^{\ast} &= 
\left(
\varepsilon _{\nu \downarrow }-\varepsilon _{\mu \uparrow }\right) \psi
_{\nu \downarrow }^{\ast }\psi _{\mu \uparrow } \notag\\
&+\nabla (\psi _{\mu \uparrow
}\nabla \psi _{\nu \downarrow }^{\ast }-\psi _{\nu \downarrow }^{\ast
}\nabla \psi _{\mu \uparrow }) \,.
\label{iden}
\end{align}
Substituting Eq.~\eqref{iden} into Eq.~\eqref{empty} one has
\begin{equation}
2(\chi _{0}^{+-}B_{xc})({\bf r,r}^{\prime },\omega )=m({\bf r)\delta }({\bf %
r-r}^{\prime }{\bf )}-\omega \chi _{0}^{+-}({\bf r,r}^{\prime },\omega )
\label{iden1}
\end{equation}
where we used the completeness condition
\begin{equation}
\sum_{\mu }\psi _{\mu \sigma }^{\ast }({\bf r)}\psi _{\mu \sigma}({\bf r}%
^{\prime }{\bf )=\delta }({\bf r-r}^{\prime }{\bf )} \,.
\label{complet}
\end{equation}
Substituting Eq.~\eqref{iden1} into Eq.~\eqref{transverse} we can
transform the latter expression to the following form 
\begin{align}
\widehat{\chi}^{+-} &= \widehat{\chi}_{0}^{+-}+\widehat{\chi}_{0}^{+-}\frac{%
2B_{xc}}{m}\widehat{\chi}^{+-} \notag\\
&= \widehat{\chi}_{0}^{+-}+\widehat{\chi}%
^{+-}-\omega \widehat{\chi}_{0}^{+-}\frac{1}{m}\widehat{\chi}^{+-}+\frac{%
\widehat{\Lambda }}{m}\widehat{\chi}^{+-}  
\label{iden2}
\end{align}
or, equivalently,
\begin{equation}
\widehat{\chi }^{+-}=m\left[\omega -\left( \widehat{\chi _{0}}^{+-}\right) ^{-1}%
\widehat{\Lambda }\right]^{-1}  \label{chi_lambda}
\end{equation}
where
\begin{align}
&\Lambda ({\bf r,r}^{\prime },\omega )=\sum_{\mu \nu }\frac{f_{\mu
\uparrow }-f_{\nu \downarrow }}{\omega-\varepsilon _{\mu \uparrow
}+ \varepsilon _{\nu \downarrow }}\psi _{\mu \uparrow }^{\ast
}({\bf r)}\psi _{\nu \downarrow }({\bf r)}  \notag\\
& \star \nabla \left[\psi _{\mu
\uparrow }({\bf r}^{\prime
}{\bf )}\nabla \psi _{\nu \downarrow }^{\ast }({\bf r}^{\prime }%
{\bf )}-\psi _{\nu \downarrow }^{\ast }({\bf r}^{\prime }{\bf
)}\nabla \psi _{\mu \uparrow }({\bf r}^{\prime }{\bf )}\right] \,.
\label{lambda}
\end{align}
Using Eqs.~\eqref{empty} and~\eqref{chi_lambda} we come to the final expression 
\begin{equation}
\widehat{\chi }^{+-}=\left( m+\widehat{\Lambda }\right) \left( \omega -I_{xc}%
\widehat{\Lambda }\right) ^{-1}.  \label{chi_final}
\end{equation}
Let us emphasize that the transformation from Eq.~\eqref{transverse} to Eq.~\eqref{chi_final} is exact. The latter however is more convenient to study the magnon spectrum. 

The susceptibility, expressed in Eq.~\eqref{chi_final}, has poles at the condition 
\begin{equation}
\omega=\Omega({\bf r,r}^{\prime },\omega ) \equiv I_{xc}\Lambda ({\bf
r,r}^{\prime },\omega ) \,. \label{Aeq}
\end{equation}
Solutions to Eq.~\eqref{Aeq} allows us to find a real-valued expression for the magnon spectrum. The imaginary part of $\Omega$ describes Stoner damping of magnons, that appear in metals. 
Note that there are many practical calculations of exchange interactions and magnon dispersion of real material, using the dynamical susceptibility~\cite{callaway1981magnetic, cooke1985new, savrasov1998linear, gorni2018spin, ke_electron_2021, PhysRevB.96.075108, PhysRevB.66.174417, costa2005ground, lounis2010dynamical}. 

The last step we describe in this section, and which allows a crucial result, is to restore effective exchange integrals from Eq.~\eqref{chi_lambda}. This procedure cannot be made in a unique way; there are, at least, two different definitions of exchange integrals which are both reasonable, but unfortunately not identical. 

First, we can try to fit interatomic exchange parameters to the poles of the susceptibility, that is, to the magnon spectrum. To do this explicitly we need a bit more transformations. Substituting Eq.~\eqref{iden} into Eq.~\eqref{lambda} one obtains the expression
\begin{align}
&\Lambda ({\bf r,r}^{\prime },\omega )=\sum_{\mu \nu }\frac{f_{\mu
\uparrow }-f_{\nu \downarrow }}{\omega - \varepsilon _{\mu
\uparrow }+\varepsilon _{\nu \downarrow}}  \notag \\
&\star \psi _{\mu \uparrow
}^{\ast }({\bf r)}\psi _{\nu \downarrow }\left[{\bf }2B_{xc}({\bf
r}^{\prime })-\varepsilon _{\nu \downarrow}+\varepsilon _{\mu
\uparrow }\right]\psi _{\nu \downarrow }^{\ast }({\bf r}^{\prime
}{\bf )}\psi _{\mu \uparrow }({\bf r}^{\prime }{\bf )} \,.
\label{lambda1}
\end{align}
Therefore, one may write
\begin{align}
&\Omega({\bf r,r}^{\prime },\omega )=\frac{4}{m({\bf r)}}J({\bf
r,r}^{\prime },\omega ) +
I_{xc}({\bf r)} \sum_{\mu \nu
}\frac{f_{\mu \uparrow }-f_{\nu \downarrow }}{\omega-\varepsilon
_{\mu \uparrow }+\varepsilon _{\nu \downarrow }} \notag \\
& \star 
\left(
\varepsilon _{\mu \uparrow}-\varepsilon _{\nu \downarrow }\right) \psi _{\mu \uparrow }^{\ast }({\bf r)}\psi _{\nu \downarrow
}({\bf r)}\psi _{\nu \downarrow }^{\ast }({\bf r}^{\prime }{\bf
)}\psi _{\mu \uparrow }({\bf r}^{\prime }{\bf )} \,.
\label{A1}
\end{align}
It is reasonable to identify the quantity
\begin{align}
\label{J}
&J({\bf r,r}^{\prime },\omega )=\sum_{\mu \nu
}\frac{f_{\mu \uparrow }-f_{\nu \downarrow }}{\omega-\varepsilon
_{\mu \uparrow }+ \varepsilon _{\nu \downarrow}}  \notag \\
& \star \psi _{\mu
\uparrow }^{\ast }({\bf r)}B_{xc}({\bf r)}\psi _{\nu \downarrow
}({\bf r)}\psi _{\nu \downarrow }^{\ast }({\bf r}^{\prime }{\bf
)}B_{xc}({\bf r}^{\prime }{\bf )}\psi _{\mu \uparrow }({\bf
r}^{\prime }{\bf )}  
\end{align}
as frequency-dependent interatomic exchange parameters. If one sets $\omega=0$ in this expression, one arrives at Rudderman-Kittel-Kasuya-Yosida (RKKY) type, indirect interactions~\cite{vonsovsky1974magnetism,Yosida_Magnetism}. As will be shown later in this review, these expressions are exactly equivalent to those from Refs.~\cite{liechtenstein1984exchange,liechtenstein1987local,Liechtenstein_1995_LDAU}. In fact, these expressions are more general, since they do not assume a rigid-moment approximation and they take into account the full coordinate dependence of the wave functions. 
Using the identity~\eqref{complet} one can also show that 
\begin{equation}
\Omega({\bf r,r}^{\prime },0)=\frac{4}{m({\bf r)}}J({\bf r,r}^{\prime },0)-2B_{xc}
({\bf r)\delta }({\bf r-r}^{\prime }) \,.  \label{A2}
\end{equation}

The other way to evaluate interatomic exchange interactions is to connect exchange parameters to the energy of spin spiral configurations, that is, with the {\it static} magnetic susceptibility $\widehat{\chi }^{+-}\left( 0\right)$. The latter can be
rewritten as
\begin{equation}
\widehat{\chi }^{+-}\left( 0\right) =m\left(
\widehat{\Omega}^{-1}-\frac{1}{2}{B_{xc}^{-1}}\right)  \label{static}
\end{equation}
which corresponds to the renormalized spin-wave energy
\begin{equation}
\widehat{\widetilde{\Omega}}=\widehat{\Omega}\left(
1-\frac{1}{2}B_{xc}^{-1}\widehat{\Omega}\right) ^{-1}  \label{bruno}.
\end{equation}
Note that this expression corresponds to the definition of exchange parameters in terms of the energy of static spin configurations~\cite{bruno2003exchange,Antropov_2003,Logan_1998}. As we will show below, this corresponds to the exchange parameters from Refs.~\cite{liechtenstein1984exchange,liechtenstein1987local,Liechtenstein_1995_LDAU}, normalized by taking into account constraints of the density functional~\cite{bruno2003exchange}. 

Thus, strictly speaking one cannot map the density functional susceptibility onto an effective Heisenberg model with interatomic exchange parameters in a unique way. The formal reason is the renormalization of the numerator, that is, the residue of the susceptibility at the magnon pole in Eq.~\eqref{chi_final}. There are however two important limits where this difference disappears.

\begin{figure}
\includegraphics[width=0.95\linewidth]{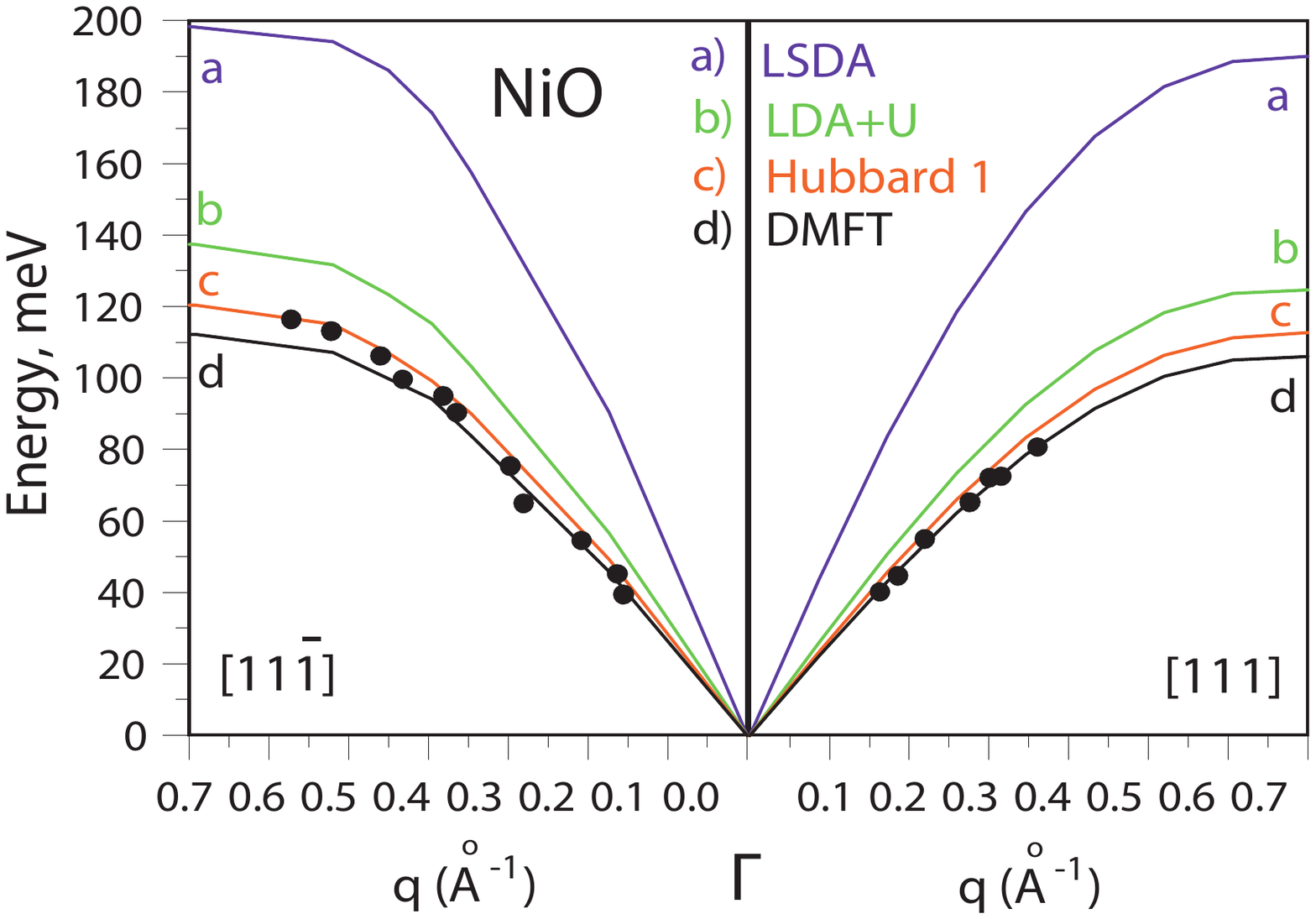}
\caption{(Color online) Calculated and measured magnon dispersion of NiO. Note that several levels of approximation for the theory are shown (solid lines) together with experiments (solid circles). Figure redrawn after Ref.~\onlinecite{PhysRevLett.97.266403}.
\label{fig1sect2}
}
\end{figure}

First, if we pass to the Fourier representation with the wave vector ${\bf q}$ and consider the limit ${\bf q} \rightarrow 0$ then, due to the Goldstone theorem, $\Omega \rightarrow 0$, the renormalization of the magnon spectrum~\eqref{bruno} disappears. This means that the expression for the spin-wave stiffness constant~\cite{liechtenstein1984exchange} determining magnon spectrum at ${\bf q} \rightarrow 0$ is well-defined and exact within the local spin density approximation. 
Second, if typical magnon energies are much smaller than the Stoner splitting, $B_{xc}^{-1}\widehat{\Omega}$ is small and the two definitions of exchange integrals coincide. This corresponds to an adiabatic approximation for magnons (note that magnon energies are much smaller than typical electron energies) which should be, of course, clearly distinguished from the adiabatic approximation in the sense of TDDFT. This is the case where the mapping of a full quantum mechanical description to the effective spin model is possible. In the sections below, we will focus on this case. 

In Fig.~\ref{fig1sect2} we highlight the results of Ref.~\onlinecite{PhysRevLett.97.266403}, using the expressions of exchange parameters discussed above. The figure shows results of a calculation for NiO, and after performing a Fourier transform from real-space $J({\bf r,r}^{\prime })$, from Eq.~\eqref{J} to reciprocal space, $J({\bf q})$, the magnon dispersion was calculated. The figure also shows experimental data and one may note that the agreement between observation and calculation is satisfactory, if the correct level of approximation is used for solving the Kohn-Sham equation~\eqref{firstKS}. For NiO dynamical mean field theory, LDA+U and the Hubbard 1 approximation are all found to reasonably well reproduce experiments. This will be discussed further in Section~\ref{examples}.

\section{Mapping electronic energies to an effective spin Hamiltonian}
\label{basicconcepts}

In the previous section we touched on the central aspect of this review; to extract from calculations of the electronic structure, parameters that accurately describe magnetic excitations. In this section we outline the basic principles of a method to do this, as was originally proposed in Ref.~\onlinecite{liechtenstein1984exchange}. A more detailed description of this method, with its extension for non-collinear spin systems when spin-orbit coupling (SOC) is also considered, will be presented in Section~\ref{detailsLKAG}. We emphasize that unless stated explicitly, we are only concerned with parameters that describe the coupling between spin moments. We start by a section that contains the essential aspect of Refs.~\onlinecite{liechtenstein1984exchange, liechtenstein1987local}, that involves how to connect changes of the energy of a spin Hamiltonian (such as the one in Eq.~\eqref{eqn1}) with changes of the grand canonical potential that contains energies of the electronic sub-system. 

\subsection{Basic assumptions}

We start by making a central assumption; that it is possible to identify well-defined regions of a material where the magnetisation density is more or less unidirectional and sizeable only close to an atomic nucleus. This implies the existence of local atomic magnetic moments (atomic spins), as is illustrated in Fig.~\ref{fig3}, with ferromagnetic, anti-ferromagnetic (AFM) or non-collinear interactions between atomic spin moments. As discussed in connection to Fig.~\ref{fig3}, very few materials, if any, are not 
accurately described in this way. 

An atomic spin moment is here chosen to be described with a direction, $\vec{e}_{i}$, quantified as:
\begin{equation}
\vec{e}_{i}= \left(\sin(\theta_{i}) \cos(\phi_{i}), \sin(\theta_{i}) \sin(\phi_{i}), \cos(\theta_{i}) \right) \;,
\label{ndef}
\end{equation}%
where $\theta_{i}$ and $\phi_{i}$ stand for the polar and azimuthal angles, respectively, of the atomic spin moment at site $i$. The rigid-spin approximation~\cite{phariseau2012electrons} is also assumed, where upon rotation of atomic spins, the length is not changed. 

The method of Ref.~\onlinecite{liechtenstein1984exchange} is an explicit method for calculations of interatomic exchange interactions, which relies on a formalism of the Green function of the electronic sub-system~\cite{Gyorffy_1985, kubler2017theory}. The basic idea is that an effective spin Hamiltonian describes accurately the energy of different atomic spin configurations, that are close to the magnetic ground state. We will below refer to the energy of the spin Hamiltonian as $\mathcal{H}$, and one needs to make sure that variation of $\mathcal{H}$, when the spin configuration is modified slightly, follows closely changes of the true total energy (the grand canonical potential, $\Omega$, described below) as provided by the electronic sub-system. 
This then allows us to map energies of the electron sub-system, as provided by e.g. density functional theory, to energies of an effective spin Hamiltonian, as given in Eq.~\eqref{eqn1}.
Practically, this mapping is based on the magnetic force theorem, which states that the variation of total energy of the electronic sub-system can be expressed in terms of variations only of occupied single particle energies~\cite{AndersenSkriverNohlJohansson+1980+93+118, mackintosh1980electrons, methfessel1982bond, liechtenstein1984exchange}. 
Recently, comparison of different mapping procedures for calculation of exchange interactions in various classes of magnetic materials was presented in Ref.~\cite{PhysRevB.103.104428}.
More details of the argumentation and its extension for correlated systems will be discussed in Subsections~\ref{secmft} and~\ref{sec:5K}, respectively.

\subsection{The mapping scheme}

In making the mapping between energies of the spin Hamiltonian and energies of the electronic sub-system, one considers as a reference state the atomic spin arrangement of the ground state, with the energy $\mathcal{H}$. Then the orientation of one atomic spin moment, at site $i$, is rotated with an infinitesimally small angle, keeping the length of the spin vector conserved (see Fig.~\ref{figone}). The variation of the direction of the spin, due to this rotation is denoted $\delta \vec{e}_{i}$ and the new direction of the perturbed spin can be written as
\begin{align}
	\vec{e}^{\prime}_i\rightarrow\vec{e}_i+\delta \vec{e}_i\,.
	\label{newpspindir}
\end{align}
The energy of this system, that can be seen as having a small perturbation from the ground state, can be written as $\mathcal{H}^{\prime }=\mathcal{H}^{\prime } (\delta \vec{e}_{i})$, where
\begin{equation}
\mathcal{H}^{\prime}= \mathcal{H}+\delta \mathcal{H}_{i}^\mathrm{one}\;.
\label{Eone}
\end{equation}%
As a second step, one considers a system with two atomic spin moments rotated, at the site $i$ and $j$. One can then express the energy of this spin arrangement as
\begin{equation}
\mathcal{H}^{\prime \prime}= \mathcal{H}+ \delta \mathcal{H}_{i}^\mathrm{one}+\delta \mathcal{H}_{j}^\mathrm{one} + \delta \mathcal{H}_{ij}^\mathrm{two} \;,  
\label{Etwo}
\end{equation}%
where $\mathcal{H}^{\prime \prime}=\mathcal{H}^{\prime \prime} (\delta \vec{e}_{i},\delta \vec{e}_{j})$ stands for the energy of a spin system with two atomic moments rotated with an infinitesimal amount (see Fig.~\ref{figtwo}). 

One may assume that the same procedure can be done for the grand canonical potential variation of the electronic system, where the value of the single site rotated system is
\begin{equation}
\Omega^{\prime}= \Omega +\delta \Omega_{i}^\mathrm{one}\;
\label{Omone}
\end{equation}%
and for the two-site rotated system is
\begin{equation}
\Omega^{\prime \prime}= \Omega + \delta \Omega_{i}^\mathrm{one}+\delta \Omega_{j}^\mathrm{one} + \delta \Omega_{ij}^\mathrm{two} \;. 
\label{Omtwo}
\end{equation}%
The next step is to derive explicit expressions for both $ \delta \Omega_{i}^\mathrm{one}$ and $\delta \Omega_{ij}^\mathrm{two}$ and to make a comparison with the $\delta \mathcal{H}_{i}^\mathrm{one}$ and $\delta \mathcal{H}_{ij}^\mathrm{two}$, respectively. The limit when the SOC is neglected, and the spins are arranged collinearly along a global quantisation axis (e.g. the $z$-direction) will here be referred to as the LKAG-limit (LKAG is after the authors of Refs.~\onlinecite{liechtenstein1984exchange,liechtenstein1987local}). A typical case for a small deviation from the collinear state with atomic moments along the $z$-axis that will be considered here is: $\delta \vec{e}_{i} \simeq \left(\delta \theta_{i}, 0, -1/2 \left( \delta \theta_{i} \right)^{2} \right)$. 

\subsection{Excitation of the spin model} 

\begin{figure}[t!]
\includegraphics[width=0.865\linewidth]{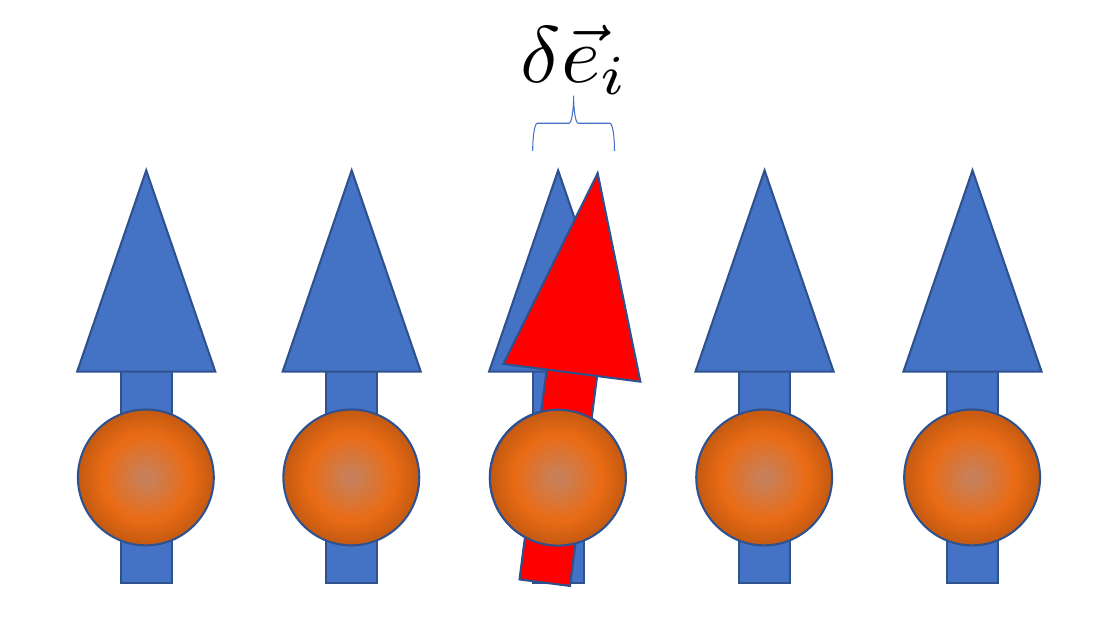}
\caption{(Color online) A schematic figure for the one-site spin rotation when the unperturbed system is collinear, ferromagnetic. An atomic spin at site $i$ is being rotated with an infinitesimal vector $\delta \vec{e}_i$. This process costs the energy $\delta \mathcal{H}_{i}^\mathrm{one}$ due to the fact that the spin interacts with every other spin in the rest of the spin system.}
\label{figone}
\end{figure}

The classical Heisenberg spin Hamiltonian has been introduced in Eq.~\eqref{neweqn1}. First, as shown in Fig.~\ref{figone}, we derive the one-site spin rotation variation, $\delta \mathcal{H}_{H,i}^\mathrm{one}$. Let us denote the non-perturbed spin configuration by the set of $\left\{ 
\vec{e}_{l}\right\} $ vectors and a perturbed system by the set of $\left\{ \vec{e}_{l}+\delta _{il}\delta \vec{e}_{i}\right\}$ where $\delta \vec{e}_{i}$ stands for an infinitesimal variation of the spin direction due to a rotation at site $i$ with the angle $\delta \theta_{i} $. One then finds 
\begin{equation}
\begin{split}
\mathcal{H}_{H}^{\prime}& =\sum_{\left<lk\right>}  J_{lk} \left( \vec{e}
_{l}+\delta _{il}\,\delta \vec{e}_{i}\right) \cdot \left( \vec{e}_{k}+\delta _{ik}\,\delta \vec{e}_{i}\right) \\
& =\mathcal{H}_{H}+  \frac{1}{2} \sum_{l\left( \neq i\right) } J_{li}\, \vec{e}_{l} \cdot \delta \vec{e}_{i}+\frac{1}{2}\sum_{k\left( \neq i\right) } J_{ik}\,\delta\vec{e}_{i} \cdot \vec{e}_{k}\;,
\label{firstperturbed}
\end{split}
\end{equation}
where the origin of the factor of $\frac{1}{2}$'s has been explained in Subsection \ref{nomenclature}. Note that Eq. (\ref{firstperturbed}) can be simplified to describe the energy gain due to the rotation as
\begin{equation}
\delta \mathcal{H}_{H,i}^\mathrm{one} =   \,\sum_{l\left( \neq i\right) } J_{li}\, \vec{e}_{l} \cdot \delta \vec{e}_{i} ,
\label{dHHone}
\end{equation}
since the interaction is symmetric, $J_{il}=J_{li}$.
This means that the energy variation of the one-site spin rotation is an energy cost resulted by the interaction of the rotated spin and its environment formed by the non-rotated spins as shown in Fig.~\ref{figone}. 

\begin{figure}[t!]
\includegraphics[width=1.\linewidth]{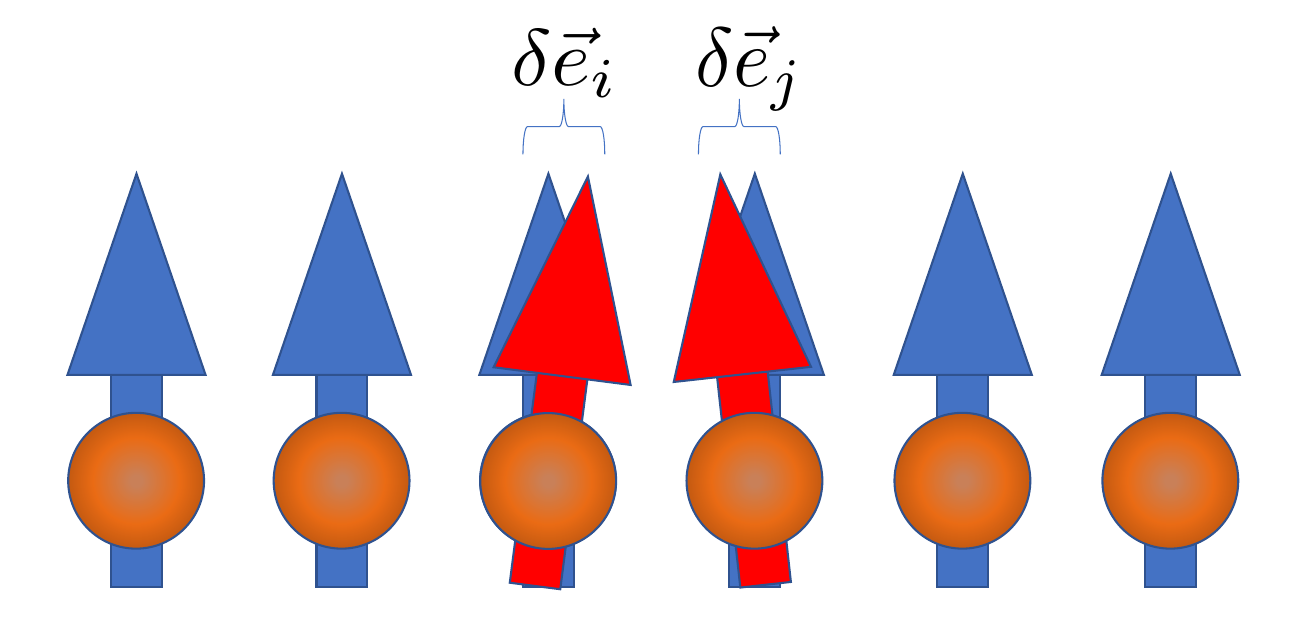}
\caption{(Color online) A schematic figure for the two-site spin rotation when the unperturbed system is collinear, ferromagnetic. Atomic spins at site $i$ and $j$ are rotated with the infinitesimal vector $\delta \vec{e}_i$ and $\delta \vec{e}_j$, respectively. This process costs the energy $\delta \mathcal{H}_{i}^\mathrm{one}+\delta \mathcal{H}_{j}^\mathrm{one}+\delta \mathcal{H}_{ij}^\mathrm{two}$ where $\delta \mathcal{H}_{i}^\mathrm{one}$ and $\delta \mathcal{H}_{i}^\mathrm{one}$ stand for the energy cost of a one-site rotations (shown in Fig.~\ref{figone}) while the interacting term, $\delta \mathcal{H}_{ii}^\mathrm{two}$, see text, characterizes the exchange energy between the spins located at site $i$ and $j$.}
\label{figtwo}
\end{figure}

If the non-perturbed configuration is now collinear ferromagnetic, i.e., $\left\{\vec{e}_{j}\right\}= \left\{(0,0,1) \right\}$ for all $j$ in the spin system with the energy $\mathcal{H}_{H}$, one obtains that $\vec{e}_{j} \cdot \delta \vec{e}_{i}=\delta e^{z}_{i}$, which is proportional to $\left(\cos \delta \theta_{i}-1 \right)$, i.e., approximately to $-(1/2)\left(\delta \theta_{i} \right)^{2}$. Therefore, in the ferromagnetic limit, one can demonstrate that
\begin{equation}
\delta \mathcal{H}_{H,i}^\mathrm{one} \simeq - \frac{1}{2}
\sum_{j\left( \neq i\right) } J_{ji}  \left(\delta \theta_{i} \right)^{2}\;.
\label{two}
\end{equation}

Next, we simultaneously rotate two spins at site $i$ and $j$ with $\delta \theta_{i} $ and $\delta \theta_{j}$, respectively. As shown in Fig.~\ref{figtwo}, the perturbed system for the two-site spin rotation is given by the set of $\left\{ \vec{e}_{l}+\delta _{il}\delta \vec{e}_{i}+\delta _{jl}\delta \vec{e}_{j}\right\}$ and its energy is 
\begin{equation}
\mathcal{H}_{H}^{\prime \prime}=\mathcal{H}_{H}+\delta \mathcal{H}_{H,i}^\mathrm{one}+ \delta \mathcal{H}_{H,j}^\mathrm{one}+ J_{ij} \delta \vec{e}_{i} \cdot \delta \vec{e}_{j}\;.
\label{Hpp}
\end{equation}
Comparing Eq.~\eqref{Hpp} to Eq.~\eqref{Etwo}, we obtain \footnote{A factor of $2$ would appear in the last term of Eq. \ref{Hpp}, but it is canceled by a factor of $\frac{1}{2}$ (that is explained in Subsection \ref{nomenclature}).}
\begin{equation}
\delta \mathcal{H}_{H,ij}^\mathrm{two}=J_{ij}\,\delta \vec{e}%
_{i} \cdot \delta \vec{e}_{j}\;.  \label{dHHtwo}
\end{equation}
In the LKAG limit when ${\delta \vec{e}_{i}=(\delta \theta_{i},0,0)}$, ${\delta \vec{e}_{j}=(\delta \theta_{j},0,0)}$ and $\delta \theta_{i}=-\delta \theta_{j}=\delta \theta$, i.e., the spins are rotated in the opposite directions, it can be shown that
\begin{equation}
\delta \mathcal{H}_{H,ij}^\mathrm{two}=J_{ij}\,\delta \theta_{i} \delta \theta_{j}=-J_{ij}\,\left(\delta \theta \right)^{2} \;.
\label{three}
\end{equation}

One can in a more general way consider a spin Hamiltonian with a tensorial coupling between the spins as follows,
\begin{equation}
\mathcal{H}_{T}=\sum_{<ij>} \vec{e}_{i} \cdot \mathbb{J}_{ij} \cdot \vec{e}_{j}\;, \label{Htens}
\end{equation}
where $\mathbb{J}_{ij}=\{J_{ij}^{\mu \nu };\mu,\nu\in\{x,y,z\}\}$. This is needed even in a collinear system when the SOC is present~\cite{udvardi2003first}. Note that $\mathcal{H}_{T}$ can be rewritten as  
\begin{equation}
\mathcal{H}_{T}=\mathcal{H}_{H}+\mathcal{H}_{anis}+\mathcal{H}_{DM}\;,
\label{tensor1}
\end{equation}
where

\begin{equation}
\mathcal{H}_{anis}= \sum_{<ij>} \vec{e}_{i} \cdot \mathbb{A}_{ij} \cdot \vec{e}_{j}\;
\label{tensor2}
\end{equation}
is the symmetric anisotropic interaction tensor and $\mathcal{H}_{H}$ and $\mathcal{H}_{DM}$ have been introduced by Eqs.~\eqref{neweqn1} and~\eqref{neweqn2}, respectively. More precisely, the $3\times 3$ tensorial interaction is given by
\begin{align}
   &\mathbb{J}_{ij}= \nonumber\\
   &\left(\begin{array}{ccc}
       J_{ij}+A^{xx}_{ij} & D^{z}_{ij}+A^{xy}_{ij} & -D^{y}_{ij}+A^{xz}_{ij} \\
      -D^{z}_{ij}+A^{xy}_{ij} & J_{ij}+A^{yy}_{ij} &  D^{x}_{ij}+A^{yz}_{ij} \\
         D^{y}_{ij}+A^{zx}_{ij} &  -D^{x}_{ij}+A^{yz}_{ij} & J_{ij}+A^{zz}_{ij} 
   \end{array}\right).
   \label{TensComps}
\end{align}
 Such a $3\times 3$ tensor can be decomposed by symmetry into three independent tensor terms; a symmetric scalar or rank 0, $\mathbb{S}$, an asymmetric vector or rank 1, $\mathbb{V}$, and lastly a symmetric rank 2 tensor term, $\mathbb{T}$, respectively. These are defined as
\begin{align}
    \mathbb{S}&=\frac{1}{3}\,\mathrm{Tr}\,\mathbb{J}\\
    \mathbb{V}&=\vec{D}\\
    \mathbb{T}&=\mathbb{A}-\frac{1}{3}\,\mathrm{Tr}\,\mathbb{A}.
\end{align}
 
While we have referred to the Heisenberg interaction as the term where the explicit magnetic interaction is a scalar, there is an alternative view that the Heisenberg interaction is an interaction that is effectively a scalar, i.e.,~$\mathbb{S}$. Such an approach ensures that the other interactions are traceless.
This means that the Dzyaloshinskii-Moryia interaction is unique, but the exact Heisenberg and second rank tensor is kind of a matter of choice.

The one- and two-site energy variations of $\mathcal{H}_{T}$ can be given as the sum of the variations of $\mathcal{H}_{H}$, $\mathcal{H}_{anis}$ and $\mathcal{H}_{DM}$, i.e., 
\begin{align}
\delta \mathcal{H}_{T,i}^\mathrm{one} = &  \sum_{j\left( \neq i\right) }
\Big(J_{ij}\,\delta \vec{e}_i \cdot \vec{e}_j \notag\\
&+\vec
{D}_{ij} \cdot \left( \delta \vec{e}_i\times\vec{e}_j \right) +\delta \vec{e}_i \cdot  \mathbb{A}_{ij} \cdot \vec{e}_j\Big)
\label{1stHamT}
\end{align}
and
\begin{align}
\delta\mathcal{H}^\mathrm{two}_{T,ij} = &  J_{ij}\, \delta \vec{e}_i \cdot \delta \vec{e}_j \notag\\
&+\vec
{D}_{ij} \cdot \left( \delta \vec{e}_i\times \delta \vec{e}_j \right) +\delta \vec{e}_i \cdot \mathbb{A}_{ij} \cdot \delta \vec{e}_j \,, 
\label{2ndHamT}
\end{align}
respectively. The expressions of energy variations of the spin Hamiltonian, in Eqs.~\eqref{1stHamT} and~\eqref{2ndHamT}, must now be compared to similar expressions for the grand canonical potential variations of the electrons. Before we make this connection, a few important aspects of electronic structure theory need to be reviewed, which is what the following section attempts to do. 

\section{Basic concepts of electronic structure theory}
\label{estruct}

In this section we introduce a few central concepts of electronic structure theory, such as the one-electron Green function and (integrated) density of states that will be needed in Section~\ref{detailsLKAG}, where we present the details of the derivation of the generalized interatomic exchange formulas.

First of all, we need an expression for the electronic energy and its variations under a perturbation, such as the rotations in Figs.~\ref{figone} and~\ref{figtwo}. The grand canonical ensemble is used for this purpose, where energy and particles of the system considered can be exchanged with a reservoir, implying that the chemical potential ($\mu$) and temperature ($T$) are relevant thermodynamic variables. The grand canonical potential can be calculated as
\begin{equation}
\Omega=E-TS-\mu N\;,
\label{eqn:grandpot}
\end{equation}
where $E$ is the energy given by the equation
\begin{equation}
E=\int\limits_{-\infty }^{\infty}d\varepsilon\, \varepsilon f( \varepsilon )n(\varepsilon ),
\end{equation}
$S$ is the entropy of the band electrons
\begin{align}
 S=-\int\limits_{-\infty }^{\infty}d\varepsilon\,n(\varepsilon )\left\{f( \varepsilon )\ln f( \varepsilon )+[1-f( \varepsilon )]\ln \left[1-f( \varepsilon )\right]\right\},
\end{align}
and $N$ is the number of electrons in the valence band. Note that $f( \varepsilon )$ is the Fermi-Dirac distribution function and $n(\varepsilon )$ denotes the density of states (DOS). The exact conditions that have proven crucial in constraining and constructing accurate approximations for ground-state DFT are generalized to finite temperature, based on the work of Mermin \cite{PhysRev.137.A1441}, can be found in Ref. \cite{PhysRevLett.107.163001}.  

\subsection{Grand canonical potential at zero temperature}

Considering that the Fermi-energy, $\varepsilon_{F}$, usually is much higher than the critical (Curie or Neel) temperature, it is for most cases enough to work in the $T=0$ approach (i.e. $f( \varepsilon )$ is a step function). In this case $\Omega=E- \varepsilon_{F} N$, i.e.,
\begin{equation}
\Omega= \int\limits_{-\infty }^{\varepsilon_{F}}d\varepsilon\, \varepsilon\, n(\varepsilon )-  \varepsilon_{F} N = - \int\limits_{-\infty }^{\varepsilon_{F}}d\varepsilon N(\varepsilon ),
\label{four}
\end{equation}
where partial integration has been used. Here the number of states function (or integrated density of states-IDOS) is introduced, $N(\varepsilon )$, and one finds that the grand canonical potential can be calculated as an integral of this function. This means that one has to determine the variations of IDOS to get the variations of the grand canonical potential. A practical way to do this is to employ the so-called Lloyd formula, that will be described in Section~\ref{detailsLKAG}. Note that the corresponding formula of Eq.~\eqref{four} for cases when the energy argument is in the complex plane, is presented at the of Subsection~\ref{secgreen}.

\subsection{Green function}
\label{secgreen}

Since the derivation of the interatomic exchange formulas relies on a Green function formalism of the electronic structure, we summarize here the most central aspects needed. A full account may be seen in Ref.~\onlinecite{economou2006green}. The Green function (or resolvent) of the electronic Hamiltonian, $H$, is defined as
\begin{equation}
G(z)=\left(z-H\right)^{-1}\;, 
\label{five}
\end{equation}
where $z \in \mathbb{C}$. This implies that $G\left(z^{*}\right)=G^{\dagger}(z)$. If both sides of the equation $(z_{2}-H)-(z_{1}-H)=z_{2}-z_{1}$ are multiplied by $G(z_{1})G(z_{2})$ and one sets that $z_{2}=z+dz$, $z_{1}=z$ and considers the limit $dz \to 0$, then the equation
\begin{equation}
\frac{dG(z)}{dz}=-G^{2}(z) 
\label{six}
\end{equation}
can be obtained. 

Next, we consider an electronic Hamiltonian, $H$, with a discrete spectrum\footnote{Our conclusions would be the same for continuous spectrum.}, with solutions $H\varphi_{\mu}=\varepsilon_{\mu} \varphi_{\mu}$. Note that $\langle \varphi_{\mu} | \varphi_{\nu}\rangle=\delta_{\mu \nu}$ and the solutions to $H$ form a complete set. The spectral resolution of the Green function can then be obtained from the so-called Lehmann representation,
\begin{equation}
G(z)=\sum_{\mu} \frac{|\varphi_{\mu} \rangle \langle \varphi_{\mu}|}{z-\varepsilon_{\mu}}  \;.
\label{green} 
\end{equation}
This implies that on the basis of the eigenfunction of $H$ the Green function could be represented as ${G_{\mu \nu} (z)=\delta_{\mu \nu} \frac{1}{z-\varepsilon_{\mu}}}$. In addition, $G(z)$ is obviously undefined for $z=\varepsilon_{\mu}$. However, considering $z$ in the complex plane, just above or below the real axis ($z=\varepsilon \pm i \delta $), allows us to define\footnote{Note that $G^{\pm}(\varepsilon)=\left(G^{\mp}(\varepsilon)\right)^{\dagger}$.}:
\begin{equation}
G^{\pm}(\varepsilon) = \lim_{\delta \to 0^{+}}G(\varepsilon \pm i \delta) \;.
\label{pm}
\end{equation}
One should note that a lattice site-dependent Green function, $G_{ij}(z)$, is relevant here, and it is obtained as 
\begin{equation}
G(z)=\sum_{ij\mu} \frac{|\phi_i \rangle\langle \phi_i|\varphi_{\mu} \rangle \langle \varphi_{\mu}|\phi_j \rangle\langle \phi_j |}{z-\varepsilon_{\mu}}
=\sum_{ij} |\phi_{i} \rangle G_{ij}(z) \langle \phi_{j} |.
\label{greenij} 
\end{equation}
with local functions $|\phi_i\rangle$ at site $i$.

\subsection{Grand canonical potential at finite temperature}

To derive the grand canonical potential at finite temperature, it is useful to find a relationship between the IDOS, DOS and the Green function, and one may note that in a system of independent fermions, the expectation value of a one-particle observable, $A$, is given as
\begin{equation} 
\langle A \rangle = \sum_{\mu} p_{\mu} \langle \varphi_{\mu} |A |  \varphi_{\mu} \rangle \;,
\label{A}
\end{equation}
where $p_{\mu}=f (\varepsilon_{\mu})$, i.e. the Fermi-Dirac distribution function. One can evaluate this expression with the help of Cauchy's theorem, which states that for a closed contour oriented clock-wise the integration of a function $g(z)/(z-a)$ is equal to $-2 \pi i g(a)$ if $a$ is within the contour (otherwise the result is zero). With the help of Cauchy's theorem and Eq.~\eqref{green}, Eq.~\eqref{A} can be simply given by $G^{+}(\varepsilon)$ as follows\footnote{More details: http://newton.phy.bme.hu/$\sim$szunyogh/Elszerk/Kkr-slides.pdf.},
\begin{equation} 
\langle A \rangle =-\frac{1}{\pi} \Im \int\limits_{-\infty }^{\infty}d\varepsilon f(\varepsilon)\, \mathrm{Tr}_{L \sigma} AG^{+}(\varepsilon) 
\end{equation}
where the trace is taken over both the orbital ($L$) and spin ($\sigma$) spaces. If $A$ is the identity operator one obtains the expression
\begin{equation} 
N=-\frac{1}{\pi} \Im \int\limits_{-\infty }^{\infty}d\varepsilon f(\varepsilon) \mathrm{Tr}_{L \sigma} G^{+}(\varepsilon)  \;.
\label{N2}
\end{equation}
This allows us to identify a relationship between the DOS and the Green function,
\begin{equation}
n(\varepsilon)=-\frac{1}{\pi} \Im \, \mathrm{Tr}_{L \sigma} G^{+}(\varepsilon)  \;.
\label{seven}
\end{equation}
In the rest of the paper we consider the limit of the upper part of the complex plane (Eq.~\eqref{pm}) and the ``$+$ symbol'' will be omitted for brevity for functions of real energies.

\begin{figure}[t!]
\includegraphics[width=0.75\linewidth]{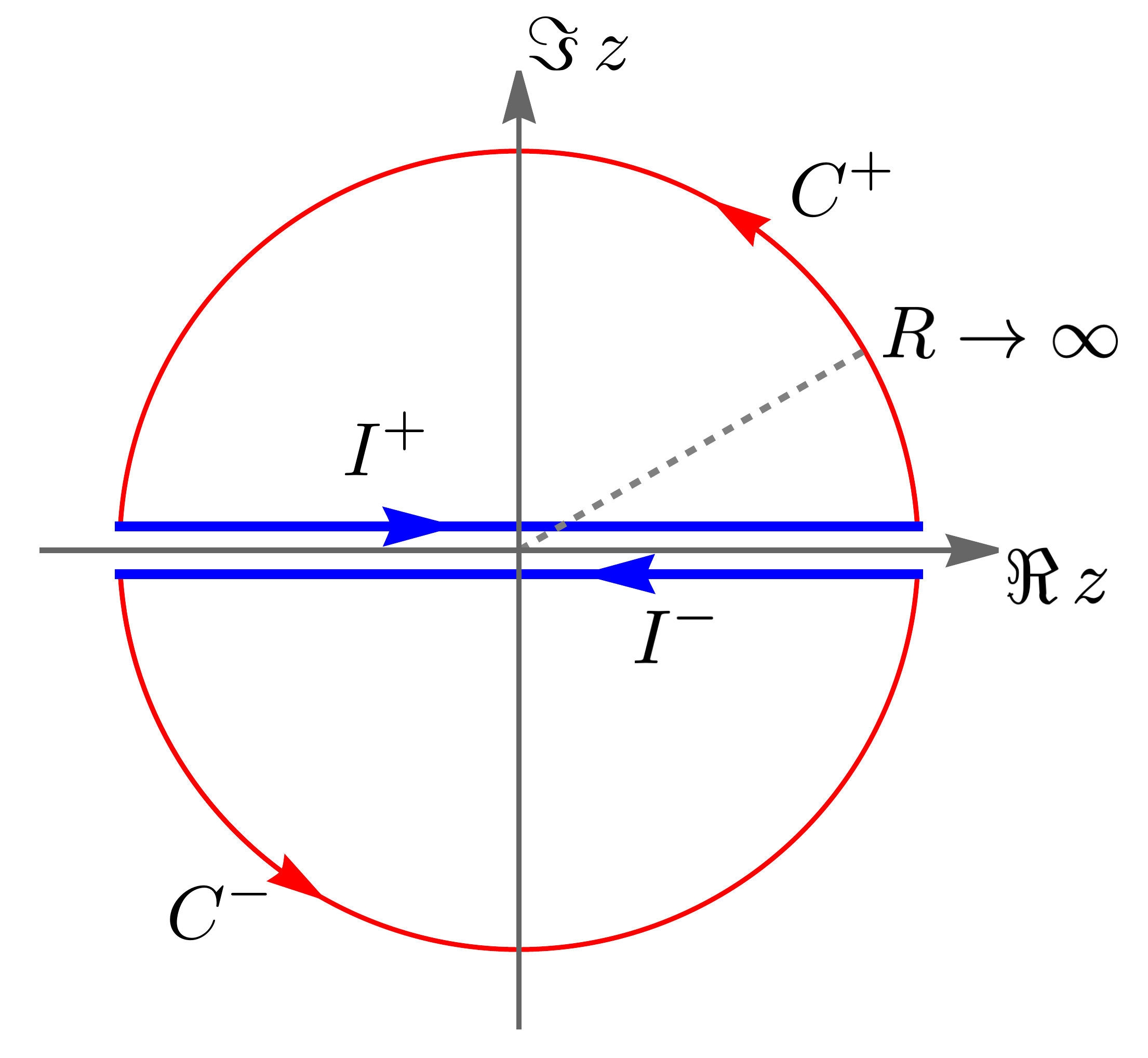}
\caption{(Color online) Integration paths in the complex plane.}
\label{Contour-GF}
\end{figure}

One can recognize that in Eq.~\eqref{N2} the integral is taken along the real axis, which is not always convenient, i.e., it is preferable to transform such integrals to the complex plane. We proceed with the realization that the DOS can equally well be calculated with the help of $G^{-}(\varepsilon)$, or since $\Re\,G^{+}(\varepsilon)=\Re\,G^{-}(\varepsilon)$ one can use the expression
\begin{equation}
n(\varepsilon)=-\frac{1}{2\pi i} \mathrm{Tr}_{L \sigma} \left\{G^{+}(\varepsilon) - G^{-}(\varepsilon)\right\} \;.
\label{dosfromg+g-}
\end{equation}
With the latter choice, the number of particles of Eq.~\eqref{N2} can be reformulated as
\begin{align} 
N&=-\frac{1}{2\pi i} \int\limits_{-\infty }^{\infty}d\varepsilon f(\varepsilon) \mathrm{Tr}_{L \sigma} \left\{G^{+}(\varepsilon) - G^{-}(\varepsilon)\right\}=\nonumber\\
&=-\frac{1}{2\pi i}\mathrm{Tr}_{L \sigma} \left\{\int\limits_{-\infty }^{\infty}d\varepsilon f  G^{+} +\int\limits_{\infty }^{-\infty}d\varepsilon f  G^{-}\right\}
\;.
\label{N3}
\end{align}
Referring to the two integrals as $I^+$ and $I^-$, one can view them as the two path integrals illustrated with thick blue lines in Fig.~\ref{Contour-GF}. A closed contour integral can be obtained by adding the two paths labeled $C^+$ and $C^-$, respectively, shown with thin red line, that both give vanishing contributions since the energies of this part of the path can be chosen to lie infinitely far away from the poles of the Green functions. Then since the integrand is analytical within these contours, these integrals can be evaluated by summing the residues that arise from the Fermi-Dirac distribution, $\mathrm{Res}\,(f,\mu+i\omega_n)=-T$, due to its poles
at the Matsubara energies $z=\mu+i\omega_n$, where $\omega_n=(2n+1)\pi T$ and $T$ is the temperature\cite{Auerbach_book}. Hence
\begin{align} 
N&=\left(I^++C^+\right)+\left(I^-+C^-\right)=\nonumber\\
&=-\frac{1}{2\pi i}\left\{\oint_+dz f  G +\oint_-dz f  G\right\}=\nonumber\\
&=-\frac{1}{2\pi i}\mathrm{Tr}_{L \sigma} \sum_{n=-\infty}^{\infty}\,(2\pi i)\,\mathrm{Res}\,(fG,\mu+i\omega_n)=\nonumber\\
	&=\,T\sum_{n=-\infty}^{\infty}\,\mathrm{Tr}_{L \sigma}\,G(\mu+i\omega_n) \,.
\label{N4}
\end{align}

\section{Detailed derivation of the exchange formulas}
\label{detailsLKAG}

In this section, we present the details of the mapping of the electronic Hamiltonian to the spin Hamiltonian given by the $\mathbb{J}_{ij}$-tensor as shown in Eq.~\eqref{Htens}. The derivation is general in the sense that we consider a non-collinear spin arrangement when the SOC interaction is present. Hence, we will give explicit expressions for the Heisenberg $J_{ij}$, the DM vector $\vec{D}_{ij}$ and the symmetric anisotropic exchange term $\mathbb{A}_{ij}$ in general, and the interpretation of the results in the LKAG limit.

\subsection{Magnetic local force theorem}
\label{secmft}

As mentioned in Section~\ref{basicconcepts}, the mapping of the electronic Hamiltonian to the spin Hamiltonian is based on the magnetic force theorem, since one can always consider small variations from the ground states, i.e, a mapping to an effective Hamiltonian is locally\footnote{See the discussion of local vs global spin model later in this Section.} possible (close enough to the magnetic ground state). 

Let us write the grand canonical potential as
\begin{equation}
\Omega=\Omega_{sp}-\Omega_{dc} \;.
\label{mlft}
\end{equation}
In this expression the subscript $sp$ stands for single particle, and $\Omega_{sp}$ simply represents the integral in Eq. (\ref{four}). In addition, $\Omega_{dc}$ stands for the interaction or "double-counting" term. Then one can calculate the first-order change in $\Omega$ when the system is under some perturbation. In deriving the magnetic force theorem small rotations are considered as perturbations. These changes are assumed to be described by some set of parameters\cite{methfessel1982bond}. As a first step, the potential is held fixed which leads to a variation in the single particle energy $\delta^{*} \Omega_{sp}$. Then, in a second step, the parameters that characterize the changes are held constant and the potential is allowed to relax to self-consistency. This leads to
variations $\delta_{1} \Omega_{sp}$ and $-\delta \Omega_{dc}$ in the single particle energies and the double counting term, respectively. However, these two contributions, $\delta_{1} \Omega_{sp}$ and $-\delta \Omega_{dc}$ cancel each other as shown in Refs.~\onlinecite{AndersenSkriverNohlJohansson+1980+93+118, mackintosh1980electrons, methfessel1982bond, liechtenstein1984exchange}. In summary, the magnetic force theorem indeed shows that the variation of total energy of the electronic sub-system can be expressed in terms of variations only of occupied single particle energies. Note that the magnetic local force theorem strictly holds only for first order variations.

Finally, we close this subsection by mentioning that it is a challenge to find the ground state due to many local minima in the DFT-total-energy landscape. The problem of finding a list of initial magnetic configurations for the practical calculations has been tackled by Refs.~\onlinecite{ZHENG2021107659, PhysRevX.11.011031}.

\subsection{Energy variation from non-collinear Kohn-Sham Hamiltonian}

Let us start this subsection with a general non-collinear state with spin moments $\{\vec{e}_i\}$ and the Kohn-Sham Hamiltonian defined by Eq.~\eqref{firstKS}. For simplicity we introduce ${\vec{B}}({\bf r})$ as ${\vec{B}}({\bf r})={\vec{B}}_{xc}({\bf r})+\vec{B}_{ext}({\bf r})$ and in condensed form we can express the spin-dependent interaction as $B\equiv{\vec{B}}({\bf r}) \cdot \vec{\sigma}$. Then one can let the directions of the local moments rotate away slightly from given magnetic configuration. Instead of the case when only one spin is rotated in the spin system, as shown in Fig.~\ref{figone} and by Eq.~\eqref{newpspindir}, we allow in principle a $\delta \vec{e}_{i}$ change at all possible sites. As we will see this is more than a sum of one-site rotations because of the intersite interactions. However, the corresponding perturbation in the electronic potential $\delta V$, which is purely spin-dependent, can be divided into local changes of the spin polarised potential in a given region around the atomic sites where the moments are varied
\begin{align}
\delta V=-\sum_i \,\delta \vec{B}_i \cdot\vec{\sigma} = -\sum_i {B}_i\,\delta \vec{e}_i \cdot\vec{\sigma}\,,\label{dyson-eq}
\end{align}
where $\vec{B}_{i}\equiv B_{i} \vec{e}_{i}$. Having the perturbation $\delta V$, one can write for the perturbed Green function, $G^{\prime}$ (omitting for simplicity the energy argument), that 
\begin{align}
G^{\prime}&=G+G\,\delta V\,G^{\prime} \, \nonumber\\ 
G&=G^{\prime}-G\,\delta V\,G^{\prime}=\left(1-G\,\delta V\right)G^{\prime} \,, \label{exactGF}\end{align}
where $G$ stands for the unperturbed Green function. 

Now, from Eq.~\eqref{six} one can deduce that 
\begin{align}
    G=-\frac{\partial \ln G}{\partial\varepsilon}
\end{align}
 which means that IDOS, which is the primitive function to the DOS of Eq.~\eqref{seven}, is given by the equation
\begin{align}
    N &=-\frac{1}{\pi}\Im\, \mathrm{Tr}_{iL \sigma} \left(-\ln G \right)
\end{align}
and the change in IDOS, is then given by the Lloyd formula~\cite{lloyd1967wave}, i.e.
\begin{align}
\delta N
&=-\frac{1}{\pi}\Im\, \mathrm{Tr}_{iL \sigma}\,\left\{-\ln G^{\prime} +\ln \left(1-G\,\delta V\right) G^{\prime} \right\} \notag\\
&= -\frac{1}{\pi}\Im\, \mathrm{Tr}_{iL \sigma}\,\ln \left(1-G\,\delta V\right) \,.
\label{Lloyd1}
\end{align}
This means that one does not have to deal with the exact Green function $G^{\prime}$ in order to calculate $\delta N$. One can also expand the logarithm in a series as long as $\delta V$ is small, which yields the expression
\begin{equation}
	\delta N=\frac{1}{\pi}\Im\, \mathrm{Tr}_{iL \sigma}\, \sum_{k=1}^{\infty}	\, \frac{(\delta V\, G)^k}{k} \,,
	\label{Lloyd}
\end{equation}
where the order of the two factors can be altered due to the properties of the trace.
Note that $G$ is the Green function corresponding to the electronic Hamiltonian, Eq.~\eqref{firstKS}. It can be decomposed to inter site terms, $G_{ij}$, according to Eq.~\eqref{greenij}, that can further be decomposed into spin-components into a form where
\begin{align}
	G_{ij}=G^0_{ij}+\vec{G}_{ij}\cdot\vec{\sigma}\,,\label{GF-s}
\end{align}
where $\vec{G}_{ij}$ is a vector with the components of $G_{ij}^{x}$, $G_{ij}^{y}$ and $G_{ij}^{z}$. We introduce here the notation $G^{\eta}_{ij}$ where the index $\eta$ enumerates both the scalar spin-independent Green function as well as the components of the spin-dependent vector Green function of Eq.~\eqref{GF-s}, i.e.~$\eta$ can be either $0$, $x$, $y$ or $z$. Note that $G_{ij}$ is defined in both the spin and orbital spaces while $G^{\eta}_{ij}$ is represented only in the orbital space. In other words, $G_{ij}$ can be represented by a 18x18 matrix while $G_{ij}^{\eta}$ is a 9x9 matrix when $spd$ orbitals are used in a practical calculation. We refer here to  $G^0_{ij}$ as the charge part and to $\vec{G}_{ij}$ as the spin part of the Green function and the physical interpretation of the decomposition will be discussed in Subsection~\ref{decomp}. Note that in the LKAG limit the vector $\vec{G}_{ij}$ has only a $z$-component, and in Subsection~\ref{seccomp} we will define the up and down spin channels with the help of $G^0_{ij}$ and $G^z_{ij}$. It should also be noted that the trace in Eqs.~\eqref{Lloyd1} and~\eqref{Lloyd} is over the atomic sites $i$, the local basis functions $L$, as well as the spin components $\sigma$.

Based on Eq.~\eqref{four}, the variation in grand canonical potential (Eq.~\eqref{eqn:grandpot}), due to the moment rotations, is obtained through integration of the change in number of states function, i.e., 
\begin{equation}
\delta \Omega=- \int\limits_{-\infty }^{\infty}d\varepsilon\, \delta N(\varepsilon )\,f(\varepsilon ),
\label{mft}
\end{equation}
where Eq.~\eqref{Lloyd} can be used for $\delta N$. The corresponding grand canonical potential variation formula at finite temperature can be expressed as
\begin{align}
    \delta \Omega= T\sum_{n=-\infty}^\infty \,\pi\,\delta N(\mu+i\omega_n )\,\label{eq:matsubara-GCpot}
\end{align}
where $\delta N$ (that along the real axis is given by Eq.~\eqref{Lloyd1}) is generalised to an expression in the complex plane 
\begin{align}
    \delta N(z)&=-\frac{1}{\pi}\,\mathrm{Tr}_{iL \sigma}\, \ln \left( 1-\delta V\, G(z)\right),\label{eq:intN-AnalCont}
\end{align}
 with the limit
\begin{align}
  \Im\,  \lim_{\Im z\rightarrow 0^+} 
    \delta N(z)=\delta N(\varepsilon)\,.
\end{align}
With Eq.~\eqref{eq:intN-AnalCont} one can rewrite the sum over Matsubara frequencies in Eq.~\eqref{eq:matsubara-GCpot} as a trace
\begin{align}
    \delta \Omega=  -\mathrm{Tr}\, \ln \left( 1-\delta V\, G\right)=\mathrm{Tr}\,\sum_k	\, \frac{(\delta V\, G)^k}{k}\,
\end{align}
which is a short notation for
\begin{align}
    \mathrm{Tr}&=\mathrm{Tr}_{\omega iL \sigma}=T\sum_{n=-\infty}^{\infty}\,\mathrm{Tr}_{iL \sigma} \,.\label{Tr-Matsubara-DFT}
\end{align}
Using analytical continuation from the Matsubara space to the real frequencies we can get the following relation~\cite{katsnelson2000first}
\begin{align}
    \mathrm{Tr}_{\omega iL \sigma}=-\frac{1}{\pi} \int\limits_{-\infty }^{\infty}d\varepsilon\, f(\varepsilon)\, \Im\, \mathrm{Tr}_{iL \sigma} \,.
\label{Tr1-Matsubara-DFT}
\end{align}

\subsection{Perturbation to first order}

One may directly conclude that whenever the lowest order in Eq.~\eqref{Lloyd} is not vanishing it will dominate providing torques on some of the local moments. Therefore we first analyse the first order term, which can be described as a sum of one-site rotations, i.e.
\begin{equation}
\delta N^\mathrm{one}=\frac{1}{\pi}\Im\, \mathrm{Tr}_{iL \sigma}\, \delta V\,G =\sum_{i} \delta N^\mathrm{one}_{i}
\label{dNfirstdefi}
\end{equation}
where the site local variation of the number of states is
\begin{align}
	\delta N^\mathrm{one}_{i}&=
-\frac{1}{\pi}\Im\,  \,\delta \vec{e}_i\cdot \mathrm{Tr}_{L \sigma}\,{B}_i\,\vec{\sigma}\,G_{ii} =\nonumber\\
&=-\frac{2}{\pi}\Im\,  \,\delta \vec{e}_i\cdot \mathrm{Tr}_{L} \,{B}_i\,\vec{G}_{ii}\,,\label{1st-order}
\end{align}
and the factor 2 arises from the trace over spin variables. The grand canonical potential variation ($\delta \Omega^\mathrm{one}_i$) due to one-site rotation (Fig.~\ref{figone}) is based on the expression $\delta N^\mathrm{one}_{i}$ (given by Eq.~\ref{1st-order}) and the details of the derivation will be presented in Subsection~\ref{seconesitederiv}. Note that Fig.~\ref{figone} shows a collinear (ferromagnetic) case. However, Eq.~\eqref{1st-order} also holds for cases when the rotation, $\delta \vec{e}_{i}$, appear in a non-collinear background of atomic moments.

\subsection{A sum rule}
\label{generalsumrule}

While $\delta N^\mathrm{one}_{i}$ (and therefore $\delta \Omega^\mathrm{one}_i$) can be obtained by direct calculation based only on onsite quantities as shown by Eq.~\eqref{1st-order}, we prefer to deepen the analysis by taking an algebraic step which allows us to express this first order term as a bilinear intersite magnetic interaction which eases the understanding of these magnetic interactions.
Since the local Green functions arise from a self-consistent solution of a magnetically ordered state one can derive an explicit expression for it in the following way. One may consider a solution as obtained from a well defined non-magnetic system with a Hamiltonian in the form of the right hand side of Eq.~\eqref{firstKS}, more precisely, $V({\bf r})=V^{\mathrm{nm}}_0$. In this case ${\vec{B}}_{xc}({\bf r})={\vec{B}}_{ext}({\bf r})=0$. Note that the non-spin-polarized potential $V^{\mathrm{nm}}_0$ for this non-magnetic state in general will not be equivalent to the corresponding spin-independent part of the potential, $V_0$, for a magnetic state. The Green function of the magnetic state is related to the Green function of the non-magnetic state, $G_{\mathrm{nm}}$, through Dyson's equation as follows (omitting for simplicity the energy argument)
\begin{align}
G&=G_{\mathrm{nm}}+G_{\mathrm{nm}}\,\Delta V\,G
\end{align}
or
\begin{align}
\left(G^{-1}-G^{-1}_{\mathrm{nm}}\right)_{ij}&=\left\{V^{\mathrm{nm}}_0-V_0+\vec{B}_j\cdot\vec{\sigma}\right\}\delta_{ij}\,,\label{MS} 
\end{align}
where the spin polarized fields can be written $\vec{B}_j=B_j\,\vec{e}_j$ and $\Delta V=V_0-\vec{B}_j\cdot\vec{\sigma}-V^{\mathrm{nm}}_0$.

In order to arrive at a suitable expression one makes use of the fact that this magnetic state has to be degenerate with the corresponding time reversed state, i.e.,~the state with all moments switched and the direction of a charge current is reversed.
The Green function for this time reversed problem, $\widetilde{G}$, is given by
\begin{align}
	\left(\widetilde{G}^{-1}-G_{\mathrm{nm}}^{-1}\right)_{ij}=\left\{V^{\mathrm{nm}}_0-V_0-\vec{B}_j\cdot\vec{\sigma}\right\}\delta_{ij}\,.\label{MS-TR} 
\end{align}
The difference between Eqs.~\eqref{MS} and~\eqref{MS-TR} gives
\begin{align}
	\left({G}^{-1}-\widetilde{G}^{-1}\right)_{ij}=2\vec{B}_j\cdot\vec{\sigma}\,\delta_{ij}\,.\label{MS-diff}
\end{align}
By letting $\widetilde{G}$ and $G$ act on Eq.~\eqref{MS-diff} from either side in a symmetric fashion, we arrive at a sum rule for the local Green functions
\begin{align}
	\widetilde{G}_{ii}-G_{ii}=\sum_j \left(G_{ij}\,\vec{B}_j\cdot\vec{\sigma}\,\widetilde{G}_{ji}+\widetilde{G}_{ij}\,\vec{B}_j\cdot\vec{\sigma}\,{G}_{ji}\right)\,.\label{TR-diff}
\end{align}

\subsection{Further decomposition of Green function and their physical interpretation}
\label{decomp}

To be able to utilize the relation of Eq.~\eqref{TR-diff} one can further decompose the components of the Green function in Eq.~\eqref{GF-s} into terms that are either even or odd under space reversal~\cite{fransson2017microscopic}. This can be done by introducing $G_{ij}^{\eta \kappa}$ where the first index $\eta$ has been introduced and explained after Eq.~\eqref{GF-s} while the second index $\kappa$ can be viewed as an indicator whether the terms that are space reversal invariant and those are not, i.e, $\kappa$ can be $0$ or $1$. This decomposition of the Green function can be summarised as,
\begin{equation}
G^\eta_{ij}=G^{\eta 0}_{ij}+G^{\eta 1}_{ij} 
\label{GF-st}
\end{equation}
where $G^{00}_{ij}$ and $\vec{G}^1_{ij}$  are time reversal invariant while $G^{01}_{ij}$ and $\vec{G}^0_{ij}$ are not. Sometimes it is  convenient to write the $x$, $y$ or $z$ dependent components of the Green function as vectors, i.e.~$\vec{G}^\kappa$. This decomposition also plays a useful role in how the Green function behaves under site exchange, since in a real local basis~\cite{fransson2017microscopic} we have the expression
\begin{align}
    G^{\eta \kappa}_{ij}=(-1)^{\kappa}{G^{\eta\kappa}_{ji}}^T \,.
    \label{etakappatrafo}
\end{align}
In fact, it has been shown~\cite{fransson2017microscopic} that these two-index Green functions are decomposed in terms that produce local charge-, $G^{00}$ or spin-densities, $\vec{G}^{0}$, and charge-, $G^{01}$, and spin-currents $\vec{G}^{1}$, respectively, an aspect we will come back to.

We can express both the Green function and its time reversed version as a superposition of these two index decomposed Green functions, as
\begin{align}
{G}&=G^{00}+G^{01}+\vec{G}^0\cdot\vec{\sigma}+\vec{G}^1\cdot\vec{\sigma}\\
\widetilde{G}&=G^{00}-G^{01}-\vec{G}^0\cdot\vec{\sigma}+\vec{G}^1\cdot\vec{\sigma} \,.
\label{tauGF-st}
\end{align}
These decomposed Green functions are then inserted in Eq.~\eqref{TR-diff}, which leads to the expression
\begin{widetext}
\begin{align}
	\widetilde{G}_{ii}-G_{ii}=\sum_j& \left\{
	\left(G^{00}+G^{01}+\vec{G}^0\cdot\vec{\sigma}+\vec{G}^1\cdot\vec{\sigma}\right)_{ij}\,\vec{B}_j\cdot\vec{\sigma}\,\left(G^{00}-G^{01}-\vec{G}^0\cdot\vec{\sigma}+\vec{G}^1\cdot\vec{\sigma}\right)_{ji}\right.\nonumber\\
	+&\left.\left(G^{00}-G^{01}-\vec{G}^0\cdot\vec{\sigma}+\vec{G}^1\cdot\vec{\sigma}\right)_{ij}\,\vec{B}_j\cdot\vec{\sigma}\,\left(G^{00}+G^{01}+\vec{G}^0\cdot\vec{\sigma}+\vec{G}^1\cdot\vec{\sigma}\right)_{ji}\right\} \,,
	\label{TR-diff-explicit}
\end{align}
which for the spin dependent and time reversal odd part, $\vec{G}^0_{ii}$, of $\frac{1}{2}(G_{ii}-\widetilde{G}_{ii})=G^{01}_{ii}+\vec{G}^0_{ii}\cdot\vec{\sigma}$
 allows us to identify the expression \cite{cardias2020dzyaloshinskii}
\begin{align}
	\vec{G}^0_{ii}
	=&-\sum_j\,\Bigg\{\left(G^{00}_{ij}\,\vec{B}_j\,G^{00}_{ji}-G^{01}_{ij}\,\vec{B}_j\,G^{01}_{ji}\right) 
	+i\left(\vec{G}^{1}_{ij}\times\vec{B}_j\,G^{00}_{ji}+G^{00}_{ij}\,\vec{B}_j\times\vec{G}^{1}_{ji}
	\right)
	-i\left(\vec{G}^{0}_{ij}\times\vec{B}_j\,G^{01}_{ji}+G^{01}_{ij}\,\vec{B}_j\times\vec{G}^{0}_{ji}
	\right) \notag\\
	&+\left(\vec{G}^{1}_{ij}\cdot\vec{B}_j\,\vec{G}^{1}_{ji}-\vec{G}^{0}_{ij}\cdot\vec{B}_j\,\vec{G}^{0}_{ji}\right)  
	-\left(
	\left(\vec{G}^{1}_{ij}\times\vec{B}_j\right)\times\vec{G}^{1}_{ji}
	-\left(\vec{G}^{0}_{ij}\times\vec{B}_j\right)\times\vec{G}^{0}_{ji}
	\right) \Bigg\}.
	\label{spin-sumrule-diff}
\end{align}
\end{widetext}
Note that the expression in Eq. (\ref{spin-sumrule-diff}) is general, despite that we arrived at it from considerations of the Green function of its normal and spin reversed state. Hence Eq. (\ref{spin-sumrule-diff}) can be used also for small angle rotations of moments\footnote{Note that $\Delta V$ that we consider in Subsections \ref{generalsumrule} and \ref{decomp} is not the same as $\delta V$ that stands for perturbations due to different kind of infinitesimally small spin rotations in the rest of Subsection \ref{detailsLKAG}.}, which will be utilized in Subsection \ref{seconesitederiv}.  

To give a physical interpretation for the charge- and spin-densities and charge- and spin-currents, it is useful to study the decomposition of the Green function in real space , $G(\mathbf{r},\mathbf{r}';\varepsilon)$, into eight independent two indexed contributions, i.e., to consider the expression
\begin{align}
	G(\mathbf{r},\mathbf{r}';\varepsilon)&=\sum_{\eta\in\{0,x,y,z\}}\sum_{\kappa=0}^1 \sigma_\eta\,G^{\eta\kappa}(\mathbf{r},\mathbf{r}';\varepsilon)\,,\label{rsGF}
\end{align}
where $\sigma_0$ represents the identity matrix. Note that the second index, $\kappa$, of the Green function in Eq.~\eqref{rsGF}, indicates whether the function is even (0) or odd (1) under the exchange of spatial coordinates ($\mathbf{r}\leftrightarrow \mathbf{r}'$) 
\begin{align}
	G^{\eta\kappa}(\mathbf{r}',\mathbf{r};\varepsilon)=(-1)^\kappa\,G^{\eta\kappa}(\mathbf{r},\mathbf{r}';\varepsilon)\,,\label{rs-asym}
\end{align}
where the $\kappa$ decomposition is defined through
\begin{align}
    G^{\kappa}(\mathbf{r},\mathbf{r}';\varepsilon)=\frac{G(\mathbf{r},\mathbf{r}';\varepsilon)+(-1)^\kappa G(\mathbf{r}',\mathbf{r};\varepsilon)}{2}.
\end{align}
The meaning of this two-index decomposition can be summarised in the equation
 \begin{align}
	G(\mathbf{r},\mathbf{r}';\varepsilon)&=\sum_{\eta\in\{0,x,y,z\}}\sum_{\kappa=0}^1\,(-1)^\kappa \sigma_\eta\,G^{\eta\kappa}(\mathbf{r}',\mathbf{r};\varepsilon)\,.\label{rsGF-rev}
\end{align}
The four different Green function, two scalar and two vector valued, as discussed above, all have a direct physical property as in the local and non-relativistic limit they give rise to charge and spin density (scalar and vector) and charge and spin current densities (vector and tensor), respectively, through the formulas
\begin{align}
	n(\mathbf{r})=&-\frac{1}{\pi}\Im \lim_{\mathbf{r}'\rightarrow\mathbf{r}}\int G^{00}(\mathbf{r},\mathbf{r}';\varepsilon)\,\mathrm{d}\varepsilon\label{n-CD}\\
	\vec{m}(\mathbf{r})=&-\frac{1}{\pi}\Im \lim_{\mathbf{r}'\rightarrow\mathbf{r}}\int \vec{G}^{0}(\mathbf{r},\mathbf{r}';\varepsilon)\,\mathrm{d}\varepsilon\label{m-SD}\\
	\mathbf{j}(\mathbf{r})=&-\frac{1}{\pi}\Re \lim_{\mathbf{r}'\rightarrow\mathbf{r}}\int \boldmath{\nabla} G^{01}(\mathbf{r},\mathbf{r}';\varepsilon)\,\mathrm{d}\varepsilon\label{j-CC}\\
	\vec{\mathbf{q}}(\mathbf{r})=&-\frac{1}{\pi}\Re \lim_{\mathbf{r}'\rightarrow\mathbf{r}} \int \boldmath{\nabla} \vec{G}^{1}(\mathbf{r},\mathbf{r}';\varepsilon)\,\mathrm{d}\varepsilon\label{q-SC}\,.
\end{align}
These four independent density quantities are important in the case of magnetic materials and is known to appear in many other approaches such as in general Hartree-Fock theory~\cite{Fukutome1981unrestricted}.

Expanding the Green function represented in real space in a local basis (Eq.~\eqref{greenij}) results in the expression
\begin{align}
	G^{\eta\kappa}(\mathbf{r},\mathbf{r}';\varepsilon)&=
	\phi_i^T(\mathbf{r})\,G_{ij}^{\eta\kappa}\,\phi_j(\mathbf{r}'),\label{grr2gij} \,
\end{align}
where space is divided into regions around the atomic sites, such that the site $i$ is specified by the position $\mathbf{r}$ and the vector of basis functions $\phi_i(\mathbf{r})$ is uniquely defined. Then the condition of Eq.~\eqref{rs-asym} leads to the relationship
\begin{align}
G^{\eta\kappa}(\mathbf{r}',\mathbf{r};\varepsilon) 
&= \phi_j^T(\mathbf{r}')\,G_{ji}^{\eta\kappa}\,\phi_i(\mathbf{r}) \notag\\
&=(-1)^\kappa\,\phi_i^T(\mathbf{r})\,G_{ij}^{\eta\kappa}\,\phi_j(\mathbf{r}') \notag\\
&=(-1)^\kappa\,\left(\phi_j(\mathbf{r}')^T\,\left\{G_{ij}^{\eta\kappa}\right\}^T\phi_i(\mathbf{r})\right)^T \notag \\ &=(-1)^\kappa\,\phi_j(\mathbf{r}')^T\,\left\{G_{ij}^{\eta\kappa}\right\}^T\phi_i(\mathbf{r})\,,
\end{align}
where the outer transpose  is superfluous since it is acting on a scalar.
This leads to the relation for the Green function matrices expanded in a real basis shown in Eq.~\eqref{etakappatrafo}, which illustrates that the decomposed Green functions that stem from currents, $\kappa=1$, are asymmetric in the direction of the propagation in contrast to those that stem from densities, $\kappa=0$.

\subsection{Bilinear interaction parameters due to one-site spin rotation}
\label{seconesitederiv}

One can generally express the variation of the grand potential as a series of  contributions coming from different orders of perturbation, as
\begin{align}
    \delta \Omega= \delta \Omega^{\mathrm{one}}+\delta \Omega^{\mathrm{two}}+\ldots\,. 
\end{align}
It is relevant to express both these two first terms in the series in terms of bilinear interaction parameters. In the case of one-site spin rotation one then has to express the one-site grand potential variation, $\delta \Omega^\mathrm{one}$, in terms of intersite Green functions, which corresponds to inserting Eq.~\eqref{1st-order} into Eq.~\eqref{mft} where Eq.~\eqref{1st-order} is given with the onsite Green function, $\vec{G}_{ii}$. However, it is only the time reversal odd spin-dependent Green function, $\vec{G}^{0}_{ii}$, that will give rise to a non-zero product $\mathrm{Tr}_{L}\,B_i\vec{G}_{ii}$ in Eq.~\eqref{1st-order}, where $\vec{G}^{0}_{ii}$ in turn can be expressed in terms of intersite Green functions due to the sum rule of Eq.~\eqref{spin-sumrule-diff}. Hence, one can express the first order term as a superposition of different pair interactions, using the expression
\begin{align}
& \delta \Omega^\mathrm{one}=	\frac{2}{\pi}\Im\, \sum_i \,\delta \vec{e}_i\cdot \int \mathrm{Tr}_{L}\,{B}_i\,\vec{G}^{0}_{ii}(\varepsilon)\,\mathrm{d}\varepsilon =\nonumber\\
&= 2 \sum_{<ij>} \delta\vec{e}_i\cdot\mathbb{J}^{(1)}_{ij}\cdot\vec{e}_j +\delta\Omega^\mathrm{one}_\mathrm{loc} \,,
\label{1st-bilinear}
\end{align}
where the tensor $\mathbb{J}^{(1)}_{ij}$ has the same form as given by Eqs. (\ref{Htens})-(\ref{TensComps}) with the exchange parameter $J^{(1)}_{ij}$, the DM vector $\vec{D}^{(1}_{ij}$, and the symmetric anisotropic interaction tensor $\mathbb{A}^{1}_{ij}$. Note that comparing Eq. (\ref{1st-bilinear}) to Eq. (\ref{1stHamT}) allows us to identify the exchange parameter $J^{(1)}_{ij}$ from the expression
\begin{align}
	J^{(1)}_{ij} = &-\frac{2}{\pi}\Im\int \mathrm{Tr}_{L }\Bigl(B_i\, G^{00}_{ij}\,B_j\,G^{00}_{ji}-B_i\, G^{01}_{ij}\,B_j\,G^{01}_{ji} \notag\\
	&+\sum_\nu B_i\,{G}^{\nu 0}_{ij}\,B_j\,{G}^{\nu 0}_{ji}-\sum_\nu B_i\, {G}^{\nu 1}_{ij}\,B_j\,{G}^{\nu 1}_{ji}\Bigr)\,\mathrm{d}\varepsilon,\label{defJ}
\end{align}
while the components of the vector $\vec{D}^{(1)}_{ij}$ and the tensor $\mathbb{A}^{(1)}_{ij}$ are given from the equation
\begin{equation}
	{D}_{ij}^{(1)\nu}=-\frac{4}{\pi}\Re\int \mathrm{Tr}_{L }\left(B_i G^{00}_{ij} B_j {G}^{\nu1}_{ji}-B_i  G^{01}_{ij} B_j\,{G}^{\nu0}_{ji}\right) \mathrm{d}\varepsilon
	\label{defD}
\end{equation}
and
\begin{equation}
  A_{ij}^{(1)\nu\mu}=-\frac{4}{\pi}\Im \int \mathrm{Tr}_{L }\left(B_i {G}^{\nu 1}_{ij} B_j {G}^{\mu 1}_{ji}-B_i {G}^{\nu 0}_{ij} B_j {G}^{\mu 0}_{ji}  \right)\mathrm{d}\varepsilon\,,\label{defA}
\end{equation}
respectively, where $\mu$ and $\nu$ can be $x$, $y$ or $z$. Note that the index (1) in $J^{(1)}_{ij}$, $\vec{D}^{(1)}_{ij}$ and $\mathbb{A}^{(1)}_{ij}$ refers to the fact that these parameters are derived from \textit{one}-site spin rotation. We also note that the prefactor 2 in the second line in Eq. \ref{1st-bilinear} arises for the same reason as why the one-site term enters twice in Eq.~\eqref{Omtwo}. The second term in Eq. \ref{1st-bilinear} is given by the formula
\begin{align}
	\delta\Omega^\mathrm{one}_\mathrm{loc}=
 \sum_{i} \delta\vec{e}_i\cdot\mathbb{J}^{(1)}_{ii}\cdot\vec{e}_i,
\end{align}
which may play roles for the magnetic anisotropy \cite{PhysRevB.52.13419} or the longitudinal exchange couplings \cite{PhysRevB.72.104437}, i.e., we have arrived at a more general model, beyond what is usually considered in bilinear spin models, as~Eq.~\eqref{1stHamT}. However, in the collinear--non-relativistic limit, this term can be shown to be cancelled by a similar local term in the second order interaction \cite{liechtenstein1987local}, which we will come back to in Subsection~\ref{seccomp}. Nevertheless, considering the intersite terms in the second line in Eq. (\ref{1st-bilinear}), a local mapping can always be made with the Heisenberg exchange parameter $J_{ij}=J^{(1)}_{ij}$, the DM vector $\vec{D}_{ij}=\vec{D}^{(1}_{ij}$ and the symmetric anisotropic interaction tensor $\mathbb{A}_{ij}=\mathbb{A}^{(1)}_{ij}$, around the magnetic order of the reference state.

\subsection{Bilinear interaction parameters due to two-site spin rotations}

Whenever the first order term vanishes the second order perturbation plays a role. This is for instance the case for a collinear state in the absence of SOC where the first order contribution is identically zero. This term might also be of importance when one aims to calculate collective excitations, i.e.~spin waves, in linear spin wave theory, where the spin Hamiltonian has to be bilinear in the variations of the magnetic moments \cite{Toth_2015}. The second order term in Eq.~\eqref{Lloyd} can be written as $\delta N^\mathrm{two}$ analogously to $\delta N^\mathrm{one}$ given by Eq. (\ref{dNfirstdefi}). Inserting $\delta N^\mathrm{two}$ into Eq.~\eqref{mft} leads to the grand potential variation $\delta \Omega^\mathrm{two}$, which corresponds to simultaneous rotations at site $i$ and $j$ as illustrated in Fig.~\ref{figtwo} and is naturally bilinear. This term also contains a local term, $\delta\Omega^\mathrm{two}_\mathrm{loc}$, that again we will ignore. Then one can obtain that
\begin{align}
& \delta \Omega^\mathrm{two}-\delta\Omega^\mathrm{two}_\mathrm{loc}=	 \nonumber\\
&=-\frac{1}{2\pi} \sum_{<ij>}\,\int \Im\,\mathrm{Tr}_{L \sigma}\, \delta \vec{e}_i \cdot \vec{\sigma} \,B_i\, G_{ij} \, \delta \vec{e}_j \cdot \vec{\sigma}\, B_j\, G_{ji}\,\mathrm{d}\varepsilon \,.
\label{deltaomegatwo}
\end{align}
Note that Fig.~\ref{figtwo} shows a collinear (ferromagnetic) case, but Eq.~\eqref{deltaomegatwo} also holds for the general non-collinear case. One should note that Eq.~\eqref{deltaomegatwo} can be simplified in a similar fashion as the first order contributions; first decompose the Green functions and then sum out the spin degrees of freedom after manipulating the matrix product by means of Pauli algebra. A comparison with Eq.~\eqref{2ndHamT} leads to the expression
\begin{equation}
 \delta \Omega^\mathrm{two}-\delta\Omega^\mathrm{two}_\mathrm{loc}= \sum_{<ij>}\delta \vec{e}_i\cdot\mathbb{J}^{(2)}_{ij}\cdot \delta \vec{e}_j,
 \label{2nd-bilinear}
\end{equation} 
where $J^{(2)}_{ij}$ is defined as
\begin{align}
	J^{(2)}_{ij} = &-\frac{{2}}{\pi}\Im\int \mathrm{Tr}_{L }\Bigl(B_i\, G^{00}_{ij}\,B_j\,G^{00}_{ji}+B_i\, G^{01}_{ij}\,B_j\,G^{01}_{ji} \notag\\
	&-\sum_\nu B_i\,{G}^{\nu 0}_{ij}\,B_j\,{G}^{\nu 0}_{ji}-\sum_\nu B_i\, {G}^{\nu 1}_{ij}\,B_j\,{G}^{\nu 1}_{ji}\Bigr)\,\mathrm{d}\varepsilon\,,
	\label{defJ2}
\end{align}
while the components of $\vec{D}^{(2)}_{ij}$ and $\mathbb{A}^{(2)}_{ij}$ are given from the expression
\begin{equation}
	{D}_{ij}^{(2)\nu}=-\frac{4}{\pi}\Re\int \mathrm{Tr}_{L }\left(B_i\, G^{00}_{ij}\,B_j\,{G}^{\nu1}_{ji}+B_i\, G^{01}_{ij}\,B_j\,{G}^{\nu0}_{ji}\right)\mathrm{d}\varepsilon
	\label{defD2}
\end{equation}
and
\begin{equation}
  A_{ij}^{(2)\nu\mu}=-\frac{4}{\pi}\Im \int \mathrm{Tr}_{L }\left(B_i{G}^{\nu 1}_{ij}B_j{G}^{\mu 1}_{ji}+B_i{G}^{\nu 0}_{ij}B_j{G}^{\mu 0}_{ji}  \right)\mathrm{d}\varepsilon\,,
  \label{defA2}
\end{equation}
respectively, where the superscript (2) in $J^{(2)}_{ij}$, $\vec{D}^{(2)}_{ij}$ and $\mathbb{A}^{(2)}_{ij}$ refers to the fact that these parameters are derived from \textit{two}-site spin rotations. This is an alternative mapping since $\mathbb{J}^{(2)}_{ij} \neq \mathbb{J}^{(1)}_{ij}$, i.e., the mapping procedures based on the one- and two-site spin rotations lead to different results in general. Their comparison and physical interpretations are discussed in Subsection \ref{seccomp}.

\subsection{Explicit symmetric or asymmetric interactions}

With a relation in hand for the decomposed Green function we observe that the interactions are explicitly determined as symmetric or asymmetric.
For example for the Dzyaloshinskii-Moryia interaction of Eq.~\eqref{defD} we can, since the trace of the transpose of a matrix is equal to the trace of the matrix and the fact that the trace of a product is invariant under cyclic permutation of the factors, derive its asymmetric property explicitly due to the property of Eq.~\eqref{etakappatrafo} as follows,
\begin{align}
{D}_{ij}^{(1)\nu} &= -\frac{4}{\pi}\Re\int \mathrm{Tr}_{L }\left(B_i G^{00}_{ij}B_j{G}^{\nu1}_{ji}-B_i G^{01}_{ij}B_j{G}^{\nu0}_{ji}\right)^T\mathrm{d}\varepsilon \nonumber\\
&=\frac{4}{\pi}\Re\int \mathrm{Tr}_{L }\left(B_j {G}^{\nu1}_{ji} B_i  G^{00}_{ij}-B_j {G}^{\nu0}_{ji} B_i  G^{01}_{ij}\right)\mathrm{d}\varepsilon \nonumber\\
&=-{D}_{ji}^{(1)\nu}\,.
\end{align}
In general we can conclude that pair interaction terms that include an even number of asymmetric Green functions become symmetric, while those that include an odd number are asymmetric.
Then it is clear that it is only the Dzyaloshinskii-Moriya interaction that is asymmetric among the bilinear interactions of Eqs.~\eqref{defJ}--\eqref{defA}. Note that the argumentation presented here holds for ${D}_{ij}^{(2)\nu}$ (see Eq.~\eqref{defD2}) as well.

\subsection{Comparison the interaction parameters obtained from one- and two-site variations}
\label{seccomp}

Since we have reformulated the first order interactions in a bilinear form (Eq.~\eqref{defJ}), it can be directly compared with the second order interactions that are naturally bilinear (Eq.~\eqref{defJ2}). There are clear differences in the two expressions, which might not be very surprising as they reflect different quantities, the first order interaction describes the local torques on the magnetic moments while the second order interaction mainly describe the interaction of rotated moments. 
However, in the LKAG limit, i.e.~with collinear order and negligible spin-orbit coupling, it has been observed \cite{liechtenstein1984exchange} that they actually give rise to identical interaction parameters.
When studying the details of this limit it turns out that 
this is slightly fortuitous.  The mapping to the spin Hamiltonian $\mathcal{H}_{T}$ based on one-site spin rotation, resulted in the exchange parameters $J^{(1)}_{ij}$, $\vec{D}^{(1)}$ and $\mathbb{A}^{(1)}$ while a similar mapping based on two-site spin variations led to the parameters $J^{(2)}_{ij}$, $\vec{D}^{(2)}$ and $\mathbb{A}^{(2)}$. In the LKAG limit there is no spin and charge current present and we can choose a global coordinate system in which the non-perturbed spin-arrangement will all point to the $z$ direction and one has the freedom to restrict the small rotations to the $xz$ plane. 
 
Let us start with the exchange parameters obtained from two-site variations. In this case we can see that the $J^{(2)}_{ij}$ parameter, defined by Eq.~\eqref{defJ2}, is reduced to the expression
\begin{equation}
	J^{(2)}_{ij}=-\frac{2}{\pi}\Im\int \mathrm{Tr}_{L }\Bigl(B_i\, G^{00}_{ij}\,B_j\,G^{00}_{ji}- B_i\,{G}^{z 0}_{ij}\,B_j\,{G}^{z 0}_{ji}\Bigr)\,\mathrm{d}\varepsilon \,. \label{defJ2LKAG}
\end{equation}
In the LKAG limit, $\vec{D}^{(2)}$, defined by Eq.~\eqref{defD2}, vanishes and the symmetric anisotropic interaction tensor $\mathbb{A}^{(2)}$, defined by Eq.~\eqref{defA2}, will only have one  non-vanishing component, $A_{ij}^{(2)zz}$ with a collinear magnetic order along the $z$-direction. However since the variation $\delta{e}_i^z=-(\delta\theta_{i})^2/2$ is quadratic in the small rotation angle $\delta \theta_i$ this term gives a variation of fourth and not second  order in the rotation angles
\footnote{When $\delta\theta_{i}=-\delta\theta_{j}= \delta \theta$ then 
$\delta{e}_i^z \delta{e}_j^z$ is proportional to $\left( \delta \theta \right)^4$.}. 
This means that only the first Heisenberg term of Eq.~\eqref{2nd-bilinear} is relevant, i.e.~of second order in the variation angle. We introduce the notation $G^{\uparrow}_{ij}=G^{00}_{ij}+G^{z0}_{ij}$ and $G^{\downarrow}_{ij}=G^{00}_{ij}-G^{z0}_{ij}$, then the LKAG exchange expression will be given in its well-known form:
\begin{equation}
    J^{(2)}_{ij}=-\frac{2}{\pi}\Im\int \mathrm{Tr}_{L}\Bigl(B_i\, G^{\uparrow}_{ij}\,B_j\,G^{\downarrow}_{ji} \Bigr)\,\mathrm{d}\varepsilon\,.
\label{LKAGnewest}
\end{equation}
We note here that substituting Eq.~\eqref{green} into Eq.~\eqref{LKAGnewest} and integrating over energy one arrives at an expression which is equivalent to Eq.~\eqref{J}~\cite{antropov1997exchange}. We also note that the leading term in the corresponding variation in the grand potential becomes
\begin{align}
	\delta\Omega^\mathrm{two}\approx \frac{1}{2}\sum_{ij}J^{(2)}_{ij}  \delta \theta_i \delta \theta_j \,,
 \label{twoforallsites}
\end{align}
i.e., only the onsite $i$-$i$ term will have a factor 1/2 and the intersite terms will be given as shown in Eq. (\ref{three}).

Next, we focus on the parameters obtained from one-site variation in the absence of SOC. For a collinear order along $z$ one only has to deal with the component $\delta e^{z}_{i}$ of the variation and the second line in Eq. (\ref{1st-bilinear}) is reduced to the expression 
\begin{align}
\delta\Omega^\mathrm{one}\approx -\sum_{ij} \left( J^{(1)}_{ij}+A_{ij}^{(1)zz} \right) {\left(\delta \theta_i \right)^{2}}/{2}.
\end{align} 
In the non-relativistic limit a global spin rotation, i.e.~all $\delta \theta_i=\delta \theta$, is always a symmetry operation, which is now seen to appear as a non-trivial cancellation of the first and second order interactions from the consideration that
\begin{align}
	& \delta\Omega=\delta\Omega^\mathrm{one}+\delta\Omega^\mathrm{two}+\ldots\approx\nonumber\\
 &\approx-\frac{1}{2}\sum_{ij} \left( J^{(1)}_{ij}+A_{ij}^{(1)zz} -J^{(2)}_{ij}\right)\left(\delta \theta\right)^{2} =0,
\end{align}
that are justified by inspection of 
Eqs.~\eqref{1st-bilinear} and \eqref{defJ2LKAG}, considering the vanishing intersite Green functions $\vec{G}^1=G^{01}=0$ in the LKAG limit. Another case when there is a cancellation between first and second order is the case of rotation of the moment at a single site $i=0$. Then we note that in the LKAG limit the sum over all intersite exchange parameters, $J_0=\sum_{<0i>}J_{0i}^\mathrm{(2)}$, is determined by
a cancellation \cite{liechtenstein1987local} of $\delta\Omega^\mathrm{one}_\mathrm{loc}$ and $\delta\Omega^\mathrm{two}_\mathrm{loc}$ in the total variation of the grand potential resulting in the expression\footnote{Note that the expressions in the second and the third line of Eq. (\ref{J0defnewest}) are proportional to $-\,{(\delta\theta_0)^2}$. This leads to a non-trivial expression for $J_{0}$ depending exclusively on onsite Green functions, which can be utilized in testing of code implementations.}
\begin{align}
\delta\Omega^\mathrm{one}&+\delta\Omega^\mathrm{two}_\mathrm{loc}=\nonumber\\
=-&\frac{2}{\pi}\Im\int \mathrm{Tr}_{L}\Bigl(B_0 \frac{G^{\uparrow}_{00}-G^{\downarrow}_{00}}{2}\,\frac{(\delta\theta_0)^2}{2}+\nonumber\\
&+\frac{1}{2} B_0\, G^{\uparrow}_{00}\,B_0\,G^{\downarrow}_{00} \,(\delta\theta_0)^2 \Bigr)\,\mathrm{d}\varepsilon=\nonumber\\
=&-\sum_{<0i>}2\left(J_{0i}^\mathrm{(1)}+A_{0i}^{(1)zz}\right)\,\frac{(\delta\theta_0)^2}{2}=\nonumber\\
=&-\sum_{<0i>}J_{0i}^\mathrm{(2)}\,{(\delta\theta_0)^2}=-J_0\,{(\delta\theta_0)^2}
\,.
\label{J0defnewest}    
\end{align}

It is also worth noting that for a collinear state with SOC included, the symmetric interactions still vanish in  first order while the asymmetric DM interaction will be finite in absence of inversion symmetry. This non-vanishing torque leads to instabilities of collinear order, e.g.~ferromagnetic states that are unstable towards cycloidal order \cite{PhysRevB.96.104416} or anti-ferromagnetic order that are unstable towards tilting which might give rise to a weak ferromagnetic order \cite{solovyev1996crucial,PhysRevB.71.184434}.

Finally, we note the components of the DM vectors $\vec{D}^{(1)}_{ij}$ and $\vec{D}^{(2)}_{ij}$ are sums of two independent terms. Both the terms are mediated by products of Green functions such that one factor is density related and the other current related, as indicated by the 0 and 1 site exchange symmetry indices $\kappa$ defined in \ref{etakappatrafo}, in Eqs.~\eqref{defD}--\eqref{defD2}. This implies that for the trivial topology with collinear spin arrangment the DM term will vanish in absence of SOC as then the current contributions are prohibited, while for general non-collinear order these interactions are non-vanishing even in absence of SOC. For the second term of the symmetric anisotropic interaction parameters $\mathbb{A}^{(1)}_{ij}$ and $\mathbb{A}^{(2)}_{ij}$ defined by Eqs.~\eqref{defA}--\eqref{defA2} is mediated by density related spin polarised Green functions that exist for all magnetic order even in absence of SOC.This term was investigated and discussed as an anisotropy anomaly in Ref.~\onlinecite{PhysRevB.82.180404}. 

We end this subsection with a comment for practical reasons, we will in Section~\ref{examples} give numerical examples of exchange interactions that are mostly based on the equations obtained from the two-site energy variations. An exception is for the results in Fig.~\ref{turz}, where the first derivative of the grand potential with respect to angle is shown.

\subsection{Local versus global spin models}
\label{Sec-5K-LocalGlobal}

We make a comment here on the distinction between local and global spin models proposed by Ref.~\onlinecite{streib2021exchange}. Here we have focused on spin models that are obtained within a generalization of the LKAG approach. This approach is still based on the fact that there is a perturbation that consists of small rotations of local moments in an already magnetic reference state. The generalization of LKAG is that the magnetic state is now allowed to have any non-collinear order and that 
relativistic effects, i.e.~mainly spin-orbit coupling, are included, but only in a weak enough limit such that the local moments are still well defined as spin moments. In such an approach the reference state will incorporate composed Green functions, $\vec{G}^{0}$, $\vec{G}^{1}$ and $G^{01}$, that are directly dependent on the magnetic order. 
Hence the mapped spin model is only valid locally on the energy vs.~configuration curve, i.e.~it is only relevant for small magnetic variations around the reference state. This is in contrast to the concept of global spin models that are supposed to be valid for all magnetically ordered states and the full curve of energy vs.~configuration.

The fact that the models are local implies that they do not have to fulfill global symmetry requirements. A magnetic state dependence of the interaction coefficients arises naturally for local models due to their dependence on the reference state~\cite{cardias2020dzyaloshinskii,streib2021exchange}. If the state dependence is taken into account for a local spin model, all global symmetries are of course recovered. 

One way to avoid the reference state dependence is to start with a non-magnetic reference state for which the Green functions of course are independent of any magnetic state. In this approach \cite{PhysRevResearch.2.033240, brinker2019chiral} there will be extensions of the formulas beyond bilinear interactions, which involve e.g.~biquadratic effects with coupling terms like $\sum_{<ij>} {\mathcal J}^{BQ}_{ij} ({\vec{S}}_i \cdot {\vec{S}}_j)^2$ and generalizations of it, i.e., 
$\sum_{<ijkl>} {\mathcal J}^{Ring}_{ijkl} ({\vec{S}}_i \cdot {\vec{S}}_j)({\vec{S}}_k \cdot {\vec{S}}_l)$, where $i,j,k,l$ are site indices. In such an approach the perturbations are proportional to the full spin dependent potentials and these larger perturbations in the series of Eq.~\eqref{Lloyd} will in general be slowly convergent so higher orders play a role.

These two approaches, with non-magnetic respectively magnetic reference states, are in a sense complimentary. While one approach includes the effects in terms of multi-spin interactions \cite{drautz2004spin, PhysRevB.101.174401} the other approach includes the same effects within the composite Green functions mediating the interaction. The first approach will have a large validity range, in favorable cases maybe even global, but will be less accurate for any given magnetic state, while the second approach can calculate the interaction parameters accurately for any general magnetic order but only for one local magnetic state at a time.

It is also important to realize that the existence of global spin models for itinerant-electron systems
is not guaranteed, since there is no way to prove that the magnetic degrees of freedom
can be globally described by a Hamiltonian dependent solely on spin operators. 
At the same time, at least for small frequencies and small wave-vectors, any ferromagnetic system should be described by the macroscopic Landau-Lifshitz
equation \cite{akhiezer1968spin,vonsovsky1974magnetism,aharoni2000introduction}.
This means that at least the expression for the spin-wave stiffness constant,
based on small variations from the ferromagnetic ground state, is always
meaningful \cite{liechtenstein1984exchange}. Moreover, within the local approximations such as dynamical 
mean-field theory (see Subsection \ref{sec:5K}) the expression
for the spin-wave stiffness constant derived from magnetic force theorem
becomes exact \cite{lichtenstein2001magnetism}.

\subsection{Exchange interactions in correlated system}
\label{sec:5K}

In order to calculate the effective exchange interaction parameters for correlated
magnetic systems the Dynamical Mean-Field Theory (DMFT) approach has been explored, with a local frequency dependent self energy. The implementation of DMFT into DFT-based first-principle calculations~\cite{PhysRevB.57.6884, kotliar-DMFT} is based on the mapping to multiband Hubbard-like model.
It assumes knowledge of effective parameters characterising local Coulomb interactions (the problem of Hubbard U).
The state-of-the-art way includes taking into account screening effects via the so-called constrained Random Phase Approximation (c-RPA)~\cite{Ferdi_CRPA_2004}. Within this approach no arbitrary parameters are introduced and calculations remain fully first-principles. Note that in the early days of this method, U was frequently used as a fitting parameter. The historical 
developments  of the method and its relations to the previous LDA+U formalism can be found in Ref.\onlinecite{kotliar-DMFT}.
What is important here is the statement that in principle DFT and DMFT can be combined in a fully ab-initio way.
The remaining questions on the applicability of c-RPA for realistic situations was recently analysed in details in Ref.~\onlinecite{Wehling_CRPA_2021}

First of all, let us prove the analog of the local force theorem in the DMFT-like theory~\cite{katsnelson2000first}.
Instead of working with
the thermodynamic potential $\Omega $ as a {\it density} functional we have
to start from its general expression in terms of an exact Green function~\cite{luttinger1960ground,kotliar-DMFT}, i.e. 
\begin{align}
\Omega &= \Omega _{sp}-\Omega _{dc} \notag\\
\Omega _{sp} &= -\mathrm{Tr}\,\left\{ \ln \left[ \Sigma -G_0^{-1}\right] \right\} 
\nonumber \\
\Omega _{dc} &= \mathrm{Tr}\,\Sigma G-\Phi
\label{first}
\end{align}
where $G,G_0$ and $\Sigma $ are an exact Green function, its bare value and
self-energy, correspondingly; $\Phi $ is the Luttinger generating functional
(sum of the all connected skeleton diagrams without free legs), $%
\mathrm{Tr}=\mathrm{Tr}_{\omega iL\sigma }$ is the sum over Matsubara frequencies $\mathrm{Tr}_\omega
...=T\sum\limits_\omega ...,$ $\omega =\pi T\left( 2n+1\right) ,$ $n=0,\pm
1,...,$ and $T$ is the temperature. Furthermore, $iL\sigma $ are site numbers ($i$),
orbital quantum numbers ($L={l,m}$) and spin projections $\sigma $,
correspondingly. Both Green functions are related via the Dyson equation: 
\begin{equation}
G^{-1}=G_0^{-1}-\Sigma  \label{DYSON}
\end{equation}
with the important variational identity
\begin{equation}
\delta \Phi =\mathrm{Tr}\,\Sigma \delta G.  \label{var}
\end{equation}
We represent the expression Eq.~\eqref{first} as a difference of ``single
particle'' ($sp$) and ``double counted'' ($dc$) terms as it is usual in the
density functional theory. When neglecting the quasiparticle damping, $%
\Omega _{sp}$ is nothing but the thermodynamic potential of ''free''
fermions but with exact quasiparticle energies. Suppose we change the
external potential, for example, by small spin rotations. Then the variation
of the thermodynamic potential can be written as 
\begin{equation}
\delta \Omega =\delta ^{*}\Omega _{sp}+\delta _1\Omega _{sp}-\delta \Omega
_{dc} , \label{var2}
\end{equation}
where $\delta ^{*}$ is the variation without taking into account the change
of the ``self-consistent potential'' (i.e. self energy) and $\delta _1$ is
the variation due to this change of $\Sigma $. Taking into account Eq.~\eqref{var} it can be easily shown (cf. Ref.~\onlinecite{luttinger1960ground,kotliar-DMFT}) that one may identify the expression
\begin{equation}
\delta _1\Omega _{sp}=\delta \Omega _{dc}=\mathrm{Tr}G\delta \Sigma  \label{var3}
\end{equation}
and hence 
\begin{equation}
\delta \Omega =\delta ^{*}\Omega _{sp}=-\delta ^{*}\mathrm{Tr}\,\ln \left[ \Sigma
-G_0^{-1}\right],  \label{var4}
\end{equation}
which is an analog of the ``local force theorem'' in the density functional 
theory~\cite{AndersenSkriverNohlJohansson+1980+93+118,mackintosh1980electrons, methfessel1982bond, liechtenstein1984exchange}.

In the DMFT scheme, the self energy is local, i.e., it is diagonal in site
indices. Let us write the spin-matrix structure of the self energy and Green
function in the following form 
\begin{align}
\Sigma_i &= \Sigma _i^c+{\vec \Sigma}_i^s{\vec { \sigma }} \notag\\
G_{ij} &= G_{ij}^c+{\vec G}_{ij}^s{\vec { \sigma }}
\label{spin}
\end{align}
where $\Sigma _i^{\left( c,s\right) }=\frac 12\left( \Sigma _i^{\uparrow
}\pm \Sigma _i^{\downarrow }\right)$, ${\vec \Sigma}_i^s=\Sigma _i^s{\vec e}_i,
$ with ${\vec e}_i$ being the unit vector in the direction of effective
spin-dependent potential on site $i$ and in the local moment approximation not depending on frequency (discussed furter in Section~\ref{Section9}), $G_{ij}^c=\frac 12\mathrm{Tr}_\sigma (G_{ij})$ and $%
{\vec G}_{ij}^s=\frac 12\mathrm{Tr}_\sigma (G_{ij} {\vec {\sigma}})$.

Then following the general idea of infinitesimal rotation of local magnetic potential/self-energy the effective exchange interactions in correlated magnetic system can be obtained by
rewriting all equations in this section with a substitution of $\Sigma_i^s$ for $B_i$, leading to the expression~\cite{katsnelson2000first}
\begin{equation}
J_{ij}= 2\mathrm{Tr}_{\omega L}\left( \Sigma _i^sG_{ij}^{\uparrow }\Sigma
_j^sG_{ji}^{\downarrow }\right) ,  \label{Jij}
\end{equation}
to be compared with Eq.~\eqref{Tr1-Matsubara-DFT}.
In the strong coupling limit for half filled Hubbard model this expression reduced to the standard Anderson kinetic exchange $t_{ij}^2/U$~\cite{stepanov2021spin}

\section{Beyond kinetic exchange}
\label{beyond_kinetic_exchange}

Let us now return to a general discussion of exchange interactions within the,
formally rigorous, scheme of time-dependent density functional presented in
Section~\ref{linearresp}. In this approach, the whole dynamics of the many-electron system is
described in terms of the time-dependent
one-particle density matrix $\rho _{\alpha \beta }\left( \mathbf{r,r,}%
t\right) =\left\langle \eta _{\beta }^{+}\left( \mathbf{r},t\right) \eta
_{\alpha }\left( \mathbf{r},t\right) \right\rangle ,$ where $\eta _{\alpha
}\left( \mathbf{r},t\right) $ is the annihilation operator for the electron
at the point $\mathbf{r}$ with spin projection $\alpha$ at the instant time
$t$. Equivalently, one can introduce the charge $n\left( \mathbf{r,}t\right)
=\mathrm{Tr}_{L \sigma}\rho \left( \mathbf{r,r},t\right) $ and magnetization $\vec{m} \left( \mathbf{r,}t\right)=\mathrm{Tr}_{L \sigma}\rho \left( \mathbf{r,r},t\right) \vec{\sigma}$ densities (also obtained in the time-independent case from Eqs.~\eqref{magn} and~\eqref{magnOE1}). In the adiabatic approximation, the spin and charge densities are expressed
in terms of Kohn-Sham spinor eigenfuctions $\psi _{\nu \alpha }\left( \mathbf{r,}%
t\right)$ and the corresponding eigenenergies $\epsilon _{\nu }\left( t\right)$
satisfying the Kohn-Sham equation of Eqs.~\eqref{schroed} and~\eqref{sham}.
The Kohn-Sham wave functions and the
corresponding energies depend here on time due
to the time-dependence of the densities and external field (the latter is
supposed to be slowly varying in time in comparison to the characteristic electron energies).

Very importantly, even in the local-density approximation there is a nonlocality
in the kinetic term in the total density functional, via nonlocality of the kinetic-energy
term $T[\hat{\rho]}$, due to nonlocality of Kohn-Sham states. 
The total effective magnetic field can be represented as
\begin{equation}
\vec{B}_{tot}(\mathbf{r})
=-\frac{\delta T}{\delta \vec{m}(\mathbf{r})}-\frac{\delta E_{xc}}{\delta
\vec{m}(\mathbf{r})}+\vec{B}_{ext}(\mathbf{r}),  \label{c345}
\end{equation}%
and the first term in the r.h.s. of Eq.~\eqref{c345} depends on $\vec{m}(\mathbf{r'})$
at $\mathbf{r'} \neq \mathbf{r}$ even if the exchange-correlation term $\vec{B}_{xc}(\mathbf{r})$ is
local. This leads to exchange interactions, i.e., a connection between magnetization
direction in different points of space. In this sense, exchange
parameters discussed until now all correspond to kinetic, or indirect, exchange. Note that despite GW-approach formally goes beyond locality it deals with the nonlocality in charge density only and not in spin density. This means that within GW theory one has only kinetic  exchange as well.

As will be discussed in Section~\ref{examples}, the whole experience of calculations of exchange
parameters via the LKAG formula, or its extensions, is that for many classes of systems it reproduces experimental data with good accuracy. This means that in most of the cases indirect, that is, a kinetic
contribution, to exchange interactions is dominant. There is nevertheless a natural question as to what is exactly neglected in this approach~\cite{katsnelson2003spin}. To answer
this question one needs to go beyond the local spin-density approximation and study
the nonlocality of $\vec{B}_{xc}[\vec{m}]$.

There are many works on a general analysis of noncollinear magnetism within density
functional without local spin-density approximation~\cite{kleinman1999density,capelle2001spin,capelle2003exploring,katsnelson2003spin,sharma2007first,peralta2007noncollinear,scalmani2012new,eich2013transverse,eich2013transverse1,bulik2013noncollinear,ullrich2018density,sharma2018source, kubler2017theory, kubler1988density, nordstrom1996noncollinear,heine1983magnetic}. Here we focus only on one aspect of this activity, namely, the applicability of the local spin-density approximation to the calculations of exchange parameters. To study this issue we need to investigate the origin of nonlocality in the exchange-correlation functionals.

At the construction of the local spin-density approximation, one starts with the calculation of the exchange-correlation energy for a homogeneous electron gas, from a given charge and spin density. A natural step in studying its nonlocality is to replace this reference system by the simplest nonuniform state, namely, the electron gas in a spin-spiral state. This approach was suggested by Kleinman~\cite{kleinman1999density} at the level of the Fock approximation and by Katsnelson and Antropov~\cite{katsnelson2003spin} at the level of the random phase approximation (RPA). The latter was developed further and used in electronic structure calculations, e.g. as published in Refs.~\onlinecite{sharma2007first,peralta2007noncollinear,scalmani2012new,eich2013transverse,eich2013transverse1,bulik2013noncollinear,ullrich2018density,sharma2018source}. To illustrate the basic idea and some simple estimations we will follow here the presentation of Ref.~\onlinecite{katsnelson2003spin}.   

Let us consider a homogeneous electron gas in the spin-density-wave (SDW) state.
The latter is characterized by anomalous averages $s_{\mathbf{p}%
}=\left\langle c_{\mathbf{p}+\mathbf{Q}/2\uparrow }^{+}c_{\mathbf{p}-\mathbf{%
Q}/2\downarrow }\right\rangle $ where $c_{\mathbf{p}\sigma }^{+}$,$c_{%
\mathbf{p}\sigma }$ are the creation and annihilation operators of
electrons with momentum $\mathbf{p}$ and spin projection $\sigma$. To consider a spin-density wave, it is convenient to use a spinor representation of the creation and annihilation operators, similar to the Gorkov-Nambu formalism in the theory of superconductivity~\cite{schrieffer_book,kurmaev_book}.
To this end, we introduce the spinor operator $\eta _{%
\mathbf{p}}=(c_{\mathbf{p}+\mathbf{Q}/2\uparrow }^{+},c_{\mathbf{p}-\mathbf{Q%
}/2\downarrow })$. Then the Hamiltonian of the homogeneous electron gas takes the form
\begin{equation}
H=\sum_{\mathbf{p}}\eta _{\mathbf{p}}h_{\mathbf{p}}\eta _{\mathbf{p}}+\frac{1%
}{2}\sum_{\mathbf{q\neq }0}\sum_{\mathbf{pp}^{\prime }}v_{c}\left( \mathbf{q}%
\right) \left( \eta _{\mathbf{p+q}}^{+}\eta_{\mathbf{p}}\right) \left( \eta
_{\mathbf{p}^{\prime }\mathbf{-q}}^{+}\eta _{\mathbf{p}^{\prime }}\right),
\label{A3}
\end{equation}%
where $v_{c}\left( \mathbf{q}\right) =4\pi e^{2}/\mathbf{q}^{2}V,$ $V$ is a
volume, $h_{\mathbf{p}}=\theta _{\mathbf{p}}+\tau _{\mathbf{p}}\sigma
_{z}-\Delta _{\mathbf{p}}\sigma _{x}$ and
\begin{align}
\theta _{\mathbf{p}}& =\frac{1}{2}\left( \varepsilon _{\mathbf{p}+\mathbf{Q}%
/2}+\varepsilon _{\mathbf{p}-\mathbf{Q}/2}\right) =\mathbf{p}^{2}/2+\mathbf{Q%
}^{2}/8-\mu ,  \notag\\
\tau _{\mathbf{p}}& =\frac{1}{2}\left( \varepsilon _{\mathbf{p}+\mathbf{Q}%
/2}-\varepsilon _{\mathbf{p}-\mathbf{Q}/2}\right) =\mathbf{pQ}/2,
\label{A5}
\end{align}%
where $\varepsilon _{\mathbf{p}}=\mathbf{p}^{2}/2-\mu $ is the energy of the
free electron and $2\Delta _{\mathbf{p}}$ is the antiferromagnetic gap,
related to the formation of the spin-density wave. Note that in this Section we use the units
$\hbar = m =1$. In the Fock approximation the gap equals to
\begin{equation}
\Delta _{\mathbf{p}}=\sum_{\mathbf{p}^{\prime }}v_{c}(\mathbf{p}-\mathbf{p}%
^{\prime })s_{\mathbf{p}^{\prime }}.  \label{A6}
\end{equation}%
To simplify the consideration as much as possible one can replace $v_{c}$ by and effective Stoner exchange splitting $%
I=\left( V_{exc}^{\uparrow }-V_{exc}^{\downarrow }\right) /\left(
n_{\uparrow }-n_{\downarrow }\right) ,$ where $V_{exc}^{\sigma }=\partial
\left( n\varepsilon _{exc}\right) /\partial n_{\sigma }.$ Then, Eq.~\eqref{A6}
can be replaced by\ $\Delta =I\left( n_{\uparrow }-n_{\downarrow }\right) /2$%
, where $\Delta $ does not depend on $\mathbf{p}.$

To calculate the correlation contribution to the energy of the homogeneous electron gas one can restrict oneself to the simplest meaningful approximation, namely, the RPA corresponding to the summation of all ``bubble'' diagrams~\cite{mahan_book,vignale_book}.
The ``bare'' Green function in the Matsubara representation has the form
\begin{equation}
G\left( i\omega _{m},\mathbf{p}\right) =\frac{1}{i\omega _{m}-h_{\mathbf{p}}}%
=\frac{i\omega _{m}-\theta _{\mathbf{p}}+\tau _{\mathbf{p}}\sigma
_{z}-\Delta _{\mathbf{p}}\sigma _{x}}{\left( i\omega _{m}-\xi _{\mathbf{p}%
_{\uparrow }}\right) \left( i\omega _{m}-\xi _{\mathbf{p}_{\downarrow
}}\right) },  \label{A9}
\end{equation}%
where $\xi _{\mathbf{p}\uparrow ,\downarrow }=\theta _{\mathbf{p}}\mp E_{%
\mathbf{p}}$ is a quasiparticle spectrum for SDW with $E_{\mathbf{p}}=\sqrt{%
\tau _{\mathbf{p}}^{2}+\Delta ^{2}}$. From Eq.~\eqref{A9} one can find the
occupation number matrix
\begin{equation}
2N_{\mathbf{p}}=\left( 1+\frac{\tau _{\mathbf{p}}\sigma _{z}-\Delta \sigma
_{x}}{E_{\mathbf{p}}}\right) f_{\mathbf{p}\uparrow }+\left( 1-\frac{\tau _{%
\mathbf{p}}\sigma _{z}-\Delta \sigma _{x}}{E_{\mathbf{p}}}\right) f_{\mathbf{%
p}\downarrow },  \label{A10}
\end{equation}%
where $f_{\mathbf{p}\sigma }=f\left( \xi _{\mathbf{p}\sigma }\right) $ is a
Fermi function. Then for the Fock contribution to the exchange-correlation
energy one has
\begin{align}
E_{Fock}&=-\frac{1}{2}\sum_{\mathbf{pp}^{\prime }}v_{c}\left( \mathbf{p}-%
\mathbf{p}^{\prime }\right) Tr\left[ N(\mathbf{p})N(\mathbf{p}^{\prime })%
\right] \notag\\ 
&=E_{Fock}^{(1)}+E_{Fock}^{(2)}\,,  
\label{A11}
\end{align}
where
\begin{align}
E_{Fock}^{(1)}& =-\frac{1}{4}\sum_{\mathbf{pp}^{\prime }\sigma }v_{c}\left(
\mathbf{p}-\mathbf{p}^{\prime }\right) f_{\mathbf{p}\sigma }f_{\mathbf{p}%
^{\prime }\sigma }\left( 1+\frac{\tau _{\mathbf{p}}\tau _{\mathbf{p}^{\prime
}}+\Delta ^{2}}{E_{\mathbf{p}}E_{\mathbf{p}^{\prime }}}\right) , \notag \\
E_{Fock}^{(2)}& =-\frac{1}{2}\sum_{\mathbf{pp}^{\prime }}v_{c}\left( \mathbf{%
p}-\mathbf{p}^{\prime }\right) f_{\mathbf{p}\uparrow }f_{\mathbf{p}^{\prime
}\downarrow }\left( 1-\frac{\tau _{\mathbf{p}}\tau _{\mathbf{p}^{\prime
}}+\Delta ^{2}}{E_{\mathbf{p}}E_{\mathbf{p}^{\prime }}}\right) .  
\end{align}%
Further, one may consider the case of small $\mathbf{Q}$ only, which is sufficient for the calculation of the contributions to the spin-wave stiffness constant. The RPA-based calculations without this restriction were first performed in Ref.~\onlinecite{eich2013transverse}. Expansion of Eq.~\eqref{A11} up to $\mathbf{Q}^2$ leads to the corrections of the chemical potential (from the conservation of the number of particles)
\begin{equation}
\delta \widetilde{\mu }=\widetilde{\mu }_{\mathbf{Q}}-\widetilde{\mu }_{%
\mathbf{Q}=0}=-\frac{\mathbf{Q}^{2}}{8F\left( n_{\uparrow },n_{\downarrow
}\right) }  \label{A13}
\end{equation}%
and to the total energy 
\begin{align}
&\frac{E_{Fock}}{V} = -\frac{e^{2}}{8\pi ^{3}} 
\Bigg\{ 
\left( p_{F\uparrow
}^{4}+p_{F\downarrow }^{4}\right) \notag\\
&- Q^{2}\left[ \left( \frac{1}{2F}-\frac{2}{3}\right) 
\left( p_{F\uparrow }^{2}+p_{F\downarrow }^{2}\right) 
+\frac{\left(p_{F\uparrow }+p_{F\downarrow }\right)^{2}}{12F^{2}} \right] 
\Bigg\}. 
\label{A14}
\end{align}
where $F=(p_{F\uparrow }+p_{F\downarrow })I(n_{\uparrow },n_{\downarrow
})/2\pi ^{2}$ is a dimensionless Stoner enhancement factor, $p_{F\sigma
}=(6\pi ^{2}n_{\sigma })^{1/3}.$

To treat the correlation effects, one may use RPA and sum up the bubble diagrams~\cite{mahan_book,vignale_book}. The corresponding expression is expressed in terms of the empty-loop polarization operator
\begin{equation}
\Pi \left( i\omega ,\mathbf{q}\right) =-Tr\sum_{\mathbf{p}%
}T\sum_{\varepsilon _{n}}G\left( \mathbf{p}+\mathbf{q},i\varepsilon
_{n}+i\omega _{n}\right) G\left( \mathbf{p},i\varepsilon _{n}\right) .
\label{A15}
\end{equation}%
The corresponding contribution to the $\Omega$-potential equals
\begin{eqnarray}
\Omega_{corr}&=&\sum_{\mathbf{q}}\int\limits_{-\infty }^{\infty }\frac{%
d\omega }{4\pi }
\Bigg\{ 
\ln \left[ \frac{1+v_{c}\left( \mathbf{q}\right) \Pi
\left( i\omega ,\mathbf{q}\right) }{1+v_{c}\left( \mathbf{q}\right) \Pi _{%
\mathbf{Q}=0}\left( i\omega ,\mathbf{q}\right) }\right] \notag  \\
&-&v_{c}\left( \mathbf{%
q}\right) \left[ \Pi \left( i\omega ,\mathbf{q}\right) -\Pi _{\mathbf{Q}%
=0}\left( i\omega ,\mathbf{q}\right) \right] 
\Bigg\} 
,  \label{A16}
\end{eqnarray}
where only $\mathbf{Q}$-dependent part of the correlation energy was considered.
Substituting Eq.~\eqref{A9} into Eq.~\eqref{A15} one finds
\begin{align}
&\Pi \left( i\omega ,\mathbf{q}\right) =\frac{1}{2}\sum_{\mathbf{p},\sigma}\left(
1+\frac{\tau _{\mathbf{p}}\tau _{\mathbf{p}+\mathbf{q}}+\Delta ^{2}}{E_{%
\mathbf{p}}E_{\mathbf{p}+\mathbf{q}}}\right) \frac{f_{\mathbf{p%
}\sigma }-f_{\mathbf{p}+\mathbf{q}\sigma }}{i\omega +\xi _{\mathbf{p}%
\mathbf{q}\sigma }-\xi _{\mathbf{p}\sigma }}    \notag \\
&+2\sum_{\bf p}\left( 1-\frac{\tau _{\mathbf{p}}\tau _{\mathbf{p}+\mathbf{q}}+\Delta
^{2}}{E_{\mathbf{p}}E_{\mathbf{p}+\mathbf{q}}}\right) \frac{f_{\mathbf{p}%
\uparrow }-f_{\mathbf{p}+\mathbf{q}\downarrow }}{i\omega +\xi _{\mathbf{p}+%
\mathbf{q}\downarrow }-\xi _{\mathbf{p}\uparrow }}. 
\label{A17}
\end{align}
The corresponding exchange-correlation addition to the spin wave
spectrum at finite $\mathbf{Q}$ can be written as
\begin{equation}
\delta \omega _{\mathbf{Q}}=\frac{4}{M}\left[ E_{SDW}(\mathbf{Q})-E_{SDW}(0)%
\right],  \label{A23}
\end{equation}
where $E_{SDW}(\mathbf{Q})$ is the total energy of the spin spiral and $M$
is the magnetic moment of the unit cell.

The next step is to restore the expression of the exchange-correlation functional
corresponding to Eq.~\eqref{A23}. The simplest rotational invariant expression 
has the form
\begin{equation}
E_{exc}=\int d\mathbf{r}\left\{ n\varepsilon _{exc}\left( n_{\uparrow
},n_{\downarrow }\right) +\lambda \left( n_{\uparrow },n_{\downarrow
}\right) D\right\} ,  \label{C45}
\end{equation}%
where $D=\left( \nabla _{\alpha }e_{\beta }\right) \left( \nabla _{\alpha
}e_{\beta }\right) =\left( \nabla \theta \right) ^{2}+\sin ^{2}\theta \left(
\nabla \varphi \right) ^{2}$ is the rotational invariant of lowest order.
Here $\vec{e}=\vec{m}/\left\vert \vec{m}\right\vert \equiv \left(
\sin \theta \cos \varphi ,\sin \theta \sin \varphi ,\cos \theta \right) .$
More detailed analysis of the functional dependence in the general density
functional can be found in Refs.~\onlinecite{sharma2007first,scalmani2012new,eich2013transverse,eich2013transverse1,bulik2013noncollinear,ullrich2018density}.
Based on the analysis of Fock and RPA expressions for the total energy of the spin-spiral state, the following expression for $\lambda$ was suggested in Ref.~\onlinecite{katsnelson2003spin}:
\begin{align}
\lambda \left( n_{\uparrow },n_{\downarrow }\right) = &-\frac{e^{2}}{16\pi ^{2}%
}\left( \frac{1}{F}-\frac{4}{3}\right) \left( V_{exc}^{\uparrow
}p_{F\uparrow }-V_{exc}^{\downarrow }p_{F\downarrow }\right) \notag \\
&+ \frac{e^{2}}{%
96\pi ^{3}F^{2}}.  
\label{A18}
\end{align}

To evaluate the importance of the non-locality of the exchange-correlation functional for
the exchange parameters one can calculate the corresponding contribution to the
spin-wave stiffness constant, which can be expressed as
\begin{equation}
D=\frac{4}{M}\left[ \lim_{\mathbf{Q}\rightarrow 0}\frac{E_{SDW}(\mathbf{Q}%
)-E_{SDW}(0)}{\mathbf{Q}^{2}}\right] .  \label{A21}
\end{equation}
Namely, Eq.~\eqref{C45} gives:
\begin{equation}
\delta D=\frac{4}{M}\int dr\lambda \left( n_{\uparrow },n_{\downarrow
}\right) ,  \label{A22}
\end{equation}%
with integration over the whole elementary cell. 
The numerical calculations for the case of Fe and Ni, performed in Ref.~\onlinecite{katsnelson2003spin}, led to the following results: whereas the standard local-spin-density approximation gave the values 239 meV{\AA}$^{2}$ and 692 meV{\AA}$^{2}$ for $D$ in bcc Fe and fcc Ni, respectively, the corrections, Eq.~\eqref{A22}, for $\delta D$ were equal to 13 meV{\AA}$^{2}$ and 45 meV{\AA}$^{2}$, respectively. Hence, the total $D$ became 253 meV{\AA}$^{2}$ and 735 meV{\AA}$%
^{2}$ for bcc Fe and fcc Ni, respectively. Thus, for these materials, that serve as important systems for testing theoretical models, the indirect (kinetic) contributions are much larger than the direct contributions from the non-locality of the exchange-correlation functional.

In the model (e.g., tight-binding) approach direct exchange enters the Hamiltonian straightforwardly, via the matrix elements 
\begin{align}
J_{ij} &= \left\langle ij\left| v\right| ji\right\rangle \notag\\
&=\int d{\bf r}d{\bf r}%
^{\prime }\psi _{i}^{\ast }({\bf r})\psi_{j}^{\ast }({\bf r}^{\prime
})v\left( {\bf r-r}^{\prime }\right) \psi_{j}({\bf r})\psi _{i}({\bf r}%
^{\prime }),  
\label{coulomb}
\end{align}
where $v\left( {\bf r-r}^{\prime }\right)$ is the effective potential of electron-electron interaction (in the simplest approximation, just Coulomb interaction). In most of the cases, this contribution is supposed to be irrelevant but in some cases it is claimed that this interaction is important and can even change the calculated magnetic ground state (e.g., transform a spin-spiral state into a ferromagnet). Examples include single-side hydrogenated graphene~\cite{PhysRevB.94.214411} and half-metallic CrO$_2$~\cite{PhysRevB.92.144407}. The direct exchange interaction is also relevant in single-side fluorinated graphene~\cite{PhysRevB.94.214411} and fourth-group adatoms at the surface of Si(111)~\cite{PhysRevB.94.224418} and SiC(0001)~\cite{PhysRevB.98.184425}. Whereas sp-bonded magnets may be considered as an exotic exception, the example of CrO$_2$ demonstrates that the issue is not completely clear even for conventional 3d-electron magnets and requires a careful investigation. 

\section{Numerical examples of interatomic exchange}
\label{examples}

In this section we provide examples of numerical calculations of interatomic exchange interactions as well as magnetic moments, for several classes of materials.  Reviews of theoretical results of magnetic materials have been published before, albeit with different focus than the present article. However, it is noteworthy that in Refs.~\onlinecite{mohn2006magnetism, kubler2017theory, eriksson2017atomistic}, a comparison between experiment and theory regarding bulk magnetic moments was made, with some of the results shown in Fig.~\ref{fig2}. In general, DFT calculations reproduce experimental magnetic moments with an error that seldom exceeds 5\%, in particular for transition metal elements and their intermetallic compounds. Since reviews of magnetic moments have been published before, we focus in this section on results of the interatomic exchange. The work in Ref.~\onlinecite{kubler2017theory} also reviews results of interlayer exchange interactions of magnetic multilayers, as well as magnon dispersion from spin-spiral calculations~\cite{kubler1988density,Sandratskii_1991,PhysRevB.58.293,PhysRevB.66.134435,doi:10.1080/000187398243573,jakobsson2015first}. Results for thin films were reviewed in Ref.~\onlinecite{Etz_2015}, where magnon measurements based on spin-polarized electron energy loss spectroscopy (SPEELS) were compared to adiabatic magnon spectra evaluated from \textit{explicit} calculations of interatomic exchange. Finally, we note that in Ref.~\onlinecite{sato2010first} a full review was published of the magnetic properties, including \textit{explicit} calculations of interatomic exchange, of diluted magnetic semiconductors. We also note here that the most direct comparison between experiment and theory of interatomic exchange interactions is likely to be the magnon dispersion. This is in contrast to, e.g., estimates of the Curie temperature, that in principle also reflects the strength of the interatomic exchange. However, most of the DFT calculations of interatomic exchange are carried out at low (in fact zero) temperature, which challenges a comparison for results at finite temperature. If the exchange interaction was independent on temperature (or magnetic configuration), a comparison to experimental results at finite temperature, such as the ordering temperature, would be unproblematic. Although most materials do have interatomic exchange that depends on temperature, there have been progress also in calculations of configuration dependent exchange, and magnetic properties at finite temperature, as discussed in Section \ref{detailsLKAG}. Before entering details of materials specific results of interatomic exchange, we note that since we give examples from previously published works, there will be a mixture of units presented. In particular, energy is in some works given in eV and sometimes in Ry. We have however been consistent with the definition of the spin Hamiltonian introduced in Section \ref{intro}, which means that a negative value of the interatomic exchange corresponds to a ferromagnetic coupling.

Early implementations of the {\it explicit} method, i.e. Eq.~\eqref{LKAGnewest}, were incorporated in the linear muffin-tin orbital (LMTO)~\cite{PhysRevLett.53.2571} and Korringa-Kohn-Rostoker (KKR)~\cite{KORRINGA1947392,PhysRev.94.1111} electronic structure methods. Both approaches were first formulated either within the muffin-tin (MT) or the atomic sphere approximation (ASA), where the potential inside each sphere is assumed to be spherically symmetric.
For closed-packed systems this is a reasonable approximation and the results were consistent.
However, with the development of so-called full-potential electronic structure methods, which are free from geometrical constraints of the self-consistent density and potential, it quickly became clear that for more loosely-packed, or low-dimensional, systems, this level of approximation is needed.
There are several ab-initio implementations, using different basis functions, that employ a full-potential approach. However, it should be noted that the computationally much more efficient ASA calculations are still being pursued with good accuracy, especially for close packed systems. 

The biggest advantage of ASA-based codes is the compact representation of the basis functions, which are atom-centered and have a well-defined angular momentum character. 
This is very convenient for implementation of the magnetic force theorem, which operates with quantities, which have a site index $i$ attached (section \ref{detailsLKAG}).
In the full-potential codes the basis set is more extended and in general a minimal basis set is avoided.
In this case, the problem of defining a good representation of the local basis (see Eq.\ref{greenij}) becomes less obvious and in general it does not have a unique solution.
This issue sometimes hinders a proper quantitative comparison between the results obtained with various codes or even implementations within a given code. 

When one evaluates the interatomic exchange interaction between two atoms, the resulting values may depend on the choice of orbitals, which represent these atoms (see e.g. Ref.~\onlinecite{PhysRevB.97.125132,Han2004}).
This issue was discussed in detail in Ref.~\onlinecite{PhysRevB.91.125133,Herrmann2015}, where the comparison between the $J_{ij}$'s obtained with the projection on the muffin-tin sphere and L{\"o}wdin-orthogonalized orbitals were presented.
Overall, the results for fcc Ni and hcp Gd were very consistent, but in general it is found that depending on system there may be an unwanted sensitivity to the projection.
Moreover, strong covalent bonding between $3d$ and ligand states also calls for either perturbing the spins of the \textit{hybrid} orbitals or for explicit treatment of ligand spins as a standalone entity~\cite{Logemann-2017, PhysRevMaterials.2.073001, PhysRevB.103.104428}.

One commonly used choice is to use Wannier functions to obtain a localized basis for the $J_{ij}$ calculations~\cite{PhysRevB.88.081405, PhysRevB.91.224405, PhysRevMaterials.2.073001,PhysRevB.101.064401}.
In particular, maximally localized Wannier functions~\cite{RevModPhys.84.1419} form an appealing basis set, which is well-defined for a given set of bands and thus enables the comparison of the magnetic interactions obtained with different DFT codes.
There are a couple of versatile softwares, which allow one to apply the present formalism for an arbitrary tight-binding Hamiltonian independent of the chosen projection scheme~\cite{Jx-software,TB2J}.

Having these issues in mind, we now proceed with a discussion of calculated results of interatomic exchange for several classes of magnetic materials.

\subsection{Elemental transition metals}
\label{elementals}

\begin{figure}[t!]
\includegraphics[width=0.9\linewidth]{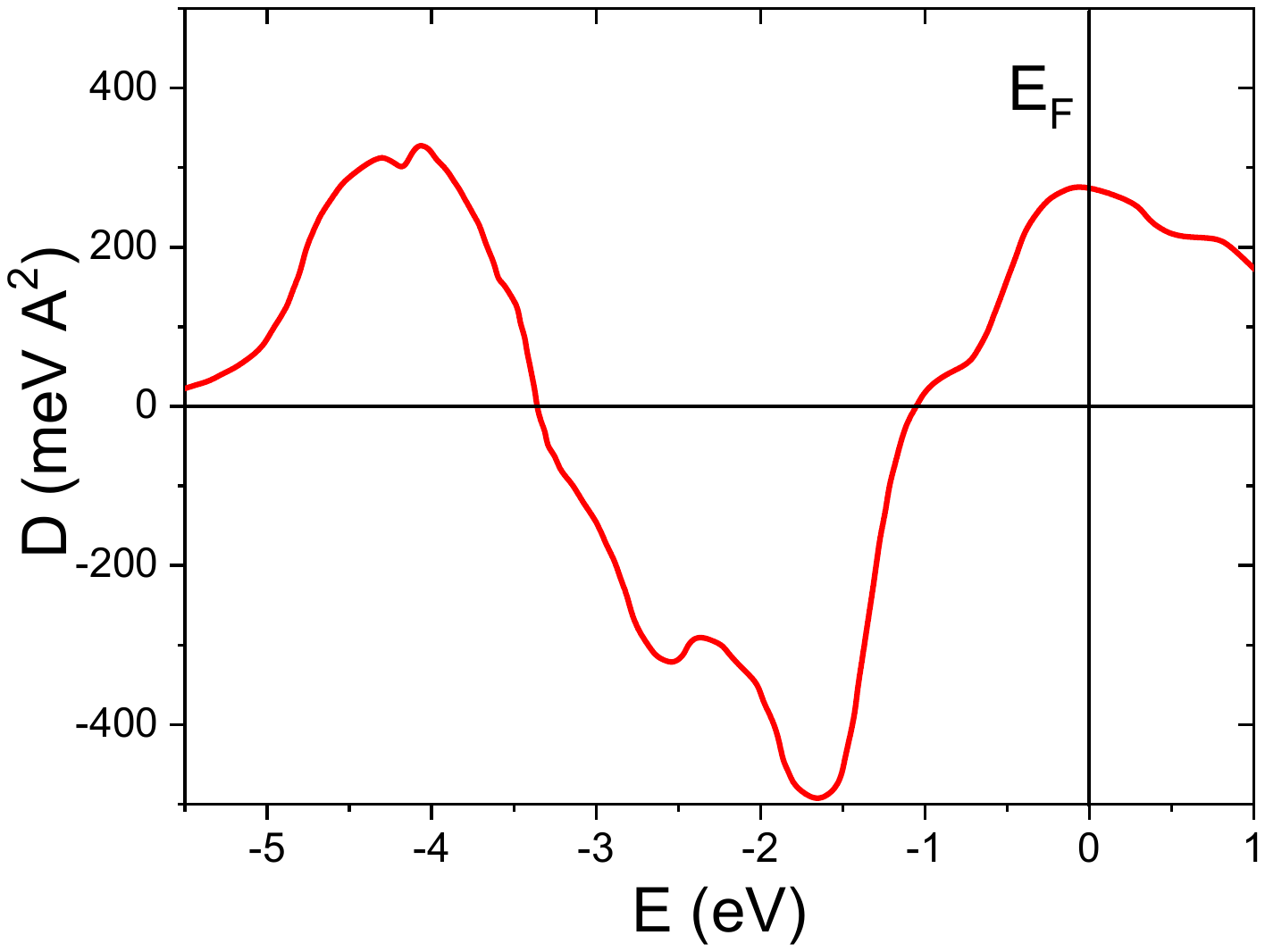}
\caption{(Color online) Spin-wave stiffness in bcc Fe as a function of the upper integration limit. Figure adopted from Ref.~\onlinecite{liechtenstein1984exchange}.}
\label{s7-fig1}
\end{figure}

One of the most important test cases for {\it explicit} calculations of interatomic exchange, is the ability to quantitatively reproduce magnetic properties such as spin-wave dispersion and ordering temperature of the three ferromagnetic 3d elements; bcc Fe, hcp Co and fcc Ni. 
The spin-wave stiffness, $D$, of bcc Fe was evaluated in the original articles of {\it explicit} calculations of interatomic exchange~\cite{liechtenstein1984exchange,liechtenstein1987local}.
In these works, the interaction between the first two coordination shells was calculated for bcc Fe. The dominant, nearest-neighbour (NN) coupling was found to be ferromagnetic (FM), while the next NN coupling was found to be antiferromagnetic, and much smaller.
The obtained value of the spin-wave stiffness, $D$, was 294 meV$\text{\AA}^2$ for bcc Fe, which is in good agreement with experimental values that range from 305 meV$\text{\AA}^2$ (Ref.~\onlinecite{PhysRevLett.44.1282}) to 314 meV$\text{\AA}^2$ (Ref.~\onlinecite{Stringfellow-1968}).
This initial result proved the formalism described in detail in Section~\ref{detailsLKAG} to be highly promising.
The formula for calculating the Heisenberg exchange, $J_{ij}$, also allowed the authors of Ref.~\onlinecite{liechtenstein1984exchange, liechtenstein1987local} to evaluate $D$ as a function of the upper integration limit, which can be viewed as the position of the Fermi level (see Fig.~\ref{s7-fig1}).
This provides valuable information of how $D$ can be affected by doping of the material.
In particular, as one changes the Fermi level to arrive at the half-filled $3d$-shell, at around -1 -- -3 eV in Fig.~\ref{s7-fig1}, the $D$ takes negative values, indicating that the FM reference state becomes unstable.

In Refs.~\onlinecite{liechtenstein1984exchange,liechtenstein1987local} it was argued that the NN exchange coupling primarily determines the value of spin-wave stiffness. 
The interactions with the neighbours beyond 2nd coordination shell were not computed, as their contribution to $D$ was expected to be negligibly small due to their oscillatory sign~\cite{oguchi1983magnetism}. 
However, later it was shown that the magnetic interactions in elemental transition metals are, in fact, extremely long-ranged~\cite{doi:10.1063/1.370495, ANTROPOV1999148, PhysRevB.62.5293} 
and obtaining a well-converged value of the spin-wave stiffness was indeed found to be extremely difficult~\cite{doi:10.1063/1.370495, ANTROPOV1999148}.

\begin{figure}[t!]
\includegraphics[width=0.85\linewidth]{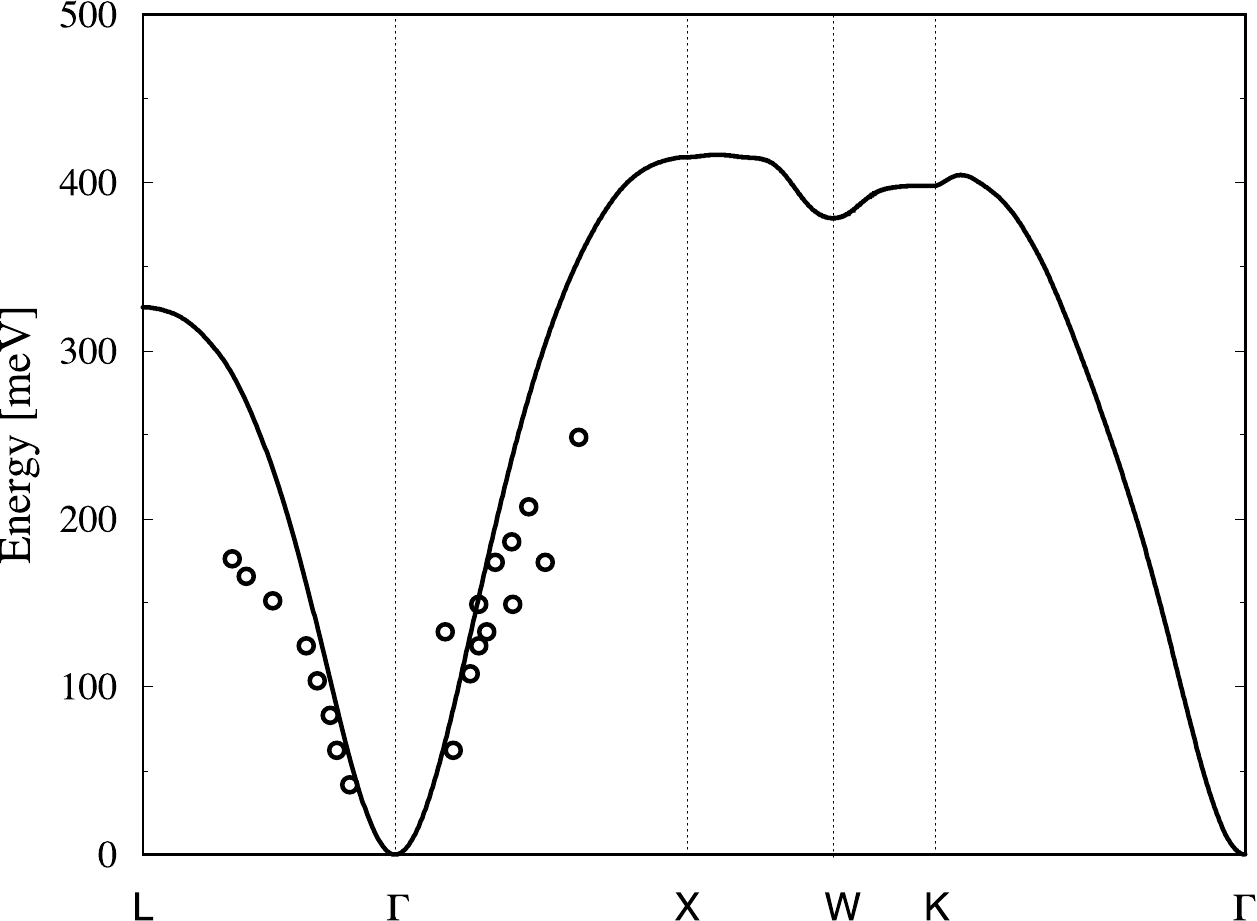}
\caption{Calculated spin-wave dispersion relation of fcc Ni from Ref.~\onlinecite{PhysRevB.64.174402}. Experimental data are taken from Ref.~\onlinecite{PhysRevLett.54.227}.}
\label{s7-fig2}
\end{figure}

Pajda and co-workers made a substantial advancement in that direction, by performing a thorough study of spin-waves and ordering temperatures, calculated from {\it explicit} values of $J_{ij}$, for bcc Fe, fcc Co and fcc Ni~\cite{PhysRevB.64.174402}. 
Their calculations were performed using a tight-binding LMTO method~\cite{PhysRevLett.53.2571}.
This work was done using the full set of valence states ($spd$ basis) and a very fine k-point mesh. For a magnetic material with one-atom per unit cell the spin-wave dispersion is governed by the exchange couplings, $J_{ij}$, in the following way:
\begin{eqnarray}
\omega(\mathbf{q})= \frac{4}{M} \sum_j J_{ij}(1-\exp{(i \mathbf{q} \cdot \mathbf{R}_{ij})}), 
\end{eqnarray}
where $M$ is the value of the saturated magnetic moment. 
Since the real space values of $J_{ij}$'s are involved, the summation has to be truncated.
The authors of Ref.~\onlinecite{PhysRevB.64.174402} considered interactions with first 195 and 172 shells for bcc and fcc metals, respectively, in order to ensure that the spin-wave dispersions are converged.
The obtained dispersion for fcc Ni is shown in Fig.~\ref{s7-fig2}.
The experimental data obtained with inelastic neutron scattering is also shown for comparison. Since the experimental spin waves become damped for higher values of $q$, it is only possible to compare experiments and theory in a region around the zone-center, and as Fig.~\ref{s7-fig2} shows, in this regime the agreement between theory and experiment is impressive. Results of similar accuracy were obtained from spin-spiral calculations~\cite{kubler2017theory}, and it is reassuring that DFT calculations of interatomic exchange obtained from different methods give similar results. In fact a direct comparison between the two methods was made for bcc Fe, with very similar results~\cite{bergqvist2005electronic}. 

\begin{figure}[t!]
\includegraphics[width=0.95\linewidth]{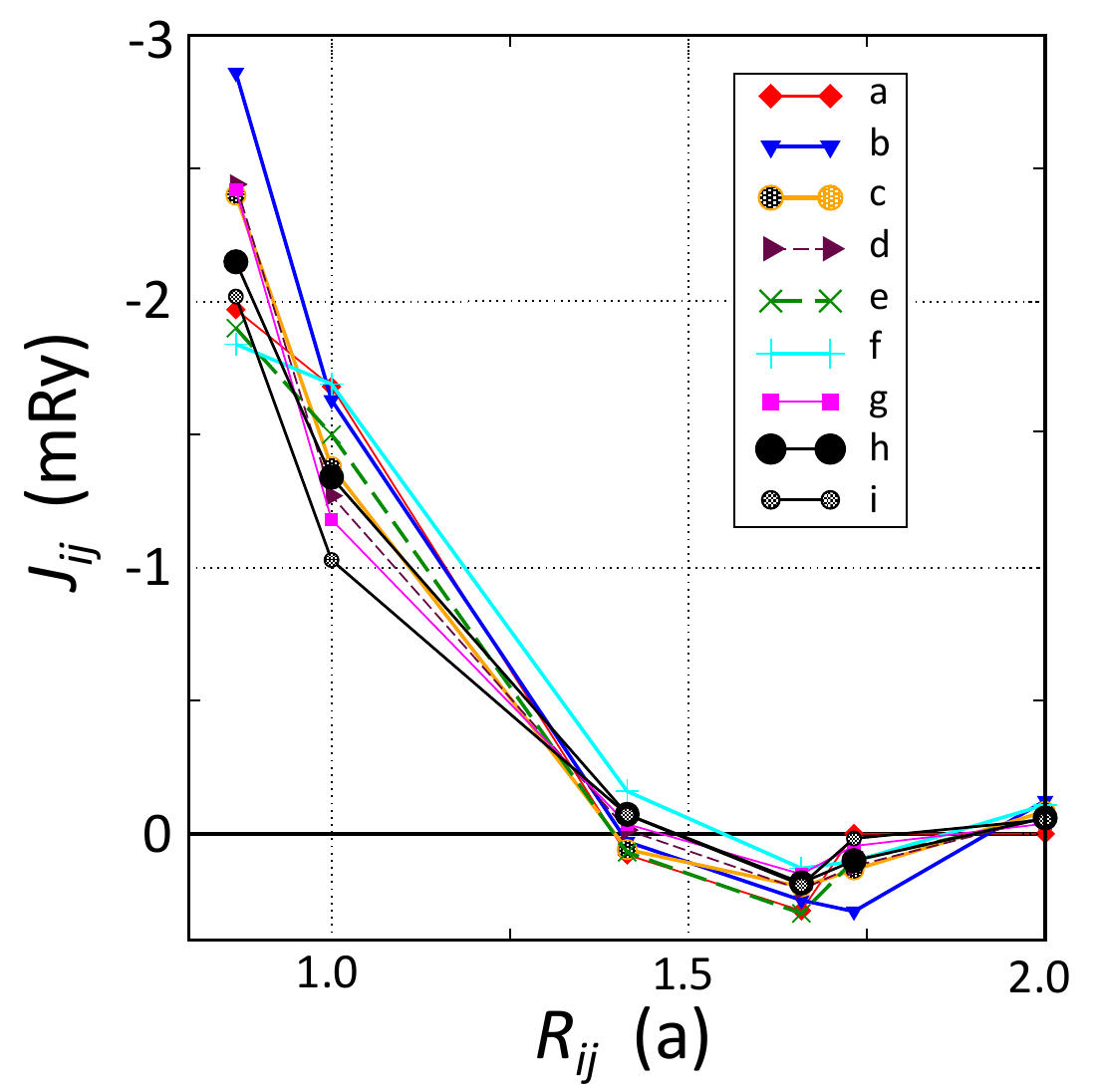}
\caption{(Color online) Interatomic exchange parameters in bulk bcc Fe calculated from Eq. (\ref{LKAGnewest}) with different code implementations. In case a) denoted by red squares a tight-binding (TB) LMTO-ASA method was used by Mor\'an et al. \cite{PhysRevB.67.012407}. In case b) another TB LMTO method was used by Pajda et al. \cite{PhysRevB.64.174402}. The case c) is obtained by the use of real space LMTO-ASA code by Frota-Pessoa et al. \cite{PhysRevB.62.5293}. The cases d) and e) are also real space LMTO-ASA calculations by Antropov et al. \cite{ANTROPOV1999148} and Schilfgaarde et al. \cite{doi:10.1063/1.370495}, respectively. The case f) is a real space tight-binding framework used by Spi\v{s}\'ak et al. \cite{SPISAK1997257}. In case of g) an LDA++ approach was used for the first time by Katsnelson et al. \cite{katsnelson2000first}. In case of h) a full-potential, relativistic calculation (RSPt) was used with extended basis and in case of i) RSPt was used by with a minimal basis (unpublished). Note that $J_{1}$=-1.9 mRy was found with a KKR calculation by Mankovsky et al. \cite{PhysRevB.102.134434}.}
\label{bccFecomparsion}
\end{figure}

In Fig.~\ref{s7-fig2} experimental data are only shown for fcc Ni. This is primarily due to the fact that it is difficult to measure inelastic scattering of polarized neutrons of Co, due to the strong self-absorption effect. In addition, the crystal structure of bulk Co is hcp, not fcc. However, as reviewed in Ref.~\onlinecite{Etz_2015}, experimental results of the magnon dispersion have been published for thin films of Co (in the fcc structure), as an overlayer of, e.g., Cu (001). Also here can one find good agreement between theory and experiments. 
The results for thin films of fcc Co were also reported in Refs.~\onlinecite{PhysRevB.53.12166, doi:10.1063/1.1689774, PhysRevLett.91.147201, Balashov-2014}, with good agreement between theory and observations.

Since this review article focuses on the \textit{explicit} method, in Fig. \ref{bccFecomparsion} we compare selected, calculated interatomic exchange parameters of bcc Fe -- which is a common test-material in case of code implementations. The exchange parameters in Fig. \ref{bccFecomparsion} are calculated by Eq. (\ref{LKAGnewest}), however, the actual electronic structure methods used, energy functionals as well as details of the implementations differ in the different investigations reported. This causes some differences between the different investigations. The first-nearest neighbor couplings (using the form of the Heisenberg Hamiltonian in Eqn.\ref{neweqn1}) were obtained as -1.97, -2.86, -2.40, -2.44, -1.90 and -1.90 mRy in Refs. \cite{PhysRevB.67.012407}, \cite{PhysRevB.64.174402}, \cite{PhysRevB.62.5293}, \cite{ANTROPOV1999148}, \cite{PhysRevB.102.134434}, and \onlinecite{PhysRevLett.116.217202}, respectively. These data, together with interactions at longer distance, are shown in Fig. \ref{bccFecomparsion}. One may note from the figure that the general behaviour of the interatomic exchange interaction, as a function of distance between atoms, is very similar for all reported studies. The strongest interactions are clearly between nearest neighbours, followed by that from next nearest neighbours, while longer range interactions are in all published studies much weaker. Fig. \ref{bccFecomparsion} also shows that differences in the value of interatomic exchange varies between the published results, which reflects the sensitivity of this parameter with respect to computational details (basis set, energy functional etc). Another relevant parameter that is extracted from a set of interatomic exchange is the total exchange value, $J_0=\sum_{<0i>}J_{0i}$ given by Eq. (\ref{J0defnewest}). Values for $J_{0}$ were found as -10.00, -11.03, -12.20 and -13.58 mRy in Refs. \cite{sakuma1999first}, \cite{PhysRevB.62.5293}, \cite{katsnelson2000first} and \cite{PhysRevB.64.174402}, respectively. These values vary with approximately the same amount as the values in 
Fig. \ref{bccFecomparsion}, which seems natural. 
Note that all numerical values we report here are adjusted\footnote{In many cases one can find $-\frac{1}{2}J_{ij}$ values in the literature where the nomenclature differs.} to the spin Hamiltonian of Ref.(\ref{neweqn1}). More details on this issue can be read in Subsection \ref{nomenclature}.

In Ref.~\onlinecite{PhysRevB.64.174402} the long-ranged character of the oscillations was discussed in great detail. Using stationary phase approximation and the asymptotic behaviour of the inter-site Green function, the long-range character of the $J_{ij}$'s was shown to be of the following form:
\begin{equation}
J_{ij} \propto \Im \frac{\exp{[i((\mathbf{k}^{\uparrow}_F+\mathbf{k}^{\downarrow}_F) \mathbf{R}_{ij} + \Phi^{\uparrow} + \Phi^{\downarrow})]}}{R_{ij}^3},
\label{Jij-long-range-exp}
\end{equation}
where $\mathbf{k}_F$ is the wave vector of energy $E_F$ having the direction such that the associated group velocity is parallel to $\mathbf{R}_{ij}$, $\Phi$ is an additional phase factor, while $\uparrow$ and $\downarrow$ denote spin projections.
For weak itinerant-electron ferromagnets, which have both spin-up (majority) and spin-down (minority) bands partially occupied, the Fermi wave vectors are real and one recovers the oscillatory exchange interaction, known as Rudermann-Kittel-Kasuya-Yosida (RKKY) mechanism of indirect exchange~\cite{PhysRev.96.99}.
At the same time, if one of the spin channels is completely empty or filled, the Fermi wave vector becomes imaginary $\mathbf{k}_F=i\mathbf{\kappa}_F$, which in turn results in the evanescence of the $J_{ij}$'s.
Thus, in weak ferromagnets one can expect more long-ranged magnetic interactions than in e.g. half-metals or strong ferromagnets, that have a filled majority-band.
This result also provides an explanation for why bcc Fe, being a weak ferromagnet, shows much more pronounced Kohn anomalies in the spin-wave spectra compared to Co and Ni \cite{PhysRevB.58.293}.

Ref.~\onlinecite{PhysRevB.64.174402} also demonstrated that the interactions with very distant neighbors must be taken into account when calculating the spin-wave stiffness.
However, by considering interactions between very distant atoms (which are more than 6 lattice constant apart), the value of $D$ keeps oscillating as one takes more coordination shells in the summation.
The reason for this is that the expression for the $D$ includes a term $R_{ij}^2$ (see Eq.~\eqref{stiffness-eta}. The $J_{ij}$'s have at worst (from a summation point of view) an $R_{ij}^{-3}$ dependence (Eq.~\eqref{Jij-long-range-exp}).
As a result, the numerical convergence of $D$ is quite a problematic.
One solution to this problem was also proposed in Ref.~\onlinecite{PhysRevB.64.174402}.
It was suggested that the expression for spin-wave stiffness can be regularized by introducing an additional decay factor $\eta$, which ensures its convergence at large distances.
So the $D(\eta)$ is then defined as:
\begin{equation}
D(\eta)=\lim_{R_{\text{max}}\rightarrow \infty } \frac{2}{3M} \sum_{R_{ij} \le R_{\text{max}}} J_{ij}R^2_{ij} \text{e}^{\left(-\eta\frac{ R_{ij}}{a}\right)}
\label{stiffness-eta}
\end{equation}
and finally the spin-wave stiffness is calculated by taking the limit of $\eta$ going to zero:
\begin{equation}
D = \lim_{\eta \rightarrow 0} D(\eta).
\label{stiffness-limit}
\end{equation}
The so obtained values are shown in Table~\ref{s7-table1}. 
They show systematically good agreement with experimental data, measured by different techniques.
In Table~\ref{s7-table1} we also show the experimental results for hcp Co. The magnetic interactions in hcp Co were calculated in several studies~\cite{doi:10.1063/1.370495, https://doi.org/10.1002/pssb.200301671, PhysRevB.92.134422}.
In one of the more recent works the spin-wave excitations and the $T_c$ were calculated by means of atomistic spin dynamics simulations and excellent agreement with experiment for both properties was reported~\cite{PhysRevB.95.214417}.

The magnetic ordering temperatures of the three FM metals were also calculated in a number of publications, from several different approaches: mean-field approximation (MFA), Tiablikov's decoupling scheme (also known as random phase approximation (RPA)~\cite{tiablikov2013methods}), classical Monte Carlo simulations or atomistic spin dynamics~\cite{antropov1995ab,antropov1996ab,evans2014atomistic, eriksson2017atomistic, turzhevskii1990degree, shirinyan2019self}, while the latter uses unsupervised machine learning.
The MFA values
for bcc Fe and fcc Co were reported to be in reasonably good agreement with experiment.
For instance, for Fe it was found to be $\sim$ 1400 K, while the experimental value is 1045 K~\cite{PhysRevB.64.174402}.
Given the fact that MFA is known to overestimate the estimates by roughly 30$\%$ as compared with a more accurate Monte Carlo method~\cite{binder2010theoretical}, the calculated value is close to what one should expect.

\begin{table}[t!]
\caption{Calculated and measured values of spin-wave stiffness in elemental ferromagnets in the units of meV$\text{\AA}^2$. }
\begin{tabular}{ccc}
    \hline
Metal & $D_{\text{theo}}$ (Ref.~\onlinecite{PhysRevB.64.174402}) & $D_{\text{exp}}$ \\
    \hline
Fe (bcc) & 250 $\pm$ 7 & 281$^{c}$, 266$^a$, 256$^a$  \\
Co (fcc) & 663 $\pm$ 6 & 384$^a$, 371$^a$, 466$^{d}$, 435$^{e}$, 580$^{f}$   \\ 
Ni (fcc) & 756 $\pm$ 29 & 374$^a$, 403$^a$, 555$^{b}$  \\
\hline
\end{tabular} \\
$^a$ - Ref.~\onlinecite{PhysRev.156.623} and references therein. \\
$^b$ - Ref.~\onlinecite{PhysRevLett.30.556} \\
$^c$ - Ref.~\onlinecite{doi:10.1063/1.2163453} \\
$^d$ - Ref.~\onlinecite{PhysRevB.53.12166} (thin films) \\
$^e$ - hcp Co, Ref.~\onlinecite{PhysRevB.53.12166} \\
$^f$ - hcp Co, Ref.~\onlinecite{pauthenet1982experimental} \\
\label{s7-table1}
\end{table}

The calculations for fcc Ni suggested a $T_c$ of about 397 in MFA and 350 K in RPA~\cite{PhysRevB.64.174402}, which is much smaller than the experimental value of about 630 K.
This striking underestimation was already reported in Refs.~\onlinecite{liechtenstein1987local,doi:10.1063/1.370495}.
However, the spin-wave stiffness is, on the contrary, overestimated as compared with experiment. 
This suggests that the inconsistency of the results for Ni can not be circumvented by a simple re-scaling of the exchange integrals.

The problem of describing magnetic excitations in fcc Ni has been addressed for a long time.
Bruno suggested that the corrections to the LKAG formula due to transverse constraining fields become substantial when the exchange splitting is small and becomes comparable with magnon energies~\cite{bruno2003exchange}, as discussed also in Section~\ref{linearresp} of this review.
This is indeed the case of fcc Ni, whose saturated magnetic moment amounts to roughly 0.6 $\mu_B$ per atom, indicating that the splitting between spin-up and spin-down bands is the smallest among three elemental magnets as shown in Ref.~\onlinecite{PhysRevB.71.214435}. Using \textit{renormalized} values of exchange parameters, it was shown that the MFA-based $T_c$ estimates can be substantially improved~\cite{bruno2003exchange}. At the same time, the employed corrections were shown not to modify the magnon spectrum~\cite{katsnelson2004magnetic}, such that the good agreement between theory and experiment remained (Fig.~\ref{s7-fig2}).

However, as discussed above, the case of Ni also raises questions whether the small moment of Ni can be treated classically. In addition, the values of the magnetic moments in fcc Ni depend significantly on the magnetic configuration, and this dependence is much more pronounced than in, e.g., bcc Fe~\onlinecite{turzhevskii1990degree,PhysRevB.55.14975,ANTROPOV1999148}. It seems that the best gauge for estimating the accuracy of interatomic exchange of fcc Ni is to compare magnon dispersion, as opposed to the Curie temperature. 

Since both the calculated magnetic moments and the interatomic exchange integrals depend on the reference state, it is reasonable to expect that the spin stiffness should be better described by the set of $J_{ij}$'s extracted from the ordered magnetic ground state, while the $T_c$ should be estimated using
a magnetic configuration found at the ordering temperature~\cite{PhysRevB.70.125115, PhysRevB.72.104437}. The problem is that representing such a state in DFT calculations is not straightforward.
In the so-called disordered local moment picture that we will be discussed in Subsection \ref{DLM} in detail, the magnetic moments experience a completely spin-disordered environment introduced via the coherent potential approximation (CPA)~\cite{PhysRev.156.809,RevModPhys.46.465, PhysRevB.45.7196}. 
However, in these calculations, the local moment in fcc Ni collapses to zero~\cite{PhysRevB.72.104437}, in contrast to observations.
A generalized Heisenberg model, which takes into account not only the short-range order effects~\cite{PhysRevB.72.140406}, but also allows the magnetic moments to change their magnitude, i.e. introducing longitudinal spin fluctuations, were proposed in Refs.~\onlinecite{PhysRevB.55.14975,ruban2007temperature, PhysRevB.78.184419}, and from this model
the calculated $T_c$'s of bcc Fe and fcc Ni are in good agreement with experimental values.

\begin{figure}[t!]
\includegraphics[width=0.80\linewidth]{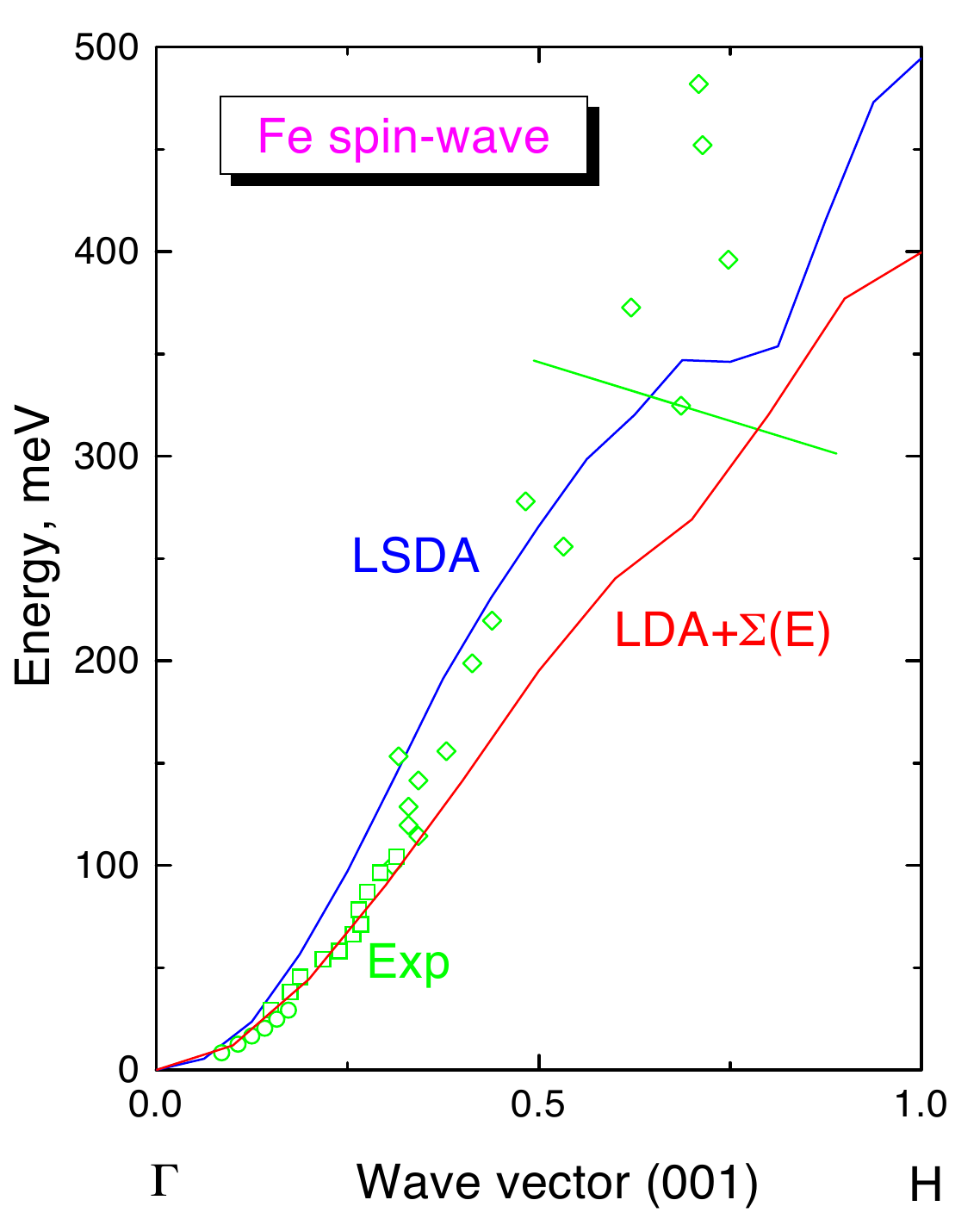}
\caption{(Color online) Spin-wave dispersion in bcc Fe as obtained from DFT+DMFT (referred to as LDA+$\Sigma(E)$) and spin-polarized DFT (LSDA) calculations~\cite{katsnelson2000first}.}
\label{s7-JFe-QMC}
\end{figure}

In general, interatomic exchange is a quantity that depends critically on the details of the electronic structure.
The results discussed so far were obtained employing LSDA or the similar, spin polarized generalized gradient approximation (GGA).
Electron correlations beyond LSDA/GGA can be captured by means of a combination of DFT and dynamical mean field theory~\cite{RevModPhys.68.13, PhysRevB.57.6884, kotliar-DMFT}.
In Ref.~\onlinecite{katsnelson2000first} this method was used to calculate interatomic exchange. 
It was shown that taking into account local correlations of bcc Fe will influence both the local magnetic moment and the $J_{ij}$'s. Subsequent work~\cite{PhysRevB.62.5293} basically confirmed this result. 

The results shown in Fig.~\ref{s7-JFe-QMC} indicate that the calculation of the spin-wave stiffness in bcc Fe, obtained using LSDA, is different from results of DFT+DMFT (by roughly 20$\%$).
We note that the starting point for these calculations was non-magnetic DFT solution, and therefore the local exchange splitting emerges purely from DMFT and is governed by the Hubbard $U$ term.
However, it was shown that if one starts from magnetic DFT and performs DMFT calculations on top of it, then the differences between LSDA and LSDA+DMFT results are quite modest~\cite{PhysRevB.91.125133}. This is partly related with the fact that the exchange splitting is introduced by LSDA and does not change much after $U$ is explicitly added to consideration.
In the case of moderate correlation strength, the overall differences in the total exchange interaction $J_{0}$ are related to the quasiparticle's mass renormalization, brought by electron-electron interactions~\cite{PhysRevB.88.085112}.
However, since the orbitals of different symmetry have different effective masses, the overall impact of dynamical correlations on each individual $J_{ij}$ is more sophisticated.
In Ref.~\onlinecite{borisov2021heisenberg} it was shown that dynamical correlations, as described by DMFT, can produce up to
30 \% variation of the leading Heisenberg and DM exchange
interactions. This was exemplified by a study of intermetallic compounds such as CoPt and FePt,
MnSi and FeGe, as well as transition metal bilayers; Co/Pt(111) and Mn/W(001).
Furthermore, non-local correlations, modelled on the GW level, have also been made for Fe, Co and Ni~\cite{QSGW-MFT-2019}, albeit with marginal changes in the Heisenberg exchange.

\begin{figure}[t!]
\includegraphics[width=\linewidth]{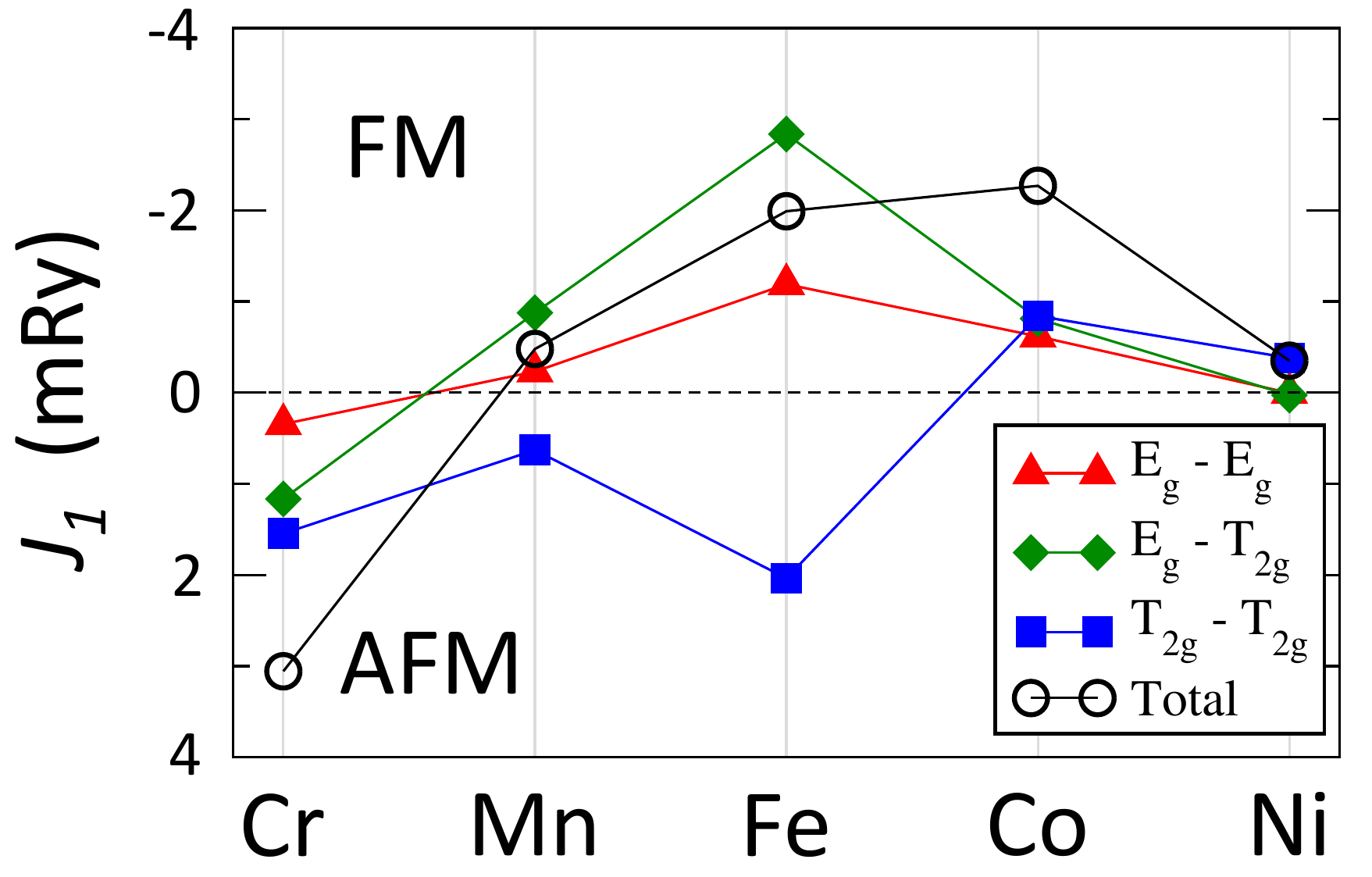}
\caption{(Color online) Calculated orbital decomposed first-nearest neighbour exchange interaction in elemental $3d$ metals in the bcc structure from Ref.~\onlinecite{PhysRevLett.116.217202}. The calculations were made with the use of the RS-LMTO-ASA method.}
\label{s7-BS}
\end{figure}

\begin{figure}[t!]
\includegraphics[width=1.\linewidth]{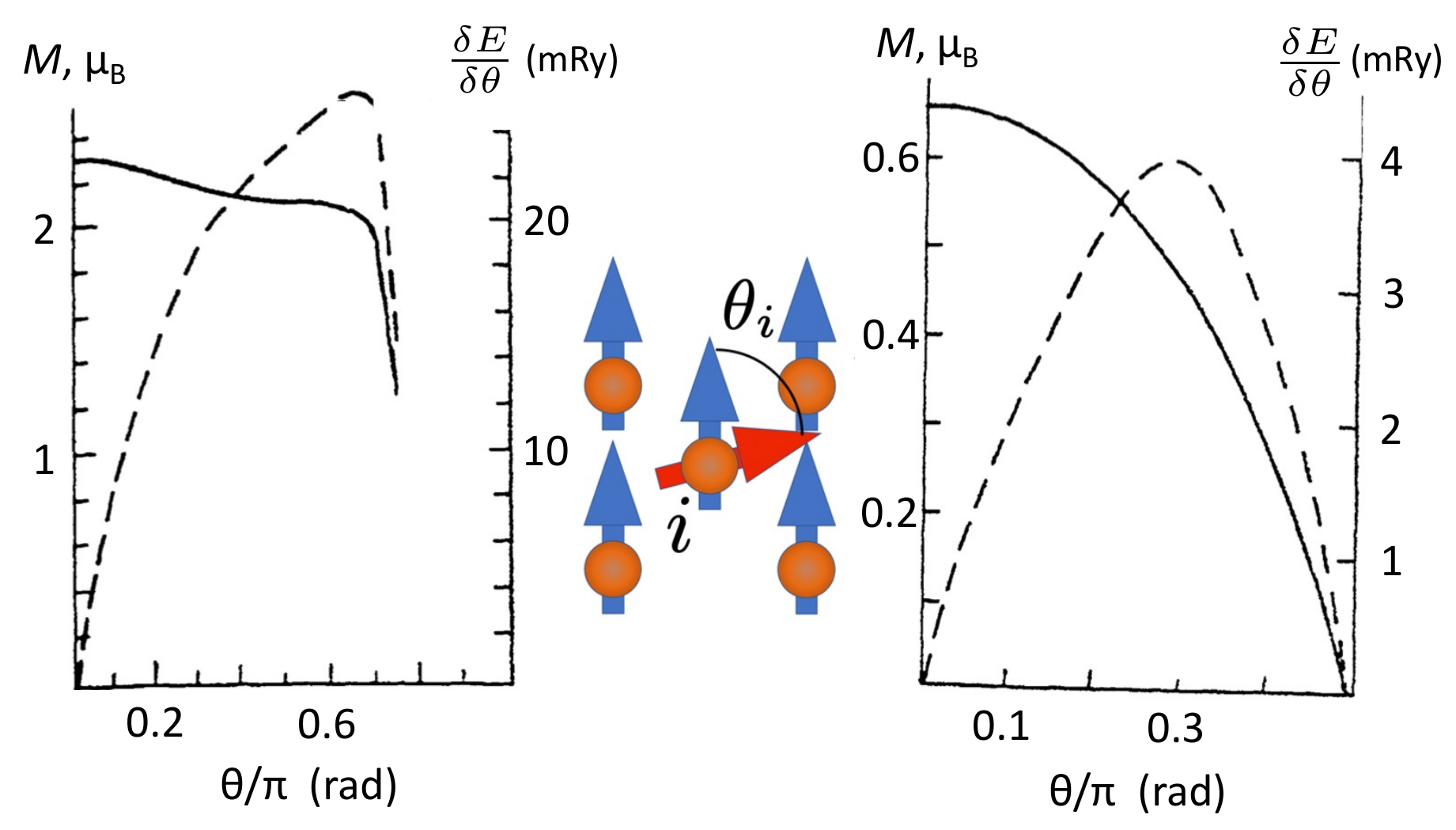}
\caption{(Color online) Magnetic moment in $\mu_{B}$'s (solid lines) and the first derivative of the energy ($\Omega$) with the respect to angle $\theta_{i}$ (dashed lines) for the case of bcc Fe (left) and fcc Ni (right) shown on the left and right sides, respectively, when one spin is rotated with a \textit{finite} $\theta_{i}$ in a ferromagnetic background~\cite{turzhevskii1990degree} as shown by the added schematic figure.}
\label{turz}
\end{figure}

A great advantage of the formalism of {\it explicit} calculation of the $J_{ij}$ parameters, is that one can perform orbital-by-orbital decomposition of each magnetic coupling. This decomposition is possible since in the LKAG formula (given by Eq.~\eqref{LKAGnewest}) the $Tr_{L}$ can first be taken over just a part of the orbitals, i.e., one can analyze the individual orbital contributions of the exchange parameter.  In a cubic material one can then follow the coupling between different irreducible representations of the 3d orbitals ($E_g$ and $T_{2g}$). This turns out to be a powerful tool for obtaining a microscopic understanding of the nature of magnetic interactions.
To be specific, if the material has cubic symmetry, the $d$ orbitals split into $E_g$ and $T_{2g}$ manifolds. 
In the basis of cubic harmonics the local exchange splitting becomes a diagonal matrix and the exchange interaction can be represented as a sum of orbital contributions $J^{mm'}_{ij}$, where an orbital $m$ on the site $i$ is coupled with each orbital $m'$ on the site $j$.
In the cubic system it is therefore natural to group these terms into three contributions:
\begin{eqnarray}
J_{ij} = J^{E_g-E_g}_{ij} + J^{E_g-T_{2g}}_{ij} + J^{T_{2g}-T_{2g}}_{ij},
\label{eq-orbjij}
\end{eqnarray}
which combine the individual orbital contributions according to the symmetry of the $d$ orbitals involved.
In Ref.~\onlinecite{PhysRevLett.116.217202} such orbital decomposition of the NN exchange integral was performed for a series of transition metal alloys in the bcc structure.
The results, also shown in Fig.~\ref{s7-BS}, reveal that in case of Mn and Fe there is a strong competition between different terms having opposite (FM and AFM) signs. 
This balance is most intricate for bcc Fe, where all three terms in Eq.~\eqref{eq-orbjij} are of comparable size.
Most interestingly, it was shown that thanks to this decomposition it was possible to identify the microscopic exchange mechanisms to each of these three channels, revealing a combination of RKKY, double- and super-exchange~\cite{PhysRevLett.116.217202}.

Overall, the sign of the NN coupling in all elemental $3d$ follows the famous Bethe-Slater curve, but it is governed by a complex interplay between different orbital contributions~\cite{cardias_bethe-slater_2017}.
This result paves the way towards designing magnetic interactions in metallic $3d$ systems in general, and allows for a deeper analysis of interatomic exchange interaction. One way to continue the analysis is to calculate the symmetry-decomposed interaction parameters between further neighbors as was done for bcc Fe \cite{PhysRevLett.116.217202} and for other $3d$ elements~\cite{cardias_bethe-slater_2017}. One of the most important conclusions in case of bcc Fe is that the exchange between the $T_{2g}$ orbitals is Heisenberg-like and long-ranged while it is relatively short-ranged with a substantial non-Heisenberg behavior in case of the $E_{g}$-$E_{g}$ and the mixed ($E_{g}$-$T_{2g}$) channel~\cite{PhysRevLett.116.217202}.

Note that the non-Heisenberg behavior of bcc Fe and especially of fcc Ni have been discussed for a long time \cite{turzhevskii1990degree}, from calculations that considered $\delta \Omega_{i}^\mathrm{one}$ when a spin is rotated by a finite $\theta_{i}$, as shown in Fig.~\ref{turz}. The results of the figure are clear; a strong configuration dependence can be observed for the magnetic moment and the angular dependence of the energy variation does not follow a sine function, especially for angles far from the ground state.

\begin{figure}[t!]
\includegraphics[width=\linewidth]{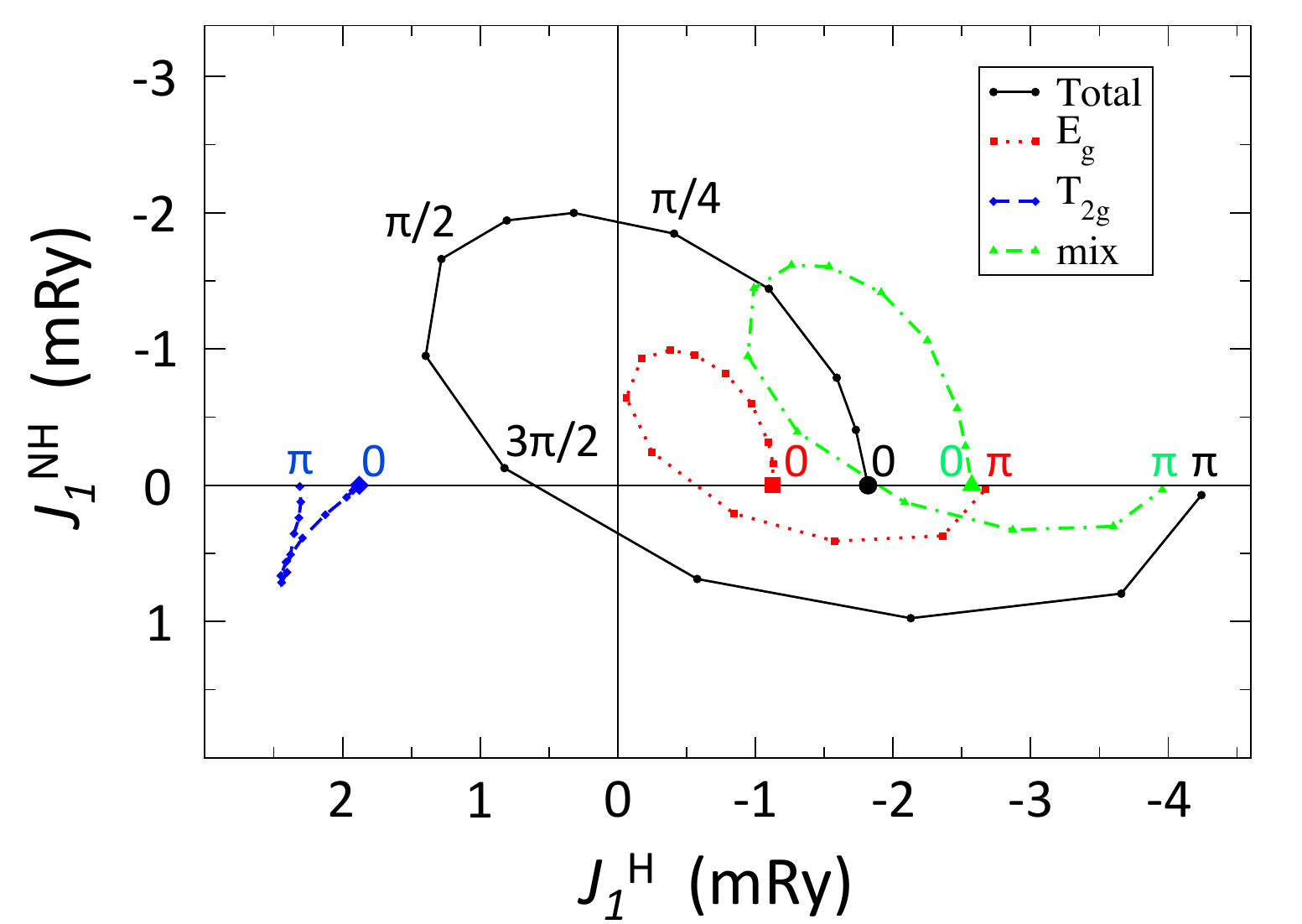}
\caption{(Color online) First-nearest neighbour Heisenberg and non-Heisenberg interatomic exchange parameters in bcc Fe when one spin is rotated by a finite $\theta_{i}$ running from zero to $\pi$ at site $i$ in a ferromagnetic background in case of bcc Fe~\cite{szilva2017theory}. $J_{1}^{H}=J_{1}^{(2)}+A^{(2)xx}_{1}$ and  $J_{1}^{NH}=-2A^{(2)zx}_{1}$, see Eqs.~\eqref{defJ2} and~\eqref{defA2}. The black (solid) curve stands for the total value while the red (dotted), blue (dashed), and green (dash-dotted) lines show its symmetry decomposition in the d channel defined by Eq.~\ref{eq-orbjij}.}
\label{spiral}
\end{figure}

In Ref.~\onlinecite{szilva2017theory} a similar system was considered, i.e., one spin was rotated by a finite $\theta_{i}$ at site $i$ on a bcc Fe lattice when all other spins formed a ferromagnetic background and $\theta_{i}$ ran from 0 to $\pi$. In this study the two-site energy variation was the main focus. Note that in general one formally gets for the two-site energy variation (in the lack of SOC) that \begin{equation}
\delta \Omega_{ij}^\mathrm{two}=- \left(J_{ij}^{H} \cos{\theta_{i}}+J_{ij}^{NH}\sin{\theta_{i}}\right) \left(\delta \theta \right)^{2}  \,,  
\end{equation}
where the terms which are proportional to a cosine and a sine function are referred to as the Heisenberg (H) term and the non-Heisenberg (NH) term, respectively and $J_{ij}^{H}=J_{ij}^{(2)}+A^{(2)xx}_{ij}$ and  $J_{ij}^{NH}=-2A^{(2)zx}_{ij}$ according to Eqs.~\eqref{defJ2} and~\eqref{defA2}. In the discussion of this paragraph, when the first-nearest neighbour couplings are considered, the $ij$ indices are replaced by the index 1. The calculated Heisenberg and non-Heisenberg results for different values of $\theta_{i}$ are shown by the solid black line in Fig.~\ref{spiral}. The figure shows that in a general, non-collinear case the non-Heisenberg contribution can be significant. However, the symmetry decomposition proves that in the $T_{2g}$ channel the system is more Heisenberg-like and the non-Heisenberg behavior originates from the $E_{g}$ and the mixed channel. This is in good agreement with the conclusions based on collinear formalism presented in Refs.~\onlinecite{PhysRevLett.116.217202, cardias_bethe-slater_2017}.

\subsection{Itinerant magnets based on 3d metal alloys and compounds}
\label{alloys3d}

\begin{figure}[t!]
\includegraphics[width=0.85\linewidth]{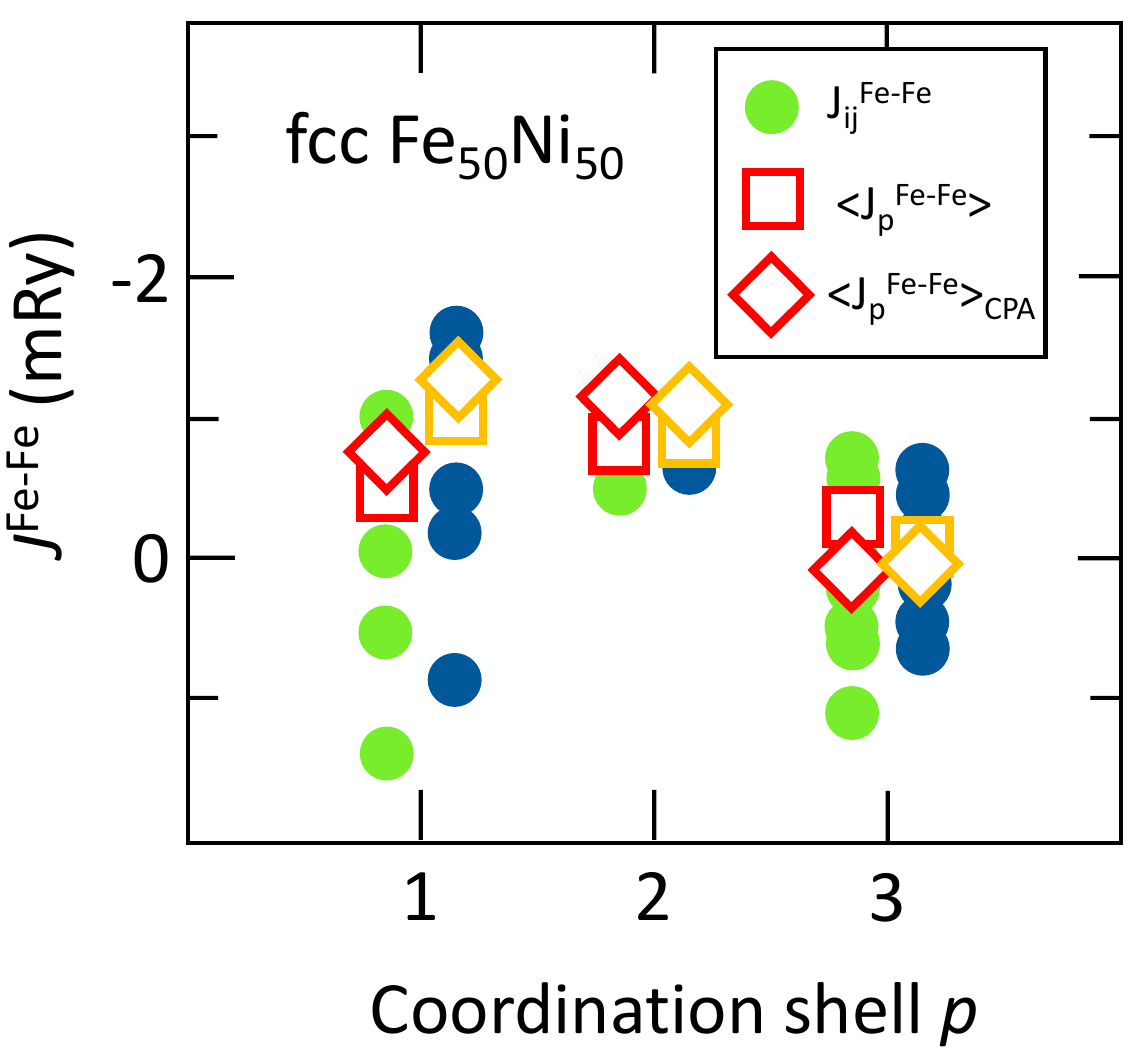}
\caption{(Color online) Calculated Fe-Fe exchange interactions with first 3 coordination shells in fcc Fe$_{0.5}$Ni$_{0.5}$ for two different unit cell volumes ($V$)~\cite{PhysRevB.71.054402}. The 16-atom supecell-based results for $V$=73.6~a.u.$^3$ and $V$=70.3~a.u.$^3$ are shown with blue (dark grey) and green (light grey) circles, respectively. Supercell- and CPA-averaged $J_{ij}$'s are shown for comparison.}
\label{s7-FeNi-Invar}
\end{figure}

The \textit{explicit} method of calculating exchange has been widely applied to study $3d$-based alloys and compounds\cite{doi:10.1080/14786430500504048,Ebert_2011}, and we describe in this subsection some selected examples.
According to the Slater-Pauling curve, the maximal magnetization per atom in $3d$ metal alloys is achieved for the Fe$_{1-x}$Co$_{x}$ family.
In the entire composition range, these alloys are ferromagnetic~\cite{doi:10.1063/1.2710181}.
Ref.~\onlinecite{doi:10.1063/1.2710181} suggested that all pairs of Fe-Fe, Fe-Co and Co-Co interactions are FM and the NN $J^{\text{Fe-Co}}$ have the highest value.
The latter result was also reported earlier for an ordered B2-FeCo system~\cite{doi:10.1063/1.370036}, highlighting the fact that the efficient hybridization between Fe and Co states results in the enhancement of both the saturated magnetization and the $T_c$.
Interestingly, for $x$>0.17, an experimental value of $T_c$ of the bcc phase is unknown, since the structural bcc-fcc transition occurs before the bcc structure reaches a Curie temperature. The temperature of the bcc-fcc transition sets a lower value of the expected $T_c$ of the bcc structure, and it is very high. In fact, MFA-based estimates predict values of 1600K for x=0.5~\cite{doi:10.1063/1.2710181}, which is consistent with expectations.
An interesting feature of this family of alloys is that by changing concentration, one gradually transforms the electronic structure, to achieve a transition from weak to strong ferromagnetism.
As a result, depending on Co concentration, the magnetic interactions (and hence the $T_c$'s) have very different sensitivity to, e.g., volume changes~\cite{doi:10.1063/1.2710181}.

Iron-nickel alloys form in the fcc crystal structure, and are celebrated thanks to the Invar effect; a vanishing thermal expansion at room temperature, which is in an intrinsic relation with the temperature dependence of the magnetic configuration~\cite{van_schilfgaarde_origin_1999}.
In Ref.~\onlinecite{PhysRevB.71.054402}, the magnetic interactions were calculated in Fe$_{0.5}$Ni$_{0.5}$ and Fe$_{0.65}$Ni$_{0.35}$. They were compared with those in (fcc) $\gamma$-Fe, and it was found that although both types of systems are frustrated, the physical picture is drastically different.
In fcc Fe, the frustration comes from the competition between FM NN exchange coupling and that with more distant neighbours, having long-ranged oscillatory character.
In contrast, the Fe-Ni alloys are characterized by highly dispersive interactions already with the first coordination shell, as one can see in Fig.~\ref{s7-FeNi-Invar}.
Although CPA-based results agree well with the averaged $J_{ij}$'s obtained from the supercell approach, it is clear that the latter captures more details and reveals strong influence of the local environment, which infers why the magnetic order of these alloys is so complex.
Note that fcc-based Fe-Mn alloys have a similar tendency to AFM coupling and non-collinearity~\cite{doi:10.1143/JPSJ.69.3072}.
Generally, for Ni-based alloys, it was found that the \textit{renormalized}~\cite{bruno2003exchange} $J_{ij}$'s provide better estimates of the $T_c$'s~\cite{PhysRevB.77.224422}, which is again related with relatively small exchange splitting of its $3d$ states.

\begin{figure}[t!]
\includegraphics[width=0.95\linewidth]{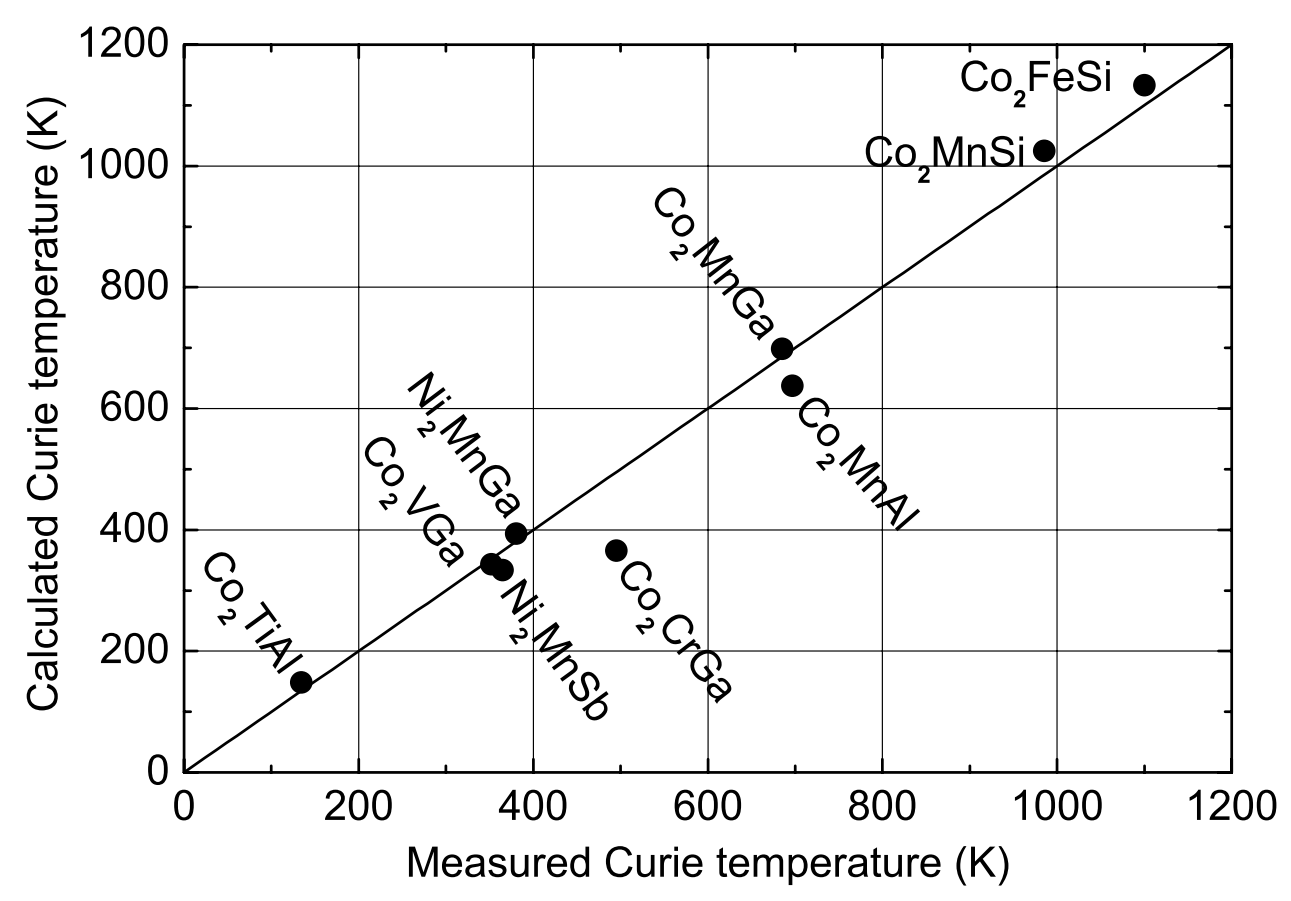}
\caption{Calculated versus measured $T_c$'s in the series of L2$_1$ Heusler alloys~\cite{Thoene_2009}.}
\label{s7-Heuslers}
\end{figure}

Heusler alloys have been intensely studied with the {\it explicit} formalism of exchange interaction~\cite{PhysRevB.71.014425,PhysRevB.73.214412, PhysRevB.78.184427,Thoene_2009,PhysRevB.81.094411,PhysRevB.90.214420,PhysRevB.91.094432, PhysRevB.89.094410,PhysRevB.92.054438,PhysRevB.93.214439}. For instance, in Ref.~\onlinecite{Thoene_2009}, a systematic study of magnetic interactions, spin wave dispersion and $T_c$ was done for the series of Heusler compounds with L2$_1$ structure.
The results shown in Fig.~\ref{s7-Heuslers} demonstrate that $T_c$'s calculated from $J_{ij}$'s combined with a mean field (MFA) estimate of the ordering temperature, are overall in excellent agreement with experiment.
Given all the approximations of this work, such as a MFA for $T_c$ estimation and the neglection of local correlations, one may regard this excellent result as somewhat fortuitous. However, it is still impressive that the theory is able to correctly reproduce the experimental trend so well.

Heusler alloys attract great attention partially due to the half-metallic character, which is observed in some of them.
However, there are many other half-metals, such as Cr- and Mn- compounds, with zinc blende structure, which were also successfully modelled by the formalism presented here~\cite{PhysRevB.68.054417, PhysRevB.81.054446, PhysRevB.82.094435}.
An overall review of the calculated $J_{ij}$'s in half-metallic magnets can be found in  Ref.~\onlinecite{katsnelson2008half}.
As expected from the earlier considerations (Eq.~\eqref{Jij-long-range-exp}), the $J_{ij}$'s in half-metals are relatively short-ranged.

\subsection{Alloys with 4\emph{d} and 5\emph{d} elements}
\label{alloys4d}

The $4d$ and $5d$ metals are typically non-magnetic due to relatively more pronounced band dispersion, which makes it difficult for the Stoner criterion to get satisfied.
However, when placed in proximity to $3d$ metals, these elements can get quite substantial induced magnetic moments~\cite{Mohn-Schwarz-1993}.
The problem of coexisting intrinsic- and induced moments was addressed in several works on FePt and CoPt alloys with L1$_0$ structure~\cite{mryasov2004magnetic,mryasov-ferh}.
It was suggested that the size of the induced moments of $5d$ elements is defined by an effective Weiss field, produced by the surrounding $3d$ magnetic moments.
This idea was later elaborated on, where a generalized Monte Carlo-based scheme was suggested, which dynamically updates the induced magnetic moments for each magnetic configuration during the simulation~\cite{PhysRevB.82.214409}.
Application of this scheme to the series of Fe$_x$Pd$_{1-x}$ and Co$_x$Pt$_{1-x}$ alloys was shown to deliver a systematically good agreement with experimental values of $T_c$.
In Ref.~\onlinecite{PhysRevB.93.024423} it was pointed out that such treatment of the induced moments effectively leads to the emergence of higher-order (biquadratic) exchange interactions between $3d$ metal moments.
Indeed, such interactions were suggested~\cite{mryasov-ferh} to play a key role in explaining the intriguing metamagnetism of FeRh~\cite{PhysRevB.92.094402}.
In ordered FePd$_3$, the biquadratic interactions were also suggested to stabilize the non-collinear 3Q phase under pressure~\cite{PhysRevB.86.174429} and they were needed to get a consistent model of magnetism in ferropnictides \cite{wysocki2011consistent}.

Alloying $3d$ metals with heavier elements can also boost the effective strength of the spin-orbit coupling. Indeed, the SOC constant of Pt $5d$ states is one order of magnitude larger than that of Fe $3d$ states, and can therefore be used to enhance anisotropic magnetic interactions and the magnetocrystalline anisotropy (MAE). Indeed, the results for Pt-doped $3d$ metals~\cite{PhysRevB.52.13419} showed that the MAE is to a large extent defined by non-local scattering of electrons from the SOC potential of Pt states.
Below, we will see how these ideas become particularly useful for inducing large magnetocrystalline anisotropy and DM interaction in low-dimensional systems.

\subsection{Results from the disordered local moment approximation}
\label{DLM}
So far we have focused most of the discussion on theoretical calculations of the electronic structure, and the mapping of these results to the Hamiltonian of Eqns. \ref{neweqn1} and \ref{neweqn2}. However, the electronic structure can have a strong configuration dependence, which was clearly demonstrated in a sequence of papers \cite{Gyorffy_1985, Staunton_1985, STAUNTON198415}. In these works finite temperature effects were introduced, by separating the variables into slow and fast, and the concept of ''temporarily broken ergodicity'' was introduced, as mentioned in connection to Fig.\ref{fig3}. 
A central aspect of these works was the description of the electronic structure above an ordering temperature by means of the  disordered local moment (DLM) model \cite{Gyorffy_1985, Staunton_1985, STAUNTON198415, PhysRevB.20.4584, doi:10.1143/JPSJ.46.1504, Edwards_1982, oguchi1983magnetism, Pindor_1983}, in which the electronic structure is evaluated from a single site approximation of the coherent potential approximation. This implies that the electronic structure at finite temperature is represented by an atom with a potentially finite magnetic moment in an environment with ''spin-average'' scattering properties. Hence there is no short range order in this model, which seems at variance with experimental results from e.g. muon-spin resonance, with significant amount of short range order also at elevated temperatures. The fluctuating local band (FLB) model\cite{PhysRevB.16.4032, correlations1981magnetism, capellmann1979theory} also builds on short range magnetic order at or even above the ordering temperature, and in fact the early works of the FLB model express the basic principles behind non-collinear electronic structure theory. In this discussion it becomes relevant to also note early works of Hubbard, who argued for a theory
that builds on itinerant electron states, but with a local exchange field that varies in direction and strength from atom
to atom (see Refs. \cite{PhysRevB.23.5974, hubbard1981magnetism}). The probability of finding a system in a given configuration of local exchange field was evaluated by an energy expression together with a Boltzman factor, allowing for calculations of magnetism at finite temperature. This theory resulted in a Curie temperature of 1840 K for Fe and 1200 K for Ni. Both
values are significantly larger than the experimental values.

\begin{figure}[t!]
\includegraphics[width=0.7\linewidth]{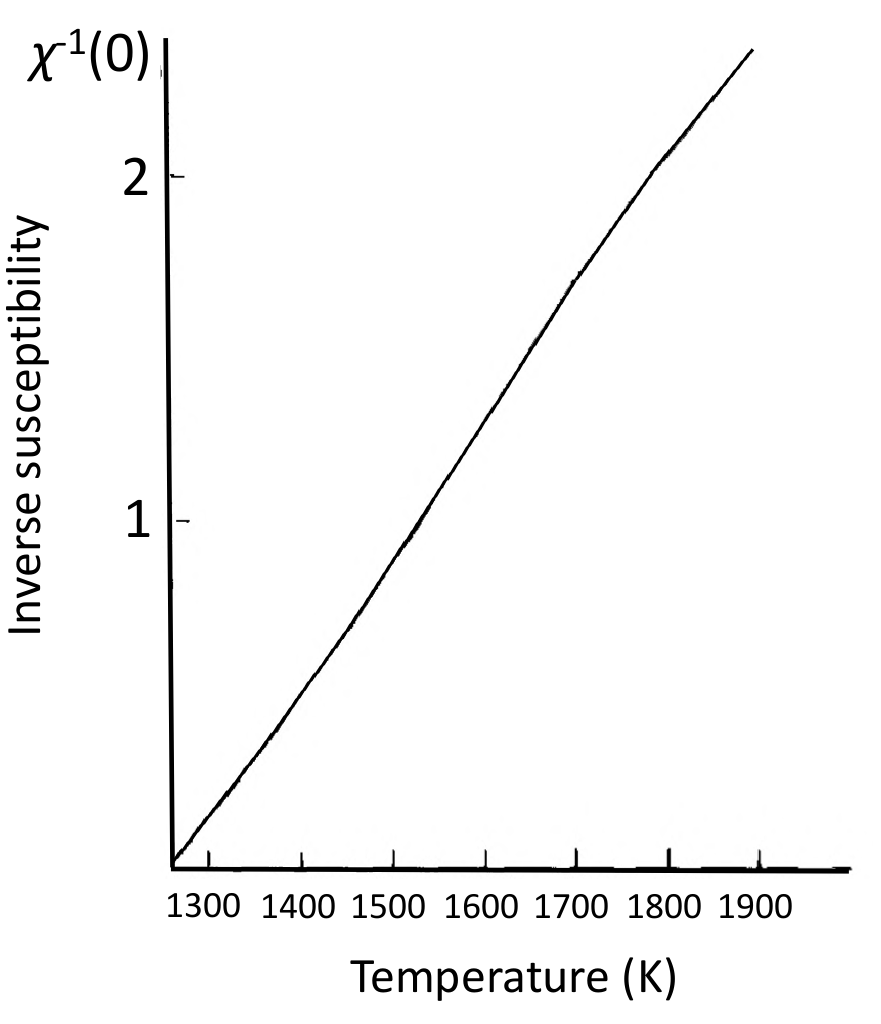}
\caption{Calculated inverse susceptibility of bcc Fe in units of $10^{-2}\mu_{B}^{-2}\left(\mathrm{Ry}/\left(a_{0}/2 \pi \right)^{2}\right)$ (where $a_{0}$=2.789 Å), from DLM electronic structure theory (see text), evaluated as a function of temperature. Figure redrawn after Ref.\cite{STAUNTON198415}.}
\label{s7-DLMnew}
\end{figure}

Although many materials show short range order also above the ordering temperature, the DLM approach, that neglects short range order, has given very encouraging results (see e.g. Refs. \cite{PhysRevB.85.174407, Khmelevskyi_2007, PhysRevB.86.045126, PhysRevB.96.174415}). As an example of this method, we show in Fig. \ref{s7-DLMnew} the inverse of the susceptibility of bcc Fe, evaluated as a function of temperature, in a calculation that builds on the DLM model \cite{STAUNTON198415}. As seen in the figure, the susceptibility diverges at 1260 K, corresponding to the ordering temperature, which is in good agreement with the experimental Curie temperature of 1040 K. There are several examples of calculations of Heisenberg exchange from the DLM approach, e.g. the works quoted above in this subsection, and the relativistic extension of the DLM approach makes possible to calculate the temperature dependence of magnetic anisotropy as well \cite{PhysRevB.74.144411}. We also note that an excellent review of critical dynamics of magnets above and below the transition temperature can be found in Ref. \cite{doi:10.1080/00018739400101535}.

\subsection{Multilayers and atoms on metallic surfaces}

With the development of epitaxial growth techniques, it is now possible to produce extremely thin layers of magnetic materials with good control of the structural homogeneity. 
The magnetic interactions in such low-dimensional magnets bring many surprises and opportunities for applications, e.g. in spintronics  and magnonics.

\begin{figure}[t!]
\includegraphics[width=0.95\linewidth]{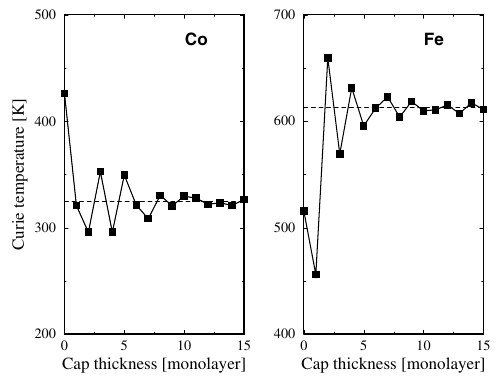}
\caption{RPA-derived estimates of the $T_c$ of Co (left) and Fe (right) monolayers on Cu(001) substrate, covered by Cu-layer of varying thickness~\cite{PhysRevLett.85.5424}.}
\label{s7-Tc-oscillations}
\end{figure}

For thin-film systems, SPEELS serves as a very accurate experimental tool for observing magnon excitations~\cite{PhysRevLett.91.147201}. 
In a number of works, the adiabatic magnon spectra, calculated using $J_{ij}$'s, are directly compared against measured spectra, with in general a good agreement~\cite{PhysRevB.89.174404, PhysRevB.90.174437, zakeri_unconventional_2021}.
In order to incorporate finite temperatures into the theory, atomistic spin dynamics simulations have also been widely used to model the surface magnons (for a review, see Ref.~\onlinecite{Etz_2015}).
Among the studied materials one observes the Co/Cu(111), Co/Cu(001), Fe/Cu(001) and Fe/W(110) systems~\cite{PhysRevB.87.144401}.

Exchange interactions in multilayers of elemental transition metals were investigated in many studies. 
Ref.~\onlinecite{Vaz_2008} provides a comprehensive overview of calculated spin-wave stiffnesses, obtained using different electronic structure methods.
An interesting result was obtained in Refs.~\onlinecite{PhysRevLett.85.5424, BRUNO2002346}, where Fe and Co monolayers on Cu(001) was considered. Depending on the thickness of the capping Cu layer, the $T_c$'s were shown to have oscillatory character.
This results, also shown in Fig.~\ref{s7-Tc-oscillations}, was suggested to be caused by the interference effects in the capping layer.
Such oscillations have actually been observed in Co/Cu/Ni trilayers~\cite{PhysRevB.59.R3938} and also explained using the {\it explicit} approach of calculating exchange interactions~\cite{PhysRevB.65.024435}.

\begin{figure}[t!]
\includegraphics[width=0.95\columnwidth]{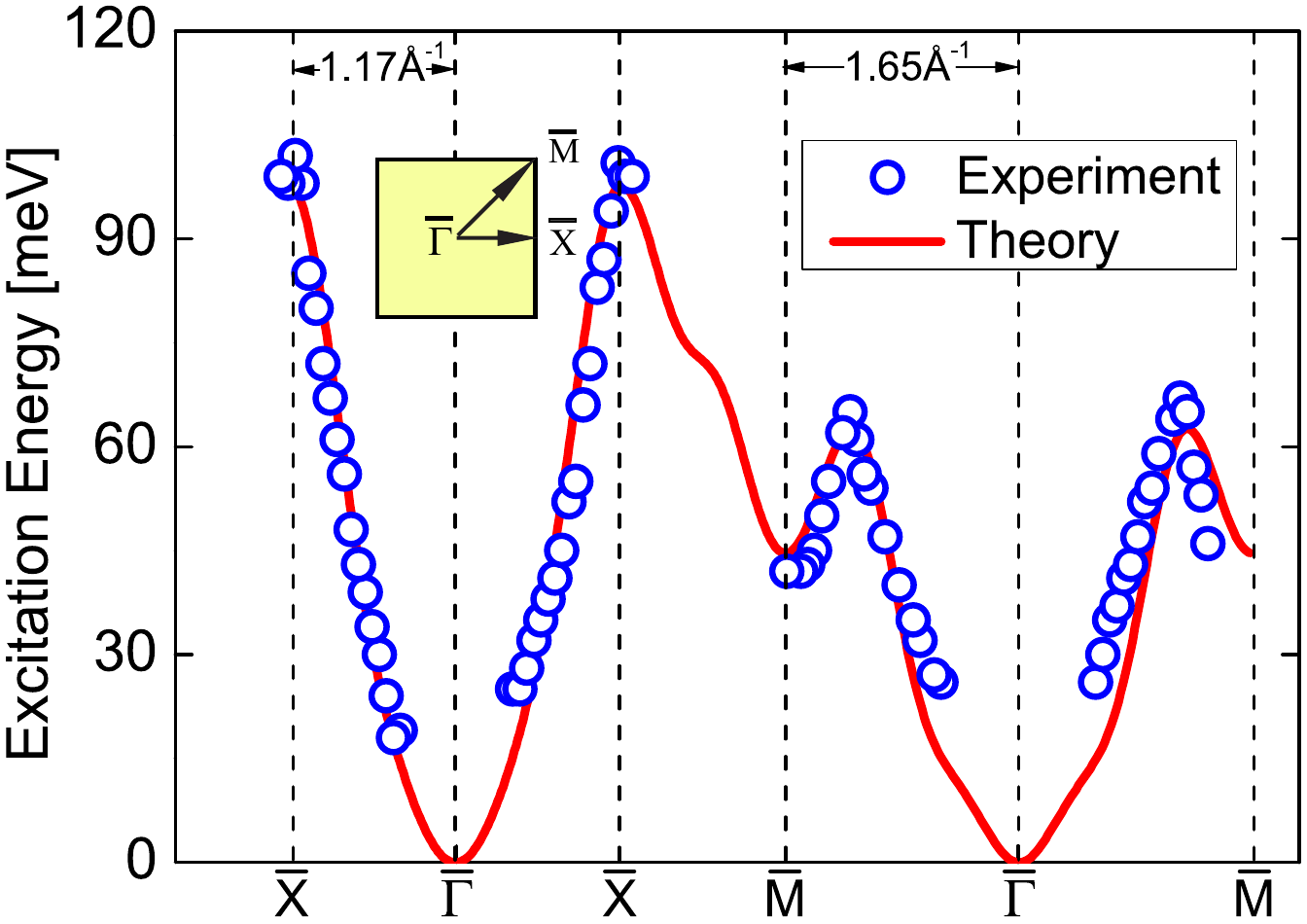}
\caption{(Color online) Computed and measured acoustic magnon dispersions in Fe/Rh(001) from Ref.~\onlinecite{PhysRevB.90.174437}. Inset shows the parts of the Brillouine zone used in the plot.}
\label{s7-Fe-on-Rh001}
\end{figure}

Multilayers of $3d$ metals on the substrates of heavier elements get even more unpredictable behaviour. 
This is partially related to substantial exerted strain as well as a modification of the bandwidth of electron states.
For instance, in Ref.~\onlinecite{PhysRevB.90.174437} the study of Fe/Rh(001) revealed a pronounced softening of acoustic magnons at the $M$ point, as seen as a dip in the dispersion in Fig.~\ref{s7-Fe-on-Rh001}.
Usually in layered systems the lowest magnon branch originates from the spins subject to the smallest effective Weiss field (defined by the total exchange interaction). In this work it was demonstrated that Fe atoms at the interface have a strong tendency to AFM coupling and therefore give the main contribution to the lowest acoustic magnon mode.
This is an unexpected result, given that bulk bcc Fe has such a pronounced NN and next NN FM interaction.
In fact, this tendency was also reported for pure Fe surface~\cite{PhysRevB.92.165129} and is related with the changes of density of states of the surface Fe atoms.
A similar tendency to AFM Fe-Fe interactions were reported for Fe/Ir(001)~\cite{PhysRevB.80.064405,PhysRevB.89.174404,zakeri_direct_2013}. We note that in a similar system, in a monolayer Fe on Rh(111), an up-up-down-down double-row-wise antiferromagnetic magnetic ground state was directly observed in Ref. \cite{PhysRevLett.120.207202}. We also note that the occurrence of a novel type of atomic-scale spin lattice in an Fe monolayer on the Ir(001) surface was predicted in Ref. \cite{ PhysRevB.92.020401}.

\subsection{Influence of spin-orbit coupling}\label{Sec:Infl-SOC}

Although several LKAG-inspired approaches for calculating relativistic interactions have been proposed, DM interactions have attracted most attention~\cite{solovyev1996crucial,PhysRevB.71.184434,udvardi2003first,PhysRevB.79.045209,katsnelson2010correlated,secchi2013non,PhysRevB.96.104416,Ebert2021}. As described here, DM parameters can be extracted using first-order or second-order variation in the spin rotation angles, depending on the situation.

The first-order approach has been utilized to calculate the instability of a ferromagnetic state towards a formation of a cycloid configuration in Ref. \cite{PhysRevB.96.104416}, as well as for the so-called weak ferromagnets (which are weakly ferromagnetic due to uncompensated antiferromagnetism, in contrast to the weak itinerant electron ferromagnets \cite{PhysRevB.71.184434, katsnelson2010correlated} discussed in Subsection \ref{elementals}). As regards weak ferromagnets this leads to good agreement with experimentally observed canting angles for both La$_2$CuO$_4$~\cite{katsnelson2010correlated} and FeBO$_3$~\cite{our-nphys-sign-DMI}. This approach relies on the fact that the canting angle is small and a collinear magnetic state, subject to a finite torque acting on the magnetic moments, is not far from the true one. Interestingly, in this approach one can rotate spin and orbital momenta separately and for both studied systems the latter contributed significantly to the total DM interaction value. Similar calculations of finite torques on collinear magnetic moments due to symmetry allowed DM interactions are the calculation of the small tiltings due to lattice distortions in LaMnO$_3$ \cite{solovyev1996crucial} and the instability of the ferromagnetic state of the B20 alloy Fe$_{1-x}$Co$_{x}$Ge towards a cycloidal spin density wave \cite{PhysRevB.96.104416}. This latter instability is the origin of the formation of skyrmion lattices in this system\cite{heinze_spontaneous_2011}.

\begin{figure}[t!]
\includegraphics[width=0.5\linewidth]{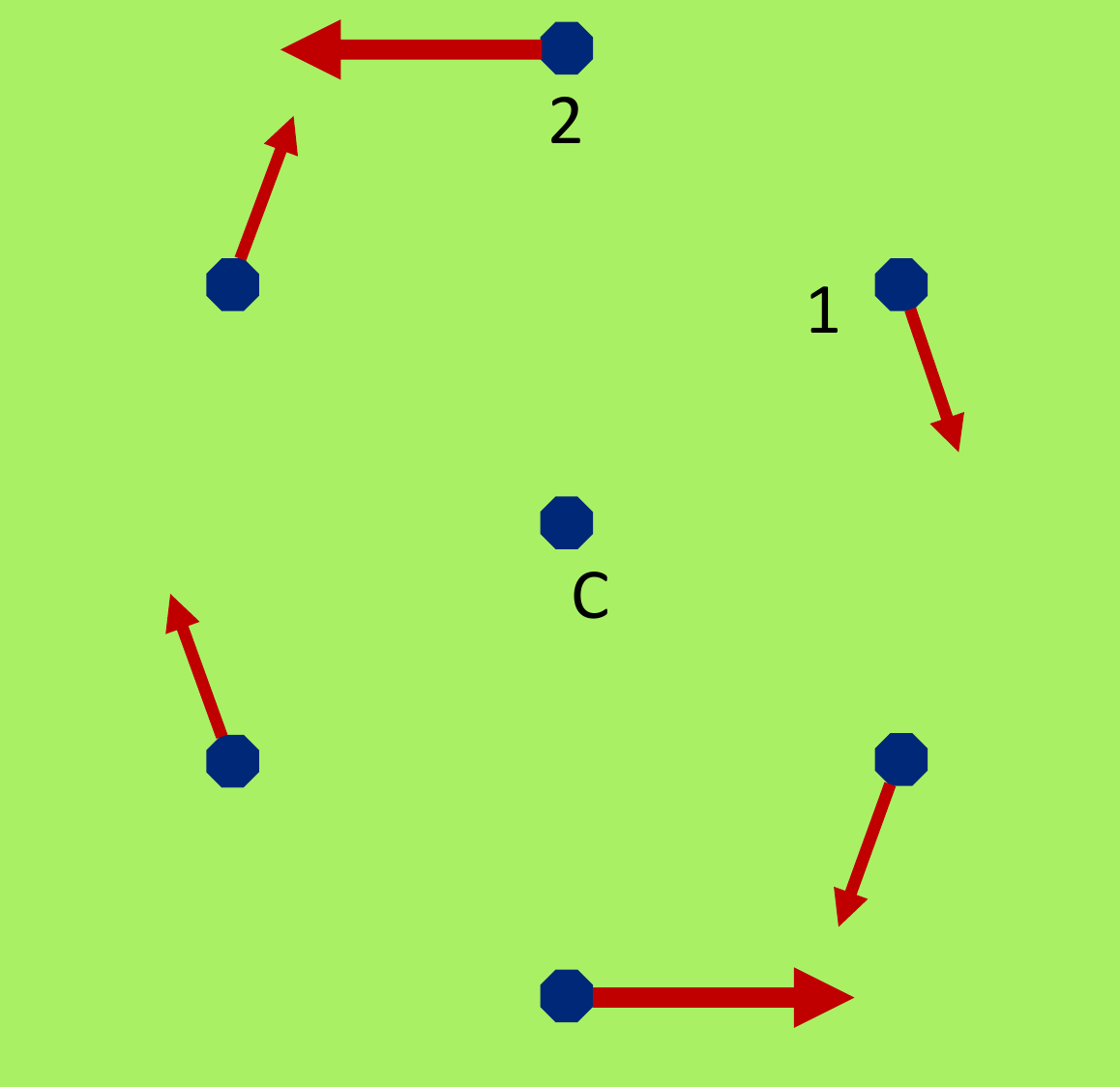}
\caption{(Color online) Schematic representation of the calculated DM interactions in Fe/W(110) between central iron atom (C) and its NN and next NN, denoted as 1 and 2, respectively. Figure is taken from Ref.~\onlinecite{PhysRevLett.102.207204}. The DM vectors are seen to obey twofold rotational symmetry.}
\label{FeW110-DM}
\end{figure}

The second order approach is most appropriate for DM interactions that are used for spin wave spectra.
The relativistic effects on the excitation spectra was demonstrated by a systematic comparison of relativistic exchange couplings calculated for Fe/Cu(001) and Fe/Au(001), in Ref.~\onlinecite{udvardi2003first}. The authors showed that strong SOC of Au-$5d$ states gives rise to substantially different magnon spectra for the in-plane and out-of-plane orientation of the magnetization. 
Currently, experimental efforts are concentrated on the studies of DM interaction in such systems~\cite{PhysRevLett.104.137203}).
Indeed, DM interactions can be effectively enhanced on the surfaces of heavy elements due to the combined effect of narrow surface states and substrate induced, large SOC.
By means of {\it explicit} calculations, it was shown that sizeable DM interaction exists between Fe atoms on a W(110) surface~\cite{PhysRevLett.102.207204}.
The so obtained DM vectors are shown in Fig.~\ref{FeW110-DM}. Due to the symmetry of the system, the DM vectors are oriented strictly in the plane of the surface, such that the z-component of the DM vector is zero. 

\begin{figure}[t!]
\includegraphics[width=\columnwidth]{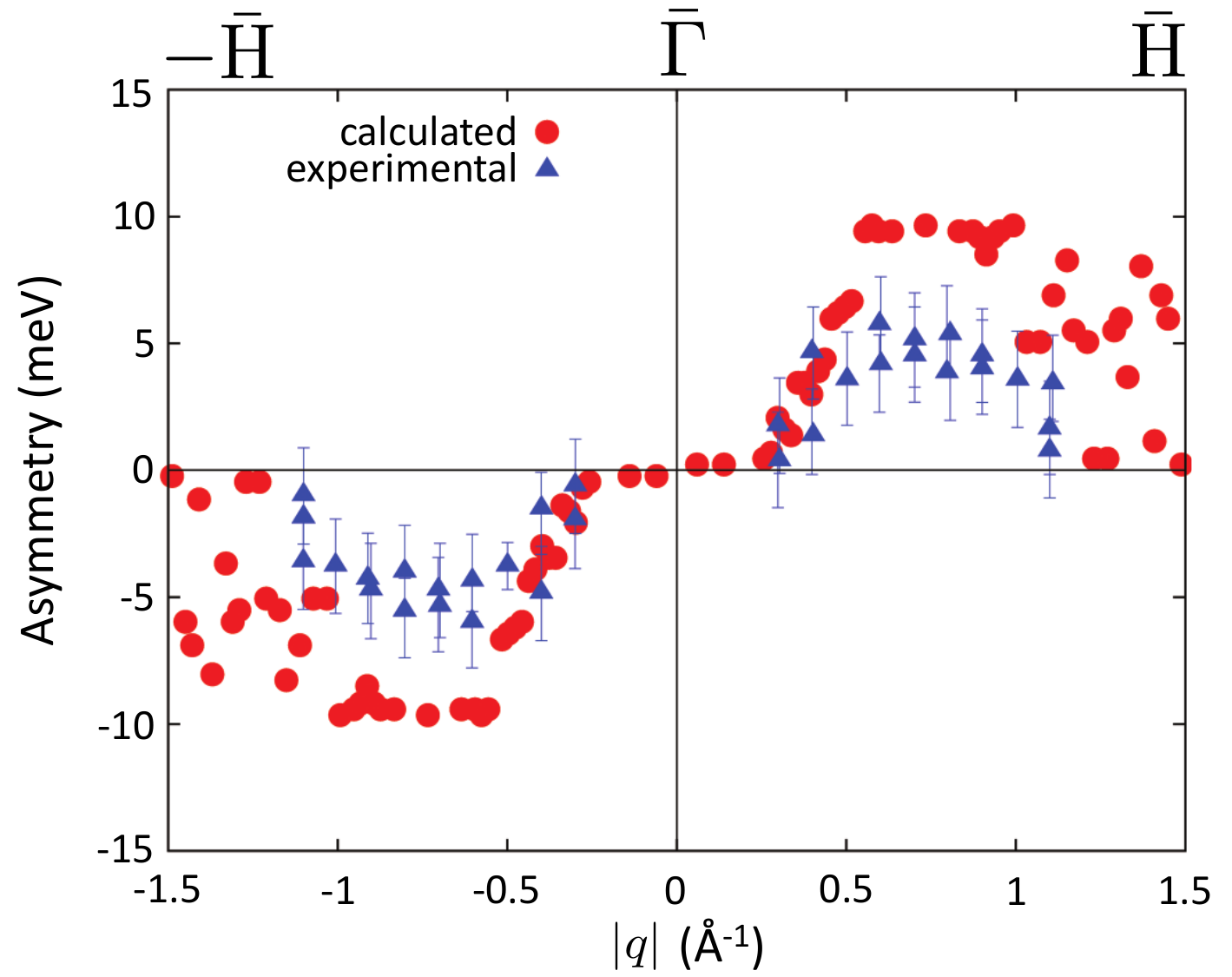}
\caption{(Color online) Experimental~\cite{PhysRevLett.104.137203} and theoretical~\cite{PhysRevLett.102.207204} chiral asymmetry of magnon spectrum of bilayer Fe/W(110) (from Ref.~\onlinecite{PhysRevB.87.144401}).}
\label{chiral-asymmetry}
\end{figure}

Moreover, in Ref.~\onlinecite{PhysRevLett.102.207204} these interactions were predicted to give rise to an asymmetry of the magnon dispersion, i.e. a preferred chirality, with an asymmetry energy defined as $\Delta E=E(\mathbf{q})- E(-\mathbf{q})$. This asymmetry was later confirmed experimentally, in Ref.~\onlinecite{PhysRevLett.104.137203}.
The comparison between computed and measured asymmetry energy for the Fe/W(110) system, also shown in Fig.~\ref{chiral-asymmetry}, was done in Ref.~\onlinecite{PhysRevB.87.144401}.
Without DM interaction, $\Delta E$ is strictly zero for all $q$-vectors.
Thus, $\Delta E$ can be effectively used for quantifying DM couplings in this class of systems, partially due  to high resolution of SPEELS-based experiments and partially due to the theory of evaluating the $D_{ij}$'s.
In this respect, the relativistic interactions between transition metals deposited on Pt(111) have attracted particular attention~\cite{PhysRevB.80.014422,PhysRevB.94.214422,PhysRevB.97.134405,PhysRevB.99.214426}.

As demonstrated, e.g., in Ref.~\onlinecite{PhysRevLett.102.207204}, a Hamiltonian with a $3 \times 3$ tensorial coupling between the spins can be considered where the $x$-component of the moment on atomic site $i$ can interact with the $y$-component of the moment on atomic site $j$, as shown in Eqs.~\eqref{Htens} and~\eqref{tensor1}. These interactions come in a form that is antisymmetric under interchange of $x$- and $y$-indices, which leads to the DM interaction discussed above. However, there is also a symmetric component to the anisotropic exchange interaction, as shown in Eq.~\eqref{tensor2}, that in some cases is significant. As an example we note a recent calculation of symmetric and antisymmetric exchange of CoPt, where the two interactions were found to be of similar size~\cite{borisov2021heisenberg}. As a final remark to this subsection, we note that more references on calculations of DM interactions by various first principles methods  can be found in a recent review focused on this topic \cite{YangDMINature}.

\subsection{Clusters of atoms on surfaces}

\begin{figure}[t!]
\includegraphics[width=\columnwidth]{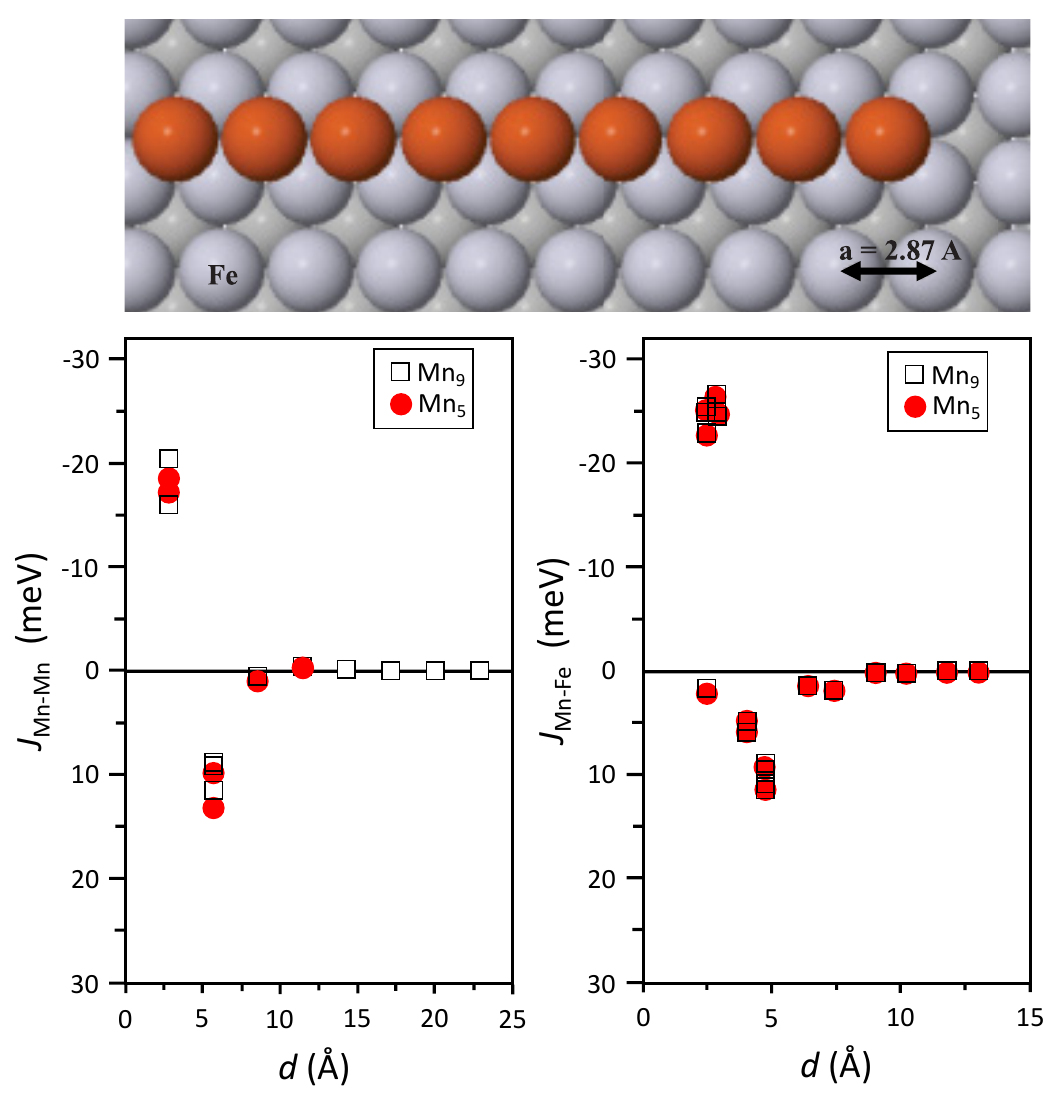}
\caption{(Color online) Geometry of Mn chain shown by orange (dark grey) spheres on a bcc Fe surface (001 orientation) with Fe atoms as light grey spheres (upper panel). Calculated exchange interactions between Mn-Mn pairs and between Fe-Mn pairs (lower panel). Data from Ref.~\onlinecite{igarashi2012first}.}
\label{realspacefig1}
\end{figure}

With the invention of real space methods for calculations of electronic structures~\cite{PhysRevLett.53.2571, haydock1975electronic}, it has become possible to study magnetic exchange interactions of systems without periodic boundary conditions. This is the case when clusters or defects are embedded into a solid or at a surface with the use of LMTO~\cite{PhysRevLett.53.2571} or KKR~\cite{KORRINGA1947392,PhysRev.94.1111} methods. In Fig.~\ref{realspacefig1} we give an example when the real-space LMTO-ASA method was used~\cite{igarashi2012first}, since its implementation is built on a Green function formalism, and the expressions of interatomic exchange (Section~\ref{detailsLKAG}) are, more or less, straightforward to implement. This has, e.g., been published in a series of works~\cite{PhysRevB.62.5293, bergman2007magnetic, ribeiro2011collinear, igarashi2012first, szilva2013interatomic, bezerra2013complex, cardias2016magnetic, szilva2017theory, carvalho2021complex}. The results shown in Fig.~\ref{realspacefig1} are obtained from a calculation based on the LSDA, for an isolated chain of Mn atoms (5 and 9 Mn atoms in the chain was considered) on top of a bcc Fe (001) surface. The results of Fig.~\ref{realspacefig1} show that the interactions are dominantly short ranged between all atom types. In addition, the interactions between Mn-Mn pairs as well as Mn-Fe pairs, are both ferromagnetic and antiferrmagnetic, depending on distance between the atoms. This competition between interactions is responsible for the complex, non-collinear magnetic structures found in this system.

\subsection{\emph{f}-electron systems}
\label{Sect:f-systems}

Unpaired electrons of transition metal $d$ states is the most common source of magnetism, but not the only one.
Many elements with partially filled electronic $f$ shells also exhibit intrinsic magnetic ordering.
Modelling magnetism of such systems is quite challenging, since the $f$ electrons are governed by a sophisticated interplay between strong local correlations, spin-orbit coupling, crystal field effects and hybridization.
Capturing all these ingredients on equal footing is a huge challenge for first-principles electronic structure calculations.

An advantage of the rare-earth elements is that their $4f$ wavefunctions are extremely localized and hybridization effects can be neglected (with few exceptions; La and Ce)~\cite{jensen1991rare}.
Indeed, although $4f$ electrons are responsible for the formation of the local magnetic moments, they do not explicitly participate in the formation of magnetic interactions~\cite{PhysRev.96.99}.
Instead, the $4f$ electrons locally spin-polarize valence $6s6p5d$ orbitals, which mediate the exchange couplings (see e.g. Ref.~\onlinecite{PhysRevB.61.4070}).

Turek \textit{et al.} have shown that by treating the $4f$ electrons as a non-interacting spin-polarized core, a very good description of magnetic interaction can be achieved for hcp Gd from calculations of a FM state~\cite{Turek_2003}.
Gadolinium orders ferromagnetically with an observed total magnetic moment of about 7.6 $\mu_B$ per atom, where 7 $\mu_B$ come from half-filled $f$-shell ($S$=7/2)~\cite{jensen1991rare}. This is reproduced by theory~\cite{colarieti2003origin}. In the work of Ref.~\onlinecite{Turek_2003}, the MFA-based estimate of the $T_c$ was 334 K, which is in excellent agreement with experiment (293 K)~\cite{jensen1991rare}.
Subsequent studies treated the paramagnetic phase of Gd by means of the DLM approach~\cite{Khmelevskyi_2007}.
Although the calculated values of NN $J_{ij}$'s were different from the FM-derived ones, a similar $T_c$ estimate was obtained.

A systematic study of the entire series of late rare-earths was published in Ref.~\onlinecite{PhysRevB.94.085137}.
It was here shown that the calculations incorporating local $4f$ correlations on Hubbard-I  level of approximation (HIA) \cite{PhysRevB.57.6884} are capable of reproducing both electronic valence band excitation spectra, showing well pronounced atomic multiplets, and magnetic interactions of these systems.
Also, full charge self-consistency in DMFT was shown to be of utter importance in order to correctly describe the exchange couplings. Fortunately, the $J_{ij}$'s can already be well described with the $4f$-as-core approach, which is much less computationally demanding.

As Ref.~\onlinecite{PhysRevB.94.085137} shows, the best possible approach to the electronic structure of the rare-earths is the HIA approximation. It reproduces measured electronic structures (both occupied and unoccupied states) and results in realistic magnetic properties. 
As discussed in Ref.~\onlinecite{PhysRevB.94.085137} LDA+U has a significantly worse performance for elemental rare earths. This is shown explicitly in Fig. 14 of Ref.~\onlinecite{PhysRevB.94.085137}, where the valence band of HIA calculations is compared to LDA+U calculations. The latter is seen to not capture experiments while the former does. Also, for compounds such as TbN, HIA gives a much better description of the total energy, equilibrium lattice constant and bulk modulus 
than LDA+U~\onlinecite{PhysRevB.89.205109}.
At the same time, as argued in Ref.~\onlinecite{PhysRevB.94.085137}, a poor-man's treatment of the 4f electrons is to consider them as non-hybridizing core states with a spin-moment constrained according to LS-coupling, an approach also considered in Ref.~\onlinecite{Turek_2003} that successfully reproduced experimental moments and exchange interactions.
In the case where non-local interaction effects are important, that is intersite Coulomb interactions, a reasonable alternative could be the self-interaction corrected local spin density (SIC-LSD) approximation~\cite{Temmerman_2007,Temmerman_1993}. 

\begin{figure}[t!]
\includegraphics[width=\linewidth]{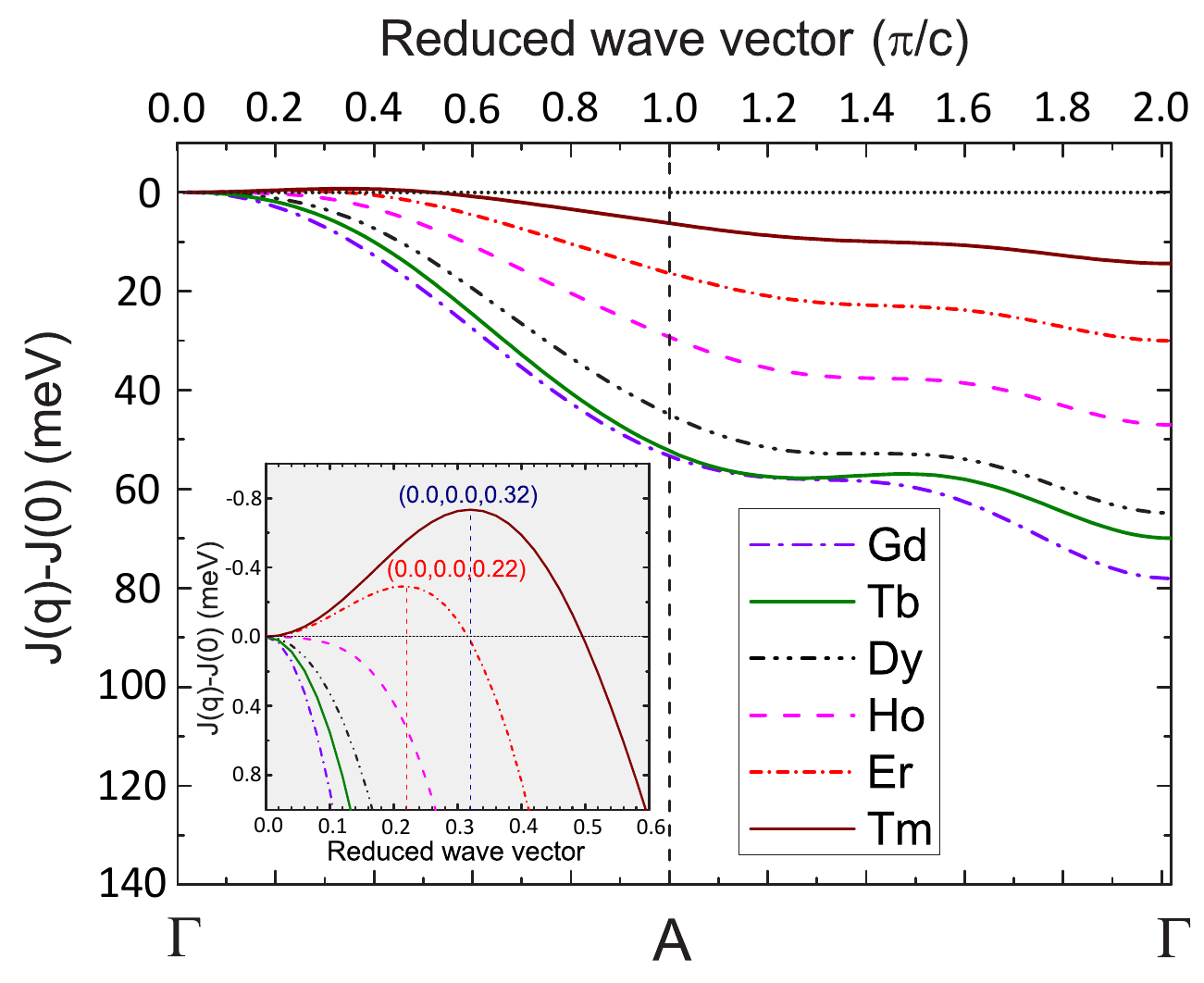}
\caption{(Color online) Fourier transform of the exchange interaction $J(\vec q) - J(0)$ in heavy elemental lanthanides~\cite{PhysRevB.94.085137}. If the minimum correspond to $\Gamma$ point, the ferromagnetic order is preferable.
\label{s7-REs-Jq}
}
\end{figure}

The Fourier transform of the obtained $J_{ij}$'s of the heavy rare-earths, calculated in Ref.~\onlinecite{PhysRevB.94.085137}, is shown in Fig.~\ref{s7-REs-Jq} (as $J(\vec q) - J(0)$).
The minimum value of this curve indicates the ground state magnetic ordering $q$-vector. The results show that Er and Tm have a tendency to have non-collinear magnetic order (similarly to Eu~\cite{PhysRevB.68.224431}).
Holmium is also on the border to have a finite-$q$ maximum, which is compatible with experiments~\cite{jensen1991rare}. In fact, Ho has just passed the border between ferromagnetism and non-collinearity and measurements demonstrate a finite spin-spiral vector. Calculations based on non-hybridizing core states with a 4f spin-moment constrained according to LS-coupling reproduce this experimental finding quite accurately \cite{LNordstr_2000}.

In Ref.~\onlinecite{PhysRevB.94.085137} the ordering temperatures were calculated from the spin Hamiltonian (Eq.~\eqref{neweqn1}) using {\it explicit} calculations of the obtained $J_{ij}$'s. Combined with Monte Carlo simulations this allowed for estimates of the ordering temperature, which was shown to be in good agreement with experiments for all studied, heavy rare earths.
Overall, the ordering temperature was found to decrease linearly with the number of electrons in the $4f$ shell. Also, an intriguing self-induced spin glass state was recently observed experimentally for elemental Nd~\cite{kamber2020self, benj2022verl}.
Ab-initio calculations of the exchange parameters of this element revealed that this is related to the unique exchange interactions of the crystal structure of Nd (dhcp), with competing FM and AFM interactions of equal strength. 

Numerous lanthanide-based systems~\cite{PhysRevB.71.174408,doi:10.1063/1.3365594, PhysRevB.86.104429, PhysRevB.96.100404,Gutfleisch2019} were successfully modelled with the methods reviewed here to calculated interatomic exchange, by treating the $4f$ electrons as core states.
Alternatively, HIA (which also neglects the hybridisation effects) was also used in several works~\cite{PhysRevB.78.060401, PhysRevB.83.205201}. Both theoretical methods to treat the 4f shell have been used to analyse the magnetism of intermetallic compounds containing 4f elements. This class of materials, often referred to as hard magnets, is of particular importance for electromagnetic applications, e.g. the conversion of mechanical energy to electricity or as key components in electrical engines (see e.g. Refs. \cite{coey2010magnetism,Skomski2021, skomskicoey, PhysRevB.48.15812}). The most established permanent magnet is Nd$_2$Fe$_{14}$B (see e.g. Refs. \cite{croat1984high, sagawa1984new, PhysRevB.29.4176}), a material that has had its electronic structure and magnetic properties investigated with DFT \cite{LNordstrom_1993, PhysRevB.41.9697}. In these earlier theories of the electronic structure of compounds containing lanthanides, the 4f shell was treated as a non-hybridized part of a spin-polarized core, where the magnetic state was confined to follow LS coupling, and in general good agreement between theory and observations was found. Calculations using the HIA have also been published for hard magnets, e.g. for SmCo$_5$ \cite{graanas2012charge} where the electronic structure and magnetic properties were found to be in good agreement with experiments \cite{PhysRevB.43.8593}. The reason why calculations based on ''4f as core'' and HIA both reproduce the experimental magnetic properties is connected to the fact that both are faithful to the standard model of the lanthanides \cite{jensen1991rare}, in which the 4f shell basically is an atomic like, non-hybridized entity. In more recent years the theory connected to HIA has been developed to also enable calculations of crystal field splittings of the 4f shell \cite{PhysRevMaterials.6.084410, PhysRevB.101.214433} an important achievement in the field, since the 4f crystal field splitting is connected to the magneto crystalline anisotropy of these systems \cite{jensen1991rare,coey2010magnetism,Skomski2021,skomskicoey}, and therefore for their excellent magnetic performance. When it comes to calculations of interatomic exchange using the LKAG formalism, fewer examples have been published. A notable recent exception is however calculations of the Heisenberg exchange of the compound Ce$_2$Fe$_{17}$ \cite{VISHINA2021161521}, a material that is considered as an alternative to Nd$_2$Fe$_{14}$B for applications as a hard magnet. Its peculiar magnetic properties was explained from electronic structure calculations coupled to the LKAG formalism of interatomic exchange \cite{VISHINA2021161521}. 

The magnetic interactions of $5f$-based compounds are much more complicated due to more pronounced hybridisation and the stronger spin-orbit coupling.
This situation often leads to strong spin-orbital mixing which in turn gives rise to high anisotropy of the spin density, so that approximating spins with dipoles does not apply any longer.
Instead, higher-order multipoles come into play, which have been extensively discussed in the context of actinide oxides~\cite{RevModPhys.81.807} as well as other actinide compounds ~\cite{PhysRevB.80.035121, Cricchio_2011}.
A new methodology has recently been applied to investigate the magnetism of UO$_2$~\cite{PhysRevB.99.094439} and NpO$_2$~\cite{Pourovskiie2025317118}.
In the former case, calculated quadrupolar exchange interactions have successfully predicted stabilization of the 3Q magnetic order in the cubic phase, which could previously only be explained by the presence of lattice distortions. In order to calculate these multipoles, a generalized many-body force theorem has been proposed in Ref.~\onlinecite{PhysRevB.94.115117}.
It relies on the assumption that the correlated states (responsible for magnetism) can be well projected onto atomic wavefunctions (calculated via HIA). Another general approach valid for less correlated actinide compounds adopts the DFT+$U$ method which treats the correlation in mean-field level \cite{PhysRevB.80.035121}. This type of approach is in agreement by large with experiments regarding magnitudes of actinide magnetic moments, with substantial orbital moments and reduced spin moments. The effect behind these calculated magnetic moments can be explained as due to the presence of large high ranked magnetic multipoles \cite{Cricchio_2011}.

\subsection{Transition metal oxides}

Transition metal oxides (TMOs) is a class of materials, which shows outstanding variety of different magnetic orders and interesting physical and chemical properties.
The magnetism of these materials can usually be explained in terms of super-exchange~\cite{PhysRev.115.2,PhysRev.100.564,KANAMORI195987,goodenough1963magnetism, kramers1934interaction} or double-exchange~\cite{PhysRev.82.403} processes. 
The competition between them is, for example, responsible for a particularly rich phase diagram of doped manganites~\cite{PhysRevLett.75.3336}.

\begin{figure}[t!]
\includegraphics[width=0.85\columnwidth]{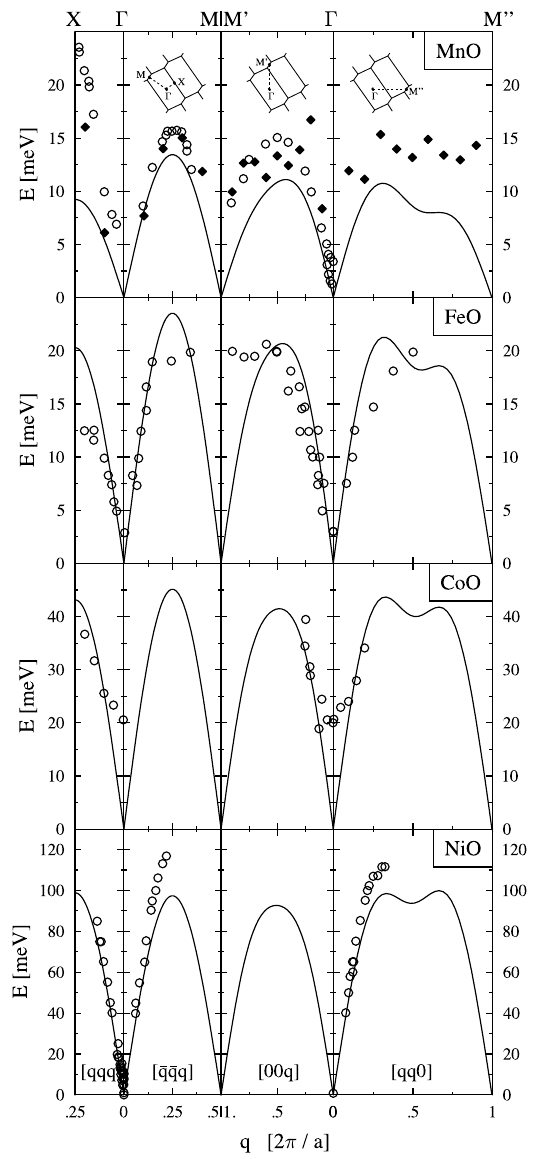}
\caption{Calculated spin-wave dispersion in MnO, FeO, CoO and NiO (solid line) from Ref.~\onlinecite{PhysRevB.80.014408}. Experimental results are shown with symbols.}
\label{s7-TMOs-magnons}
\end{figure}

Oguchi \textit{et al.} calculated magnetic interactions in transition metal monoxides MnO and NiO from first principles at an early  stage~\cite{oguchi1983magnetism,PhysRevB.28.6443}.
The obtained values were too high compared with experiment, which was most likely related to the absence of strong local correlations in the calculation.
Indeed, later it was shown that taking a Hubbard $U$ term into account for the transition metal $3d$ states, substantially improves the situation~\cite{PhysRevB.58.15496,PhysRevB.80.014408,Logemann-2017,PhysRevB.97.184404}.
As was demonstrated in Ref.~\onlinecite{PhysRevB.80.014408}, a systematic, good description of the N{\'e}el temperatures can be obtained in the whole series of transition metal monoxides using a self-interaction-corrected (SIC) version of DFT, although the valence band spectrum of these types of calculations do not agree with observations. The magnon spectra calculated from SIC theory, of MnO, FeO, CoO and NiO~\cite{PhysRevB.80.014408}, is shown in Fig.~\ref{s7-TMOs-magnons}, and one may note excellent agreement with experimental data.
Furthermore, the impact of dynamical correlations (treated in DMFT) on the $J_{ij}$'s of the transition metal monoxides was considered in Refs.~\onlinecite{PhysRevLett.97.266403,PhysRevB.91.125133}.
Although there are some quantitative differences, the results of DMFT are rather close to the LDA+U results, SIC data and values from the HIA (see Fig.~\ref{fig1sect2}).
This result might seem counter-intuitive, since the electronic structure resulting from the different approaches is substantially different~\cite{graanas2012charge}. The likely reason is probably due to the fact that wide-gap TMOs are close to the $U~>>~t$ limit, where the exchange integrals are roughly defined as $t^2/U$, which is similar in the various approaches.
The interatomic exchange extracted from a calculation of the susceptibility, using the GW approximation, also provide results of similar quality~\cite{Kotani_2008}.

Perovskite $3d$ oxides were studied very intensely by Solovyev and co-workers~\cite{PhysRevB.53.7158,solovyev1996crucial,PhysRevLett.82.2959,PhysRevLett.83.2825,PhysRevB.74.054412}.
Despite a huge variety of magnetic phases which are found in these materials, the $J_{ij}$'s are usually consistent with experimental ground state magnetic orders. Clearly this is a very rewarding result. 
Treating the electron interactions beyond  DFT  usually results in better values of the interatomic exchange interactions of these materials. Notably, LaMnO$_3$ may be an interesting exception to this rule.
In Ref.~\onlinecite{PhysRevB.53.7158} it was suggested that the Hubbard $U$ acting on the $e_g$ and $t_{2g}$ orbitals of this compound are different, due to differences in the screening of the two sets of orbitals.
It was thus suggested that having no $U$ is a better choice than adding the same $U$ on the entire set of Mn-$3d$ orbitals for LaMnO$_3$.
However, this result depends on details of the implementation, as discussed in Ref.~\onlinecite{PhysRevB.98.125126}.

Generally, TMO's are regarded as good Heisenberg magnets, in the sense that the spins are localized around $3d$ ions and the interactions are of bilinear kind without strong configuration dependence.  
However, the total energies of different magnetic orders are not always consistent with the Heisenberg model of Eq.~\eqref{neweqn1}, as was reported in several publications~\cite{doi:10.1143/JPSJ.78.054710, Logemann-2017}.
Oxygen polarization is suggested to be responsible for this inconsistency~\cite{PhysRevB.97.184404}.
Moreover, for certain oxides, like LiCu$_2$O$_2$, which have 90$^{\circ}$ superexchange, \textit{direct} exchange also plays a crucial role~\cite{PhysRevB.75.224408}.
Direct exchange interaction was first introduced in the original Heitler-London scheme~\cite{heitler_wechselwirkung_1927}. Oxides with more complex crystal structures~\cite{PhysRevB.73.014418,PhysRevB.78.195110,Jodlauk_2007,Barker_2020,PhysRevB.104.174401}, including the ones with $4d$ and $5d$ elements~\cite{PhysRevB.65.144446,PhysRevB.86.064441,PhysRevB.94.064427} have also been successfully analysed with respect to the interatomic exchange, using the method reviewed here. We also note that multi-spin interactions have been found also to be important in magnetic oxides \cite{PhysRevB.91.165122}.

Finally, it is worth pointing out that for heavy transition metals of the 5$d$ series, the large spin orbit coupling leads to strong spin-orbital mixing as in  the $j=1/2$ pseudo-spin relevant for Ir oxides \cite{PhysRevLett.101.226402}. The arguments in Sect.~\ref{Sect:f-systems} for the case where the pure spin moment has less meaning is valid also for the 5$d$-oxides.

\subsection{Novel 2D magnets}
Magnetism in layered van-der-Waals (vdW)-bonded materials was already reported in the 1960's~\cite{doi:10.1143/JPSJ.15.1664, doi:10.1063/1.1714194}.
For a long time, these materials were not in the focus of researchers, but in recent years they have attracted enormous attention~\cite{Gong2017, Huang2017}.
The discovery of intrinsic 2D magnetic order does not only challenge well-established preconceptions about 2D magnetism~\cite{PhysRevLett.17.1133}, but also offers prospects for building ultra-thin spintronic devices by combining these types of layered materials~\cite{Burch2018, Gibertini2019}.

\begin{figure}[t!]
\includegraphics[width=0.9\linewidth]{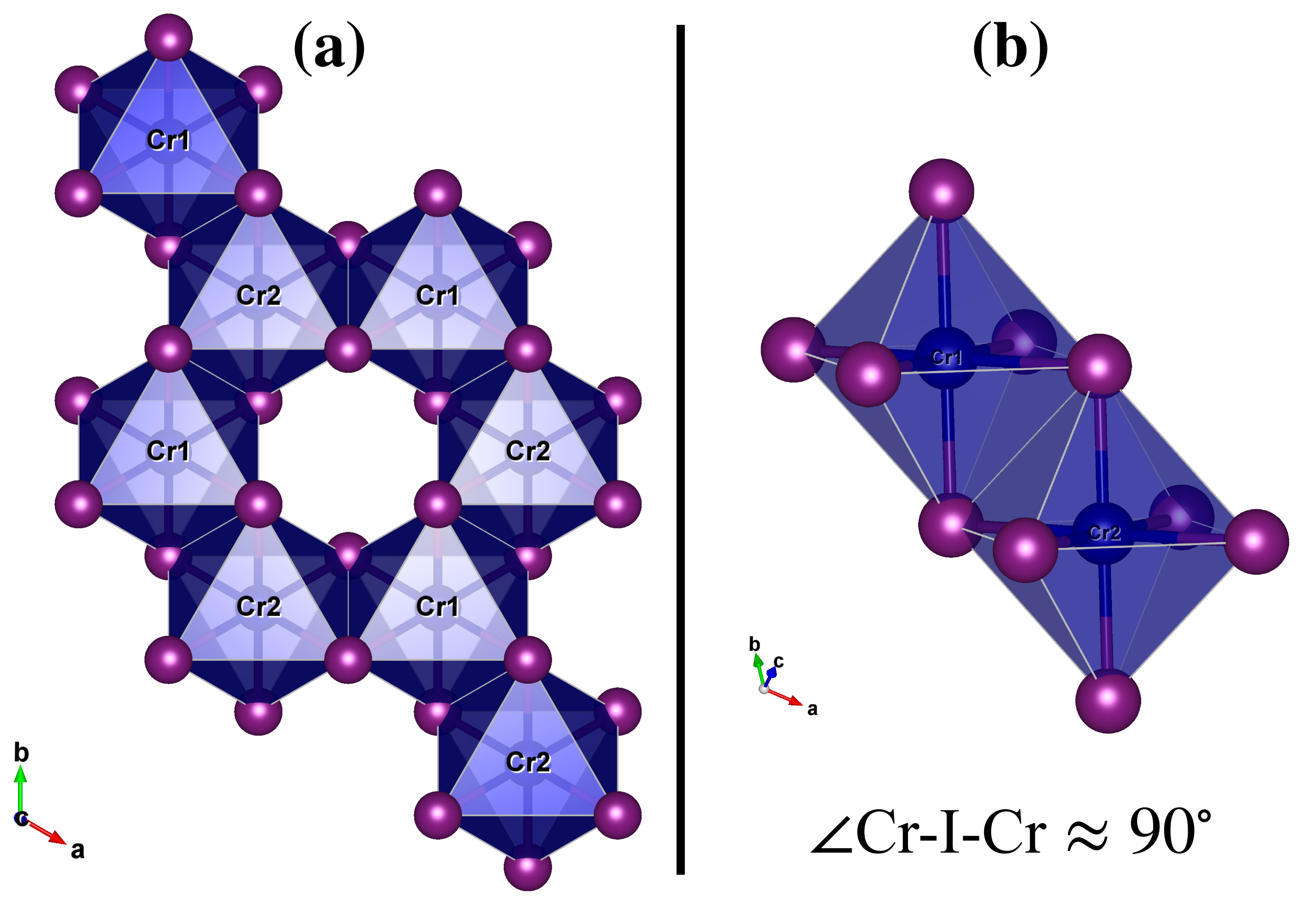}
\caption{(Color online) Left panel: Crystal structure of CrI$_3$ monolayer with I atoms shown by spheres with a light part in their centre, and Cr shown by homogeneously colorod spheres. Cr atoms form honeycomb lattice. Right panel: Local structure of the Cr-I-Cr bond.}
\label{s7-cri3-str}
\end{figure}

CrI$_3$ is the most well-studied example of 2D magnets. It is ferromagnetic and the $T_c$ of its monolayer form is 45 K, which is slightly smaller than that of the bulk form (61 K)~\cite{Huang2017}.
The crystal structure of the monolayer of CrI$_3$ is shown in Fig.~\ref{s7-cri3-str}.
Here Cr atoms are seen to form a honeycomb lattice and each of them is surrounded by six iodine atoms forming an octahedron.
Two I octahedra of the NN Cr atoms are sharing one edge, as illustrated in Fig.~\ref{s7-cri3-str}(b), which results in the Cr-I-Cr bond angle being close to 90$^{\circ}$.
The material is an insulator, so it is expected that the magnetic interactions are defined by a superexchange process involving also the $I 5p$ states.
Nominally, the Cr$^{3+}$ ions should be characterized by a d$^3$ configuration, with a half-filled $t_{2g}$ shell and with the $e_g$ states are completely empty.
In Ref.~\onlinecite{PhysRevB.99.104432}, it was shown that the $e_g$ states form very strong covalent bonds with $I 5p$ orbitals and become effectively occupied by hybridization and band broadening.
As a result, the NN $J_{ij}$'s between Cr atoms are affected by two competing contributions, namely the AFM superexchange between half-filled $t_{2g}$ orbitals and FM superexchange between $t_{2g}$ and $e_g$ states. The latter dominates and results in the overall FM sign of the NN exchange.
The same physics was confirmed to take place also in case of monolayer CrI$_3$~\cite{Kashin_2020, soriano_environmental_2021}.
Since the structure is the same in all three chromium halides, CrX$_3$ (X=$\{$Cl,Br,I$\}$), the complex nature of the NN coupling in these materials explains why its sign is so sensitive to lattice distortions and strain~\cite{PhysRevB.98.144411, PhysRevLett.127.037204, PhysRevB.105.104418}. Similar orbital analysis for the interlayer coupling~\cite{PhysRevMaterials.3.031001} has provided a microscopic description of the theoretically predicted stacking-dependent magnetic order in bilayered CrI$_3$~\cite{sivadas_stacking-dependent_2018}, which was also confirmed experimentally~\cite{bi-cri3-pressure1, bi-cri3-pressure2}. The calculated magnetic interactions in trilayer CrI$_3$ were also suggested to exhibit similar features~\cite{doi:10.1021/acs.jpcc.1c04311}.

One intriguing aspect of bulk CrI$_3$ is a large gap ($\approx$ 4 meV) between the two magnon branches, which was observed experimentally~\cite{PhysRevX.8.041028}.
There are mainly two mechanisms that have been proposed to explain this, namely a large next NN DM interaction~\cite{PhysRevX.8.041028} or NN Kitaev interactions~\cite{PhysRevLett.124.017201}.
Relativistic exchange interactions in bulk and monolayer of CrI$_3$ were studied in Ref.~\onlinecite{PhysRevB.102.115162}.
According to that work, where both conventional DFT as well as two different flavours~\cite{ldau1,lsdau} of LDA+$U$ calculations were employed, both calculated DM interaction and Kitaev terms were found to be too small to induce a substantial gap in the magnon spectrum at the $K$ point.
The work of Ref.~\onlinecite{ke_electron_2021} suggested that this is moderately correlated materials with strong non-local interaction effects and GW approximation combined with a Hubbard $U$, is needed to reproduce the magnon spectrum and most importantly that the $\approx$ 4 meV magnon gap is open by correlation enhanced interlayer coupling. More elaborate discussions on the role of non-local correlation effects and on the importance of charge self-consistency in CrX$_3$ can be found in 
Refs.~\cite{Swagata_GW_2021,Swagata_SCF_2021}. Given the relative young age of this field of magnetic materials, it is likely that other mechanisms will be discussed in the future. 

Other 2D magnets, such as Cr$_2$Ge$_2$Te$_6$~\cite{wang2019giant}, Fe$_3$GeTe$_2$~\cite{jang2020origin}, CrO$X$ ($X$=$\{$Cl,Br$\}$)~\cite{PhysRevMaterials.5.034409}, FeX$_2$~\cite{PhysRevB.103.054409} have also been studied with the help of {\it explicit} calculations of interatomic exchange. Interestingly, the class of 2D versions of Cr$_2$X$_2$Te$_6$ (X=Ge and Si) systems was predicted from ab-initio electronic structure theory~\cite{lebegue2013two} before the experimental realization. 
A rather common feature of the magnetic 2D materials is that they are characterized by relatively strong hybridization of $3d$ orbitals of the transition metals and the $p$ orbitals of the ligand states.
In this case, the choice of electronic states that should be used in the expressions of interatomic exchange (Section~\ref{detailsLKAG}), i.e. the projection scheme, becomes particularly non-trivial.
This issue has been raised by Solovyev and co-workers in Refs.~\onlinecite{PhysRevB.99.104432, wang2019giant}.

\subsection{sp-magnets}

\begin{figure}[t!]
\includegraphics[width=1.0\columnwidth]{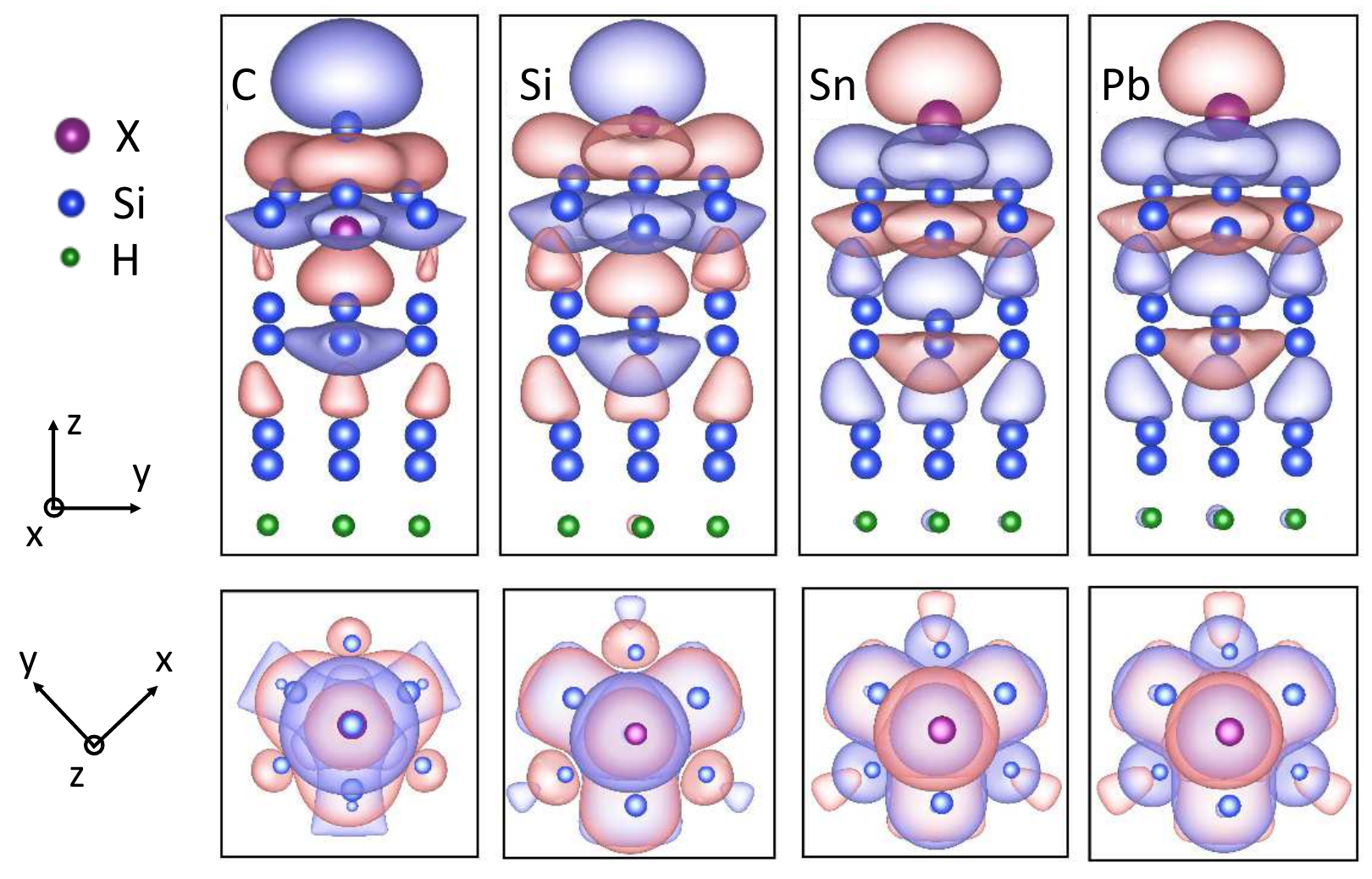}
\caption{(Color online) Maximally localized Wannier functions representing the band crossing the Fermi level in Si(111):$X$ where ${X = \{\text{Sn, C, Si, Pb}\}}$. Big (violet) spheres denote adatoms while isosurfaceses give the
different parts of Wannier functions. 
The figure is taken from Ref.~\onlinecite{PhysRevB.94.224418}.}
\label{s7-sp-WF}
\end{figure}

Another class of systems where the magnetism emerges from highly covalent states is $sp$-magnets.
One example of such materials is semi-hydrogenated or fluorinated graphene~\cite{PhysRevB.94.214411}. Another example is systems of $X$ adatoms (${X = \{\text{Sn, C, Si, Pb}\}}$) deposited periodically on silicon Si(111)~\cite{PhysRevB.60.13328, PhysRevB.68.235332, PhysRevB.71.033403, PhysRevLett.98.126401, Zhang2010, Li2013, PhysRevLett.120.196402}, germanium Ge(111)~\cite{PhysRevLett.79.2859, PhysRevB.64.075405, PhysRevB.104.045126}, or SiC(0001)~\cite{PhysRevLett.114.247602} surfaces. These systems are characterized by the presence of a single relatively narrow half-filled band crossing Fermi level, which is subject to strong local and non-local electron correlations (see, e.g., Refs.~\onlinecite{Hansmann_2013, PhysRevLett.110.166401, PhysRevB.94.224418}). Although this band originates from the $sp$-electrons of adatoms, its wavefunction is highly delocalized and has tails well inside the Si slab, as can be seen in Fig.~\ref{s7-sp-WF}. 
It has been proposed that this band leads to a magnetic instability and various exotic magnetic orders can be realized in these materials. For instance, the low-temperature ground state of Si(111):$X$ systems ranges from a 120$^{\circ}$-N{\'e}el~\cite{PhysRevB.82.035116} to a collinear row-wise~\cite{PhysRevB.83.041104} and different non-collinear chiral~\cite{PhysRevLett.120.196402, Vandelli} magnetic orders.
Moreover, formation of skyrmions is suggested to emerge upon application of a high magnetic~\cite{PhysRevB.94.224418} or a high-frequency laser~\cite{PhysRevLett.118.157201} fields.

Due to the delocalized nature of the orbitals, carrying the magnetic moments, the influence of \textit{direct} exchange mechanism is extremely pronounced.
It gives rise to a ferromagnetic exchange (as expected for wavefunctions with small overlap) and its magnitude is so strong that it may compensate 
indirect exchange contributions~\cite{PhysRevB.94.224418}.
As a result, the isotropic exchange can be effectively suppressed, which results in relatively large $\mid \vec D\mid/J$ ratio, where $\mid \vec D \mid$ is the size of DM interaction~\cite{PhysRevB.94.224418, Vandelli}. 
Since this ratio defines the period of magnetic texture, the suppression of $J$ has led to the proposition that extremely compact skyrmions can be realized~\cite{PhysRevB.98.184425}. 
Ultimately, it has been envisaged that exchange-free skyrmions can also potentially emerge~\cite{nanoskyrmions-2019}.

\subsection{Molecular magnets}

Single molecular magnets is a class of systems where transition metal atoms are embedded in an organic environment~\cite{doi:10.1126/science.265.5175.1054}.
The chemical formula of these systems are quite complicated.
For example,  K$_6$[V$_{15}$As$_6$O$_{42}$(H$_2$O)]$\cdot$8H$_2$O is one of them, which is most often referred to as V$_{15}$ for the sake of brevity. The coupling between the $3d$ magnetic moments often results in a total magnetization that is uncompensated, where the net moment is regarded as a total molecular spin.
Since the interactions between these molecular complexes are very weak their collective behaviour is similar to that of an ensemble of non-interacting point-like magnetic entities.
Thus, molecular magnets not only allow to address fundamental aspects of magnetism on the mesoscale~\cite{PhysRevLett.84.3454, PhysRevLett.84.3458}, but also find their applications in spintronics~\cite{bogani_molecular_2008,mannini_magnetic_2009}.

DFT calculations have been widely used to understand the basic electronic and magnetic properties of molecular magnets (for a review, see Ref.~\onlinecite{https://doi.org/10.1002/pssb.200541490}).
The formalism of Section~\ref{detailsLKAG}) has been widely applied to model magnetic interactions and excitation spectra in the systems, like V$_{15}$~\cite{PhysRevB.70.054417}, Mn$_4$~\cite{doi:10.1021/ic901930w} and Mn$_{12}$~\cite{boukhvalov2002molmag, PhysRevB.89.214422}.
In these works it was shown that a very good description of both electron spectroscopy and magnetic excitations is only possible if the correlation effects of the $3d$ states are taken into account via application of LDA+$U$ approach, similarly to the situation of the $3d$ oxides.
We note that the total energy difference method has also been widely used to extract the $J_{ij}$ parameters for these systems (see e.g.~\cite{https://doi.org/10.1002/chem.200500041,PhysRevB.69.014416}).

\begin{figure}[t!]
\includegraphics[angle=0,width=0.8\columnwidth]{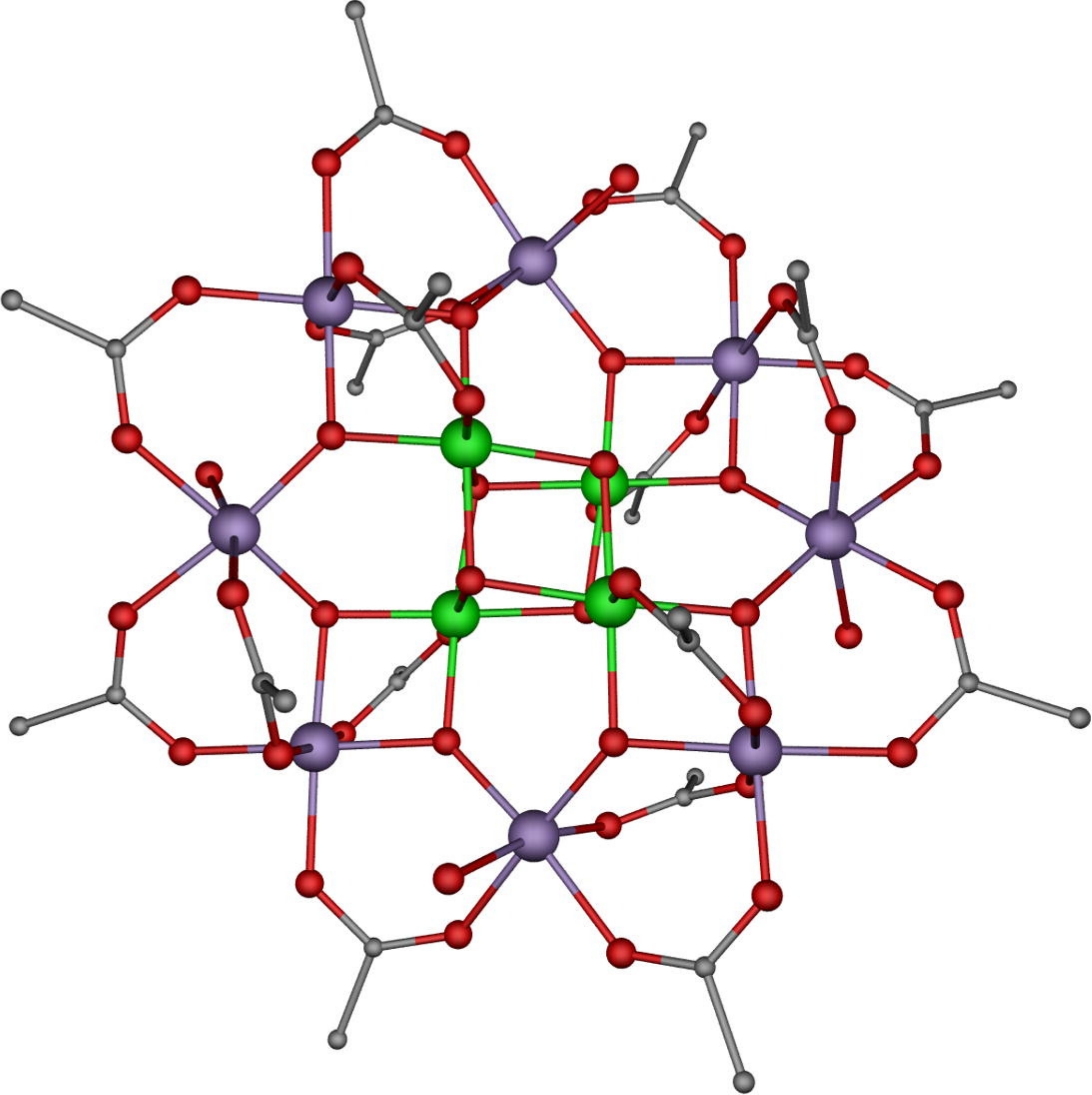}
\caption{(Color online) Crystal structure of Mn$_{12}$-acetate from Ref.~\onlinecite{ZABALALEKUONA2021213984}. Purple atoms (large dark grey spheres) represent Mn$^{3+}$ ($S$=2) ions and the green (large light gray) ones correspond to Mn$^{4+}$ ($S$=3/2). Carbon and oxygen are shown in (small light) grey and (small dark grey) red, respectively, hydrogen atoms have been omitted for the sake of clarity. }
\label{s7-mn12-str}
\end{figure}

The most complete description of exchange interactions in molecular magnets was done for Mn$_{12}$ acetate in Ref.~\onlinecite{PhysRevB.89.214422}.
The structure of this complex, shown in Fig.~\ref{s7-mn12-str}, contains two inequivalent types of Mn atoms having different oxidation states.
Eight Mn$^{3+}$ and four Mn$^{4+}$ ions are coupled antiferromagnetically, which results in the total, uncompensated spin $S$=10.
Contrary to previous works, which only addressed isotropic interactions, Ref.~\onlinecite{PhysRevB.89.214422} adds relativistic exchange interactions and single-ion anisotropy to the picture. 
Overall, the following spin Hamiltonian was considered in Ref.~\onlinecite{PhysRevB.89.214422};
\begin{equation}
\hat{\mathcal{H}}= \mathcal{H}_{DM} + \mathcal{H}_{H}
 + \sum_{i \mu \nu} \hat S^{\mu}_i A^{\mu\nu}_i \hat S^{\nu}_i ,
\label{mn12-ham}
\end{equation}
where $\{\mu,\nu\} \in \{x,y,z\}$ and $A^{\mu\nu}_{i}$
is the single site anisotropy tensor. This is hence a generalization of the sum of Eqs.~\eqref{eqn1} and~\eqref{eqn2}, since magnetic crystalline anisotropy is included.

Since transition metal ions in such molecular complexes have a relatively low-symmetric environment, the Dzyaloshinskii-Moriya interactions can take relatively large values (more than typically encountered in bulk $3d$ oxides).
This was exactly the case for Mn$_{12}$ complex, where the calculations with spin-orbit coupling revealed that the ferrimagnetic arrangement of Mn spins is canted due to the presence of DM interaction.
Combining Heisenberg exchange, DM interaction and magneto crystalline anisotropy, the authors of Ref.~\onlinecite{PhysRevB.89.214422} performed an exact diagonalization study of the complete 12-spin Hamiltonian given by Eq.~\eqref{mn12-ham}, treating all constituent spins as quantum operators.
Thanks to a very efficient realization of a parallel Lanczos algorithm, it was possible to calculate 50 lowest eigenvalues of the system, which allowed for a qualitative comparison with inelastic neutron scattering data and to assign different measured peaks to the transitions from the lowest $S$=10 to excited $S$=9 multiplets.

\section{Out of equilibrium exchange}
\label{Section8}

Femtosecond laser sources provide a unique possibility to manipulate magnetism at ultrafast time scales~\cite{RevModPhys.82.2731, Mentink_2017}. 
In particular, the light irradiation of magnetic materials allows one to modify the value of the exchange interaction~\cite{PhysRevLett.91.227403, subkhangulov2014all, mikhaylovskiy2015ultrafast}.
The idea of small spin rotations as a way to derive effective exchange interactions can be generalized to the case of time-dependent electron Hamiltonians~\cite{secchi2013non}. 
In Ref.~\onlinecite{secchi2013non} the approach was applied to the time-dependent multi-orbital Hubbard model, that is, only onsite interaction was taken into account. The Hamiltonian has the form
\begin{align}
\hat{H}(t) \equiv \hat{H}_T(t) + \hat{H}_V,
\end{align}
where $\hat{H}_T(t)$ is the time-dependent single-particle Hamiltonian
\begin{align}
\hat{H}_T(t) & \equiv \sum_{i_a \lambda_a} \sum_{i_b \lambda_b} T_{i_a \lambda_a, i_b \lambda_b}(t) \sum_{\sigma} \hat{\phi}^{\dagger}_{i_a \lambda_a \sigma} \hat{\phi}_{i_b \lambda_b \sigma}  \nonumber   \\
&  = \sum_a \sum_b T_{a b}(t) \hat{\phi}^{\dagger}_{a} \cdot \hat{\phi}_{b},
\end{align}
where we have grouped the site and orbital indexes according to $a \equiv (i_a, \lambda_a)$ and $b \equiv (i_b, \lambda_b)$, and introduced the spinor fermionic operators
\begin{align}
\hat{\phi}^{\dagger}_a = \left( \begin{matrix} \hat{\phi}^{\dagger}_{a \uparrow} & \hat{\phi}^{\dagger}_{a \downarrow} \end{matrix} \right), \quad \quad
\hat{\phi}_{b} = \left( \begin{matrix} \hat{\phi}_{b \uparrow} \\ \hat{\phi}_{b \downarrow} \end{matrix} \right) .
\end{align}
The interaction Hubbard-like Hamiltonian $\hat{H}_V$ is assumed to be time-independent:
\begin{align}
\hat{H}_V \equiv \frac{1}{2} \sum_{i} \sum_{\lambda_1 \lambda_2 \lambda_3 \lambda_4} \sum_{\sigma \sigma'} V_{\lambda_1 \lambda_2 \lambda_3 \lambda_4} \hat{\phi}^{\dagger}_{i \lambda_1 \sigma} \hat{\phi}^{\dagger}_{i \lambda_2 \sigma'} \hat{\phi}_{i \lambda_3 \sigma'} \hat{\phi}_{i \lambda_4 \sigma}.
\end{align}

The spinor field operators $\hat{\phi}_{a}$ describes both spin and charge dynamics of the interacting itinerant-electron system. To separate supposedly slow spin dynamics from the fast charge dynamics one can introduce the rotational matrices
\begin{align}
R_i(t) \equiv  \left( \begin{matrix} \sqrt{1 - \left| \xi_i(t) \right|^2} &  \xi_i^*(t) \\
                                           -\xi_i(t) & \sqrt{1 - \left| \xi_i(t) \right|^2} \end{matrix} \right),
\label{R matrix}
\end{align}
where we have introduced bosonic fields
\begin{align}
\xi_i(t) \equiv - \text{e}^{\text{i}\varphi_i(t)} \sin\left[ \theta_i(t) / 2 \right] ,
\label{def xi}
\end{align}
with $\theta_i \in \left[ 0, \pi \right[$ , $\varphi_i \in \left[ 0, 2 \pi \right[ $ being the polar angles that determine the spin axis on site $i$ at time $t$; it holds that $R_i^{\dagger}(t) \cdot R_i(t) = 1$.

The matrix $\hat{R}$ provides a transition to the new field operators $\hat{\psi}_{a}$ via the transformation
\begin{align}
& \hat{\phi}_{a}^\dagger(t) = \hat{\psi}_{a}^\dagger(t)  R^{\dagger}_{a}(t),  \nonumber \\
& \hat{\phi}_{a}(t) =  R_{a}(t)  \hat{\psi}_{a}(t), 
\label{spin rotation}
\end{align}
and we assume that in the new coordinate frame the average spin at the site $i$ at time instant $t$, $\left< 0 \left| \hat{\psi}_{a \sigma}^{\dagger}\hat{\boldsymbol{\sigma}}_a \hat{\psi}_{a \sigma} \right| 0 \right>$, is directed along the $z$ axis. Thus, all the information about instant direction of the local spin, $\left< 0 \left| \hat{\phi}_{a \sigma}^{\dagger}\hat{\boldsymbol{\sigma}}_a \hat{\phi}_{a \sigma} \right| 0 \right>$, is passed to the bosonic field, $\xi_i(t)$. 

In the approach of Ref.~\onlinecite{secchi2013non}, the problem is reformulated at the Baym-Kadanoff-Keldysh contour~\cite{KadanoffBaymBook, Rammer86, Leeuwen13, KamenevBook}, which is a common way to proceed in non-equilibrium quantum statistical mechanics. The effective action of the system is expanded, up to the second order, in the angles of spin rotations $\theta_i(t)$, and the result is compared to the effective action of the time-dependent classical Heisenberg model. As a result, we have expressions for the time-dependent exchange parameters which are expressed in terms of single-particle Green functions and electron self-energies. Both the derivation and the final expressions are quite cumbersome, and we refer the reader to the original paper~\cite{secchi2013non}. 
The procedure can be dramatically simplified if we consider electron correlations at the level of time-dependent mean-field approximation~\cite{secchi2016non}. In this case one can derive relatively compact expressions for the time-dependent magnetic susceptibility and extract the exchange parameters from them, similar to the method which we used in Section~\ref{linearresp}. The corresponding expression has the form~\cite{secchi2016non}:
\begin{widetext}
\begin{align}
J_{i j}(t)    = \, &  \mathrm{i} \, \Sigma_{i S}( t )  \lim_{\epsilon \rightarrow 0^+} \int_{0}^{\infty} \mathrm{d} \tau  \mathrm{e}^{ - \epsilon  \tau }        \,  \Sigma_{j S}( t -  \tau/2)
 \left[    \left( G^<_{\downarrow} \right)_{j , t -  \tau/2}^{i ,  t +  \tau/2}
  \left( G^>_{\uparrow} \right)^{j ,  t -  \tau/2}_{i ,  t +  \tau/2}
-   \left( G^>_{\downarrow} \right)^{i ,  t +  \tau/2}_{j ,  t -  \tau/2}   \left( G^<_{\uparrow} \right)_{i , t +  \tau/2}^{j ,  t -  \tau/2}    \right] .
\label{J EXCHANGE}
\end{align}
\end{widetext}
Here $\Sigma_{i S}( t )=\frac{1}{2}\left(\Sigma_{i \uparrow}( t )-\Sigma_{i \downarrow}( t )\right)$ is the spin part of the local self energy which is dependent only on one time, $t$, in the mean-field approximation and $(G^{<,>}_{\sigma})^{i,t}_{j,t'}$ are the corresponding components of the Keldysh two-time Green functions. 

The theoretical description of interacting electronic systems under different time-dependent perturbations, such as an applied electric field, generally requires the use of advanced many-body numerical techniques.
However, there exists a particular type of the perturbation, namely an off-resonant periodic driving, that can be addressed in a relatively simple way.
Indeed, this type of driving brings the system to a non-equilibrium steady state, and the corresponding many-body problem can therefore be solved using existing time-independent approaches.
The standard theoretical framework to describe the periodically driven system is the Floquet formalism~\cite{doi:10.1080/00018732.2015.1055918, RevModPhys.89.011004}.
This method relies on an effective time-independent Hamiltonian description of the non-equilibrium system at stroboscopic times. 
In the limiting case of a high-frequency driving, this effective Hamiltonian can be derived analytically. 
The key idea is to take advantage of a high-frequency feature of the light and use a Magnus-like perturbation expansion that allows one to reduce the time evolution of a quantum state to a time-independent eigenvalue problem with respect to the effective Hamiltoniana~\cite{ITIN2014822, PhysRevLett.115.075301}. 
This can be done as follows:
The time-periodic Hamiltonian, $H(t)$, of the initial problem obeys the time-dependent Schr\"odinger equation
\begin{align}
i\partial_{t} \Psi(\lambda,t) = H(t)\Psi(\lambda,t).
\label{eq:ShrEq_sec8}
\end{align}
One can introduce a dimensionless parameter ${\lambda = \delta{}E/\Omega}$, which compares a certain energy scale ${\delta{}E}$ of the system to the frequency $\Omega$ of the applied field. 
One the tries to find a unitary transformation ${\Psi(\lambda,\tau) = \exp\{-i\Delta(\tau)\}\psi(\lambda,\tau)}$ that removes the time dependence of the Hamiltonian. 
Here, we introduce ${\tau=\Omega{}t}$ and also impose that ${\Delta(\tau) = \sum_{n=1}^{+\infty}\lambda^{n}\Delta_{n}(\tau)}$ with $\Delta_{n}(\tau)$ being a $2\pi$ periodic function. 
Then, the Schr\"odinger equation~\eqref{eq:ShrEq_sec8} can be rewritten as
\begin{align}
i\partial_{t}\psi(\lambda,\tau) = \lambda \bar{\cal H}\psi(\lambda,\tau)
\end{align}
with an effective Hamiltonian
\begin{align}
\bar{\cal H} = e^{i\Delta(\tau)}\bar{H}(\tau) e^{-i\Delta(\tau)} - i\lambda^{-1}e^{i\Delta(\tau)}\partial_{\tau}e^{-i\Delta(\tau)} .
\end{align}
Here, the bar over the Hamiltonian means a normalization on the energy scale ${\delta{}E}$: ${\bar{H}(\tau) = H(\tau)/\delta{E}}$. 
Using the series representation ${\bar{\cal H} = \sum_{n=1}^{+\infty}\lambda^{n}\tilde{H}_{n}}$, one can determine operators $\tilde{H}_{n}$ and ${\Delta_{n}(\tau)}$ iteratively in all orders in $\lambda$. 
The zeroth order term in this representation is given by the time-average over the period of the driving ${\tilde{H}_0 = \langle \bar{H}(\tau) \rangle = \bar{H}_0}$ defined as ${\bar{H}_m = \int_{-\pi}^{+\pi} \frac{d\tau}{2\pi}e^{im\tau}\bar{H}(\tau)}$. 
The first- and the second-order terms $\lambda$ in the effective Hamiltonian are given by the following equations
\begin{align}
\tilde{H}_{1} = &- \frac12\sum_{m\neq0} \frac{\left[\bar{H}_{m}, \bar{H}_{-m} \right]}{m}, \\
\tilde{H}_{2} = &\,\frac12\sum_{m\neq0} \frac{\left[\left[\bar{H}_{m}, \bar{H}_{0}\right],\bar{H}_{-m} \right]}{m^2} ~+\notag\\
&\,\frac13 \sum_{m\neq0}\sum_{n\neq0,m} \frac{\left[\left[\bar{H}_{m}, \bar{H}_{n-m}\right],\bar{H}_{-n} \right]}{mn},
\end{align}
where the square brackets stand for a commutator.
The resulting effective time-independent Hamiltonian describes the stroboscopic dynamics of the system, whereas its evolution between two stroboscopic times is encoded into the time-dependent function $\Delta_{n}(\tau)$.
Importantly, this approach allows one to explore interesting phases of matter and to control different properties of materials through a direct tuning of model
parameters (hopping amplitudes and electronic interactions) that in Floquet theory become explicitly dependent on characteristics of the applied perturbation (see e.g. Refs.~\onlinecite{PhysRevLett.115.075301, PhysRevLett.116.125301, PhysRevB.93.241404, PhysRevB.94.174503, PhysRevB.95.024306, PhysRevLett.118.157201, PhysRevB.101.161101, PhysRevB.102.220301}). 

The introduced formalism can also be used for calculating magnetic exchange interactions under the effect of the high-frequency light irradiation~\cite{mentink2015ultrafast, PhysRevLett.115.075301,      PhysRevLett.118.157201, Mentink_2017, claassen2017dynamical, 10.21468/SciPostPhys.6.3.027}.
In particular, in a strong-coupling limit $U\gg t$, where $U$ is the Coulomb interaction and $t$ is the hopping amplitude, one can make a Schrieffer-Wolff transformation in order to map the derived effective Hamiltonian onto a Heisenberg Hamiltonian~\cite{CHAO1977163, Chao_1977, PhysRevB.37.9753, spalek2007tj}.
In the presence of an external time-dependent perturbation this transformation has been performed in Refs.~\onlinecite{PhysRevLett.116.125301, PhysRevLett.118.157201, PhysRevB.102.220301}.
The resulting isotropic symmetric exchange interaction ${J = J^{K} - J^{D}}$ contains two contributions.
The kinetic exchange interaction corresponds to a usual antiferromagnetic (AFM) superexchange $J^{K} = \tilde{t}^{2}/U$ that exists in equilibrium.
However, out-of equilibrium $J^{K}$ contains the hopping amplitude ${\tilde{t} = t{\cal J}_{0}({\cal E})}$ that is renormalised by the $m$-th order Bessel function of the first kind ${{\cal J}_{m}({\cal E})}$ due to the effect of a high-frequency light irradiation.   
The dimensionless parameter ${{\cal E} = eE_0a/\Omega}$ contains the strength of the laser field $E_0$, the elementary charge $e$, and the lattice constant $a_0$.  
The AFM exchange $J^{K}$ competes with the direct ferromagnetic (FM) exchange interaction ${J^{D} = J^{D}_{\rm bare} + J^{D}_{\rm ind}}$.
The bare part of the direct exchange $J^{D}_{\rm bare}$ stems from the non-local electronic interactions and is present already in equilibrium (see e.g. Refs.~\onlinecite{PhysRevB.75.224408, PhysRevB.78.195110, PhysRevB.88.081405, PhysRevB.94.224418, PhysRevB.94.214411}).
The second part corresponds to the contribution that is induced by the high-frequency light irradiation~\cite{PhysRevLett.115.075301, PhysRevLett.116.125301}
\begin{align}
J^{D}_{\rm ind} = 2t^2U\sum_{m=1}^{+\infty} \frac{{\cal J}^{2}_{m}({\cal E})}{m^2\Omega^2-U^2}.
\end{align}
Remarkably, for the case of a nearly resonant driving $\Omega\simeq{}U/m$~\cite{mentink2015ultrafast, PhysRevLett.115.075301} or when the bare direct exchange is sufficiently large~\cite{PhysRevLett.118.157201} the total isotropic symmetric exchange interaction can be substantially modified by the high-frequency light and can even change sign under certain conditions. 
The introduced formalism can also be extended to other types of magnetic exchange interactions, such as the Dzyaloshinskii-Moriya~\cite{PhysRevLett.118.157201}, the chiral three-spin~\cite{claassen2017dynamical}, and the biquadratic
exchange~\cite{10.21468/SciPostPhys.6.3.027} interaction that all can be tuned by high-frequency laser pulses.
In particular, the light control of magnetic interactions may dynamically induce chiral spin liquids in frustrated Mott insulators~\cite{claassen2017dynamical}.
This may also allow for creation, stabilization, and modifying the shape of skyrmions in materials where these topological spin textures do not exist at equilibrium conditions~\cite{PhysRevLett.118.157201}.  
Moreover, when the isotropic symmetric exchange interaction $J$ is completely suppressed by the light irradiation, one can access a unique phase where magnetic properties of the system are governed solely by the Dzyaloshinskii-Moriya interaction~\cite{nanoskyrmions-2019}. 

\section{Local moment formation and spin-dynamics}
\label{Section9}

Historically, the density functional theory became the standard language for the theory of magnetism and magnetic interactions.
As discussed in previous sections, in this framework exchange interactions can be obtained considering variations of the total energy with respect to
small rotations of magnetic moments starting from equilibrium ground states. 
Despite the success of this approach in describing many magnetic materials, there are several important problems that cannot be addressed using this language. 
Indeed, realistic models for magnetic materials that are derived within DFT are interacting electronic problems.
However, finding a possibility of mapping these electronic models onto Heisenberg-like spin problems is a highly nontrivial task that remains unsolved in the framework of DFT. In addition, calculating the exchange interactions using the magnetic force theorem is based on the assumption that the variation of the magnetization from the ground state magnetic configuration is small, which is frequently not the case, especially for itinerant electron systems.

The most common way to introduce an effective spin model for an interacting electronic problem is based on a Schrieffer-Wolff transformation~\cite{CHAO1977163, Chao_1977, PhysRevB.37.9753, spalek2007tj}, which, strictly speaking, is justified only at integer filling in the limiting case of a very large interaction between electrons. Already $t$-$J$ or $s\text{-}d$ exchange models~\cite{vonsovsky1974magnetism} that are frequently used to describe the physics of a doped Mott insulator cannot be easily mapped onto a pure spin Hamiltonian. 
Moreover, spin degrees of freedom in the transformed problem are described in terms of composite fermionic variables and not in terms of physical bosonic fields as would be desirable for pure spin models.
This results in a need to introduce artificial constraints in order to conserve the length of the total spin.
In addition, one also has to assume that the average value of these composite fermionic variables that define the local magnetization is nonzero.
The latter is hard to justify in a paramagnetic regime, where, generally speaking, it should also be possible to introduce a Heisenberg-like spin model.

Even though already deriving an effective spin problem for interacting electrons is not an easy task, one must do more than that and find a way to introduce a correct equation of motion for spin degrees of freedom.
For localized spins, the classical equation for the spin precession can be obtained by evaluating path integrals over spin coherent states in the saddle-point approximation~\cite{Inomata_book, Auerbach_book, Schapere_book}. 
In this approach, the kinetic term that describes the rotational dynamics of spins originates from the topological Berry phase, for which the conservation of the length of the total spin on each site is a necessary condition. 
For this reason, generalizing the formalism of spin-coherent states to itinerant electronic problems is mathematically a highly non-trivial task.
Nevertheless, finding a way to derive the equation of motion for the local magnetic moment in the framework of electronic problems is crucially important for a correct description of the full spin dynamics of the system. 
Indeed, studying classical spin Hamiltonians allows one to describe only a uniform precession of the local magnetic moment.
Taking into account dissipation effects, e.g. Gilbert damping, requires to couple classical spins to itinerant electrons~\cite{Sayad_2015, PhysRevLett.117.127201}.
In addition, considering classical spins disregards quantum fluctuations of the modulus of the local magnetic moment~\cite{doi:10.1146/annurev-conmatphys-031214-014350} that have been observed in recent experiments~\cite{PhysRevLett.100.205701, merchant2014quantum, jain2017higgs, PhysRevLett.119.067201, hong2017higgs, PhysRevLett.122.127201}.
In analogy with high-energy physics, these fast fluctuations are usually described in terms of a massive Higgs mode~\cite{PhysRevLett.13.321, HIGGS1964132, PhysRevLett.13.508, PhysRevLett.13.585}, while slow spin rotations are associated with Goldstone modes that originate from the broken rotational invariance in spin space.

The problem of describing the physics of the local magnetic moment in the framework of interacting electronic models was intensively studied in late 1970th -- early 1980th~\cite{PhysRevB.16.4032, PhysRevB.16.4048, PhysRevB.16.4058, PhysRevB.19.2626, PhysRevB.20.4584, 19791504, doi:10.1143/JPSJ.49.178, doi:10.1143/JPSJ.49.963, Hasegawa_1983, Edwards_1982, EDWARDS1983213}. 
In these works the local moments were formally introduced into the Hubbard model by using the Hubbard-Stratonovich transformation and making use of a static approximation for the introduced decoupling fields. 
Note that the static approximation in the Hubbard model is closed conceptually to the disordered local moment approach~\cite{oguchi1983magnetism, Pindor_1983, Gyorffy_1985, Staunton_1986, PhysRevLett.69.371, PhysRevB.67.235105} within the density functional theory.
As a result, initial translationally invariant system of interacting electrons is replaced by a single-particle problem of electrons moving in a random magnetic field acting on spins. Fluctuations in the direction of these fields are taken into account thus allowing to go beyond a mean-field approach. For the case of the Hubbard or $s$-$d$ exchange models at Bethe lattices, one can build the effective classical spin Hamiltonian taking into account both Anderson superexchange and Zener double exchange of essentially non-Heisenbergian character~\cite{auslender1982effective, AUSLENDER1982387}.  
This approach allowed one to go far beyond Stoner picture of itinerant-electron magnetism and clarified several important questions such as the origin of Curie-Weiss law for magnetic susceptibility above Curie temperature but it did not result in a complete quantitative theory of magnetism of itinerant electrons. In particular, it does not work at low temperatures where magnon-like dynamical excitations play a crucial role. An attempt to add these effects and to come to an unified picture in a phenomenological way was made by Moriya and collaborators which is summarized in the book~\cite{moriya_book}. 
Several important questions remained yet unsolved, 
e.g., the role of dynamical fluctuations that are known to be responsible for the Kondo effect~\cite{PhysRev.158.570} was not clarified. 

There were also many attempts to address the problem of the spin dynamics of interacting electrons. 
To get the Berry phase, one usually follows a standard route that consists in introducing rotation angles for a quantization axis of electrons~\cite{PhysRevLett.65.2462, PhysRevB.43.3790, doi:10.1142/S0217979200002430, DUPUIS2001617}.
These angles are considered as path integral variables to fulfill rotational invariance in the spin space.
In this case, the Berry phase term appears as an effective gauge field that, however, is coupled to fermionic variables instead of a spin bosonic field.
Considering purely electronic problems makes it difficult to disentangle spin and electronic degrees of freedom.
For this reason, until very recently it was not possible to connect the Berry phase to a proper bosonic variable that describes the modulus of the local magnetic moment.
For the same reason, it was also not possible to introduce a proper Higgs field to describe fluctuations of the modulus of the magnetization.
Indeed, in electronic problems this field is usually introduced by decoupling the interaction term~\cite{sachdev2008quantum, PhysRevX.8.011012, ScheurerE3665, PhysRevX.8.021048, PhysRevX.10.041057}.
First, such decoupling field does not have a clear physical meaning and its dynamics does not necessary correspond to the dynamics of the local magnetic moment.
Also, in actual calculations this effective Higgs field is usually treated in a mean-field approximation assuming that it has a non-zero average value, which is non-trivial to justify in a paramagnetic phase. 
One should also keep in mind that although the decoupling of the interaction term is a mathematically exact procedure, it can be performed in many different ways.
In particular, this fact leads to a famous Fierz ambiguity problem~\cite{PhysRevD.68.025020, PhysRevB.70.125111, Jaeckel} if the decoupling field is further treated in a mean-field approximation.

The aim of this section is to collect all previous achievements in describing spin degrees of freedom of interacting electrons and unify them in a general theory of spin dynamics and effective exchange interactions in strongly correlated systems. Below we discuss how an effective quantum spin action written in terms of physical bosonic variables can be rigorously derived starting from a pure electronic problem. Importantly, we show that this derivation can  be performed without assuming that the average magnetization is nonzero and without imposing any constraints such as artificial magnetic fields.
We illustrate that the introduced effective spin problem allows one to obtain all kinds of exchange interactions between spins and thus to establish relations between the magnetic local force approach and the standard language of response functions.
Further, we show that the corresponding equation of motion for this action correctly describes the dissipative rotational dynamics of the local magnetic moment via the Berry phase and Gilbert damping term, and also takes into account the Higgs fluctuations of the modulus of the magnetic moment. At the end, we introduce a physical criterion for the formation of the local magnetic moment in the system and show that this approach is applicable even in the paramagnetic regime. As a whole, this section provides a solid and mathematically consistent background for a complete description of spin dynamics in strongly correlated electron systems.

\subsection{Derivation of the bosonic action for the fermionic problem}

To introduce a consistent theory of spin dynamics, we will mainly follow the route presented in Refs.~\onlinecite{stepanov2018effective, stepanov2021spin} and will use the action formalism based on Feynman path integral technique as a more appropriate language for treating many-body quantum problems.
We start with a general action for a multi-orbital extended Hubbard model, as a particular example of the strongly-correlated electronic problem that possesses spin dynamics
\begin{align}
&{\cal S}_{\rm latt}[c^{(*)}] = -\int_{0}^{\beta} d\tau \sum_{jj',\sigma\sigma',ll'} \hspace{-0.1cm} c^{*}_{j\tau\sigma{}l} \left[{\cal G}^{-1}\right]^{\tau\tau{}ll'}_{jj'\sigma\sigma'}
c_{j'\tau\sigma'l'}^{\phantom{*}} \notag\\
&+\frac12\int_{0}^{\beta} d\tau \, \Bigg\{\sum_{j,\sigma\sigma',\{l\}} \hspace{-0.1cm} U^{\phantom{*}}_{l_1 l_2 l_3 l_4} c^{*}_{j\tau\sigma{}l_1} c^{\phantom{*}}_{j\tau\sigma{}l_2} c^{*}_{j\tau\sigma'{}l_4} c^{\phantom{*}}_{j\tau\sigma'{}l_3} \notag\\
&\hspace{1.75cm}+\sum_{jj',\varsigma,\{l\}} V^{jj'\varsigma}_{l_1l_2l_3l_4} \rho^{\varsigma}_{j\tau{}l_1l_2}\rho^{\varsigma}_{j'\tau{}l_4l_3} \Bigg\}.
\label{eq:s9_action_latt}
\end{align}
This action is written in terms of annihilation (creation) fermionic Grassmann variables $c^{(*)}_{j\tau\sigma{}l}$ and is considered in the lattice $j$, imaginary time $\tau$, spin ${\sigma = \{\uparrow, \downarrow\}}$, and orbital $l$ space.
The bare (non-interacting) Green function is defined by the inverse of the matrix
\begin{align}
\left[{\cal G}^{-1}\right]^{\tau\tau'll'}_{jj'\sigma\sigma'} = \delta_{\tau\tau'}\left[\delta_{jj'}\delta_{\sigma\sigma'}\delta_{ll'}(-\partial_{\tau}+\mu)-\varepsilon^{\sigma\sigma'}_{jj'll'}\right].
\label{eq:s9_bare_GF}
\end{align}
It contains the chemical potential $\mu$ and the hopping matrix $\varepsilon^{\sigma\sigma'}_{jj'll'}$.
The latter has the following form in the spin space
${\varepsilon^{\sigma\sigma'} = \varepsilon\,\delta_{\sigma\sigma'} + i\,\vec{\kappa}\cdot\vec{\sigma}_{\sigma\sigma'}}$, where the diagonal part $\varepsilon$ of this matrix corresponds to the usual hopping amplitude of electrons. 
The non-diagonal part $\vec{\kappa}$ accounts for the spin-orbit coupling in the Rashba form~\cite{Rashba, PhysRevB.52.10239}. The interacting part of the model action~\eqref{eq:s9_action_latt} consists of the local Coulomb potential $U^{\phantom{*}}_{l_1 l_2 l_3 l_4}$ and the non-local interaction $V^{jj'\varsigma}_{l_1l_2l_3l_4}$ (${V^{jj}=0}$) between electrons in the charge (${\varsigma=c}$) and spin (${\varsigma=s=\{x,y,z\}}$) channels. 
Composite fermionic variables ${\rho^{\varsigma}_{j\tau{}ll'} = n^{\varsigma}_{j\tau{}ll'} - \langle n^{\varsigma}_{ll'} \rangle}$ describe fluctuations of charge and spin densities ${n^{\varsigma}_{j\tau{ll'}} = \sum_{\sigma\sigma'} c^{*}_{j\tau\sigma{}l} \, \sigma^{\varsigma}_{\sigma\sigma'} c^{\phantom{*}}_{j\tau\sigma'l'}}$ around their average values. 

We note that the exchange interactions between spins in the bosonic problem that we aim to derive are non-local, while the dynamics of the magnetic moment is usually described by local Berry and Higgs terms.
For this reason, it would be useful to explicitly decouple local and non-local correlations in the system. The works in
Refs.~\onlinecite{stepanov2018effective, stepanov2021spin} propose to perform this decoupling by considering the local site-independent reference problem that accounts for the local part of the lattice action~\eqref{eq:s9_action_latt}
\begin{align}
&{\cal S}^{(j)}_{\rm imp}[c^{(*)}] = - \iint_{0}^{\beta} d\tau\,d\tau'
\sum_{\sigma,ll'} c^{*}_{j\tau\sigma{}l} \left[g_{0}^{-1}\right]^{ll'}_{\tau\tau'} c^{\phantom{*}}_{j\tau'\sigma{}l'} \notag\\
&+\frac12 \int_{0}^{\beta} d\tau \sum_{\sigma\sigma',\{l\}} U^{\phantom{*}}_{l_1 l_2 l_3 l_4} c^{*}_{j\tau\sigma{}l_1} c^{\phantom{*}}_{j\tau\sigma{}l_2} c^{*}_{j\tau\sigma'{}l_4} c^{\phantom{*}}_{j\tau\sigma'{}l_3},
\label{eq:s9_action_imp}
\end{align}
where 
\begin{align}
\left[g_{0}^{-1}\right]^{ll'}_{\tau\tau'} = \delta_{\tau\tau'}\delta_{ll'}(-\partial_{\tau}+\mu) - \Delta^{ll'}_{\tau\tau'}
\end{align}
is the inverse of the bare Green function of the reference system.
The action~\eqref{eq:s9_action_imp} has the form of the impurity problem of dynamical mean-field theory~\cite{RevModPhys.68.13} and is intended to describe the local correlation effects of the initial lattice action~\eqref{eq:s9_action_latt}. 
This is achieved by introducing a non-stationary hybridization function ${\Delta^{ll'}_{\tau\tau'} = \Delta^{ll'}(\tau-\tau')}$ that aims at capturing the effect of surrounding electrons on a given impurity site.
In general, the impurity problem~\eqref{eq:s9_action_imp} can be considered either in a polarized~\cite{stepanov2018effective} or in a non-polarized~\cite{stepanov2021spin} form, which corresponds to an ordered or paramagnetic solution for the problem, respectively.
At present, we stick to a non-polarized local reference system, which allows one to describe a regime of the system where the average local magnetization is identically zero $\langle n^{s}_{ll'}\rangle_{\rm imp}=0$.
In this case, the hybridization function $\Delta^{ll'}_{\tau\tau'}$ is spin independent, and can be determined from the self-consistent condition ${\frac12\sum_{\sigma}G^{\tau\tau'll'}_{jj\sigma\sigma}=g^{ll'}_{\tau\tau'}}$~\cite{stepanov2021spin} that equates the spin diagonal, local part of the interacting lattice Green function $G^{\tau\tau'll'}_{jj\sigma\sigma}$ and the interacting Green function of the local reference problem $g^{ll'}_{\tau\tau'}$.
A DMFT-like form of the reference system~\eqref{eq:s9_action_imp} allows for the exact solution of this local problem using, e.g., the continuous-time quantum Monte Carlo method~\cite{PhysRevB.72.035122, PhysRevLett.97.076405, PhysRevLett.104.146401, RevModPhys.83.349}.
This implies that corresponding local many-body correlation functions including the full interacting Green function $g^{ll'}_{\tau\tau'}$ and the susceptibility $\chi^{\varsigma\,\tau\tau'}_{l_1l_2l_3l_4}$ can be obtained numerically exact.
This drastically simplifies investigation of many physical effects that are directly related to local electronic correlations, which, in particular, includes formation of the local magnetic moment~\cite{stepanov2021spin}.
We will discuss this point in more details in the last part of this section.

After isolating the local reference system, the non-local correlations are contained in the remaining part of the lattice action ${{\cal S}_{\rm rem}[c^{(*)}] = {\cal S}_{\rm latt}[c^{(*)}] - \sum_{j}{\cal S}^{(j)}_{\rm imp}[c^{(*)}]}$.
However, the local and non-local correlation effects are not yet disentangled, because ${\cal S}_{\rm imp}[c^{(*)}]$ and ${\cal S}_{\rm rem}[c^{(*)}]$ are written in terms of the same fermionic Grassmann variables.
Calculating any physical observable using the present form of the lattice action will immediately mix these correlations up.
After that, a separation of them is possible only by a complex resummation of corresponding contributions to a Feynman diagrammatic expansion~\cite{PhysRevB.91.165134, BRENER2020168310}.  
As an alternative, there exists a simpler way to completely disentangle local and non-local correlation effects.
The idea consists in integrating out the reference system as proposed in the dual fermion (DF)~\cite{PhysRevB.77.033101, PhysRevB.79.045133, PhysRevLett.102.206401} and the dual boson (DB)~\cite{Rubtsov20121320, PhysRevB.90.235135, PhysRevB.93.045107, PhysRevB.94.205110, PhysRevB.100.165128} theories.
To this aim, we first rewrite the non-local part of the action in terms of new fermionic ${c^{(*)} \to f^{(*)}}$ and bosonic ${\rho^{\varsigma}\to\phi^{\varsigma}}$ variables by means of the Hubbard-Stratonovich transformation~\cite{stratonovich1957method, PhysRevLett.3.77}. After this transformation, the lattice action ${\cal S}_{\rm latt}[c^{*},f^{*},\phi^{\varsigma}]$ depends on two fermionic and one bosonic variables.
Original Grassmann variables $c^{(*)}$ are contained only in the local part of the lattice action, which includes the impurity problem~\eqref{eq:s9_action_imp}, and thus can be integrated out.

Before making this integration, one should recall that isolating local correlation effects should help to correctly describe dynamics of spin degrees of freedom.
In general, spin dynamics might have a non-trivial form, since it involves a combination of a slow spin precession and fast Higgs fluctuations of the modulus of the local magnetic moment.
For this reason, it is more convenient to treat these two contributions separately.
In electronic systems, the Berry phase term that describes the uniform spin precession is commonly obtained by transforming original electronic variables to a rotating frame~\cite{PhysRevLett.65.2462, PhysRevB.43.3790, doi:10.1142/S0217979200002430, DUPUIS2001617}.
This can be achieved by introducing a unitary matrix in the spin space
\begin{align}
R_{j\tau} = 
\begin{pmatrix}
\cos(\theta_{j\tau}/2) & - e^{-i\varphi_{j\tau}}\sin(\theta_{j\tau}/2) \\
e^{i\varphi_{j\tau}}\sin(\theta_{j\tau}/2) & \cos(\theta_{j\tau}/2)
\end{pmatrix}
\end{align}
and making the corresponding change of variables ${c_{j\tau{}l} \to R_{j\tau} c_{j\tau{}l}}$, where ${c_{j\tau{}l} = (c_{j\tau{}l\uparrow}, c_{j\tau{}l\downarrow})^{T}}$. Rotation angles $\Omega_{R} = \{\theta_{j\tau}, \varphi_{j\tau}\}$ are considered as site $j$ and time $\tau$ dependent variables.
Introducing an additional functional integration over them allows one to preserve the rotational invariance in the spin space.
As a consequence, the modified lattice action takes the following form; ${{\cal S}_{\rm latt}[c^{*}, f^{*}, \phi^{\varsigma},\Omega_{R}]}$.

The Berry phase arises from the local impurity problem that upon rotation becomes~\cite{stepanov2021spin}
\begin{align}
{\cal S}^{(j)}_{\rm imp}[c^{(*)}] &\to 
{\cal S}^{(j)}_{\rm imp}[c^{(*)}] +
\int_{0}^{\beta} d\tau \sum_{s,l} {\cal A}^{s}_{j\tau} \rho^{s}_{j\tau{}ll}.
\label{eq:s9_Berry_term}
\end{align}
The ${z}$ component of an effective gauge field ${\cal A}^{s}_{j\tau}$ has the desired form of the Berry phase term ${{\cal A}^{z}_{j\tau} = \frac{i}{2} \dot{\varphi}_{j\tau} (1-\cos\theta_{j\tau})}$.
To exclude other components of the gauge field from consideration, one usually assumes that the rotation angles $\Omega_{R}$ correspond to the spin-quantization axis of electrons. 
In this case, the composite fermionic variable in the spin channel $\rho^{s}$ is replaced by its $z$ component $\rho^{z}$ which is coupled to the ``correct'' component of the gauge field ${\cal A}^{z}_{j\tau}$.
Proceeding in this direction leads to several problems.
Associating rotation angles with the spin-quantization axis is non-trivial to formulate in a strict mathematical sense.
In Refs.~\onlinecite{doi:10.1142/S0217979200002430, DUPUIS2001617} it was done introducing a slave boson approximation.
However, there is no guarantee that the {\it average magnetization} on a given lattice site will also point in the $z$ direction.
Indeed, the spin-quantization axes on different sites may point in different directions, which may induce an effective mean magnetic field that will change the direction of the magnetization on a given site.
In particular, this does not allow one to replace the composite fermionic variable $\rho^{z}$ by its average value in the Berry phase term~\eqref{eq:s9_Berry_term}.
Moreover, in the paramagnetic phase this replacement does not make sense, because the average magnetization in this case is identically zero.
Finally, in Eq.~\eqref{eq:s9_Berry_term} the effective gauge field ${\cal A}^{s}_{j\tau}$ is coupled to a composite fermionic variable $\rho^{s}$ instead of a proper vector bosonic field that describes fluctuations of the local magnetic moment.
This representation of spin degrees of freedom does not conserve the length of the total spin, which is a necessary condition for a correct description of a spin precession.

We emphasize that the rotation angles cannot be associated with the direction of the newly introduced bosonic field for spin degrees of freedom $\phi^{s}$.
This field enters the lattice action as an effective quantum magnetic field that polarises the electrons~\cite{stepanov2018effective, stepanov2021spin} and is frequently associated with the Higgs field~\cite{sachdev2008quantum, PhysRevX.8.011012, ScheurerE3665, PhysRevX.8.021048, PhysRevX.10.041057}.
However, this effective bosonic field is introduced as the result of a Hubbard-Stratonovich transformation and does not have a clear physical meaning.  
Moreover, even if it would be possible to associate $\phi^{s}$ with the physical Higgs field, its dynamics would not necessarily correspond to the dynamics of the local magnetic moment.
All these observations suggest that the idea to describe the spin precession in terms of rotation angles is very appealing, but one has to find a way to relate these angles to the direction of the local magnetic moment and not to the spin-quantization axis or to the effective Higgs field.

After transforming the original electronic variables $c^{(*)}$ to a rotating frame they can finally be integrated out, which results in the, so-called, dual boson action ${{\cal S}_{\rm latt}[f^{(*)},\phi^{\varsigma},\Omega_{R}]}$~\cite{Rubtsov20121320, PhysRevB.90.235135, PhysRevB.93.045107}.
In this action, bare propagators for the fermionic $f^{(*)}$ and bosonic $\phi^{\varsigma}$ variables are purely non-local and explicitly depend on rotation angles $\Omega_{R}$~\cite{stepanov2021spin}.   
All local correlations are absorbed in the interaction part of the fermion-boson action $\tilde{\cal F}[f^{(*)},\phi^{\varsigma},\Omega_{R}]$ that consist of all possible fermion-fermion, fermion-boson, and boson-boson vertex functions of the local reference problem~\eqref{eq:s9_action_imp}.
To proceed further, we truncate the interaction at the two particle level and keep only the four-point (fermion-fermion) $\Gamma$ and three-point (fermion-boson) $\Lambda^{\varsigma}$ vertices.
This approximation is widely used in the dual fermion approach~\cite{PhysRevB.77.033101, PhysRevB.79.045133, PhysRevLett.102.206401}, the dual boson method~\cite{Rubtsov20121320, PhysRevB.90.235135, PhysRevB.93.045107, PhysRevB.94.205110, PhysRevB.100.165128}, and the recently introduced dual triply irreducible local expansion ($\text{D-TRILEX}$)~\cite{PhysRevB.100.205115, PhysRevB.103.245123, 2022arXiv220406426V}, including their diagrammatic Monte Carlo realizations~\cite{PhysRevB.94.035102, PhysRevB.96.035152, PhysRevB.102.195109} that provide results in a good agreement with the exact benchmark methods~\cite{PhysRevB.94.035102, PhysRevB.96.035152, PhysRevB.97.125114, PhysRevX.11.011058, PhysRevB.102.195109, PhysRevB.103.245123}.

Integrating out the reference system not only disentangles local and non-local correlations, but also allows one to get rid of composite fermionic variables $\rho^{\varsigma}$ that are no longer present in the dual boson action ${{\cal S}_{\rm latt}[f^{(*)},\phi^{\varsigma},\Omega_{R}]}$. 
Now, charge and spin degrees of freedom are described by a proper bosonic field $\phi^{\varsigma}$ that has a well-defined propagator and a functional integration over them. 
Moreover, in this action the gauge field ${\cal A}^{s}_{j\tau}$ is coupled (up to a certain multiplier) to the spin component of this bosonic field $\phi^{s}$~\cite{stepanov2021spin}. 
However, as discussed above, the bosonic variable $\phi^{\varsigma}$ does not have a clear physical meaning.
The way of introducing a physical bosonic variable was proposed in Ref.~\onlinecite{stepanov2018effective} and was inspired by works~\cite{doi:10.1142/S0217979200002430, DUPUIS2001617} where a similar transformation was performed for fermionic fields.
The idea consists in introducing a source field $\eta^{\varsigma}$ for the {\it original composite fermionic variable} $\rho^{\varsigma}$ that describes fluctuations of charge and spin densities. 
Then, after obtaining the dual boson action one performs one more Hubbard-Stratonovich transformation ${\phi^{\varsigma}\to\bar\rho^{\varsigma}}$ that makes $\eta^{\varsigma}$ the source field for the resulting {\it physical bosonic field} $\bar\rho^{\varsigma}$.
Further, unphysical bosonic fields $\phi^{\varsigma}$ are integrated out, which leads to the fermion-boson action ${{\cal S}_{\rm latt}[f^{(*)},\bar\rho^{\varsigma},\Omega_{R}]}$.

Importantly, the derived fermion-boson action has a simpler form compared to the dual boson action ${{\cal S}_{\rm latt}[f^{(*)},\phi^{\varsigma},\Omega_{R}]}$. 
Indeed, the interaction part of the fermion-boson action contains only the three point vertex function $\Lambda^{\varsigma}$.
The four-point vertex $\Gamma$ that is present in the dual boson action is approximately cancelled by the counterterm that is generated during the last Hubbard-Stratonovich transformation~\cite{stepanov2018effective, PhysRevB.100.205115}. 
As a result, the fermion-boson action ${{\cal S}_{\rm latt}[f^{(*)},\bar\rho^{\varsigma},\Omega_{R}]}$ takes the form of an effective $t$-$J$ or $s\text{-}d$ exchange model~\cite{vonsovsky1974magnetism} that describes local charge and spin moments $\bar\rho^{\varsigma}$ coupled to itinerant electrons $f^{(*)}$ via the local fermion-boson vertex function $\Lambda^{\varsigma}$. 
Moreover, in this action the gauge field ${\cal A}^{s}_{j\tau}$ is coupled to the spin component of the physical bosonic field $\bar\rho^{s}$ as desired for a correct description of the rotational dynamics of the local magnetic moment~\cite{stepanov2021spin}. 

We note that at this point all parameters of the fermion-boson action, including the coupling of the gauge field ${\cal A}^{s}_{j\tau}$ to the bosonic field $\bar\rho^{s}$, explicitly depend on the rotation angles $\Omega_{R}$.
From the very beginning, these angles are introduced to account for the spin precession explicitly.
For this reason, $\Omega_{R}$ should be related to the direction of the local magnetic moment, which in the fermion boson action is defined by a bosonic vector field $\bar\rho^{s}$.  
It is convenient to rewrite the latter in spherical coordinates as ${\rho^{s}_{j\tau{}ll'} = M^{\phantom{*}}_{j\tau{}ll'} e^{s}_{j\tau}}$, where $M_{j\tau{}ll'}$ is a scalar field that describes fluctuations of the modulus of the orbitally-resolved local magnetic moment. 
In this expression we assume that the multi-orbital system that exhibits a well-developed magnetic moment is characterised by a strong Hund's exchange coupling that orders spins of electrons at each orbital in the same direction.
Therefore, the direction of the local magnetic moment in the system is
defined by the orbital-independent unit vector $\vec{e}_{j\tau}$, e.g. described by a set of polar angles ${\Omega_M=\{\theta'_{j\tau},\varphi'_{j\tau}\}}$ associated with this vector.
It has been shown in Ref.~\onlinecite{stepanov2021spin} that taking the path integral over rotation angles $\Omega_{R}$ in the saddle point approximation allows one to equate these two sets of angles ${\Omega_R=\Omega_M}$ that from now on define the direction of the local magnetic moment.
After that, the remaining dependence on rotation angles can be eliminated from fermionic parts of the fermion-boson action.
This can be achieved in the adiabatic approximation that assumes that characteristic times for electronic degrees of freedom are much faster than for spin ones.

The bosonic problem that describes the behavior of charge and spin densities can be obtained integrating out fermionic fields $f^{(*)}$.
The fermion-boson action is Gaussian in terms of these fields, so this integration can be performed exactly.
The resulting bosonic action takes the following final form~\cite{stepanov2021spin}
\begin{widetext}
\begin{align} 
{\cal S}_{\rm latt}
= 
& - \mathrm{Tr} \ln \left[ \big[\tilde{\cal G}^{-1}\big]^{\tau\tau'll'}_{jj'\sigma\sigma'}  - \delta^{\phantom{*}}_{jj'} \int^{\beta}_{0} d\tau_1 \sum_{\varsigma,l_1l'_1} \, \sigma^{\varsigma}_{\sigma\sigma'} \Lambda^{\varsigma\,\tau\tau'\tau_1}_{ll'l_1l'_1} \, \bar\rho^{\varsigma}_{j\tau_1l'_1l_1} \right]
+\frac12\int_{0}^{\beta} d\tau \sum_{jj',\varsigma,\{l\}} \bar\rho^{\varsigma}_{j\tau{}ll'} \, V^{jj'\varsigma}_{ll'l_1l'_1} \, \bar\rho^{\varsigma}_{j'\tau{}l'_1l_1}
\notag\\
&- \frac12 \iint_{0}^{\beta} d\tau\,d\tau' \sum_{j,\{l\}}
\Bigg\{
\bar\rho^{c}_{j\tau{}ll'} \left[\chi^{c\,-1}\right]^{\tau\tau'}_{ll'l_1l'_1} \bar\rho^{c}_{j\tau'l'_1l_1}
+ M^{\phantom{*}}_{j\tau{}ll'} \left[\chi^{z\,-1}\right]^{\tau\tau'}_{ll'l_1l'_1} M^{\phantom{*}}_{j\tau'l'_1l_1}
\Bigg\}
+ \int_{0}^{\beta} d\tau \sum_{j} {\cal A}^{z}_{j\tau} {\cal M}^{\phantom{*}}_{j\tau}.
\label{eq:s9_Boson_action}
\end{align}
\end{widetext}
Importantly, in this action the modulus of the total magnetic moment ${{\cal M}_{j\tau} = \sum_{l}M^{\phantom{*}}_{j\tau{}ll}}$ is coupled only to the $z$ component of the effective gauge field ${\cal A}^{z}_{j\tau}$ that gives exactly the desired Berry phase term.
Other components of the gauge field disappear upon associating rotation angles with the direction of the local magnetic moment.

\subsection{Exchange interactions in many-body theory and relation to other approaches}

Before introducing the explicit expression for the exchange interaction it is worth noting that an unambiguous definition for this quantity does not exist. The exchange interactions are internal parameters of the model and thus depend on the particular form of the considered Hamiltonian.
In its turn, the latter crucially depends on the downfolding scheme used to map the interacting electronic problem onto an effective bosonic (i.e., spin) model.
For instance, it has been shown that considering small local variations from the ordered magnetic state leads to the bilinear exchange interaction that depends on the magnetic configuration, and the resulting spin Haimiltonian also contains higher-order non-linear exchange interactions that are not negligible \emph{a priori}~\cite{auslender1982effective, AUSLENDER1982387}.
On the other hand, one can try to map the interacting electronic problem onto a global Heisenberg-like spin model with only bilinear exchange interaction. In this case, the value of the bilinear exchange might be different compared to the one of the non-linear spin model.

However, both forms of the spin Hamiltonian are useful. 
The form that contains non-linear exchange interactions better reproduces the spectrum of spin waves~\cite{PhysRevB.64.174402}.  
On the other hand, the Heisenberg Hamiltonian is a standard model for atomistic spin simulations and gives reasonable thermodynamic properties of the system~\cite{eriksson2017atomistic}. In order to establish connection between different definitions for the exchange interaction, we start with the bosonic action~\eqref{eq:s9_Boson_action} derived above.
In this action local and non-local correlation effects are completely disentangled by construction of the theory.
The first line in Eq.~\eqref{eq:s9_Boson_action} describes non-local exchange interactions between charge $\bar\rho^{c}$ and spin $\bar\rho^{s}$ densities.
The first term in this expression is responsible for all possible kinetic exchange processes (including higher-order ones) mediated by electrons.
This can be illustrated by directly expanding the logarithm function to all orders in $\bar\rho^{\varsigma}$ variables.
Since this expansion is performed in terms of the bosonic variables that correspond to charge and magnetic densities, the resulting bilinear and non-linear exchange interactions are well defined.
This expansion is essentially different from the one performed in terms of rotation angles in DFT-based formalisms. Indeed, the latter is based on the magnetic force theorem (see Section~\ref{secmft}), which cannot be used however for the discussion of higher-order expansion terms in rotation angle. 
The situation is similar to that in the problem of calculations of elastic moduli of solids in density functional: whereas the first-order variations with respect to deformation are very simple and can be calculated according to the local force theorem, the second-order variations contain a lot of additional terms related to the differentiation of the double-counting contributions~\cite{zein1984}. 
At the same time, the effective bosonic action discussed here is based on formally exact transformations. 

The bilinear exchange interaction $J^{\varsigma\varsigma'}_{jj'}$ is given by the second order of the expansion
\begin{gather}
J^{\varsigma\varsigma'\tau\tau'}_{jj'll'l''l'''} = 
\int_{0}^{\beta} \{d\tau_i\} \sum_{\{\sigma_i\},\{l_i\}}
\notag\\ 
\star ~ \Lambda^{*\,\varsigma\,\tau\tau_1\tau_2}_{ll'l_1l_2} \, \tilde{\cal G}^{\tau_1\tau_3l_1l_3}_{jj'\sigma_1\sigma_3} \, \tilde{\cal G}^{\tau_4\tau_2l_4l_2}_{j'j\sigma_4\sigma_2} \, \Lambda^{\varsigma'\tau_3\tau_4\tau'}_{l_3l_4l''l'''},
\label{eq:s9_J}
\end{gather}
where a ``transposed'' three-point vertex function
${\Lambda^{*\,\varsigma\,\tau_1\tau_2\tau_3}_{l_1l_2l_3l_4} = \Lambda^{\varsigma\,\tau_3\tau_2\tau_1}_{l_4l_3l_2l_1}}$ is introduced to simplify notations.
$\tilde{\cal G}$ stands for the non-local Green function given by the difference between DMFT $G$ and impurity $g$ Green functions
\begin{align}
\tilde{\cal G}^{\tau\tau'll'}_{jj'\sigma\sigma'} = G^{\tau\tau'll'}_{jj'\sigma\sigma'} - \delta^{\phantom{*}}_{jj'}\delta^{\phantom{*}}_{\sigma\sigma'} g^{ll'}_{\tau\tau'}.
\label{eq:s9_G_dual}
\end{align}
The DMFT Green function corresponds to the bare lattice Green function~\eqref{eq:s9_bare_GF} dressed in the exact self-energy $\Sigma^{\rm imp}$ of the local reference problem~\eqref{eq:s9_action_imp}~\cite{RevModPhys.68.13}.
According to the self-consistency condition, the local part of the DMFT Green function is identically equal to the exact local Green function $g$ of the reference problem.

The diagonal part of the bilinear exchange interaction is given by the Heisenberg exchange $J^{ss}_{jj'}$ for spin~\cite{stepanov2018effective, stepanov2021spin} and the Ising interaction $J^{cc}_{jj'}$ for charge~\cite{stepanov2019effective} densities.
The latter will be discussed in details in Section~\ref{Section10}.
The non-diagonal ${J^{s\neq{}s'}_{jj'}}$ components give rise to the Dzyaloshinskii-Moriya and the symmetric anisotropic interactions (see, e.g., Ref.~\onlinecite{PhysRevB.52.10239}) that may appear in the system due to spin-orbit coupling.
These kinetic exchange interactions compete with the bare non-local electron-electron interaction $V^{\varsigma}_{jj'}$ that plays a role of a direct exchange between charge and spin densities.
This makes the total, non-local bilinear exchange interaction to have the form
\begin{align}
{\cal I}^{\varsigma\varsigma'}_{jj'} = J^{\varsigma\varsigma'}_{jj'} + \delta^{\phantom{*}}_{\varsigma\varsigma'} V^{\varsigma}_{jj'}.
\label{eq:s9_exch}
\end{align}
Importantly, the non-local interaction $V^{\varsigma}_{jj'}$ enters the bosonic problem in the same way as it was introduced in the initial lattice action~\eqref{eq:s9_action_latt}.
We also note that the direct spin-spin interaction $V^{s}_{jj'}$ usually has the opposite sign to the kinetic interaction $J^{ss}_{jj'}$. 
More involved interactions~\cite{auslender1982effective, AUSLENDER1982387}, e.g. the ring~\cite{PhysRevB.47.11329, PhysRevB.59.1468, PhysRevLett.83.5122}, the chiral three-spin~\cite{PhysRevLett.93.056402, bauer2014chiral, PhysRevB.95.014422, grytsiuk2020topological, zhang2020imprinting, PhysRevB.103.L060404} 
and the four-spin~\cite{PhysRevB.76.054427,heinze2011spontaneous, paul2020role} exchange interactions can be obtained by expanding the first term in Eq.~\eqref{eq:s9_Boson_action} to higher orders in the $\rho^{\varsigma}$ variable. For calculations of bilinear exchange interactions~\eqref{eq:s9_J} in a realistic material context see Ref.~\onlinecite{Vandelli}.

At this step we can already establish relation between bilinear exchange interactions derived using a magnetic force theorem and a quantum many-body path-integral technique. In this case it is convenient to work in the Matsubara fermionic $\nu$ and bosonic $\omega$ frequency representation. 
To simplify expressions we further omit orbital indices that can be restored trivially.
First, we note that the three-point vertex function $\Lambda^{\varsigma}$ for the zeroth bosonic frequency can be obtained from single-particle quantities 
\begin{align}
\Lambda^{s}_{\nu,\omega=0} = \triangle^{s}_{\nu} + \chi^{s\,-1}_{\omega=0}
\label{eq:s9_vertex_sigma}
\end{align}
by varying the self-energy of the local reference problem~\eqref{eq:s9_action_imp} with respect to the magnetization~\cite{stepanov2021spin}
\begin{align}
\triangle^{s}_{\nu} = \partial{}\Sigma^{\rm imp}_{\nu}/\partial{}M_{\omega=0}.
\label{eq:s9_triangle}
\end{align}
In the ordered phase, where the spin rotational invariance is broken, this variation can be approximated as 
\begin{align}
\triangle^{s}_{\nu} = \frac{\Sigma^{\rm imp}_{\nu\uparrow\uparrow} - \Sigma^{\rm imp}_{\nu\downarrow\downarrow}}{2\langle M \rangle} .
\label{eq:s9_triangle2}
\end{align}
This relation is justified by local Ward identities and the fact that in the regime of a well-developed magnetic moment the renormalized fermion-fermion interaction (four-point vertex function) does not depend on fermionic frequencies~\cite{stepanov2018effective}.
Therefore, in Eq.~\eqref{eq:s9_vertex_sigma} the $\triangle^{s}_{\nu}$ term describes the spin splitting of the self-energy due to polarization of the system. In turn, $\chi^{s\,-1}_{\omega=0}$ can be seen as a kinetic self-splitting effect, because ${\chi^{s}_{\omega} = - \langle \rho^{s}_{\omega} \rho^{s}_{-\omega} \rangle_{\rm imp}}$ is the exact spin susceptibility of the reference system.  
In magnetic materials with a relatively large value of the magnetic moment the kinetic contribution can be neglected.
Indeed, in this case the spin splitting of the self-energy is determined by the Hund's exchange coupling.
The latter is much larger than the inverse of the spin susceptibility, for which the estimation $\chi^{s\,}_{\omega=0}\sim T^{-1}$ holds due to Curie–Weiss law~\cite{moriya_book}.
Then, the static exchange interaction ${J^{ss'}_{jj'}(\omega=0) = \int d\tau'\,J^{ss'}_{jj'}(\tau-\tau')}$ (see Ref.~\onlinecite{stepanov2021spin} for discussions) reduces to the form
\begin{align}
J^{ss'}_{jj',\omega=0} = 
\sum_{\nu,\{\sigma\}} \triangle^{s}_{j\nu} \, \tilde{\cal G}^{\sigma_1\sigma_3}_{jj'\nu} \,
\triangle^{s'}_{j'\nu} \,
\tilde{\cal G}^{\sigma_4\sigma_2}_{j'j\nu}
\label{eq:s9_J_w0}
\end{align}
that under the approximation~\eqref{eq:s9_triangle2} coincides with the expression~\eqref{Jij} that for the ordered phase was derived in Section~\ref{sec:5K} using the magnetic force theorem~\cite{liechtenstein1984exchange, liechtenstein1985curie, liechtenstein1987local, katsnelson2000first, cardias2020dzyaloshinskii}.
Note that Eq.~\eqref{eq:s9_J_w0} contains the sum over spin indices $\{\sigma\}$ and for this reason does not contain the prefactor 2, that is present in Eq.~\eqref{Jij}.
The magnetic force theorem can also be applied in a paramagnetic phase.
In the HIA this was done in Ref.~\onlinecite{PhysRevB.94.115117}, and the result coincides with Eq.~\eqref{eq:s9_J_w0}, where the relation~\eqref{eq:s9_triangle} is calculated numerically exactly.
It should be emphasized that in Eq.~\eqref{eq:s9_J}, and consequently in Eq.~\eqref{eq:s9_J_w0}, the vertex function~\eqref{eq:s9_vertex_sigma} and thus the self-energy~\eqref{eq:s9_triangle} are given by the {\it local} reference system~\eqref{eq:s9_action_imp}. Moreover, the Green function~\eqref{eq:s9_G_dual} that enters the expression for the exchange interaction is also dressed only in the local self-energy. The spin splitting $\triangle^{s}$ obtained from the non-local self-energy was introduced in Ref.~\onlinecite{SECCHI2016112}. 
However, the corresponding exchange interaction is formulated in terms of bare (non-interacting) Green functions and can be derived considering only the density-density approximation for the interaction between electrons. For these reasons, the limit of applicability of this approach and the relation to other methods remain unclear.  

In addition, if the fermionic frequency-dependence in Eq.~\eqref{eq:s9_vertex_sigma} is fully neglected, the the vertex function can be approximated by the inverse of the local bare polarization $\Lambda^{s} \simeq \chi^{0\,-1}_{\omega=0}$, where ${\chi^{0}_{\omega} = \sum_{\nu}g_{\nu}g_{\nu+\omega}}$.
Then, the exchange interaction~\eqref{eq:s9_J} reduces to the form of an effective bare non-local susceptibility, as was derived in Ref.~\onlinecite{Antropov_2003},
\begin{align}
J^{ss'}_{jj',\omega=0} = \chi^{0\,-1}_{\omega=0}\,\tilde{X}^{0}_{jj',\omega=0}\,\chi^{0\,-1}_{\omega=0},
\end{align}
where ${\tilde{X}^{0}_{jj'\omega} = \sum_{\nu}\tilde{\cal G}_{jj'\nu} \,
\tilde{\cal G}_{j'j\nu+\omega}}$.

One can also establish a relation between the results of the introduced many-body theory result and the bilinear exchange interaction that can be deduced from the lattice susceptibility $X^{\varsigma\varsigma'}_{jj'}$ using the following expression:
\begin{align}
\bar{J}^{\varsigma\varsigma'}_{j\neq{}j'} = \delta_{jj'}\delta_{\varsigma\varsigma'}\left[\chi^{\varsigma} \right]^{-1} - \left[X^{-1}\right]^{\varsigma\varsigma'}_{jj'}\,.
\label{eq:s9_susceptibility}
\end{align}
This expression was used in the works in Refs.~\cite{Antropov_2003, PhysRevB.91.195123, PhysRevB.96.075108, PhysRevB.99.165134} to estimate the magnetic exchange interaction based on the DMFT approximation for the spin susceptibility~\cite{RevModPhys.68.13}.
One can find that this form for the bilinear exchange interaction~\eqref{eq:s9_susceptibility} can also be obtained from the derived above many-body theory if the non-linear action~\eqref{eq:s9_Boson_action} is approximated by the Gaussian form
\begin{align}
\bar{\cal S} = - \frac12 \iint_{0}^{\beta} d\tau\,d\tau' \sum_{jj',\varsigma\varsigma'} \bar\rho^{\varsigma}_{j\tau} \left[X^{-1}\right]^{\varsigma\varsigma',\tau\tau'}_{jj'} \bar\rho^{\varsigma'}_{j'\tau'}.
\label{eq:s9_Sapprox}
\end{align}
Since the bosonic variables $\bar{\rho}^{\varsigma}$ correspond to the charge and magnetic densities, the quantity $X^{\varsigma\varsigma',\tau\tau'}_{jj'}$ is nothing more than the lattice susceptibility~\cite{stepanov2018effective, stepanov2019effective, stepanov2021spin}.
More accurately this approximation can be done using Peierls-Feynman-Bogoliubov variational principle~\cite{PhysRev.54.918, Bogolyubov:1958zv, feynman1972}.
Comparing the two actions~\eqref{eq:s9_Boson_action} and~\eqref{eq:s9_Sapprox} shows that in this case the bilinear exchange interaction should indeed be given by the relation~\eqref{eq:s9_susceptibility}.

Effectively, this procedure corresponds to the mapping of the spin problem~\eqref{eq:s9_Boson_action} that contains all possible exchange interactions onto an effective Heisenberg problem that accounts only for the bilinear exchange.
It should be emphasised that for this reason it would be incorrect to relate two expressions for the bilinear exchange  introduced in Eqs.~\eqref{eq:s9_J} and~\eqref{eq:s9_susceptibility}. Indeed, equating these two quantities corresponds to truncating the expansion of the logarithm in the bosonic action~\eqref{eq:s9_Boson_action} at the second order in terms of $\bar{\rho}$ variables.
In other words, it means neglecting the effect of the higher-order exchange interactions on the lattice susceptibility and, consequently, on the bilinear exchange interaction $\bar{J}$.
Taking this effect into account will obviously modify the expression~\eqref{eq:s9_J} for the bilinear exchange interaction.
In particular, it will result in dressing the Green's functions $\tilde{G}$ by the non-local self-energy and in the renormalization of one of the two vertex functions, $\Lambda$, by collective non-local fluctuations in Hedin's fashion~\cite{PhysRev.139.A796}.

These observations confirm the statement that we made at the beginning of this Section, namely that the expression for the exchange interaction strongly depends on the form of the considered spin model.
If one is limited to the simplest approximation with only bilinear form of the exchange interaction, then the latter should be calculated via the Eq.~\eqref{eq:s9_susceptibility} provided that consistent calculation for the lattice susceptibility is possible. 
For instance, using the DMFT form of the susceptibility might already be questionable, because it accounts for the renormalization of the vertex function (in the ladder approximation) but disregards the non-local self-energy. 
At the same time, if a more accurate model that contains the bilinear and the non-linear exchange interactions is considered, these interactions should be computed in the form given by the action~\eqref{eq:s9_Boson_action}. 
In this case, the bilinear interaction is given by Eq.~\eqref{eq:s9_J} or its approximation~\eqref{eq:s9_J_w0}.
Calculating it via the lattice susceptibility~\eqref{eq:s9_susceptibility} would be incorrect, because it would lead to a double-counting problem for the higher-order interactions, since some contribution of them is already taken into account in the lattice susceptibility. 
The difference between the two forms for the bilinear exchange interaction can also serve as a measure of the importance of the non-linear exchange processes in the system.

\subsection{Equation of motion for the local magnetic moment}

The second line in the bosonic action~\eqref{eq:s9_Boson_action} contains only local contributions that describe dynamics of charge and spin degrees of freedom. 
The first term in this line accounts for the Higgs fluctuations of the modulus of the charge $\rho^{c}$ and spin $M$ moments around their average value.
This can be seen by formally expanding the time-dependence of the moments in powers of ${\tau-\tau'}$.
For the local magnetic moment this gives
\begin{align}
{\cal S}_{\rm Higgs} &= - \frac12 \iint_{0}^{\beta} d\tau\,d\tau' \sum_{j}
M^{\phantom{*}}_{j\tau} \left[\chi^{z\,-1}\right]_{\tau\tau'} M^{\phantom{*}}_{j\tau'} \notag\\
&\simeq - \frac12\int_{0}^{\beta} d\tau \sum_{j} \left\{\chi^{z\,-1}_{\omega=0} \, M^2_{j\tau} + \frac{\partial^{2}\chi^{z\,-1}_{\omega}}{2\,\partial\omega^2}\Big|_{\omega=0} \, \dot{M}^{2}_{j\tau} \right\}.
\label{eq:s9_Higgs}
\end{align}
The first order difference in time vanishes, because the exact local susceptibility $\chi^{\varsigma}_{\omega}$ is the even function of the frequency $\omega$.
The Lagrangian equation for this action immediately gives the standard equation of motion for a simple harmonic oscillator ${\ddot{M}_{j\tau} + \lambda^2 M_{j\tau} = 0}$, where ${\lambda^2 = - 2\chi^{z\,-1}_{\omega=0}/\left.\left(\partial^{2}_{\omega}\chi^{z\,-1}_{\omega}\right)\right|_{\omega=0}}$.
Note that in our definition the susceptibility $\chi^{\varsigma}_{\omega}$ is negative.
However, this expansion has to be performed with ultimate care.
Indeed, Higgs fluctuations of the modulus of the local magnetic moment are fast, and the spin susceptibility is strongly non-local in time~\cite{stepanov2021spin}. 
For this reason, there is no uniform justification that the Higgs fluctuations can be accurately described using an equal-time term (second line of Eq.~\eqref{eq:s9_Higgs}) instead of the full non-stationary in time local part of the lattice action (first line of Eq.~\eqref{eq:s9_Higgs}).

The last term in the bosonic action~\eqref{eq:s9_Boson_action} that contains the effective gauge field ${\cal A}^{z}_{j\tau}$ accounts for the rotational spin dynamics.
It has been shown in Ref.~\onlinecite{stepanov2021spin} that after averaging over fast Higgs fluctuations the equation of motion for the bosonic action reduces to the standard Landau-Lifshitz-Gilbert form. 
To illustrate this, we replace the scalar field $M_{j\tau}$ by its constant non-zero average value $\langle M_{j\tau}\rangle=2S$ and introduce ${\vec{S}_{j\tau} = S\,\vec{e}_{j\tau}}$.
The spin part of the action becomes
\begin{align}
{\cal S}_{\rm spin} = \int^{\beta}_0 d\tau\sum_{j} \left( i\dot{\varphi}_{j\tau}(1-\cos\theta_{j\tau}) \,S - \vec{S}_{j\tau}\cdot\vec{h}_{j\tau} \right),
\label{eq:s9_action_spin}
\end{align}
where we explicitly rewrote the gauge field in terms of rotation angles.
Components of the effective magnetic field $\vec{h}_{j\tau}$ can be expressed via the bilinear exchange interaction and the effective magnetic field that appears due to spin-orbit coupling~\cite{stepanov2021spin}
\begin{align}
h^{s}_{j\tau} = -~4\int_0^{\beta} d\tau'\sum_{j',s'}\mathcal{I}^{ss'}_{jj'}(\tau-\tau') \, S^{s'}_{j'\tau'} + h^{{\rm soc}\,s}_{j\tau} .
\label{eq:s9_h_eff}
\end{align}
In the general case, the equation of motion for the non-stationary spin action~\eqref{eq:s9_action_spin} is a complex set of integro-differential equations. 
However, one can make use of the fact that the interaction between spins is determined by the super-exchange processes mediated by electrons~\eqref{eq:s9_J} and thus decays fast on the time scales of inverse band width.
Instead, the time-dependence of the angle variables $\varphi_{j\tau}$ and $\theta_{j\tau}$ is slow, because the spin precession is slow in time~\cite{Sayad_2015, PhysRevLett.117.127201, PhysRevLett.125.086402}.
Contrary to the case of Higgs fluctuations, this allows one to expand the time-dependence of the spin variable $S^{s'}_{j'\tau'}$ in Eq.~\eqref{eq:s9_h_eff} up to the first order in powers of ${\tau-\tau'}$, which allows to write
\begin{align}
h^{s}_{j}(t) = &-4\sum_{j',s'} I^{\mathrm{R}\,ss'}_{jj'}(\Omega=0) \, S^{s'}_{j'}(t) + h^{{\rm soc}\,s}_{j}(t) \notag\\
&-4\sum_{j',s'}\frac{\partial}{\partial\Omega}\left. {\rm Im} \, I^{\mathrm{R}\,ss'}_{jj'}(\Omega)\right|_{\Omega=0}\dot{S}^{s'}_{j'}(t).
\label{eq:s9_h_stationarry}
\end{align}
With this expression for the effective magnetic field the spin problem~\eqref{eq:s9_action_spin} becomes stationary in time, and the corresponding equation of motion for this action takes the standard Landau-Lifshitz-Gilbert form
\begin{align}
\dot{\vec{S}}_{j}(t) = -\, \vec{h}_{j}(t)\times\vec{S}_{j}(t).
\label{eq:s9_eq_motion}
\end{align}
This expression can be derived by making analytical continuation that transforms the imaginary-time exchange interaction ${{\cal I}^{ss'}_{jj'}(\tau-\tau')}$ to a retarded function ${I^{{\rm R}\,ss'}_{jj'}(t-t')}$ in real time $t$. 
In turn, $I^{{\rm R}\,ss'}_{jj'}(\Omega)$ is a Fourier transform of the retarded exchange interaction to real frequency $\Omega$.
This transformation allows one to obtain the Gilbert damping, which is described by the last term in the effective magnetic field~\eqref{eq:s9_h_stationarry}.
A similar expression for the Gilbert damping was derived in Refs.~\onlinecite{Sayad_2015, PhysRevLett.117.127201} for the case of a classical spin coupled to the system of conduction electrons.
Note that the Gilbert damping cannot be obtained in the imaginary-time representation, because the exchange ${{\cal I}^{ss'}_{jj'}(\tau-\tau')}$ is an even function of time. 
Physically, this means that dissipation effects cannot be visible in the equilibrium formalism. 

There are several restrictions for the derived Landau-Lifshitz-Gilbert equation of motion that have to be discussed.  
Eq.~\eqref{eq:s9_eq_motion} describes the spin precession that is assumed to be slow in time compared to electronic processes in the system.
The corresponding effective magnetic field~\eqref{eq:s9_h_stationarry} thus takes into account only the low-frequency part of the exchange interaction.
In general, the exchange term~\eqref{eq:s9_J} has a non-trivial frequency dependence and even diverges at high frequencies, because it is given by a non-local part of the inverse of the lattice susceptibility~\eqref{eq:s9_susceptibility}. 
Non-adiabatic effects that correspond to high-frequency behavior of the exchange interaction are not taken into account by the Eq.~\eqref{eq:s9_eq_motion}. The latter can only be described using the derived bosonic action~\eqref{eq:s9_Boson_action} that has no restriction on the regime of frequencies, but is non-stationary in time.

Another important point is that the Higgs and the Berry phase terms, in the form they enter the bosonic action~\eqref{eq:s9_Boson_action}, can be obtained only after associating the rotation angles with the direction of the local magnetic moment.
As discussed above, this can be done taking the path integral over rotation angles in the saddle point approximation.
However, this approximation can be justified only for the case of a large  magnetic moment~\cite{stepanov2021spin}.
In practice, it means that the classical Landau-Lifshitz-Gilbert equation of motion is applicable only in the multi-orbital case, where the large value of the local magnetic moment is provided by a strong Hund's coupling.
If the magnetic moment is small, spin dynamics in the system is governed by quantum fluctuations.
In this case, the local magnetic moment can still be well-defined, but its behavior can no longer be described in terms of classical equations of motion.

\subsection{Local magnetic moment formation}

The Landau-Lifshitz-Gilbert equation of motion~\eqref{eq:s9_eq_motion} makes physical sense only for a non-zero value of the average magnetic moment $\langle M \rangle$. 
In the ordered phase this is ensured by a non-zero average value of the magnetization.
Defining $\langle M \rangle$ in a paramagnetic regime is much more problematic, because in this case the average magnetization is identically zero.
For this reason, the value of $\langle M \rangle$ is commonly estimated from the static (equal-time) spin susceptibility as
\begin{align}
3\chi^{z}_{\tau\tau} = \langle M^2 \rangle \simeq \langle M \rangle \big(\langle M \rangle +2 \, \big).
\label{eq:s9_XlocS}
\end{align}
However, this approximation gives quite large and almost temperature-independent value for the magnetic moment even in the high-temperature regime where the moment is not yet formed~\cite{stepanov2021spin}.
Taking into account dynamical screening effects changes the value of the average moment, but it still remains substantially larger compared to the one measured experimentally~\cite{PhysRevLett.104.197002, PhysRevB.86.064411, PhysRevLett.125.086402}.
This result can be explained by the fact that the local spin susceptibility simultaneously accounts for correlations of the local magnetic moment and for spin fluctuations of itinerant electrons.
These two contributions to the susceptibility cannot be easily disentangled.

\begin{figure}[t!]
\includegraphics[width=0.95\linewidth]{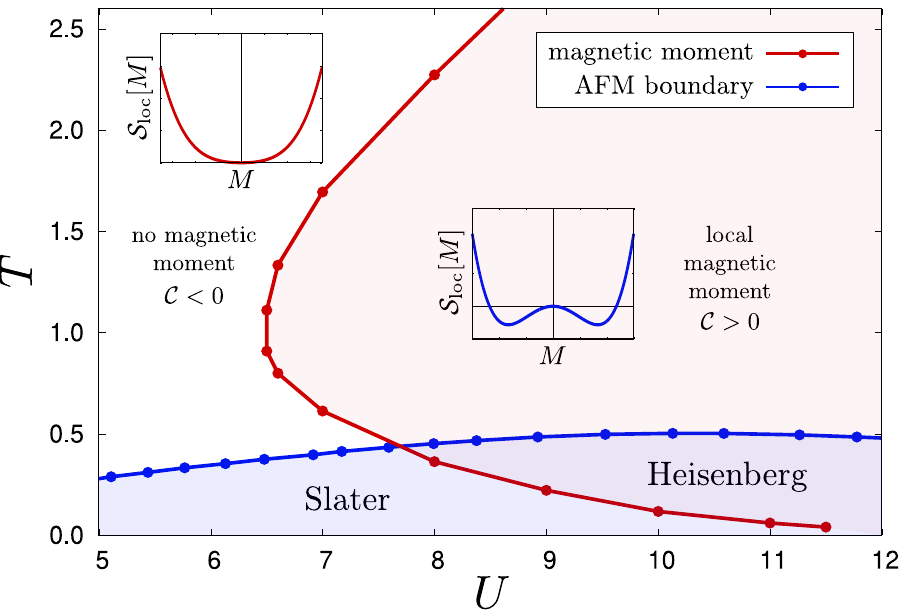}
\caption{\label{fig:Section9_phase} (Color online) Phase diagram for the 3D Hubbard model as a function of temperature $T$ and local Coulomb interaction $U$. Red (light grey) line corresponds to the criterion~\eqref{eq:s9_condition_C} for the formation of the local magnetic moment. Blue (dark gray) line depicts the N\'eel phase boundary obtained in Ref.~\onlinecite{PhysRevB.92.144409}. The insets show the local free energy~\eqref{eq:s9_action_local} as a function of the magnetic moment in two regimes, when it does not exist to the left of the red (light grey) line and where it is already formed shown by the red (light grey) shaded area. Figure is adapted from Ref.~\onlinecite{stepanov2021spin}.}
\end{figure}

In Ref.~\onlinecite{stepanov2021spin} the average value of the magnetic moment was proposed to obtain from the free energy of the local problem that describes the behavior of the magnetic moment.
The action of this local problem
\begin{align} 
{\cal S}_{\rm loc}
= 
&-\Tr\ln \left[[g^{-1}]_{\tau\tau'} \delta^{\phantom{*}}_{\sigma\sigma'} + \int^{\beta}_{0} d\tau_1 \sum_{\varsigma} \sigma^{\varsigma}_{\sigma\sigma'} \Lambda^{\varsigma}_{\tau\tau'\tau_1} \rho^{\varsigma}_{\tau_1} \right] \notag\\
&- \frac12 \iint_{0}^{\beta} d\tau\,d\tau' \sum_{\varsigma}\rho^{\varsigma}_{\tau} \left[\chi^{\varsigma\,-1}\right]_{\tau\tau'} \rho^{\varsigma}_{\tau'} 
\label{eq:s9_action_local}
\end{align}
can be derived by excluding the contribution of itinerant electrons from the local reference system~\eqref{eq:s9_action_imp}.
The resulting problem reminds of the bosonic action~\eqref{eq:s9_Boson_action}, where the non-local Green function $\tilde{\cal G}$ is replaced by the full local Green function $g$.
In the introduced local problem~\eqref{eq:s9_action_local} the magnetic moment appears as a result of a spontaneous symmetry breaking.
According to Landau phenomenology~\cite{LL_V} the latter corresponds to the change of the free energy from a paraboloid-like form with a minimum at ${\langle M \rangle=0}$ to a mexican-hat potential characterized by a continuous set of minima at ${\langle M \rangle\neq0}$ (see insets in Fig.~\ref{fig:Section9_phase}).
Remarkably, the resulting value for the average local magnetic moment appears to be substantially smaller than the one deduced from the local spin susceptibility~\eqref{eq:s9_XlocS}. 

The change of the form of the free energy can be captured by the sign change of its second variation with respect to the the local magnetic moment
\begin{align}
-\frac{\partial^2{\cal S}_{\rm loc}[\rho^{s}]}{\partial\rho^{s}_{\tau}\,\partial\rho^{s}_{\tau'}} = \left[\chi^{s\,-1}\right]_{\tau\tau'} - J^{\rm loc}_{\tau\tau'}.
\label{eq:s9_local_criterion}
\end{align}
The right-hand side of this equation can be seen as a self-exchange between the local magnetic moments, because it is given by the inverse of the local susceptibility with subtracted contribution of itinerant electrons.
The latter is described by a local analog of the kinetic exchange interaction~\eqref{eq:s9_J}
\begin{align}
J^{\rm loc}_{\tau\tau'} = 
\int^{\beta}_{0} \{d\tau_i\} \sum_{\sigma} \Lambda^{*\,s}_{\tau\tau_1\tau_2} g^{\sigma}_{\tau_1\tau_3} g^{\sigma}_{\tau_4\tau_2} \Lambda^{s}_{\tau_3\tau_4\tau'}.
\label{eq:s9_J_loc}
\end{align}
It is important to emphasize that the local magnetic moment exists only at relatively long times compared to single-electron processes.
In the static limit the moment is screened by Kondo effect or by intersite exchange-induced spin flips.
For this reason, formation of the local magnetic moment in the system corresponds to the symmetry breaking at intermediate time scales.
Consequently, as has been shown in Ref.~\onlinecite{stepanov2021spin}, the second variation of the local free energy~\eqref{eq:s9_local_criterion} changes sign at any times except ${\tau=\tau'}$.
Therefore, the formation of the local moment is not a real physical transition and should be considered as a crossover effect.
The static contribution to the local problem~\eqref{eq:s9_action_local} is contained in the inverse of the local susceptibility ${\chi^{s\,-1}_{\tau\tau'} = (\Pi^{s\,{\rm imp}}_{\tau\tau'})^{-1} - \delta_{\tau\tau'}U^{s}}$.
It is given by the bare local interaction in the spin channel ${U^{s}=-U/2}$.
In this expression $\Pi^{s\,{\rm imp}}_{\tau\tau'}$ is the exact polarization operator of the reference system~\eqref{eq:s9_action_imp}. 
The criterion for the local magnetic moment formation can thus be obtain by explicitly excluding this static contribution from Eqs.~\eqref{eq:s9_action_local} and~\eqref{eq:s9_local_criterion}.
The corresponding condition written in the frequency space is that
\begin{align}
{\cal C} = \big(\Pi^{s\,{\rm imp}}_{\omega=0}\big)^{-1} - J^{\rm loc}_{\omega=0} = 0.
\label{eq:s9_condition_C}
\end{align}
This expression illustrates that when the effective self-exchange becomes diamagnetic ${({\cal C}>0)}$ the system acquires a magnetic moment.
The derived criterion~\eqref{eq:s9_condition_C} can be approximately related to the first variation of the local electronic self-energy with respect to the magnetization. This fact suggests that the formation of the local magnetic moment is energetically favorable when this variation is negative, which minimizes the energy of electrons.

Applying the derived criterion~\eqref{eq:s9_condition_C} to interacting electronic systems shows that the local magnetic moment develops at temperatures well above the phase transition to the ordered state~\cite{stepanov2021spin}.
At the same time, the moment can be formed only above a relatively large critical value of the local Coulomb interaction $U$, which for the case of a half-filled single-orbital cubic lattice exceeds the half of the bandwidth.
The corresponding result is shown in Fig.~\ref{fig:Section9_phase}, where the blue (dark grey) line corresponds to the N\'eel phase boundary, and the red (light grey) line is obtained from the condition~\eqref{eq:s9_condition_C}.
At low temperatures the red (light grey) line determines the point at which the local magnetic moment disappears.
In the regime of large interactions this is related to Kondo screening~\cite{Hewson_book, PhysRevLett.126.056403}.
At small $U$, the local magnetic moment is destroyed by local spin fluctuations, which corresponds to the regime of valence fluctuations of the Anderson model~\cite{Hewson_book}.
The low-temperature branch of the red (light grey) line splits the ordered phase into two parts, which allows one to distinguish between Slater~\cite{PhysRev.82.538, PhysRevB.94.125144} and Heisenberg regimes of spin fluctuations.

To summarise, the path-integral formalism allows us to derive the bosonic problem~\eqref{eq:s9_Boson_action} that describes spin dynamics of itinerant electronic systems.
The non-local part of this problem gives a general form for all kinds of magnetic exchange interactions.
Upon certain approximations, the derived expression for the bilinear exchange~\eqref{eq:s9_J} reduces to the result that was originally introduced in a completely different framework of DFT.
These approximations are justified by the existence of a well-developed magnetic moment in the system and determine the limit of applicability of the DFT result. Apart from deriving the magnetic interactions, the path-integral formalism makes it possible to introduce the equation of motion for spin degrees of freedom.
It was shown that for a relatively large value of the magnetic moment its slow rotational dynamics is described by a standard Landau-Lifshitz-Gilbert equation, and fast Higgs fluctuations can be taken into account by the local non-stationary in time contribution to the bosonic problem.  
Deriving the criterion for the formation of the local magnetic moment completes the path-integral formulation of the theory of magnetism and magnetic interactions.

\section{Non-magnetic analogues of exchange interaction}
\label{Section10}

The basic idea presented and discussed in this review is an idea of coarse-grained description of collective behavior in a system of strongly interacting electrons in solids. 
The prototype example is magnetism, and ``gross'' variables in the coarse-grained description of spin degrees of freedom are angles determining directions of individual local magnetic moments. 
Technically, the main tool is the magnetic force theorem when we express the variation of the total thermodynamic potential under small spin rotations in terms of variations of single-electron Green function. 
This approach is general and can be applied to other collective phenomena than for magnetism. 
Here we consider two examples, namely, superconductors and charge-ordered systems. 
Since these subjects are auxiliary for the main aim of the review we restrict ourselves by presentation of main ideas and some illustrative results emphasizing similarities with the discussed approach to magnetic exchange interaction. 

We start with the case of superconductors; our presentation in this part will mostly follow Ref.~\onlinecite{harland2019josephson}. 
The superconductor is characterized, in the simplest case of singlet Cooper pairing, by a complex-valued order parameter meaning a wave function of condensate of the Cooper pairs. 
There is a huge literature on the subject; for a very basic introduction the text books in Refs.~\onlinecite{schrieffer_book, mahan_book, abrikosov_book} can be recommended. 

Let us consider a model of a strong-coupling superconductor with Cooper pairs relatively well localized in real space, an analog of a magnet with well-defined local magnetic moments. This is a very poor model for conventional superconductors with a typical diameter of Cooper pairs in thousands of interatomic distances~\cite{schrieffer_book, abrikosov_book} but it can be reasonably well applicable to cuprate high-temperature superconductors assuming that we consider the lattice of copper plaquettes rather than individual sites~\cite{licht_kats_cluster2000, harland2019josephson}. 
Then, the macroscopic superconductivity in the system can be described in terms of a coherence of the phase of the local Cooper pairs $\theta _{i}$ which are supposed to be all equal in the ground state (without the loss of generality, this ground-state value of the phase can be chosen as zero). 
The model that can address the issue of superconducting phase ordering, and thus macroscopic quantum properties of the superconductor, is the Josephson lattice model,
\begin{equation}
\mathcal{H}_{\mathrm{eff}}=  \sum_{<ij>} J_{ij}\cos \left( \theta _{i}-\theta _{j}\right),
\label{eq:s10_hjosephson}
\end{equation}
where $i,j$ are (super)site indices (e.g., plaquette indices for the two-dimensional Hubbard model used in the theory of superconducting cuprates). 
The Josephson coupling parameters $J_{ij}$ determine in particular superfluid density and London penetration depth~\cite{schrieffer_book, mahan_book, abrikosov_book}. 

\begin{figure}[t!]
\centering
\includegraphics[width=0.65\linewidth]{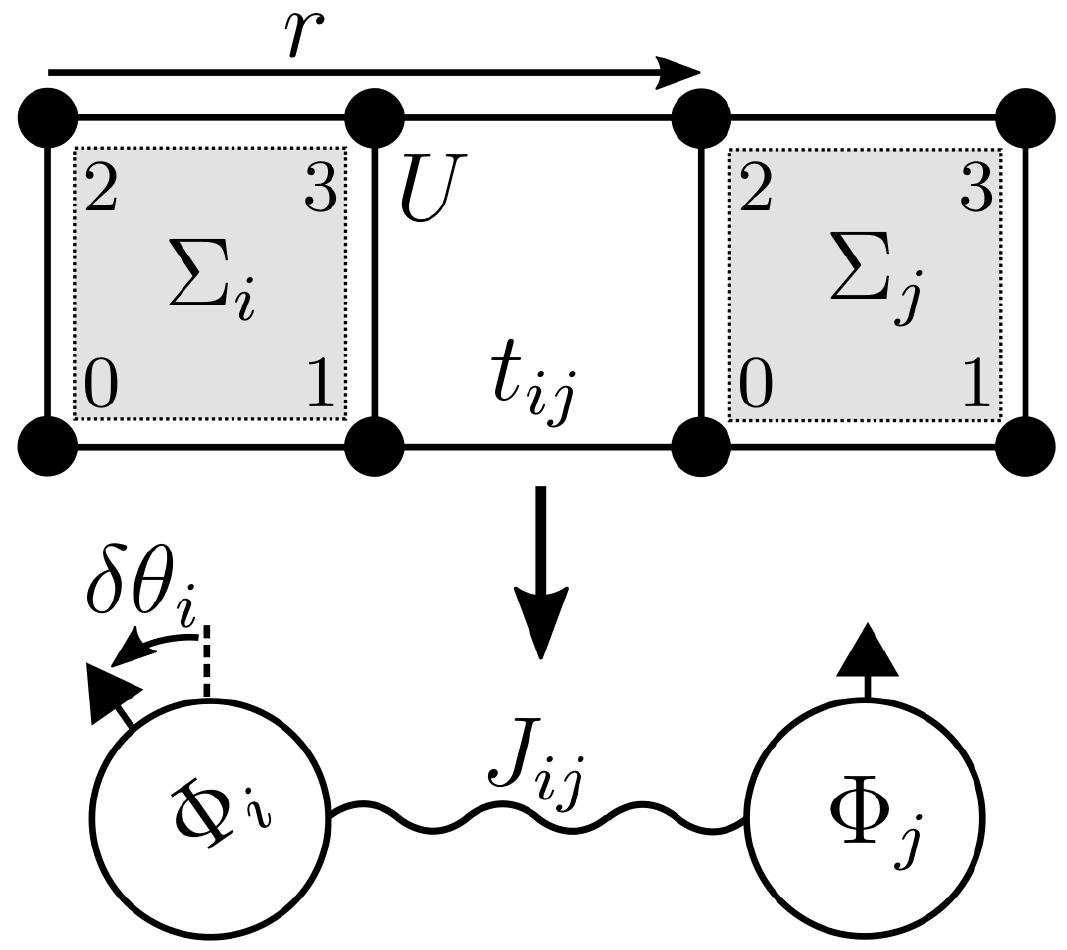}
\caption{Illustration of the Hubbard-plaquette lattice ($t_{ij}$, $U$) with lattice vector $r$, self-energies $\Sigma_i$ and plaquette sites $0,1,2,3$. It is mapped to the Josephson lattice model with effective  coupling $J_{ij}$ of plaquettes due to phase fluctuations $\delta \theta_i$ of the $d$-wave superconducting order parameter $\Phi_i$.
Figure is adapted from Ref.~\onlinecite{harland2019josephson}.
\label{fig:illu}}
\end{figure}

Instead of magnetic systems where we deal with the local rotational (or SU(2)) symmetry, for singlet superconductors we deal with the U(1) symmetry (see Fig.~\ref{fig:illu}). 
Following the general approach accepted in this review, we have to calculate the variation of the thermodynamic potential under small phase variations, and the answer will be expressed in terms of single-particle Green function. 
In the superconducting state, the latter is a supermatrix with normal and anomalous part (the so-called Nambu-Gor'kov representation)~\cite{schrieffer_book}: 
\begin{align}
\label{eq:s10_senambu}
\begin{pmatrix}
G^{p\uparrow} & F \\
F  & G^{h \downarrow}
\end{pmatrix}^{-1}_{ij}
=
\begin{pmatrix}
G^{p\uparrow}_0  & 0 \\
0 & G^{h\downarrow}_0
\end{pmatrix}^{-1}_{ij}
-
\delta_{ij}
\begin{pmatrix}
\Sigma^{p\uparrow}  & S  \\
S  & \Sigma^{h \downarrow}
\end{pmatrix}_{i},
\end{align}
where $G^{a\sigma}_0$ and $G^{a\sigma}$ are the normal parts of the bare ($G_0$) and interacting ($G$) Green's functions for an electron (${a=p}$) and a hole (${a=h}$) with the spin projection ${\sigma\in\{\uparrow,\downarrow\}}$. 
$F$ is the anomalous part of the interacting Green function, which is considered to be local in the supersite, as in Refs.~\onlinecite{licht_kats_cluster2000, harland2019josephson}. 
$\Sigma^{a\sigma}$ and $S$ are the normal and anomalous parts of the self-energy, respectively.

To obtain explicit expressions for the Josephson couplings, $J_{ij}$, we have to calculate the variation of the thermodynamic potential $\Omega$ under small variations of the superconducting phases, and compare the result with~\eqref{eq:s10_hjosephson}. Following the consideration of the exchange interactions within dynamical mean-field theory, discussed in Subsection~\ref{sec:5K}, we start with a general representation of the thermodynamic potential in terms of single-particle and double-counted contribution
with the Luttinger-Ward functional, $\Phi$, and use the local-force theorem. The result is~\cite{harland2019josephson}
\begin{equation}
\delta \Omega \simeq \sum_{ij} \Tr \left( \delta_{ij} G_{ii}\delta^\ast\Sigma_i 
+ \frac{1}{2} G_{ij} \delta^\ast\Sigma_j G_{ji} \delta^\ast\Sigma_i\right)
\label{eq:s10_omegavariation},
\end{equation}
where $\delta^\ast$ denotes the local variation of the self-energy $\Sigma$ without taking into account its variation due to the self-consistency procedure. We omit here for simplicity matrix indices of intra-plaquette and Nambu spaces. 

The variation of the self-energy under an infinitesimal change of the local phase, $\delta \theta_i$, entering Eq.~\eqref{eq:s10_omegavariation} in a homogeneous environment reads
\begin{align}
\label{eq:s10_variationsigma}
\begin{split}
\delta^\ast\Sigma_i &= e^{i\delta\theta_i \sigma_z/2} \Sigma_i e^{-i\delta\theta_i \sigma_z/2} -\Sigma_i\\
&=\begin{pmatrix} \Sigma^{p\uparrow}_i && e^{i\delta\theta_i}S_i \\ e^{-i\delta\theta_i}S_i && \Sigma^{h\downarrow}_i \end{pmatrix} - \Sigma_i\\
&\simeq \begin{pmatrix} 0 && \left(i\delta\theta_i - \frac{\left(\delta\theta_i\right)^2}{2}\right)S_i \\ \left(-i\delta\theta_i - \frac{\left(\delta\theta_i\right)^2}{2}\right)S_i && 0\end{pmatrix},
\end{split}
\end{align}
where $\Sigma^{p\uparrow}_i$, $\Sigma^{h\downarrow}_i$, and $S_i$ are electron-up, hole-down, and anomalous parts of the supersite self-energy, respectively, and the third Pauli matrix $\sigma_z$ acts in the Nambu-space. 

A straightforward calculation up to second order in $\delta\theta$ results in
\begin{widetext}
\begin{equation}
\label{eq:s10_je8}
\delta\Omega = \sum_{ij} \Tr_{\omega \alpha} \left(G^{p\uparrow}_{ij}S_jG^{h\downarrow}_{ji}S_i - \delta_{ij}  F_{ii}S_i- F_{ij}S_jF_{ji}S_i\right)\delta\theta^2_i
+\frac{1}{2}\sum_{ij} \Tr_{\omega \alpha} \left(F_{ij}S_jF_{ji}S_i  - G^{p\uparrow}_{ij}S_jG^{h\downarrow}_{ji}S_i\right)\delta\theta^2_{ij}.
\end{equation}
\end{widetext}
The trace goes over Matsubara frequencies and over the sites within the supersite ($\alpha$). 

The term $\propto \delta\theta^2_i$ vanishes, which reflects the gauge invariance of the theory, that can be checked by the direct calculation~\cite{harland2019josephson}. 
The remaining non-local term is proportional to $\delta\theta^2_{ij}$, i.e.,
\begin{equation}
\label{eq:s10_lft4}
\delta \Omega \equiv -\frac{1}{2} \sum_{<ij>} J_{ij} \delta \theta^2_{ij}.
\end{equation}
This expression should be compared with~\eqref{eq:s10_hjosephson} to find the coupling constants $J_{ij}$. The answer is an expression where
\begin{align}
\label{eq:s10_j}
J_{ij}= 2 \Tr_{\omega \alpha} \left( G_{ij}^{p\uparrow}  S_{j} G_{ji}^{h\downarrow} S_{i} - F_{ij} S_{j} F_{ji}  S_{i} \right).
\end{align}

In order to study macroscopic observables of the Josephson lattice model, we take the continuum, long-wavelength limit of~\eqref{eq:s10_hjosephson}. 
In this limit, the interaction becomes the superconducting stiffness; 
\begin{gather}
\label{eq:s10_stiffaak}
I_{ab} = -\frac{1}{\left(2\pi\right)^d}\int\! d^dk \Tr_{\omega \alpha} \\
\star \left( \frac{\partial G^{p\uparrow}(k)}{\partial k_a} S \frac{\partial G^{h\downarrow}(k)}{\partial k_b} S
-\frac{\partial F(k)}{\partial k_a} S \frac{\partial F(k)}{\partial k_b} S\right)\nonumber
\end{gather}
with the effective Hamiltonian
\begin{equation}
\label{eq:s10_hjosephsoncontinuum}
H_{\mathrm{eff}}=\frac{1}{2}\sum_{ab} I_{ab}\int d^dr\,\frac{\partial \theta }{\partial r_{a}}\frac{\partial \theta }{\partial r_{b}}.
\end{equation}
If we assume that the discussed lattice is isotropic (in two or three dimensions), we have ${I_{ab} = I \delta_{ab}}$, where the constant $I$ is related to the London penetration depth~\cite{schrieffer_book, abrikosov_book}:
\begin{equation}
\label{eq:s10_lpd}
\frac{1}{\lambda^{2}} = \frac{16\pi e^2}{\hbar^2 c^2} I.
\end{equation}

\begin{figure}[t!]
\centering
\includegraphics[width=0.95\linewidth]{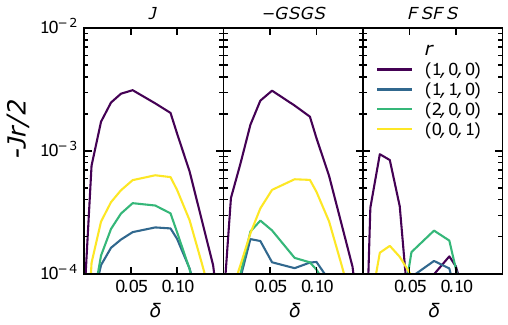}
\caption{(Color online) Josephson coupling $J_r$ (left) and its constituents, $GSGS$ (center) and $FSFS$ (right), as functions of doping $\delta$ and for different plaquette translations $r$ at $T=1/52\sim 0.02$, $t_\perp = 0.15$. Figure is adapted from Ref.~\onlinecite{harland2019josephson}.
\label{fig:jgsgsfsfs}}
\end{figure}

We present an example of the calculated Josephson couplings, $J_r$, for plaquette-translations $r$ in Fig.~\ref{fig:jgsgsfsfs}. 
The figure shows that $J_r$ reduces sharply with increasing plaquette-translation length ${|r|}$, and thus the short-range components of $J_r$ alone can give a complete description. 
The strongest coupling is $J_{100}$, followed by the interlayer coupling $J_{001}$. 
They have their maxima around $\delta = 0.05$ and ${\delta = 0.1}$, respectively. 
All couplings diminish at large dopings, ${\delta > 0.1}$.
The first term of Eq.~\eqref{eq:s10_j} ($GSGS$) is negative, and the second ($FSFS$) is positive. 
$GSGS$ is a mixed term with normal ($G$) and anomalous ($S$) contributions. 
This term provides the main contribution to $J$, that  
can be finite only if there is a superconducting gap and therefore a finite anomalous self-energy, $S$. 
Regarding the largest contributions to the nearest neighbour Josephson coupling $J_{(1,0,0)}$, $GSGS$ is about 3 times as large as $FSFS$.

Another interesting feature of correlated materials that can be potentially described by a corresponding bosonic model is charge ordering.
In electronic systems this phenomenon attracts a considerable attention since the discovery of the Verwey transition in magnetite Fe$_3$O$_4$~\cite{VERWEY1941979, doi:10.1063/1.1746466, Mott}.
Further, effects similar to the Verwey transition have been observed in many other materials, such as the rare-earth compound Yb$_4$As$_3$~\cite{PhysRevB.71.075115, 0295-5075-31-5-6-013, doi:10.1080/0001873021000057114}, transition metal MX$_2$~\cite{PhysRevB.89.235115, ritschel2015orbital, ugeda2016characterization} and rare-earth R$_3$X$_4$~\cite{FURUNO1988117, IRKHIN199047, doi:10.1080/01418638008221893} chalcogenides (${\rm M = V, Nb, Ta}$; ${\rm R = Eu, Sm}$; ${\rm X = S, Se}$), Magn\'eli phase Ti$_4$O$_7$~\cite{doi:10.1080/01418638008221887, doi:10.1080/01418638008221888, EYERT2004151, 0953-8984-18-48-022}, vanadium bronzes Na$_x$V$_2$O$_5$ and Li$_x$V$_2$O$_5$~\cite{doi:10.1080/0001873021000057114, doi:10.1080/01418638008221890}. 
In these materials the charge ordering is driven by the strong non-local Coulomb interaction and/or the electron-phonon mechanism.
Both these interactions effectively reduce the strength of the local Coulomb repulsion~\cite{PhysRevLett.111.036601, PhysRevB.94.165141, PhysRevB.52.4806, PhysRevLett.94.026401, PhysRevLett.99.146404} and may even result in an effective attraction between electrons.
Describing these effects in the framework of {\it ab initio} electronic models requires to use very advanced many-body approaches, such as the quantum Monte-Carlo technique~\cite{PhysRevB.90.085146, PhysRevB.89.205128, Buividovich:20174n}, the $GW$ method combined with the extended dynamical mean-field theory~\cite{PhysRevB.87.125149, PhysRevB.95.245130}, the dynamical cluster approximation~\cite{PhysRevB.95.115149, PhysRevB.97.115117, PhysRevB.99.245146}, or the dual theories~\cite{PhysRevB.90.235135, PhysRevB.94.205110, van2018competing, PhysRevB.102.195109, stepanov2021coexisting}.
These theoretical calculations require significant numerical efforts, which additionally motivates reformulating the original electronic problem in terms of effective bosonic variables.

Similarly to magnetism, the charge ordering is characterised by the local order parameter -- the onsite electronic density. This ordering appears as the result of a spontaneous symmetry breaking of a discrete lattice symmetry contrary to the case of a magnetic ordering, which is associated with breaking of a continuous $SU(2)$ symmetry.
For this reason, effective models formulated in terms of scalar bosonic variables are more suitable for addressing this problem.
In particular, Ising-like models are frequently used for describing the ordering in alloys~\cite{PhysRevB.70.125115, PhysRevB.72.104437, PhysRevB.79.054202, PhysRevLett.105.167208, PhysRevB.83.104203}.
In this framework, one deals with a configuration energy written in terms of effective interactions $V^{(n)}_{\alpha}$ for clusters of order $n$ and type $\alpha$.
For the case of a binary alloy $A_{c}B_{1-c}$ with the concentration $c$ the configuration energy can be written as
\begin{align}
H_{\rm conf} &= \sum_{p}V^{(2)}_{p}\sum_{i,j\in{}p}\sigma_{i}\sigma_{j} 
+ \sum_{t}V^{(3)}_{t}\sum_{i,j,k\in{}t}\sigma_{i}\sigma_{j}\sigma_{k} \notag\\
&+ \sum_{q}V^{(4)}_{q}\sum_{i,j,k,l\in{}q}\sigma_{i}\sigma_{j}\sigma_{k}\sigma_{l} + \ldots
\end{align}
where scalar variables $\sigma_{i}$ take the value $-1$ or $+1$ depending whether $A$ or $B$ atom occupies the site $i$.
Parameters for this microscopic model can be derived from {\it ab initio} energy calculations within the framework of density functional theory~\cite{PhysRevB.27.5169, 0305-4608-13-11-017, ducastelle1991order, 0034-4885-71-4-046501}.
To this aim, one can apply, e.g., a generalized perturbation theory~\cite{doi:10.1080/00318087508228689, Gautier_1975, Giner_1976, ducastelle1976generalized, Treglia_1978, Ducastelle_1980, PhysRevB.36.4630, doi:10.1080/13642819708205703}.
In this approach effective cluster interactions $V^{(n)}_{\alpha}$ can either be obtained by calculating the corresponding $n$-point correlation functions (see, e.g., Refs.~\onlinecite{PhysRevB.66.024202, PhysRevB.83.104203}) or from the single-electron energy using the force theorem~\cite{mackintosh1980electrons}.  
In the latter case, the variation of the concentration of atoms of a given kind is considered as a perturbation. This seems to be very different from consideration of small spin rotations, the primary topic of this review, that has been used successfully in the case of magnetism. Nevertheless, the resulting pair interaction between sites $j$ and $j'$ is given by the expression
\begin{align}
V^{(2)}_{jj'} = -\frac{2}{\pi}  {\Im}\int^{E_{F}}_{-\infty} dE  \Delta{}t_j \, \tilde{G}_{jj'}(E) \, \Delta{}t_{j'} \, \tilde{G}_{j'j}(E),
\label{eq:s10_exchange_alloys}
\end{align}
which very closely resembles the magnetic exchange interaction derived using the magnetic force theorem (see Section~\ref{detailsLKAG}). 
Here, ${\Delta{}t_{j} = (t^{A}_{j} - t^{B}_{j})/2}$ is the difference between single-site scattering matrices for $A$ and $B$ type of atoms, and $\tilde{G}_{jj'}(E)$ is the partial interatomic Green function of the reference system provided by a random alloy. 

As in the case of magnetism, using the force theorem does not allow one to rigorously determine limits of applicability of the theory.
In this regard, deriving effective Ising-like models in the many-body framework should be beneficial.
In the context of interacting electronic problems this has been achieved in Refs.~\onlinecite{stepanov2019effective, stepanov2021spin}.
The corresponding derivation was discussed in Section~\ref{Section9} leading to an effective bosonic problem~\eqref{eq:s9_Boson_action}.
It is important to note that introducing the bosonic model for charge degrees of freedom does not require imposing the adiabatic approximation that separates time- and energy-scales of single- and two-particle fluctuations in the magnetic case~\cite{stepanov2019effective}.

All possible interactions between the electronic densities at different lattice sites can be obtained by expanding the logarithm in Eq.~\eqref{eq:s9_Boson_action} in terms of the bosonic field $\rho^{c}$ that describes fluctuations of the charge densities $n$ around their average values.
The explicit form for the pair interaction is given by Eqs.~\eqref{eq:s9_J} and~\eqref{eq:s9_exch}. 
The tree-point vertex function $\Lambda^{c}$ that enters the kinetic exchange~\eqref{eq:s9_J} represents a remormalized local coupling between electronic and charge degrees of freedom.
Thus, this vertex can be seen as a single-site scattering matrix, which makes the many-body expression for the exchange interaction~\eqref{eq:s9_J} very similar to the pair cluster interaction derived in the context of alloys~\eqref{eq:s10_exchange_alloys}.

Mapping the quantum bosonic problem for electronic densities~\eqref{eq:s9_Boson_action} onto a classical Ising-like model can be justified only in the regime of well-developed charge fluctuations.
In a broken symmetry (charge ordered) phase, the electronic density at a given lattice site strongly differs from the average density of the system.
This allows one to replace the bosonic variable $\rho^{c}_{j}$ at each site $j$ by its average value $\langle \rho^{c}_{j} \rangle$, which reduces the quantum bosonic action~\eqref{eq:s9_Boson_action} to a classical Ising-like Hamiltonian.
In the normal phase the average density on each lattice site is uniform, which makes it difficult to introduce the corresponding classical problem and complicates determining the regime of applicability of this approach.

\begin{figure}[t!]
\includegraphics[width=0.75\linewidth]{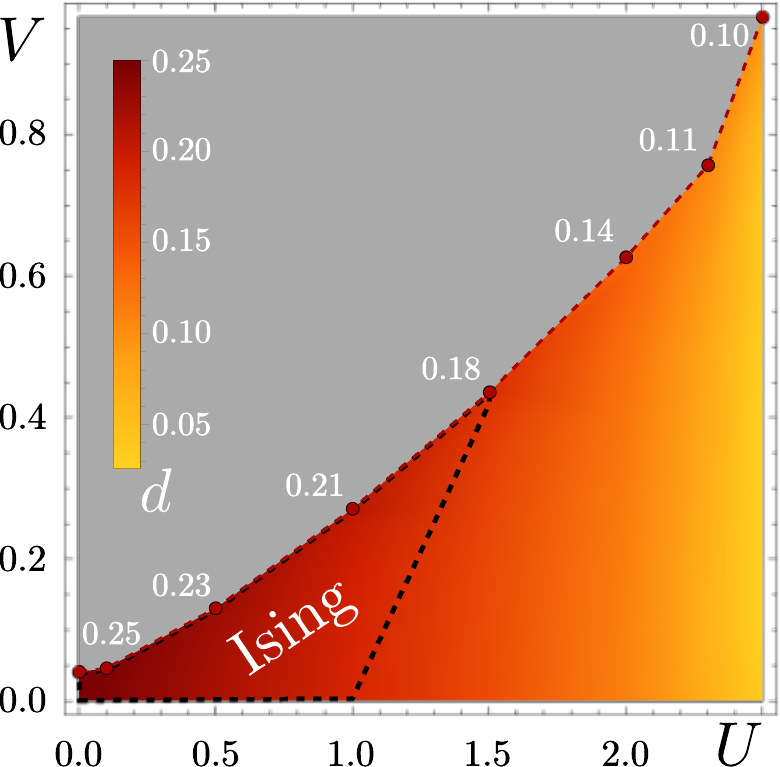}
\caption{(Color online) Double occupancy of the extended Hubbard model shown on the $U$-$V$ phase diagram. Calculations are performed in the normal phase, where the value of the double occupancy $d$ is depicted by color. The (light) gray color depicts the charge ordered phase. The black dashed line surrounds the area of the large double occupancy $d\gtrsim70\%\,d_{\rm max}$, where charge excitations can be described by an effective Ising model.
Values of Coulomb interactions $U$ and $V$ are given in units of half of the bandwidth $4t=1$, $t$ is the nearest-neighbor hopping amplitude. The inverse temperature for this calculation was set to ${T^{-1}=50}$. Figure is taken from Ref.~\onlinecite{stepanov2019effective}. 
\label{fig:Section10_phase}}
\end{figure}

In Ref.~\onlinecite{stepanov2019effective} the double occupancy $d=\langle n_{\uparrow} n_{\downarrow} \rangle$ of the lattice site was proposed as a measure of the strength of the charge fluctuations in the normal phase. 
The double occupancy for a particular case of the extended Hubbard model on a square lattice is shown in Fig.~\ref{fig:Section10_phase}.
The result is obtained at half filling, where the maximum value of the double occupancy is ${d_{\rm max}=0.25}$.
In this model the charge ordered phase (light grey area) is driven by the nearest-neighbor Coulomb interaction, $V$.
If the latter defeats the onsite Coulomb repulsion, $U$, the electronic density forms a checkerboard pattern on the lattice made of alternating doubly occupied and empty sites. 
For a given value of $U$ the maximum value of the double occupancy appears at the boundary between the normal and ordered phases, depicted by a dashed red line.
This fact confirms that the strongest charge fluctuations in the normal phase emerge in the region close to the phase transition to the ordered state.
However, the value of the double occupancy is not uniformly distributed along the phase boundary and decreases with the increase of the local Coulomb interaction. 
It has been shown in Ref.~\onlinecite{stepanov2019effective} that strong charge fluctuations drastically suppress the frequency dependence of the effective local electron-electron interaction (two-particle irreducible four-point vertex function).
The value of the double occupancy at which the effective local interaction is nearly frequency independent and coincides with the actual Coulomb interaction, $U$, was estimated as $d\gtrsim70\%\,d_{\rm max}$.
This condition defines the Ising regime of the system, depicted by the black dashed line in Fig.~\ref{fig:Section10_phase}, where charge fluctuations are indeed well-developed. Interestingly, this regime is not limited to small values of the local interaction, $U$, that for some values of $V$ exceed half of the bandwidth.

In the Ising regime of the normal phase the quantum action~\eqref{eq:s9_Boson_action} can be mapped onto an effective classical Hamiltonian.
This can be achieved by replacing the corresponding bosonic variable by an effective charge density, which is given by the square root of the double occupancy ${\rho^{c}\to\sqrt{d}}$.
Note that determining the effective charge density can be performed more accurately by finding the minimum of the local free energy in the same way as is done for estimating the value of the local magnetic moment (see discussion in Section~\ref{Section9}).
However, using the two particle correlation function (the double occupancy) to define the average density in the case of charge degrees of freedom is also well justified, contrary to the case of magnetism, where the magnetic phase corresponds to the ordering of single-particle quantities (local magnetizations). Since charge ordering is realised through the formation of double occupations, one needs to characterize this state from two-particle observables.
Ref.~\onlinecite{stepanov2019effective} shows that the effective Ising model introduced in such a simple way is able to predict the transition temperature between the normal and charge ordered phases in a good agreement with much more elaborate methods, even though the calculations are performed in the unbroken symmetry phase.

\section{Summary and outlook}

The developments that started in Refs.~\onlinecite{inoue1967interaction,lacour1974magnetic,gyorffy1980momentum, liu1961exchange, oguchi1983magnetism, PhysRevB.28.6443} culminated in Ref.~\onlinecite{liechtenstein1984exchange}, with a practical and efficient scheme of extracting exchange interactions between atomic magnetic moments of solids and molecules. This has opened up a field of research where a deeper understanding of magnetic interactions is possible. These early works on {\it explicit} calculations  of interatomic exchange enabled new dimensions of DFT and DMFT calculations, and it is now routine to extract from electronic structure calculations on one scale (involving a few atoms per unit cell) information about exchange interactions on a much larger scale (involving pair interactions between thousands of atoms), that if needed can be used to evaluate parameters of micromagnetic simulations~\cite{poluektov2016scale, poluektov2018coupling}. This represents multiscale transitions between three length scales and enables simulations of magnetic phenomena on scales equal to that of experimental sample sizes, without using experimental information as input. In addition to offering a deeper understanding of basic magnetic exchange between atoms, the method of Ref.~\onlinecite{liechtenstein1984exchange} has so far been used to calculate ordering temperatures of materials and to map out magnon dispersions (via adiabatic approaches or in spin-dynamics simulations via the dynamic structure factor, e.g as reviewed in Ref.~\onlinecite{eriksson2017atomistic}). It has also been used to address ultra fast magnetisation phenomena observed in pump probe measurements~\cite{evans2015quantitative} as well as to analyze topological magnetic states~\cite{pereiro2014topological} and spin glass formation~\cite{kamber2020self, benj2022verl}, to name a few\footnote{Developments in electronic structure theory in Uppsala with can be found here: https://www.physics.uu.se/research/\\materials-theory/ongoing-research/code-development/\\developments-in-electronic-structure-theory/. In addition, OpenMX, https://www.openmx-square.org from Tokyo, AMULET from Ekaterinburg, http://www.amulet-code.org, Artaios from Hamburg, https://github.com/molspintron, and TB2J, a python package for computing magnetic interaction parameters, https://github.com/mailhexu/TB2J should be mentioned. As we mentioned before, the exchange interaction parameters can be calculated by KKR codes as well. A corresponding link of the group of Samir Lounis is available here: https://iffgit.fz-juelich.de/kkr/jukkr. Assuming that calculations with the code dealing with periodic structures are intended, the wiki page for the calculation of exchange coupling constants can be found here: https://iffgit.fz-juelich.de/kkr/jukkr/-/wikis/jumu/jijdij.}.   

It is foreseeable that the method of Ref.~\onlinecite{liechtenstein1984exchange} will continue to be developed, to enable a more detailed and deeper understanding of the mechanisms that govern the properties of a magnetic material. An example here is the coupling of spin- and lattice degrees of freedom, where initial steps have been taken. In a recent work~\cite{mankovsky2022angular} spin-lattice parameters were calculated from an extension of the formalism of Ref.~\onlinecite{liechtenstein1984exchange}. Hence coupled motion, e.g. involving magnons and phonons, are now possible to consider in combined spin-lattice simulations~\cite{antropov1995ab, PhysRevB.99.104302}. It is foreseeable that these developments will continue to be developed, so that a natural output from electronic structure calculations are a set of interaction parameters that enable simulations of all relevant collective modes and the coupling between them.

The theories reviewed here have focused on bilinear effects, such as the ones expressed in Eq.~\eqref{eqn1}. This is natural in the spirit of the LKAG approach with perturbations corresponding to small rotations of the local moments. As the perturbations all can be considered infinitesimal, higher order than two make little sense. However, perturbational approaches \cite{brinker2019chiral} that start with a non-magnetic reference state  and where the perturbations then have to be larger, the convergence is slower and higher order terms do play a large role \cite{PhysRevResearch.2.033240,grytsiuk2020topological}. These multi-spin and multi-site interactions become cumbersome to calculate systematically in general so in most cases the interaction parameters are instead determined through fitting of the total energies.
As these two perturbational approaches lead to different descriptions and interpretation, their complementarity, discussed in Section \ref{Sec-5K-LocalGlobal}, will hopefully in the future  be utilized in order to increase the understanding of complex magnetic systems.

In these extensions, that one can expect will become under focus in the years to come, it would be of interest to also analyse the interaction terms in an orbital composed fashion, in the same way as was done for bilinear exchange~\cite{PhysRevLett.116.217202} as shown here in Fig.~\ref{s7-BS}. In connection to this analysis we mention that a similar analysis of the DM interaction is not straight forward since spin-orbit coupling mixes orbitals that otherwise would belong to separate irreducible representations. Orbital decomposed DM interaction hence becomes an issue of which basis is the most natural to use, which most likely will depend from material to material, given that spin-orbit coupling is either the weakest (for the 3d transition metals) or equal in size to other interactions of the electronic Hamiltonian (e.g. for the actinides). 

The primary focus of this review is on the magnetic dipole of an atom, as calculated from the expectation value of a spin-operator. This is natural since for the majority of materials it is the most commonly observed order parameter. However, for some solids other order parameters are of relevance, e.g. the rank 5 or triakontadipole order that has been observed in NpO$_2$~\cite{RevModPhys.81.807}. 
It would be valuable if this method could be  generalised from calculations of interactions between rank 1 spin moments, to the case of calculations of interactions of multipoles of rank $r$. This would require extensions. For instance the method of small rotations, as shown in Figs.~\ref{figone} and~\ref{figtwo} would have to be generalised to be appropriate for these multipoles and instead of the three independent type of interaction parameters of Eq.~\eqref{1st-bilinear}  one has derived expressions for $2r+1$ independent types of interaction parameters.

To illustrate that the research field reviewed here is very much a living, developing activity, we note a recent set of publications regarding details of the spin Hamiltonian in Eq.~\eqref{neweqn2}. In Refs.~\onlinecite{cardias2020dzyaloshinskii, cardias2020first} it was suggested that DM-like interaction terms can be realized for non-collinear magnetic structures, even if spin-orbit interaction is neglected (or is vanishingly small). This interpretation was criticized in Ref.~\onlinecite{dos2021proper}, who suggested that fundamental interactions of DM character have to rely on an electronic Hamiltonian with spin-orbit coupling included. Further elaborations on non-relativistic DM interaction were published in Refs.~\onlinecite{cardias2022comment, dos2022reply} without a firm consensus being reached. 

Alternative ways to extract exchange parameters have recently been suggested~\cite{streib2022adiabatic}, e.g., from tight-binding electronic structure theory and adiabatic spin-dynamics simulations, where the local Weiss field is evaluated from the so-called constraining field. In this work it was suggested that effective interatomic exchange can be evaluated (dynamically) from the energy curvature tensor of any magnetic configuration. It was  demonstrated in Ref.~\onlinecite{streib2022adiabatic} that both moment lengths
and effective exchange interactions can depend quite strongly on the magnetic
configuration. Terms obtained from such an approach, that goes beyond the weak relativistic limit, contribute to (isotropic) exchange~\cite{secchi2013non} and their relation to non-local crystal field excitations can be the subject of further studies.

Apart from magnetism of electrons in solids, there are very interesting magnetic phenomena related to ordering of nuclear spins in solid helium-3~\cite{roger1983helium}. In this case, the exchange interactions cannot be described by bilinear spin Hamiltonians, and three- and four-spin exchange interactions turn out to be highly important~\cite{roger1983helium, ceperley1995helium}. Apart from solid helium-3, monolayers of helium-3 on graphite is the other example of a system with complicated nuclear-spin-based magnetism~\cite{fukuyama2008helium}. Applications of the methods presented here to such systems seems to be an interesting direction of further development.  

The last three sections of the review (\ref{Section8}, \ref{Section9}, and~\ref{Section10}) present an alternative approach to the theory of exchange interactions, in light of contemporary quantum-many body theory with its mathematically more advanced tools, like path integrals and Feynman diagrams. Changing the language allows one to go much further than the initial formulation considering the systems out-of-equilibrium (Section~\ref{Section8}), nonmagnetic collective phenomena such as charge ordering and superconductivity (Section~\ref{Section10}), and giving a full derivation of equations of spin dynamics for itinerant-electron systems, including not only exchange interaction-related term but also dynamical, spin-precession term (Section~\ref{Section9}). These new developments are relatively recent, and their potential for applications is far from being unveiled completely. Especially, a systematic study of laser-induced nonlinear magnetic phenomena within the developed formalism seems to be an extremely promising direction.

As a final remark of the outlook section, we note that equations of the form of Eq.~\eqref{neweqn1} (and extensions of it) have been used for research outside of materials science, or even natural science. In the Ising approximation of the classical Heisenberg Hamiltonian, the atomic spins are arranged in a z graph, usually a lattice, that can be in one of two states (+1 or -1)~\cite{ising1925contribution} and the strength of the interaction is given by $J_{ij}$ in Eq.~\eqref{neweqn1}. This inspired the so-called classical voter model, and its extensions, which represents an idealized description for the evolution of opinions in a population~\cite{clifford1973model, holley1975ergodic, gleeson2013binary}. In the classic voter model, similar to the Ising model, each voter can assume two states, -1 or +1. A voter at site $i$ is selected at random and copies the state of a randomly chosen neighbor voter $j$. Another example where the Ising-model (and percolation theory) can be used is epidemics as it is shown in a comprehensive review focused on Covid-19~\cite{mello2021epidemics}. The work of Giorgio Parisi on the hidden patterns in spin glasses~\cite{mezard1987spin} should also be mentioned, since it gave an extremely important contribution to the theory of complex system, which is a quantitative, predictive and experimentally verifiable science~\cite{thurner2018introduction}. In case of complex systems a macroscopic pattern can emerge of the mutual influence of a large number of individuals~\cite{anderson1972more, principi2016self, bagrov2020multiscale} and it makes it possible to understand phenomena, not only in physics but also in other, very different areas, such as mathematics, biology, neuroscience and machine learning~\cite{castellano2009statistical, wolf2018physical, baity2018comparing}.

{\bf Acknowledgements} Valuable discussions with V. Antropov, A. Bergman, V. Borisov, R. Cardias, A. Delin, E. Delzceg, I. Di Marco, J. Fransson, O. Gr\aa na\"as, J. Hellsvik, H. Herper, J. Jonsson, A. Katanin, A. Klautau, V. Mazurenko, I. Miranda, C. S. Ong, M. Pereiro, L. Pourovskii, A. Ruban, B. Sanyal, S. Savrasov, I. Solovyev, S. Streib, D. Thonig, P. Thunstr\"am, R. Vieira, and A. Vishina are acknowledged. In particular, the critical reading and the many useful comments by A. Ruban are acknowledged. E. A. S. acknowledges support from the European Union’s Horizon 2020 research and innovation program under the Marie Skłodowska Curie Grant Agreement No. 839551– 2DMAGICS. O. E., A. I. L., and M. I. K. acknowledge sup- port from the European Research Council via Synergy Grant No. 854843 (the FASTCORR project). O. E. and L. N acknowledge support from the Swedish Research Council (VR), and O. E. also acknowledges support from the Swedish Foundation for Strategic Research (SSF), the Swedish Energy Agency (STEM), the Wallenberg Initiative Materials Science for Sustainability (WISE) funded by the Knut and Alice Wallenberg Foundation (KAW), eSSENCE, and STandUP. A. I.L. acknowledges support from the German Research Foundation through the research unit QUAST, FOR 5249, Project No. 449872909.

\bibliography{bio}

\end{document}